\documentclass[preprint,secnumarabic,preprintnumbers,aps,superscriptaddress]{revtex4}
\usepackage[pdftex]{graphicx}
\usepackage{amssymb}
\usepackage{mathptmx}

\newcommand{\be}{\begin{equation}}
\newcommand{\ee}{\end{equation}}  
\newcommand{\bea}{\begin{eqnarray}}
\newcommand{\eea}{\end{eqnarray}}
\newcommand{\noi}{\noindent}

\newcommand{\lsim}{\mathrel{\lower4pt\hbox{$\sim$}}
\hskip-11.5pt\raise1.6pt\hbox{$<$}\;}

\newcommand{\gsim}{\mathrel{\lower4pt\hbox{$\sim$}}
\hskip-11.5pt\raise1.6pt\hbox{$>$}\;}

\newcommand {\delma}{$\Delta m^{2}_{31}$\hspace{1mm}}

\newcommand{\nubar}{\mbox{$\overline{\nu}$}\hspace{1mm}}

\newcommand{\nova}{NO$\nu$A\hspace{1mm}}

\newcommand{\thetae}{\mbox{$\theta_{13}$}\hspace{1mm}}

\newcommand{\sthetae}{\mbox{sin$^2$2$\theta_{13}$}\hspace{1mm}}

\newcommand{\deltacp}{\mbox{$\delta_{CP}$}\hspace{1mm}}

\def\lsim{\mathrel{\raise.3ex\hbox{$<$\kern-.75em\lower1ex\hbox{$\sim$}}}}
\def\gsim{\mathrel{\raise.3ex\hbox{$>$\kern-.75em\lower1ex\hbox{$\sim$}}}}

\begin{document}

\preprint{Fermilab-0801-AD-E} 
\preprint{BNL-77973-2007-IR}

\title{ ~~~ \\ ~~~ \\
Report of the US  long baseline neutrino experiment study\\
}

\author{V.~Barger}
\affiliation{Department of Physics, University of Wisconsin, Madison, WI 53706, USA}

\author{M.~Bishai}
\affiliation{Physics Department, Brookhaven National Laboratory, Upton, NY 11973, USA}

\author{D.~Bogert}
\affiliation{Fermi National Accelerator Laboratory, Batavia, IL 60510, USA}

\author{C.~Bromberg} 
\affiliation{Department of Physics and Astronomy,  Michigan State University, East Lansing, MI 48824, USA}

\author{A.~Curioni}
\affiliation{Department of Physics, Yale University, New Haven, CT 06520, USA}

\author{M.~Dierckxsens}
\affiliation{Physics Department, Brookhaven National Laboratory, Upton, NY 11973, USA}

\author{M.~Diwan}
\affiliation{Physics Department, Brookhaven National Laboratory, Upton, NY 11973, USA}

\author{F.~Dufour}
\affiliation{Department of Physics, Boston University, Boston, MA 02215, USA}

\author{D.~Finley} 
\affiliation{Fermi National Accelerator Laboratory, Batavia, IL 60510, USA}

\author{B.~T.~Fleming}
\affiliation{Department of Physics, Yale University, New Haven, CT 06520, USA}

\author{J.~Gallardo}
\affiliation{Physics Department, Brookhaven National Laboratory, Upton, NY 11973, USA}

\author{J.~Heim}
\affiliation{Physics Department, Brookhaven National Laboratory, Upton, NY 11973, USA}

\author{P.~Huber}
\affiliation{Department of Physics, University of Wisconsin, Madison, WI 53706, USA}

\author{C.~K.~Jung}
\affiliation{Stony Brook University, Department of Physics and Astronomy, Stony Brook, NY 11794, USA}

\author{S.~Kahn}
\affiliation{Physics Department, Brookhaven National Laboratory, Upton, NY 11973, USA}

\author{E.~Kearns}
\affiliation{Department of Physics, Boston University, Boston, MA 02215, USA}

\author{H.~Kirk}
\affiliation{Physics Department, Brookhaven National Laboratory, Upton, NY 11973, USA}

\author{T.~Kirk}
\affiliation{Department of Physics, University of Colorado, Boulder, CO 80309, USA}

\author{K.~Lande}
\affiliation{Department of Physics and Astronomy, University of Pennsylvania, Philadelphia, PA 19104, USA}

\author{C.~Laughton}
\affiliation{Fermi National Accelerator Laboratory, Batavia, IL 60510, USA}

\author{W.Y.~Lee}
\affiliation{Lawrence Berkeley National Laboratory, Physics Division, Berkeley, CA 94720, USA}

\author{K.~Lesko}
\affiliation{Lawrence Berkeley National Laboratory, Physics Division, Berkeley, CA 94720, USA}

\author{C.~Lewis} 
\affiliation{Deparment of Physics, Columbia University, New York, NY 10027, USA} 

\author{P.~Litchfield}
\affiliation{School of Physics and Astronomy, University of Minnesota, Minneapolis, MN 55455, USA}

\author{A.~K.~Mann} 
\affiliation{Department of Physics and Astronomy, University of Pennsylvania, Philadelphia, PA 19104, USA}

\author{A.~Marchionni}
\affiliation{Fermi National Accelerator Laboratory, Batavia, IL 60510, USA}

\author{W.~Marciano}
\affiliation{Physics Department, Brookhaven National Laboratory, Upton, NY 11973, USA}

\author{D.~Marfatia} 
\affiliation{Department of Physics and Astronomy, University of Kansas, Lawrence, KS 66045, USA} 

\author{A.~D.~Marino}
\affiliation{Fermi National Accelerator Laboratory, Batavia, IL 60510, USA}

\author{M.~Marshak}
\affiliation{School of Physics and Astronomy, University of Minnesota, Minneapolis, MN 55455, USA} 

\author{S.~Menary}
\affiliation{Department of Physics and Astronomy, York University, Toronto, Ontario M3J1P3, Canada} 

\author{K.~McDonald}
\affiliation{Department of Physics, Princeton University, Princeton, NJ 08544, USA}  

\author{M.~Messier} 
\affiliation{Department of Physics, Indiana University, Bloomington, IN 47405, USA} 

\author{W.~Pariseau} 
\affiliation{Deapartment of Mining Engineering, University of Utah, Salt Lake City, UT 84112, USA} 

\author{Z.~Parsa}
\affiliation{Physics Department, Brookhaven National Laboratory, Upton, NY 11973, USA}

\author{S.~Pordes}
\affiliation{Fermi National Accelerator Laboratory, Batavia, IL 60510, USA}

\author{R.~Potenza}
\affiliation{Instituto Nazional di Fisica Nucleare, Dipartimento de Fisica e Astronomia, 
University Di Catania, I-95123, Catania, Italy} 
 
\author{R.~Rameika}
\affiliation{Fermi National Accelerator Laboratory, Batavia, IL 60510, USA}

\author{N.~Saoulidou}
\affiliation{Fermi National Accelerator Laboratory, Batavia, IL 60510, USA}

\author{N.~Simos}
\affiliation{Physics Department, Brookhaven National Laboratory, Upton, NY 11973, USA}

\author{R.~Van~Berg} 
\affiliation{Department of Physics and Astronomy, University of Pennsylvania, Philadelphia, PA 19104, USA} 

\author{B.~Viren} 
\affiliation{Physics Department, Brookhaven National Laboratory, Upton, NY 11973, USA}

\author{K.~Whisnant} 
\affiliation{Department of Physics, Iowa State University, Ames, IA 50011, USA}  	

\author{R.~Wilson}
\affiliation{Department of Physics, Colorado State University, Fort Collins, CO 80523, USA}

\author{W.~Winter}
\affiliation{Institue f\"ur theoretische Physik und Astrophysik, University of W\"urzburg,
D-97074, W\"urzburg, Germany}

\author{C.~Yanagisawa} 
\affiliation{Stony Brook University, Department of Physics and Astronomy, Stony Brook, NY 11794, USA}

\author{F.~Yumiceva}
\affiliation{The College of William and Mary, Williamburg, VA 23187, USA} 

\author{E.~D.~Zimmerman}
\affiliation{Department of Physics, University of Colorado, Boulder, CO 80309, USA}

\author{R.~Zwaska} 
\affiliation{Fermi National Accelerator Laboratory, Batavia, IL 60510, USA}

\date{\today}

\newpage

\begin{abstract} 
This report provides the results of an extensive and important study
of the potential for a U.S. scientific program that will extend our
knowledge of neutrino oscillations well beyond what can be anticipated
from ongoing and planned experiments worldwide.  The program examined
here has the potential to provide the U.S. particle physics community
with world leading experimental capability in this intensely
interesting and active field of fundamental research.  Furthermore,
this capability is not likely to be challenged anywhere else in the
world for at least two decades into the future.  The present study was
initially commissioned in April 2006 by top research officers of
Brookhaven National Laboratory and Fermi National Accelerator Laboratory
 and, as the study evolved,
it also provides responses to questions formulated and addressed to
the study group by the Neutrino Scientific Advisory Committee (NuSAG)
of the U.S. DOE and NSF.  The participants in the study, its Charge
and history, plus the study results and conclusions are provided in
this report and its appendices.  A summary of the conclusions is
provided in the Executive Summary.
\end{abstract}

\maketitle  

\tableofcontents

\newpage 

\section{Executive Summary}

This report provides the results of an extensive and important study
of the potential for a U.S. scientific program that will extend our
knowledge of neutrino oscillations well beyond what can be anticipated
from ongoing and planned experiments worldwide.  The program examined
here has the potential to provide the U.S. particle physics community
with world leading experimental capability in this intensely
interesting and active field of fundamental research.  Furthermore,
this capability is not likely to be challenged anywhere else in the
world for at least two decades into the future.  The present study was
initially commissioned in April 2006 by top research officers of
Brookhaven National Laboratory and Fermilab and, as the study evolved,
it also provides responses to questions formulated and addressed to
the study group by the Neutrino Scientific Advisory Committee (NuSAG)
of the U.S. DOE and NSF.  The participants in the study, its Charge
and history, plus the study results and conclusions are provided in
this report and its appendices.  A summary of the conclusions is
provided in this Executive Summary.

The study of neutrino oscillations has grown continuously as its key
impact on particle physics and various aspects of cosmology have
become increasingly clear.  The importance of this fundamental physics
was recognized by the National Research Council\cite{nrc} and the Office of
Science and Technology Policy\cite{pofu}, and its national budget priority has
been established in a joint OSTP-OMB policy memorandum in 2005\cite{eopm}. 
 In fact, as the present study confirms, it is now possible to design
practical experiments that are capable of measuring all the parameters
that characterize 3-generation neutrino oscillations, including the
demonstration of CP-violation for a significant range of
parameter values beyond present limits.  Also, one of the experimental approaches,
in which the detector (regardless of technology) is deployed deep underground,
considered in this study has the potential to contribute, 
 to a significant improvement of our knowledge about
nucleon decay and natural sources of neutrinos.

The two experimental approaches studied here are complex in their
detailed technical realization, comprising several detector
technologies, various specific neutrino beam designs and different
measurement strategies.  They have in common, however, the
exploitation of experimental baselines of $\sim$1000 km (a key
advantage of a U.S. based program) and both approaches make effective
use of existing Fermilab accelerator infrastructure
with modest upgrades. 
  The experimental detectors required are
very massive (in the several hundred kiloton range) because the
interaction rates are small.  The designs for
such detectors vary from already-demonstrated at a scale of 50 kTon
(Super Kamiokande) to
somewhat speculative (large liquid Argon). 
In both cases, significant R\&D is still
needed to demonstrate feasibility and  obtain a reliable cost estimate 
for the scale needed here.  The study has shown
however, that it will be feasible and practical to carry out the
desired program of important neutrino physics, perhaps together with
improved nucleon decay and natural neutrino investigations in the same
neutrino detector.

The output of the present study is twofold: 1) technical results and
conclusions that report the results of the study and address the
charge letter; 2) answers to the 15 questions posed to the study group
by the NuSAG Committee.  These two outputs comprise  more than 50 pages of
detailed commentary and they are provided in full in the body of the
report and Appendix A.  Here, we attempt to provide a somewhat
condensed version of the study results and conclusions while urging
the reader to consult the full text of the report on any points that
may appear to be questionable or unclear.  The summary results and
conclusions were discussed and agreed to at the September 17, 2006
meeting of the study group.

Results and Conclusions: 

\begin{itemize} 

\item Very massive detectors with efficient fiducial mass of
$>100$ kTon are needed for the accelerator long baseline 
neutrino program of the future.  We define efficient fiducial mass 
as fiducial mass multiplied by the signal efficiency. 
For accelerator based neutrino physics, this could correspond to 
several hundred kTon if the detector is a water Cherenkov detector
and $>100$ kTon if it is liquid argon TPC with high expected
efficiency.  
These detectors could be key shared research facilities for the future particle,
nuclear and astrophysics research programs.  Such a detector(s) could
be used with a long baseline neutrino beam from an accelerator
laboratory to determine (or bound) leptonic CP violation and measure
all parameters  of neutrino oscillations.  At the same time, if located in a
low background underground environment, it would have additional
physics capabilities for proton decay and continuous observation of
natural sources of neutrinos such as supernova or other astrophysical
sources of neutrinos.

\item  The Phase-II program will need  considerable upgrade to the 
current accelerator intensity from FNAL.  
Main Injector accelerator intensity
upgrade to $\sim$ 700 kW is already planned for Phase-I of the program
(NO$\nu$A). A further upgrade to 1.2 MW is under design and discussion  as 
described briefly in this report.  
The phase-II program could be carried out with  
these planned upgrades. Any further improvements, perhaps with a new intense
source of protons, will obviously increase the statistical sensitivity and 
measurement precision.  

\item  A water Cherenkov detector of multi-100kTon size is needed to obtain
sufficient statistical power to reach good sensitivity to CP
violation. This requirement is independent of whether one uses the
off-axis technique or the broadband technique in which the detector is
housed in one of the DUSEL sites.

\item  High signal efficiency at high energies and excellent background
reduction in a liquid argon TPC allows the size of such a detector to
be smaller by a factor of 3 compared to a water Cherenkov detector for
equal sensitivity. Such a detector is still quite large.  

\item The water Cherenkov technology is well established.  The issues of
signal extraction and background reduction were discussed and
documented at length in this study. The needed background reduction is
achievable and well understood for the broadband beam discussed in this 
report, but not yet fully optimized.  Key issues for scaling up the current
generation of water Cherenkov detectors (Super-Kamiokande, SNO, etc.) 
and locating such detectors in underground locations in DUSEL are well understood. 
The cost and schedule for such a detector could be
created with high degree of confidence. A first approximation for this
was reported to the workshop.

\item For a very large liquid argon time projection detector
key technical issues have been identified for the building of the
detector.  A possible development path includes understanding argon
purity in large industrial tanks, mechanical and electronics issues associated
with long wires, and construction of at least one prototype in the
mass range of 1 kTon.

\item In the course of this study, we have examined the surface operation
of the proposed massive detectors for accelerator neutrino physics.
Water Cherenkov detectors are suitable for deep underground locations
only.  Surface or near-surface operation of liquid argon TPCs is
possible but requires that adequate rejection of cosmic rays be
demonstrated.  Surface or near-surface operation capability is
essential for the off-axis program based on the existing NuMI beam-line
because of the geographic area through which the beam travels.

\item Additional detailed technical conclusions of the study are noted in
the Results and Conclusions section of this report.  These results
could influence the detailed design of the specific program selected.

\end{itemize} 

Detailed sensitivity estimates for the choices under consideration can be 
obtained from Section \ref{sensi}.  Here we will give a broad comparison of 
the different experimental approaches.

In the course of this year long study we have been able to draw several very clear conclusions. Regardless of which options evolve into a future program, the following will be required.
\begin{enumerate}
\item A proton source capable of delivering 1 - 2 MW to the neutrino production target.
\item Neutrino beam devices (targets and focusing horns) capable of efficient operation at high intensity.
\item Neutrino beam enclosures which provide the required level of environmental and personnel radiological protection.
\item Massive ($>>$100 kton) detectors which have have high efficiency, resolution and background rejection.
\item For each of the above items, significant investment in R and/or D is required and needs to be an important aspect of the current  program.
\end{enumerate}

We have found that the main areas of this study can be discussed
relatively simply if we divide them into two broad categories : 1) The 
neutrino beam configuration and 2) The detector technology. Further, we
are able to summarize our conclusions in two  tables which show
the pros and cons of the various options.

In Table \ref{cetable1} we compare the pros and cons of using the existing NuMI
beam and locating detectors at various locations, 
versus a new wide
band neutrino beam, from Fermilab but directed to a new laboratory
located at one of the potential DUSEL sites, i.e. at a baseline of
1300 to 2600 km.

In Table \ref{cetable2} we compare the pros and cons of constructing massive
detectors (~100 - 300 kT total fiducial mass) using either water
Cherenkov or liquid argon technology.

\begin{table}[h]
\centering
\begin{tabular}{|l|l|l|}
\hline
 & Pro & Con \\
 \hline
NuMI On-axis & Beam exists; & L $\sim$ 735 km \\
                          & Tunable spectrum; &  Sensitivity to mass hierarchy is limited \\
                          &                  & Difficult to get flux $<3$ GeV \\
\hline                          
NuMI Off-axis   & Beam exists ;                             &  L $\sim$ 800 km  \\
(1st maximum) & Optimized energy;          &  Limited sensitivity to mass hierarchy \\
                           & Optimized location  for  & \\
                           & 1st detector;                    & \\
                           & Site will exist from \nova project; & \\
\hline
NuMI Off-axis & Beam exists; &  L $\sim$ 700-800 km; \\
(2nd maximum) & Optimized energy; & Extremely low event rate;  \\   
                             & Improves mass hierarchy  &  A new site is needed; \\
                             & sensitivity if $\theta_{13}$ is large; &  Energy of events is $\sim 500 MeV$;  \\
                             &   & Spectrum is very narrow  \\
\hline
WBB to DUSEL & More optimum (longer) baseline;   & New beam construction project $>$\$100M;  \\
                             & Can fit oscillation parameters    &   Multi-year beam construction; \\   
                             & using energy spectrum;                 &  \\
                             & Underground DUSEL site for detector; & \\
                             & Detector can be multi-purpose; & \\
\hline
\end{tabular}
\caption{\label{cetable1}Comparison of the existing NuMI beam to a possible new wide band low energy (WBLE) beam to DUSEL }
\end{table}

\begin{table}[h]
\centering
\begin{tabular}{|l|l|l|}
\hline
 & Pro & Con \\
 \hline
Water  & Well understood and proven technology;      &  Must operate underground;  \\ 
Cherenkov		&  Technique demonstrated by SuperK (50kT);  &  Scale up factor is $<10$;    \\
                &                                          & Cavern stability must be assured  \\ 
                &			&  and could add cost uncertainty; 			\\
                &    New  background rejection techniques  & NC background depends on spectrum \\
	         &  available; 		& 		and comparable to instrinsic background; \\ 
                   &  Signal energy resolution $\sim10\%$;   &  Low $\nu_e$ signal efficiency (15-20\%);   \\
                                 & Underground location      &                                    \\
                                 & makes it a multi-purpose detector;      &       \\
				 & Cosmic ray rate at 5000ft is $\sim$0.1 Hz.   &                \\
 & Excellent sensitivity to $p\to \pi^0 e^+$ & Low efficiency to $p\to K^+ \bar\nu$ \\ 
\hline      
Liquid   & Technology demonstrated by & Scale up factor of $\sim$300 is needed;   \\   
Argon         &  ICARUS (0.3kT); &   \\
TPC     &    &  Needs considerable R\&D for costing; \\ 
       & Promises high efficiency and &  Not yet demonstrated  by  \\ 
                                 & background rejection;     & simulation of a large detector; \\
                                 &  Has potential to operate     &  Needs detailed safety design for   \\
                                 & on (or near) surface;          &  deep location in a cavern; \\
                                 & Could be placed on surface  & Needs detailed  demonstration            \\
			&	either at NuMI Offaxis or DUSEL;          &   of cosmic ray rejection;    \\
				&				& Surface cosmic rate $\sim$500kHz; \\ 
                                 & Better sensitivity to            & Surface operation limits \\
                                 & $p\to K^+ \bar\nu$                & physics program; \\
\hline

\end{tabular}
\caption{\label{cetable2} Comparison of Water Cherenkov to Liquid Argon detector technologies }
\end{table}

\clearpage 

\newpage

\section{Introduction}

This report details the activities and the results of a several month 
long study on long baseline neutrinos.   This workshop (named 
the US joint
study on long baseline neutrinos) was sponsored by both Fermi 
National Accelerator Laboratory and Brookhaven National Laboratory.

{\bf Charge}: 
 This study grew out of two parallel efforts. 
An earlier attempt to create a joint FNAL/BNL task force on long baseline 
neutrinos was initiated by the management of these two laboratories. 
Later the need arose to provide input to the 
neutrino scientific advisory committee (NuSAG) which was asked to 
address the APS study's recommendation for a next generation neutrino beam and 
detector configuration. The NuSAG charge is in Appendix B. 
The APS study report can be obtained from 
\underline{http://www.aps.org/neutrino/}. 
The study principals  created a charge with  specific 
scenarios for an accelerator based program. The charge from the 
chairs of the study is in Appendix C.

{\bf Membership}: Although the study group was asked 
to mainly focus on a 
next generation program within the US, participation from the world wide 
community of particle physicists was sought.  In particular, physicists 
engaged in the European equivalent of this study 
(the International Scoping Study: \underline{http://www.hep.ph.ic.ac.uk/iss/}) 
were kept abreast of our progress.   The list of physicists who 
 participated in this study by either contributing 
written material, presentations, or discussion is  at 
\underline{http://nwg.phy.bnl.gov/fnal-bnl/}. 

The membership was divided into several subgroups. The accelerator 
subgroup studied and summarized the proton intensities available mainly 
from FNAL. The neutrino beam subgroup summarized the 
neutrino beam intensities and event rates for various possibilities. 
The water Cherenkov subgroup summarized the current understanding of
the conceptual design of such a detector as well as the state of the art 
in simulating and reconstructing events in  such a detector.  
The liquid argon detector subgroup studied the capabilities of such a 
detector as well as the feasibility of building a detector large 
enough to  collect sufficient numbers of events. The results from
 each of these groups is either in  presentations,  technical 
documents prepared in the near past, or in technical documents prepared 
specifically for this study.

{\bf Scope of the work}: 
As specified in Appendix C, the scope of our work was limited to 
conventional horn focused accelerator neutrino beams from US 
accelerator laboratories.  It was asked that we study a next generation
program by placing massive detectors either off-axis on the surface 
for the NuMI beam-line at FNAL, or by building a new intense beam-line
aimed towards a new deep underground science laboratory (DUSEL)  in the 
western US. The detector technology to be considered was either a water 
Cherenkov detector or a liquid Argon time projection chamber.  
The international scoping study (ISS) on the other hand focused 
on new technology ideas such as beta-beams and muon storage ring based 
neutrino factories.

\begin{itemize}

\item In the following we will refer to the NO$\nu$A program using the NuMI 
off axis beam as Phase-I.   We will not study or comment on this phase 
extensively since it has been previously reviewed extensively, but it will 
be necessary for us to use the extensive existing material for this 
phase to study the next two items. 

\item An upgraded off-axis  program with multiple 
detectors, including a massive liquid argon detector, as Phase-II(option A).
There could be various versions of Phase-II(option A), with or without 
a liquid argon detector, with a water Cherenkov detector, and/or 
detectors at various locations off axis. We will attempt to 
elaborate on all of these.   

\item A program using a new beam-line towards DUSEL, housing a massive 
multipurpose detector, either a water Cherenkov  or a liquid argon
detector, will be called Phase-II(option B). We will provide information on 
the DUSEL candidate sites as well as the two options for a multipurpose 
detector. 

\end{itemize}

{\bf Schedule}: The study followed the schedule outlined in Appendix
\ref{sch}. 
The first meeting of the FNAL and BNL management 
that led to the study was held at BNL on November 14, 2005.  
The charge of the workshop which defined the scope of 
the work was finalized  after the  meeting on March 6-7, 2006. 
It was decided at this meeting that since the time for the report was 
short, it was best to create small subgroups to work on individual 
papers for the study. These papers would be distributed to the study group as well as the 
NuSAG committee as they were prepared. 

A set of presentations were made to the NuSAG committee on May 20,
2006. Results from on-going work was reviewed at this meeting. We
selected July 15, 2006 as a deadline for preparation of the individual
papers.  Many, but not all, papers were prepared by July 15, and were
distributed by web-site
(\underline{http://nwg.phy.bnl.gov/fnal-bnl}).

After discussion within the working group a  summary report
(this report) was commissioned. The contents of this report were
reviewed by the study group on September 16-17, 2006. The deadline for
delivering a preliminary  report to NuSAG was October, 2006.

\section{Physics goals of a Phase-II program}

There is now an abundance of evidence that neutrinos oscillate among the  
three known flavors $\nu_e$, $\nu_{\mu}$ and $\nu_{\tau}$, thus indicating
that they have masses and mix with one another\cite{ref1}.  Indeed, modulo an anomaly 
in the LSND experiment, all observed neutrino oscillation phenomena are well 
described by the 3 generation mixing

{
\be\bf
\pmatrix{|\nu_e> \cr |\nu_\mu> \cr |\nu_\tau>} =\bf U \pmatrix{|\nu_1> \cr
\bf|\nu_2> \cr\bf |\nu_3>}
\ee

\[
U = \pmatrix{\bf c_{12}c_{13} & \bf s_{12}c_{13} &\bf s_{13}e^{-i\delta} \cr
\bf-s_{12}c_{23}-c_{12}s_{23}s_{13}e^{i\delta} &\bf
c_{12}c_{23}-s_{12}s_{23}s_{13}e^{i\delta} &\bf s_{23}c_{13} \cr
\bf s_{12}s_{23}-c_{12}c_{23}s_{13}e^{i\delta} &\bf
-c_{12}s_{23}-s_{12}c_{23}s_{13}e^{i\delta} &\bf c_{23}c_{13}} \nonumber
\label{eqone}
\]

\[
c_{ij} = \cos\theta_{ij} \quad , \quad s_{ij}=\sin\theta_{ij}, ~~~ i,j=1,2,3
\]}

\noi with  $|\nu_i>,~ i=1,2,3$, the neutrino mass eigenstates.

Atmospheric neutrino oscillations are governed by a mass squared difference
$\Delta m^2_{32}=m^2_3-m^2_2 = \pm2.5\times 10^{-3} {\rm eV}^2 $\cite{superkatm}
and mixing angle $\theta_{23}\simeq 45^{\circ}$; findings that have been 
confirmed by accelerator generated neutrino beam studies  at Super-Kamiokande and MINOS\cite{k2k,minos}.

As yet, the sign of $\Delta m^2_{32}$ is undetermined.  The so-called normal
 mass hierarchy,~~~ $ m_3 > m_2$, suggests a positive sign which is also
 preferred 
by theoretical models.  However, a negative value (or inverted hierarchy) 
can certainly be accommodated, and if that is the case, the predicted rates
 for neutrino-less double beta decay will likely be larger and more easily
accessible experimentally.  Resolving the sign of the mass hierarchy is an 
extremely important issue.  In addition, the fact that $\theta_{23}$ is
large and near maximal is also significant for model building.  Measuring that 
parameter with precision is highly desirable.

In the case of solar and reactor neutrino oscillations \cite{sno,kamland,kamland2}, one finds
$\Delta m^2_{21}=m^2_2-m^2_1 \simeq 8\times 10^{-5} {\rm eV}^2 $ and 
 $\theta_{12}\simeq 32^{\circ}$.
Again, the mixing angle is relatively large (relative to the analogous
Cabbibo angle $\simeq 13^{\circ}$ of the quark sector).  In addition, 
$\Delta m^2_{21}$ is large enough, compared, to $\Delta m^2_{32}$, to make
long baseline neutrino oscillation searches for CP violation feasible and
could  yield positive results, i.e. the stage is set for a 
future major discovery (CP violation in the lepton sector).

Currently, we know nothing about the value of the CP violating phase 
$\delta$ ~ ($0 < \delta <  360^{\circ}$) and only have an upper bound \cite{chooz} on 
the as yet unknown mixing angle  $\theta_{13}$ ($\theta_{13} < 13^{\circ}$)

$$ \sin^22\theta_{13}\leq 0.2 $$

The value of $\theta_{13}$  is likely to be determined by the coming 
generation of reactor $\bar\nu_e$  disappearance and accelerator based 
$\nu_\mu\to\nu_e$ appearance experiments if $\sin^22\theta_{13}\geq 0.01$. 
Knowledge of $\theta_{13}$ and $\delta$ would complete our determination 
of the 3 generation lepton mixing matrix and provide a measure of leptonic 
CP violation via the Jarlskog invariant.

$$
J_{CP} \equiv \frac18 \sin2\theta_{12}\sin2\theta_{13} \sin2\theta_{23}
\cos\theta_{13} \sin\delta.
$$

\noi
If we use the above limit for $\theta_{13}$ then 
$J_{CP}^{Leptonic}<0.05 \times \sin \delta$,  
 which could easily turn out to be much larger than the analogous quark
degree of CP violation $J_{CP}^{Quarks}\simeq 3\times 10^{-5}$.

Based on our current knowledge and future goals, a phase II neutrino program 
should include:
\begin{itemize}

\item{Completing the measurement of the leptonic mixing matrix}, 

\item Study of CP violation,

\item {Determining the values of all parameters with high precision 
         including $J_{CP}$ as well as the sign of $\Delta m^2_{32}$ },  

\item {Searching for exotic effects perhaps due to sterile neutrino mixing,
extra dimensions, dark energy etc.}

\end{itemize}

\medskip
Of the above future neutrino physics goals, the search for and study of 
CP violation is of primary importance and should be our main objective for
several reasons which we briefly outline.

CP violation has so far only been observed in the quark sector of the 
Standard Model.  Its discovery in the leptonic sector should shed additional
light on the role of CP violation in Nature.  Is it merely an arbitrary 
consequence of inevitable phases in mixing matrices or something deeper?
Perhaps, most important, unveiling leptonic CP violation is particularly 
compelling because of its potential connection with the observed
 matter--antimatter asymmetry of our Universe, a fundamental problem at the 
heart of our existence.  The leading explanation is currently a leptogenesis
scenario in which decays of very heavy right--hand neutrinos created in the
early universe give rise to a lepton number asymmetry which later becomes
a baryon--antibaryon asymmetry via the B-L conserving 't Hooft mechanism
of the Standard Model at weak scale temperatures.

Leptogenesis offers an elegant, natural explanation for the matter--antimatter
asymmetry; but it requires some experimental confirmation of its various 
components before it can be accepted.  Those include the 
existence of very heavy right--handed neutrinos as well as lepton number and 
CP violation in their decays.

Direct detection of those phenomena is highly unlikely; however, indirect 
connections may be established by studying lepton number violation in 
neutrinoless double beta decay and CP violation in ordinary neutrino
oscillations.  Indeed, such discoveries will go far in establishing 
leptogenesis as a credible, even likely scenario.  For that reason,
neutrinoless double beta decay and leptonic CP violation
in neutrino oscillations are given very high priorities by the particle 
and nuclear physics communities.

Designing for CP violation studies in next generation neutrino programs has
other important benefits.  First , the degree of difficulty  
to establish CP violation and determine $J_{CP}^{leptonic}$  is
 demanding but doable.  It requires an intense proton beam of about
1--2~MW and a very large detector ($250 \sim 500$ kton Water Cherenkov or a
liquid argon detector of size $\sim 100$ kTon which could be    
equivalent in sensitivity due to its better performance).  
Such an ambitious infrastructure will allow very precise 
measurements of all neutrino oscillation parameters as well as the sign of
 $\Delta m^2_{32}$ via $\nu_{\mu}\to \nu_{\mu}$ disappearance and 
 $\nu_{\mu}\to \nu_{e}$ appearance studies.  It will also provide a 
sensitive probe of ``New Physics''
deviations from 3 generation oscillations, perhaps due to sterile neutrinos, 
extra dimensions, dark energy or other exotic effects.

A well instrumented 
very large detector, in addition to its accelerator based neutrino program, 
could be sensitive to proton decay which is one of the top priorities in 
fundamental science.  
Assuming that it is located underground and shielded from cosmic rays, it
can push the limits on proton decay into modes such as $p\to e^{+}\pi^{0}$ to 
$10^{35}$yr sensitivity or beyond, a level suggested by gauge boson mediated
 proton decay in super-symmetric GUTs.  Indeed, there is such a natural
 marriage between the requirements to discover leptonic CP violation and
see proton decay (i.e. an approximately 500 kTon water Cherenkov detector) that
it could be hard to imagine undertaking either effort without being able to do
the other. 

Such a large detector would also have additional physics
capabilities.  It could study atmospheric neutrino oscillations with very
 high statistics and look for the predicted relic supernova
neutrinos left over from earlier epochs in the history of the Universe, a
potential source of cosmological information.  Also, if a supernova should 
occur in our galaxy (expected about every 30 years), such a detector would see
about 100,000 neutrino events.  In addition, it could be used to look for
 signals of $n-\bar n$ oscillations in nuclei and highly penetrating GUT 
magnetic monopoles which would leave behind a trail of monopole catalyzed
 proton decays.  

The physics potential of a very large underground detector is extremely rich. 
The fact that it can also be used to determine (or bound) leptonic CP 
violation and measure all 
facets of neutrino oscillations gives such a facility 
outstanding discovery potential.
It would be an exciting, central component of the world's particle physics
program for many decades. On the other hand, a staged approach using existing 
beam facilities should also be explored to determine an optimum strategy.

\section{Strategies for the Phase-II program using a conventional beam}
\label{sfp2}

In this section we will describe the essential features of an off-axis narrow 
band beam versus an on-axis  broad band beam. We will then briefly summarize 
how these features can be used to extract the CP violation effect as well as 
all the other parameters of importance in neutrino oscillations.

Throughout this report we are concerned with 
conventional horn focused beams in the US: the existing 
NuMI beam at FNAL or a new super neutrino beam that could be optimized 
for a detector at a new deep underground national laboratory (DUSEL) with 
a possible large detector (either underground or on the surface). 
The measurement of most interest is always the appearance measurement, 
$\nu_\mu \to \nu_e$, for which the horn focused beam has 
a limitation from the irreducible background of $\nu_e$ contamination in
the beam. The level of  contamination depends on neutrino energy
and also    the beam design 
and the off-axis angle, but it is in the range of $\sim 0.5 - 1\%$ for
most practical beams. This contamination comes from 
decays of muons and kaons in the beam. These cannot be 
completely eliminated.      The second source of background is 
neutral current events that mimic electron showers. This background is 
considered reducible by detector design. In particular, a fine grained
 detector such as a liquid argon TPC detector will be capable of reducing such 
background to very small levels. Most of the remaining report will be concerned 
with the best strategy for obtaining sufficient signal events  while reducing 
these backgrounds. In this section we will not discuss the issues of 
backgrounds in detail, but give a guide to the signal spectra,
event rates and comment on the implications.

\begin{figure}
\centering\leavevmode
\mbox{\includegraphics[angle=0,width=0.49\textwidth]{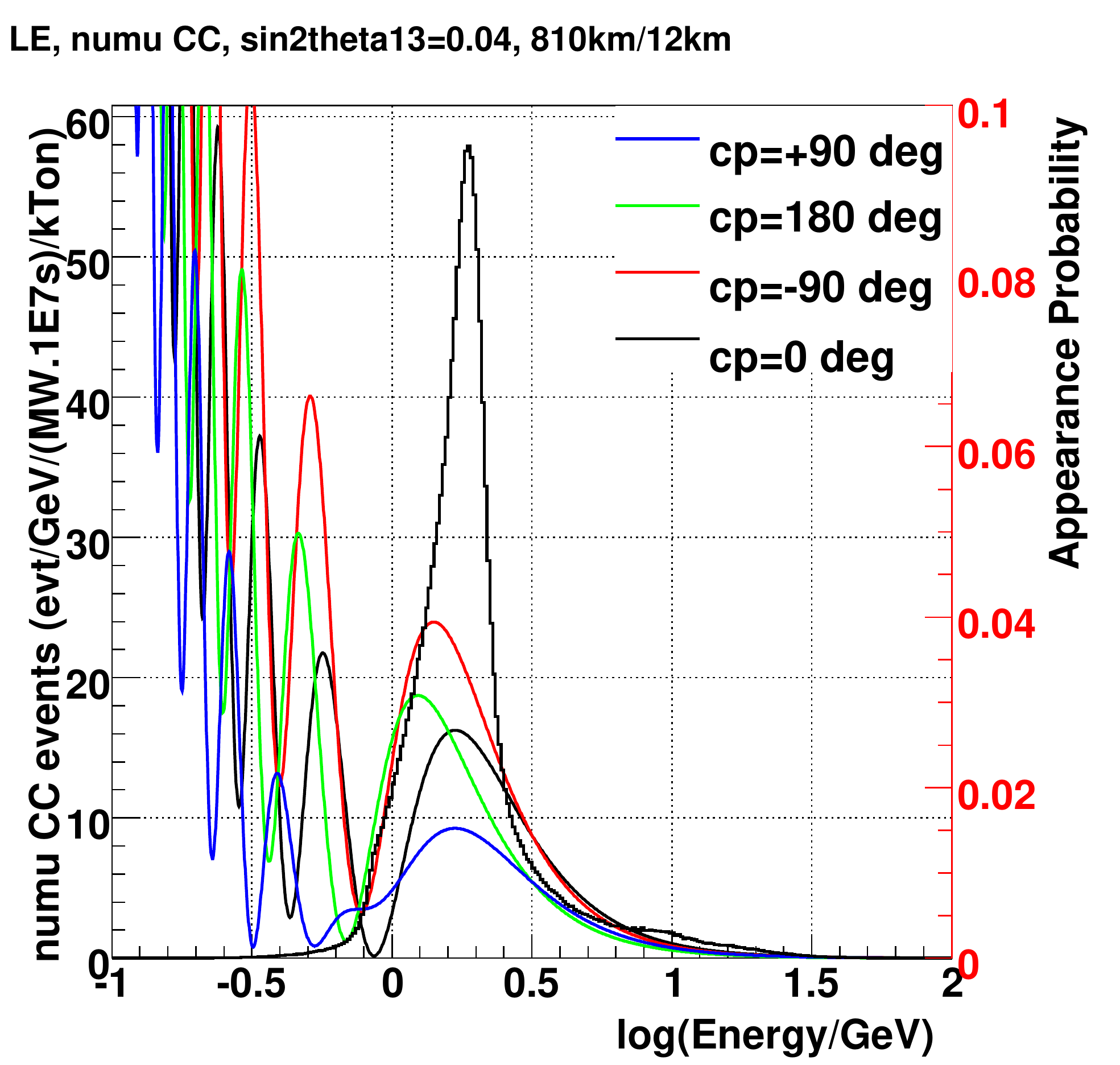}
\includegraphics[angle=0,width=0.48\textwidth]{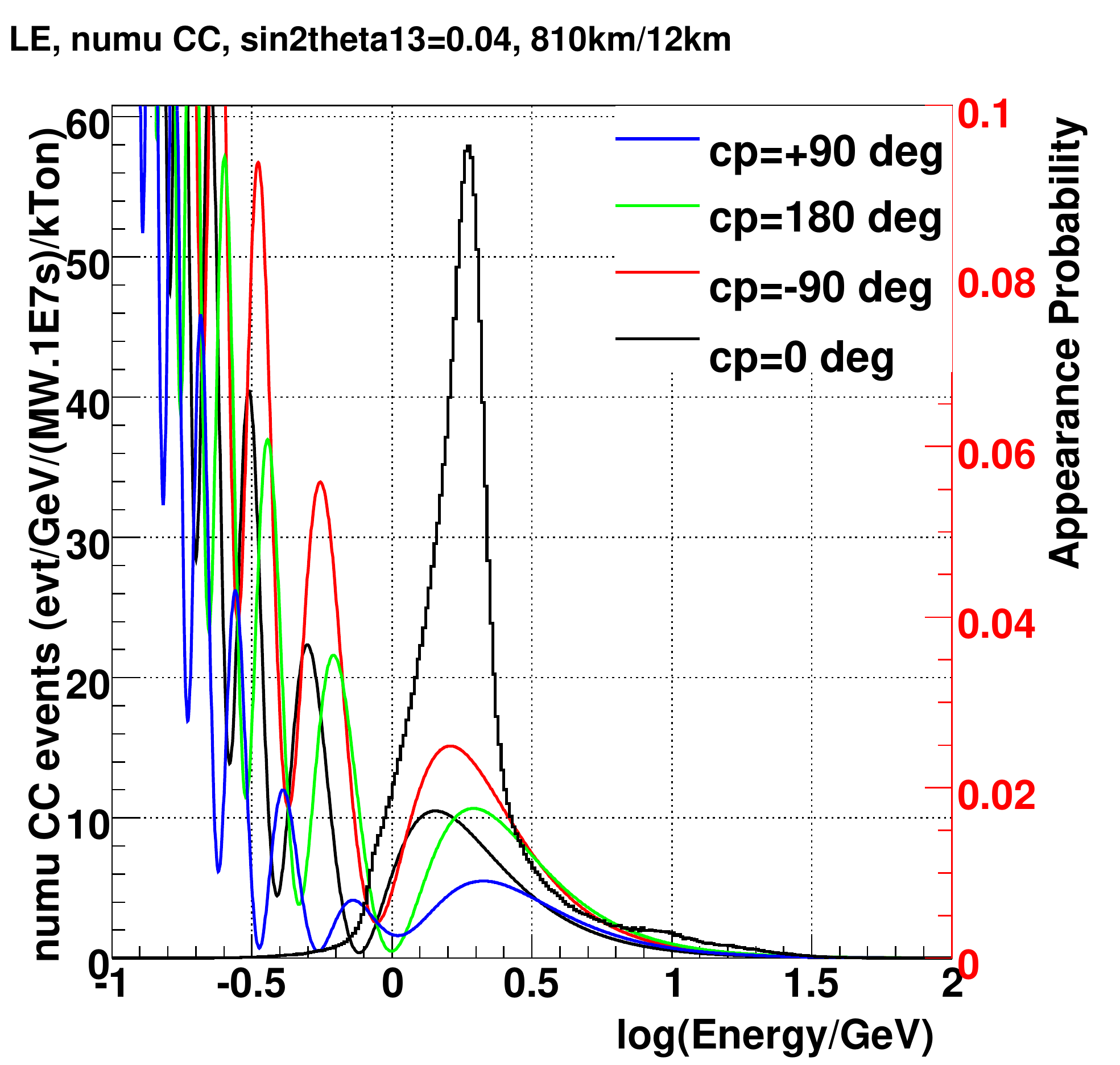}}
 \caption{(in color) Spectrum of charged current $\nu_\mu$ events at a 12 km
 off-axis location at 810 km on the NuMI beam-line. The spectrum is
 normalized per GeV per $MW\times 10^7 sec$ protons of 120 GeV. The low energy
 (LE) setting of the NuMI beam-line is used for this plot.  Overlayed
 is the probability of $\nu_\mu \to \nu_e$ conversion for $\sin^2 2
 \theta_{13} = 0.04$ with rest of the oscillation parameters as
 described in the text. The left plot is for regular mass ordering and
 right hand side is for reversed mass ordering. Figure includes no detector effects 
such as efficiencies, resolution, or backgrounds.  
  \label{f12km} }
\end{figure}

\begin{figure}
\centering\leavevmode
\mbox{\includegraphics[angle=0,width=0.49\textwidth]{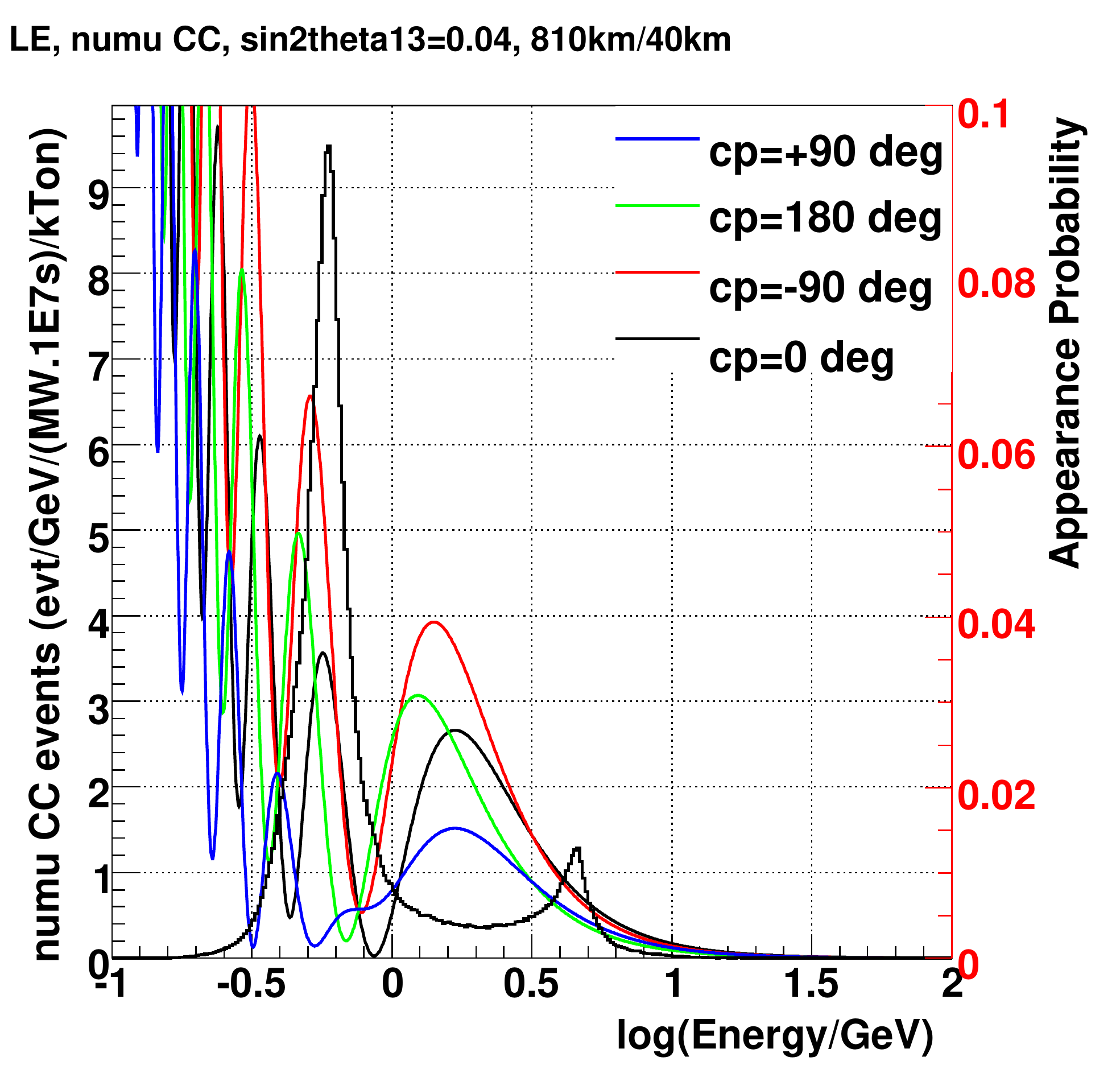}
\includegraphics[angle=0,width=0.48\textwidth]{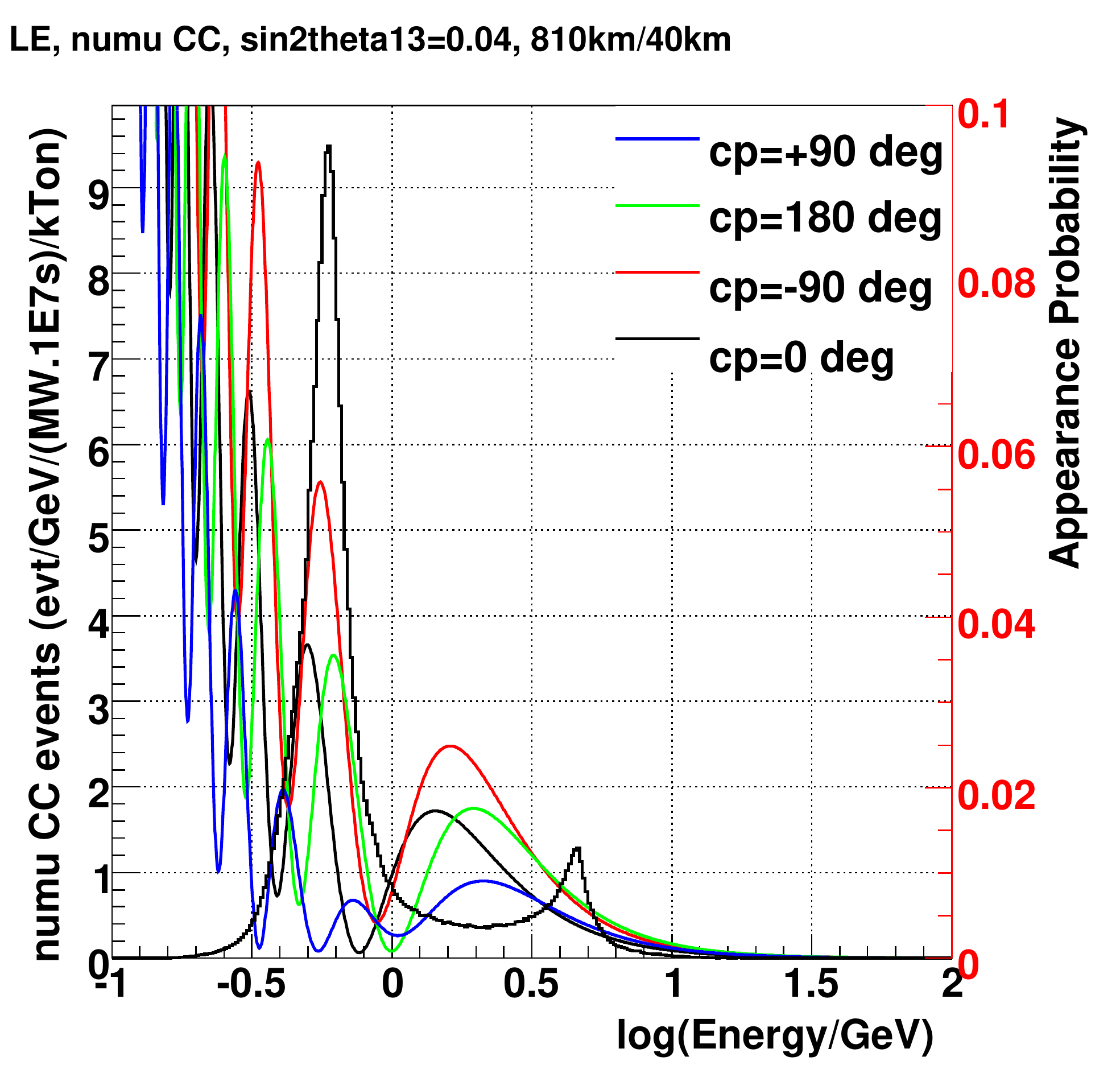}}
 \caption{(color) Spectrum of charged current $\nu_\mu$ events at a 40 km
 off-axis location at 810 km on the NuMI beam-line. The spectrum is
 normalized per GeV per $MW\times 10^7 sec$ protons of 120 GeV. The low energy
 (LE) setting of the NuMI beam-line is used for this plot.  Overlayed
 is the probability of $\nu_\mu \to \nu_e$ conversion for $\sin^2 2
 \theta_{13} = 0.04$ with rest of the oscillation parameters as
 described in the text. The left plot is for regular mass ordering and
 right hand side is for reversed mass ordering. 
Figure includes no detector effects 
such as efficiencies, resolution, or backgrounds.  
 \label{f40km} }
\end{figure}

\begin{figure}
\mbox{\includegraphics[angle=0,width=0.49\textwidth]{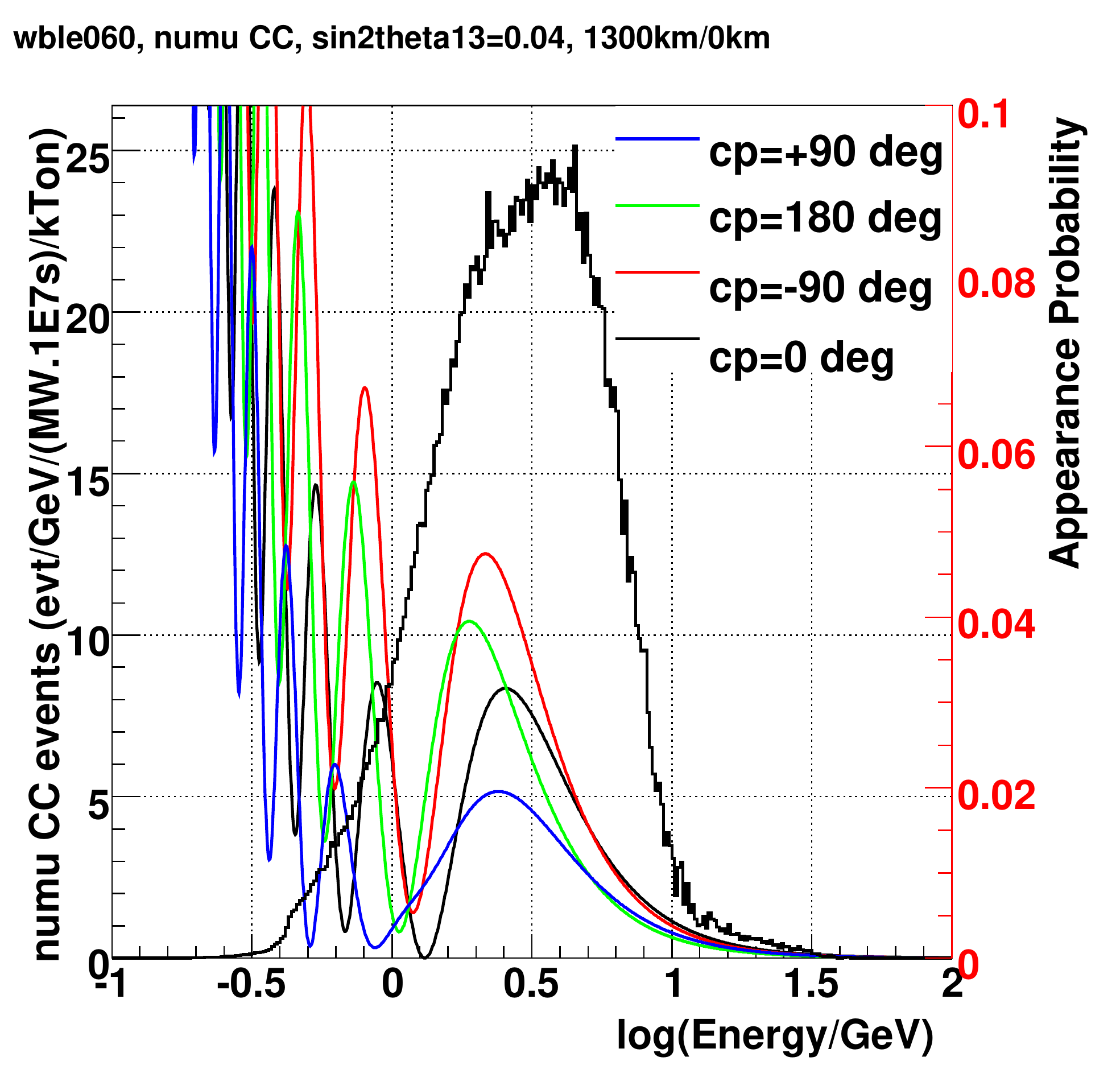}
\includegraphics[angle=0,width=0.48\textwidth]{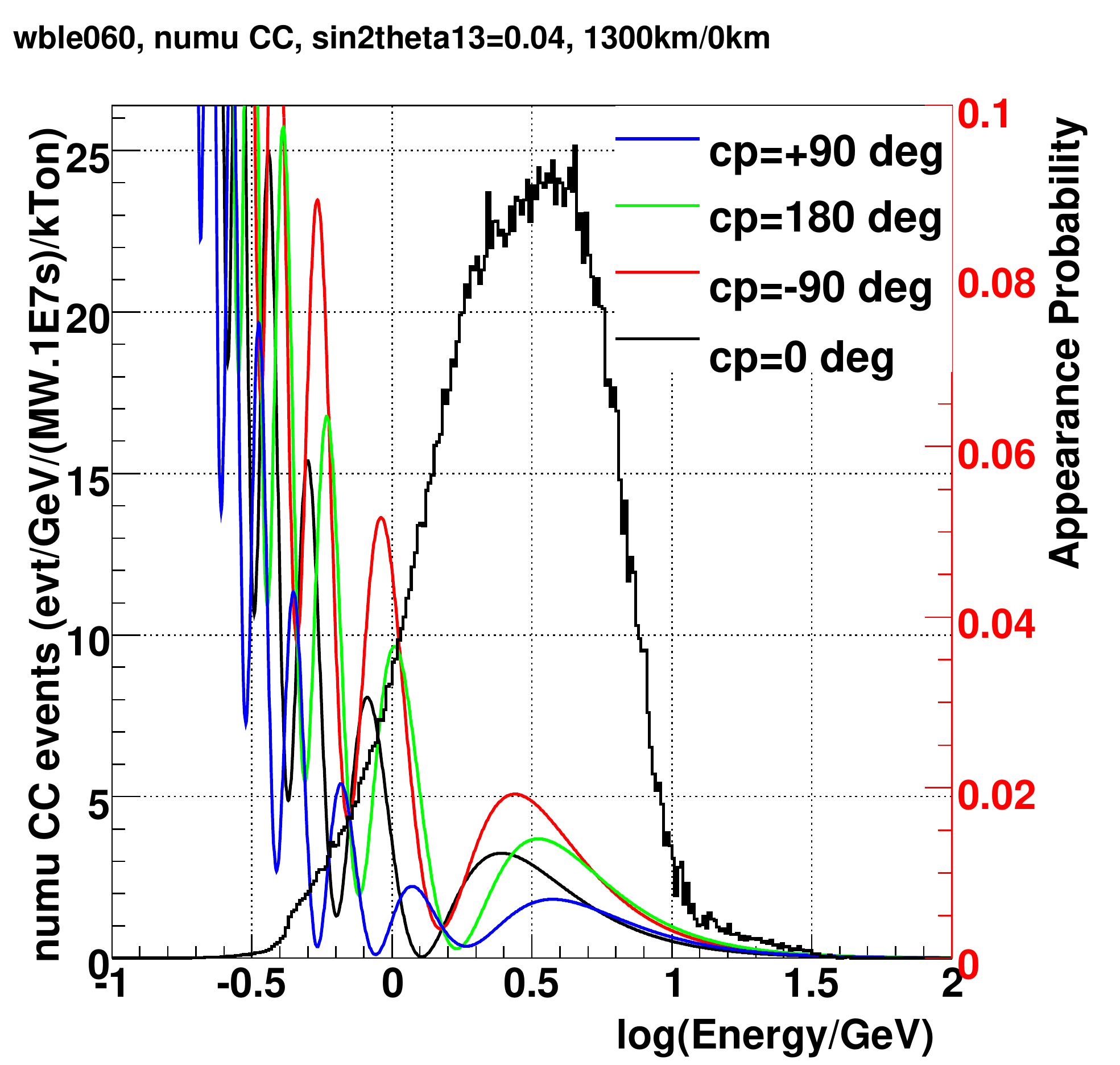}}
\centering\leavevmode
 \caption{(color) Spectrum of charged current $\nu_\mu$ events using a new
 wide band beam from FNAL to a location at 1300 km. The spectrum is
 normalized per GeV per $MW\times 10^7 sec$ protons of 60 GeV. Overlayed is the
 probability of $\nu_\mu \to \nu_e$ conversion for $\sin^2 2
 \theta_{13} = 0.04$ with rest of the oscillation parameters as
 described in the text. The left plot is for regular mass ordering and
 right hand side is for reversed mass ordering.  
Figure includes no detector effects 
such as efficiencies, resolution, or backgrounds.  
\label{fwble1300} }
\end{figure}

\begin{figure}
\mbox{\includegraphics[angle=0,width=0.49\textwidth]{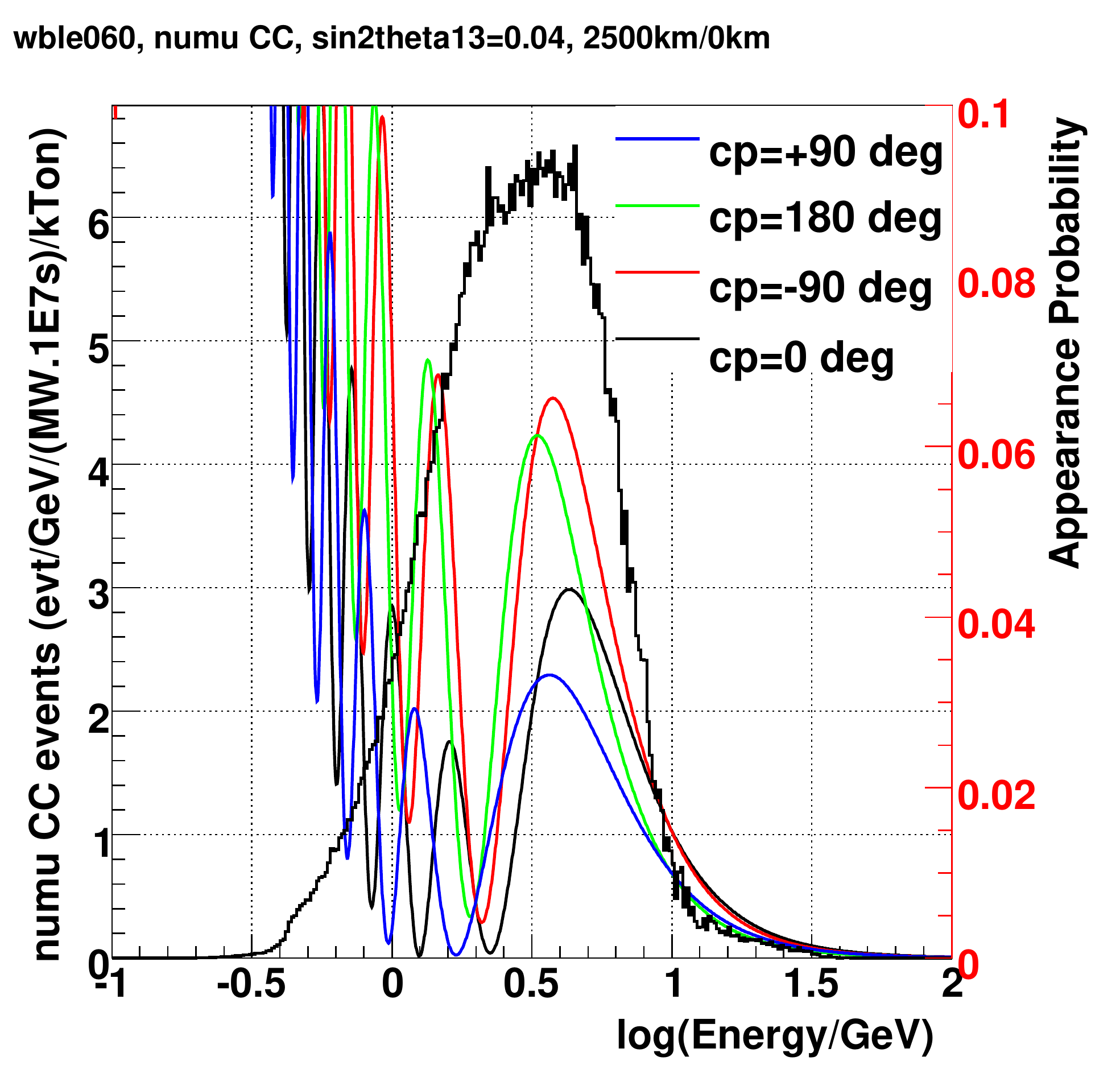}
\includegraphics[angle=0,width=0.48\textwidth]{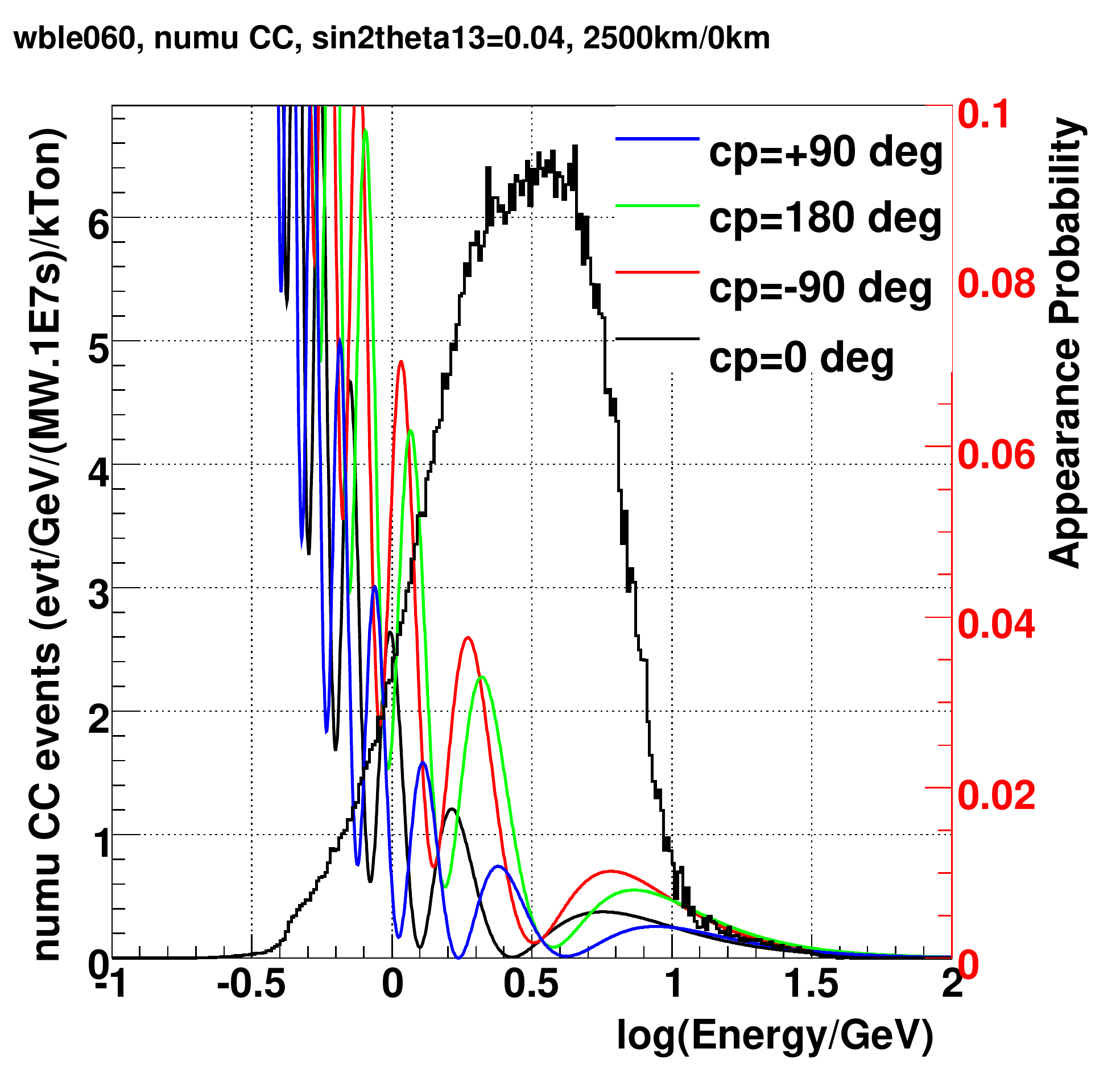}}
\centering\leavevmode
 \caption{(color) 
Spectrum of charged current $\nu_\mu$ events using a new
 wide band beam from FNAL to location at 2500 km. The spectrum is
 normalized per GeV per $MW\times 10^7 sec$ protons of 60 GeV. Overlayed is the
 probability of $\nu_\mu \to \nu_e$ conversion for $\sin^2 2
 \theta_{13} = 0.04$ with rest of the oscillation parameters as
 described in the text. The left plot is for regular mass ordering and
 right hand side is for reversed mass ordering.  
Figure includes no detector effects 
such as efficiencies, resolution, or backgrounds.  
\label{fwble2500} }
\end{figure}

Figures \ref{f12km}, \ref{f40km}, \ref{fwble1300}, and \ref{fwble2500}
show the spectra of concern.  Care is required in comparing these
plots because they are plotted on a logarithmic energy scale. The
normalization is per GeV of neutrino energy per kTon of detector mass
per $MW\times 10^7 sec$ protons of the appropriate energy on target.

These spectra were obtained by detailed simulations using the GNuMI
computer program\cite{minos}. For these figures a simple recipe was used to obtain
charged current event rate
\cite{offaxs}: a cross section of $0.8\times 10^{-38} cm^2/GeV$ 
($0.35\times 10^{-38} cm^2/GeV$ for anti-neutrinos) was
 used above 0.5 GeV and 
the quasi-elastic cross section was used below 0.5 GeV. 
There could be
small differences due to the detector target type (water, argon, etc.),
but this is a good approximation \cite{zeller}.  For figures \ref{f12km} and
\ref{f40km} we have used the low energy (LE) setting of the NuMI beam
configuration which gives a better flux at the 40 km site. Reference
\cite{offaxs} contains spectra for other choices.  For all the
off-axis spectra 120 GeV protons were used and the normalization is
for $MW\times 10^7 sec$ protons; for 120 GeV protons this corresponds to 
$5.2\times 10^{20}$ protons.

  For Figures  \ref{fwble1300} and \ref{fwble2500}, the GNuMI program was
  modified for a wide band low energy (WBLE) design for the horns as well
  as a new  decay tunnel with 4 m diameter and 400 m length; 
these are described in
  detail in \cite{wble}. For the WBLE beam, there is a choice of
  running with protons from 40 GeV to 120 GeV. For these plots we have
  chosen 60 GeV protons.  The normalization is for $MW\times 10^7sec $ protons
  of 60 GeV. The spectra shown here should not be considered 
  optimum. After thorough design and optimization there could  be
  modest improvements, but at this point we are confident that
  these numbers are sufficiently good for this review.

For Figures \ref{f12km} to \ref{fwble2500} we have superimposed the
expected probability of $\nu_\mu \to \nu_e$ conversion for the
appropriate distance and for the following oscillation parameters:
$\Delta m^2_{32} = 0.0025 eV^2$, $\Delta m^2_{21} = 8\times 10^{-5}
eV^2$, $\sin^2 2 \theta_{12} = 0.86$, $\sin^2 2 \theta_{23}=1.0$, and
$\sin^2 2 \theta_{13}=0.04$; the curves are for several choices of the
CP phase and the left and right hand side plots are for the two
different mass orderings.

In Table \ref{rate1} we have calculated the rate of electron
appearance events for various scenarios by integrating the spectrum
together with the appearance probability.  This event rate is for all
charged current events; no detector efficiency factors are applied.
A detector with efficient fiducial mass of 100 kTon is assumed with 
$\sim 10^7$ sec of running time with 1 MW of proton beam. 
No consideration for backgrounds, energy thresholds, or resolution
effects are in this table.

Also note  that Figures \ref{f12km} to \ref{fwble2500} 
 do not show the event rates from anti-neutrinos. These can be obtained from
the study web-site \cite{website}. 
We include anti-neutrino rates and spectra in later sections with more detail.  
We have included  anti-neutrino event rate in Table \ref{rate1}. 

\begin{figure}
\centering\leavevmode
\mbox{\includegraphics[angle=0,width=0.48\textwidth]{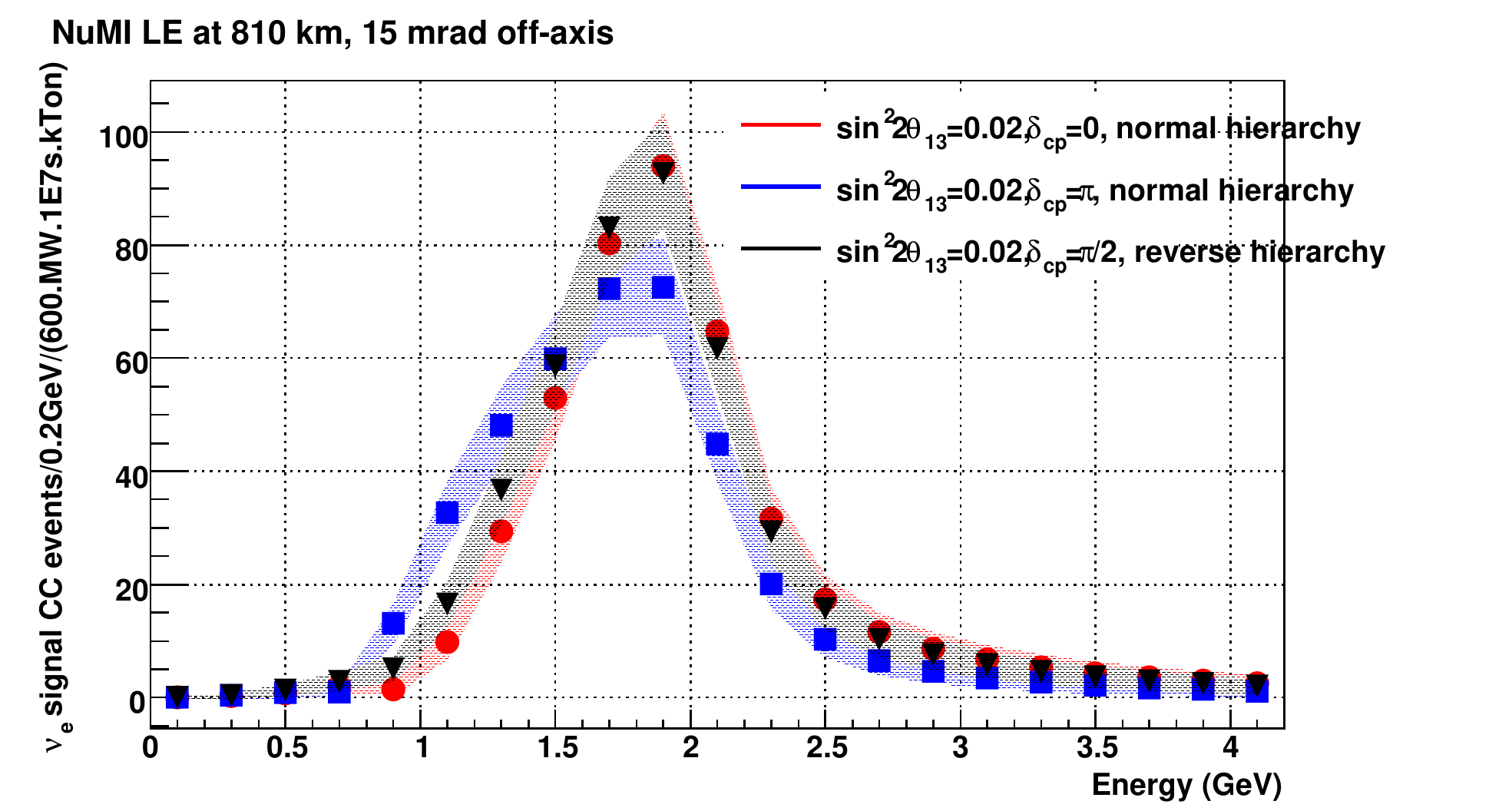}
\includegraphics[angle=0,width=0.48\textwidth]{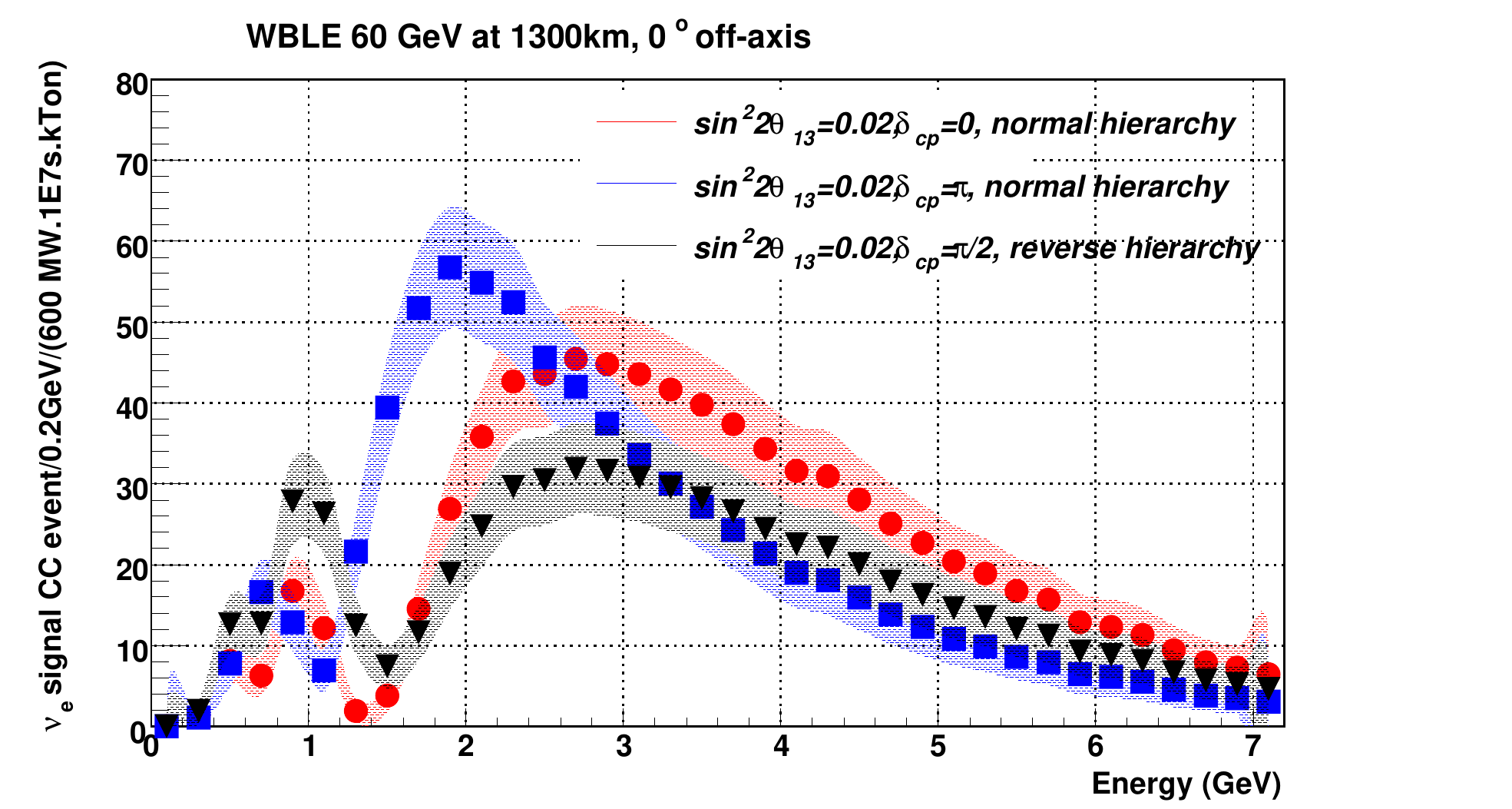}}
 \caption{(in color) 
Spectrum of charged current $\nu_\mu\to \nu_e$ events 
using the LE beam tune at 12 km off-axis 810 km location (left)
and with a new wide band beam from FNAL (using 60 GeV protons)
 to a location at 1300 km. 
The spectra are 
 normalized for  $600 MW\times 10^7 sec$ and the width of the band indicates the 
statistical error.  
The parameters used for oscillations are shown in the figure, the 
remaining parameters are as  described in the text. 
Figure includes no detector effects 
such as efficiencies, resolution, or backgrounds.  
  \label{resolvecp} }
\end{figure}

After considering the figures
 and the table we make the following observations:

\begin{itemize} 

\item  
For simplicity we look at the electron neutrino event rate at
$\delta_{CP} =0$ and compare it to $\delta_{CP} = -90^o$. In the limit
that one has resolved the mass hierarchy using the anti-neutrino data,
the modulation of the neutrino rate with $\delta_{CP}$ will give us
the CP parameter measurement that we seek. One can immediately see
that the size of the CP effect for the maximum CP ($-90^o$) is
approximately $3 \sigma$. To achieve this within a year of running
(with no consideration for efficiencies, backgrounds, etc.) the
efficient fiducial mass of the detector must be $100 kTon$ range if
the accelerator power is limited to be $\sim$1 MW. This conclusion is
regardless of the eventual choice for the beam-line.

\item The size of the CP effect (for the maximum $90^o$) increases modestly 
from $\sim 3 \sigma$ for the off axis (810 km) 12 km option to about
$4.8 \sigma$ for 2500 km.  Much of this increase can be traced to the
large CP effect at the higher oscillation nodes that become available
for the larger distances. The loss of statistics due to distance (as
$1/L^2$) is largely compensated by the increase in the strength of the
CP related signal \cite{marciano, previous, wpapers}.  By combining the 40 km
off-axis rates with the 12 km there is also a modest improvement in
the overall CP measurement. Nevertheless, for the choice of spectra in
this report, the baseline length related effects for a CP measurement
are not dramatic for the range of choices in this study.

\item   Remarkably, it
 should also be noticed that the size of the CP effect in 
the number of sigma is approximately the same for 
the different values of $\sin^2 2 \theta_{13}$. 
It has been pointed out, therefore, that for $\sin^2
2 \theta_{13} \ge 0.003$, which is the range accessible for
conventional accelerator beams, the size of the exposure (efficient
fiducial mass multiplied by the total incident beam power) needed to
obtain a good measurement of the CP parameter is independent of
$\theta_{13}$ \cite{marciano, parke, hql04}. 
This is explained by the following argument.  The asymmetry defined by
$$A\equiv \frac{P(\nu_\mu\to\nu_e) -P(\bar\nu_\mu\to\bar\nu_e)}{P
(\nu_\mu\to\nu_e) + P(\bar\nu_\mu\to\bar\nu_e)} $$
is proportional to $J_{CP}^{leptonic}$ and therefore grows linearly with 
$\sin \theta_{13}$, but $P(\nu_\mu\to\nu_e)$ is to leading order
proportional to $\sin^2 2 \theta_{13}$ and therefore the statistical 
figure of merit, the error on the asymmetry $A$ should have little
 dependence on  $\theta_{13}$. 

\item 
The size of the matter effect (the difference between the event rate for 
the two choices of mass ordering) is approximately $3\sigma$ for 
the 12 km off axis location for $\sin^2 2 \theta_{13} =0.02$ for 
neutrino rates alone.  
It is a much larger effect for longer baselines. 
 The probability curves show that  
the effect is large for the first oscillation node in all cases.
This effect will clearly compete with 
the CP effect and must be determined along with the CP effect for clarity. 
The matter effect clearly is much stronger for  larger value of 
$\theta_{13}$, and therefore for a larger value
 of $\theta_{13}$, it will be 
easier to determine the mass hierarchy.

\item 
Examination of the probability curves in Figure \ref{f12km} shows that the 
12 km off axis spectrum is sensitive mainly to the first oscillation node. 
The probability is affected not only by the CP phase, but also by 
the value of $\theta_{13}$, the mass ordering, the uncertain values of 
other parameters such as $\Delta m^2_{32}$ and $\theta_{23}$.  Also note that 
the probability curves at any particular energy have degeneracies in the
CP phase.  These degeneracies have been discussed in the 
literature \cite{barger,minakata, bmw}. Therefore, to make a clean determination of 
CP violation, one either needs very good energy resolution (to exploit the 
small energy dependence within the first node) with good statistics, or 
one needs to perform another measurement at the high oscillation node by placing 
another detector further off-axis. This is one of the options to be examined in 
this report.

\item 
Examination of the probability curves in Figures \ref{fwble1300} and \ref{fwble2500} 
shows that the energy dependence of the probability can be measured in a 
single detector  by creating a beam spectrum that matches the 
first few nodes over the $>1000$ km long baseline.    
Obviously, in such a scenario the neutrino energy must be
measured in the detector with sufficient resolution 
while suppressing backgrounds \cite{previous}.  
This is also an option to be considered in this report.  
An illustration of how the various degeneracies affect the measurement is shown in 
Figure \ref{resolvecp}. The figure illustrates the energy dependence for neutrino 
running only. It is clear that narrow band running will have additional ambiguities. 
How these can be broken 
 with additional anti-neutrino running
or with high statistics and resolution will be discussed later.

\item 
The neutrino event rate is roughly proportional 
to the total  proton beam power; the exact numbers and deviations from this 
rule will be discussed below. The total power that can be obtained from 
FNAL Main Injector after upgrades increases with the output 
 proton energy, and therefore it is important to maintain 
the highest possible proton energy for either the off-axis or 
on-axis  scenarios.  For the off-axis experiment
 the preferred running is at 
the highest, 120 GeV, proton energy. For the FNAL-to-DUSEL option, there 
could be significant advantage at running with lower proton energy. This will 
reduce the long high energy tail $>5 GeV$  of the neutrino spectrum. This tail is 
outside the  interesting  oscillation region and may contribute increased 
background in the form of neutral current events that reconstruct to have 
lower neutrino energy. The event rates given in table \ref{rate1} for WBLE 
assume running with 1 MW of power at  60 GeV. In the following we will comment on 
how  1 MW power can be obtain while maintaining the a flux with low 
high energy neutrino tail. The easiest way, of course, is by having a small off-axis 
angle. The flux that could be obtained with a $0.5^o$ off-axis 
angle to DUSEL at 1300 km is shown in Figure \ref{fwblehalf}.

\end{itemize}

\begin{figure}
\centering\leavevmode
\mbox{\includegraphics[angle=0,width=0.49\textwidth]{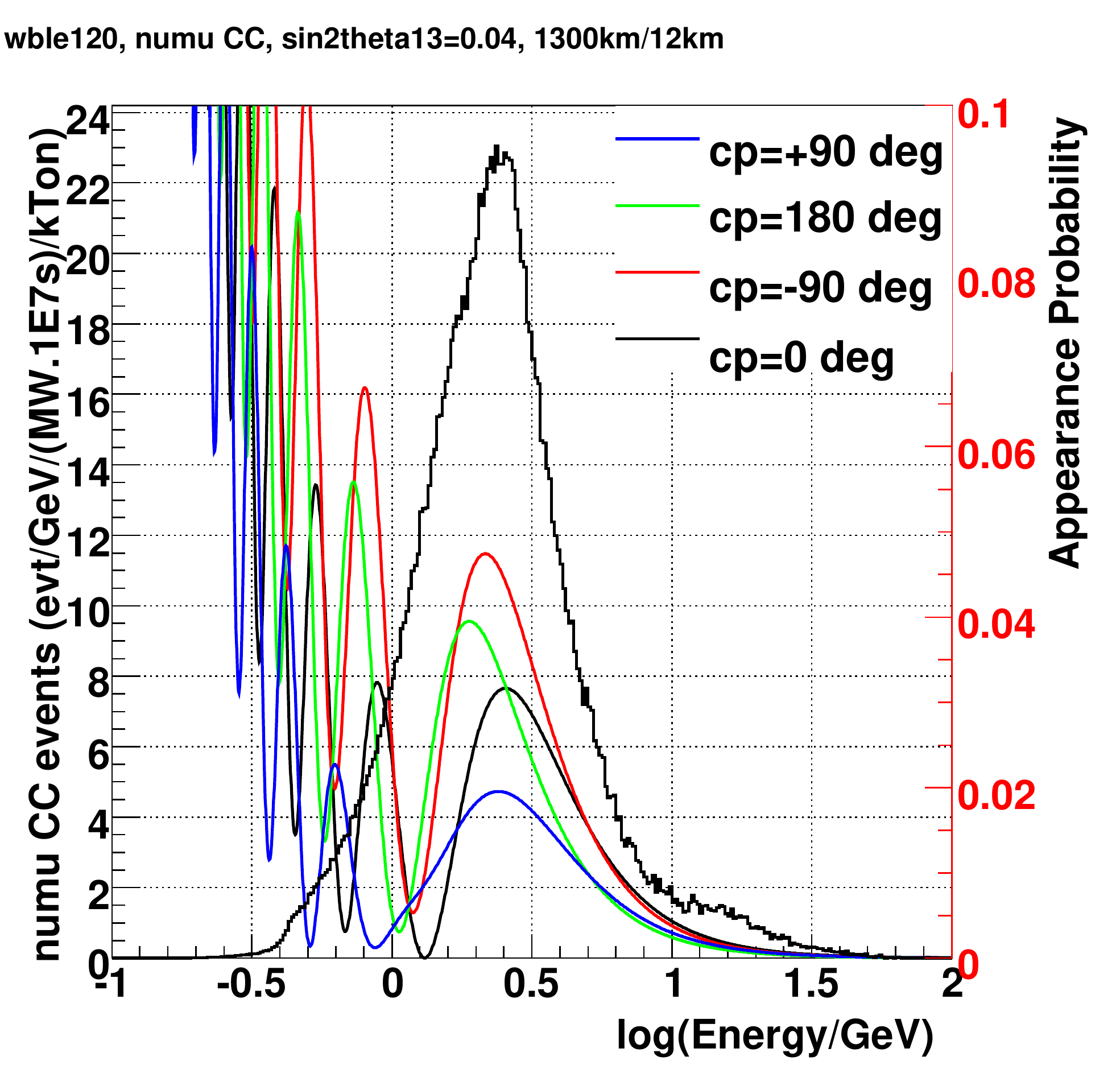}
\includegraphics[angle=0,width=0.48\textwidth]{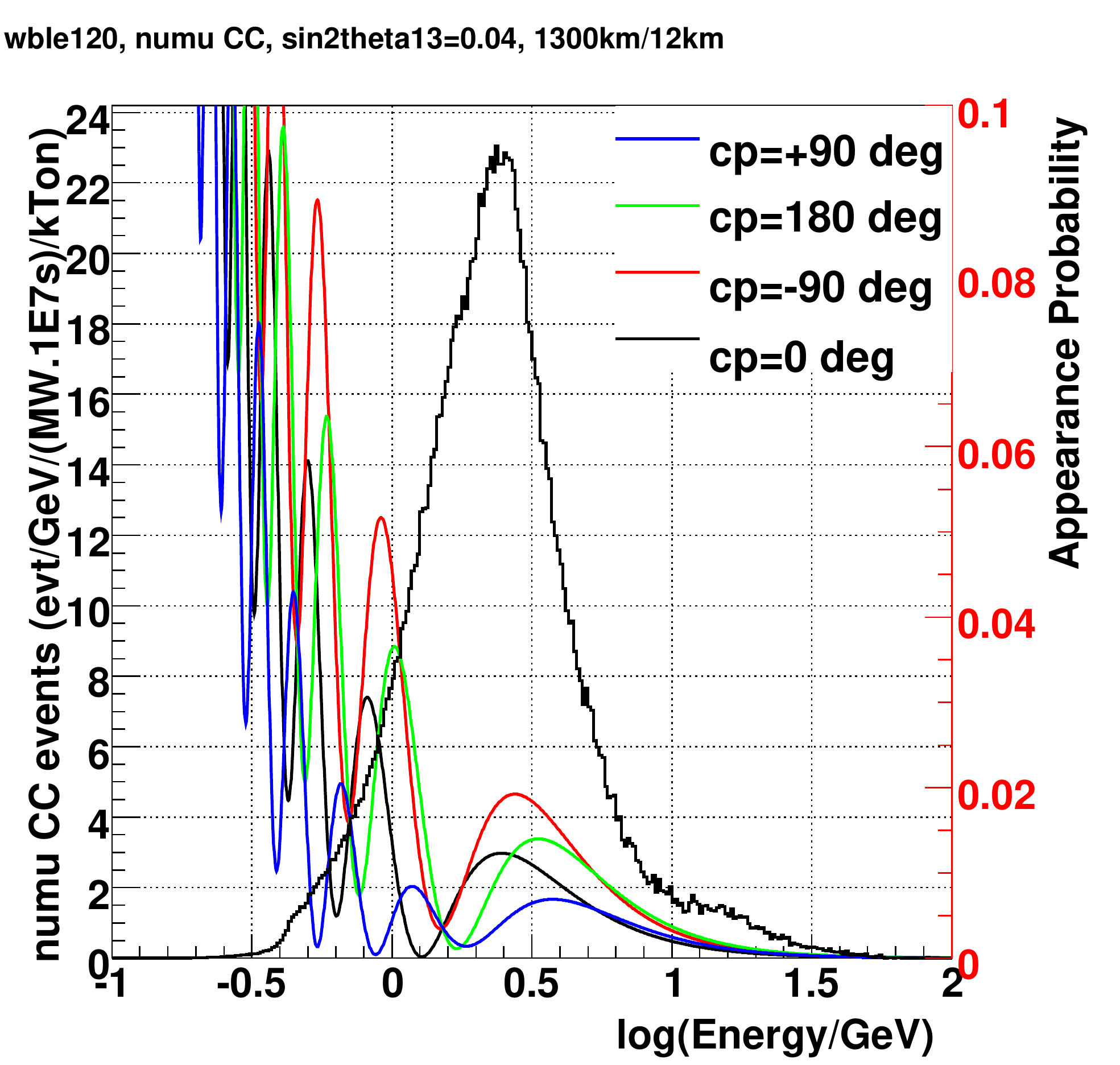}}
 \caption{(in color) 
Spectrum of charged current $\nu_\mu$ events using a new
 wide band beam from FNAL to location at 1300 km with slightly off axis location (12km)
to reduce the high energy tail.  The spectrum is
 normalized per GeV per $MW\times 10^7 sec$ protons of 120 GeV. Overlayed is the
 probability of $\nu_\mu \to \nu_e$ conversion for $\sin^2 2
 \theta_{13} = 0.04$ with rest of the oscillation parameters as
 described in the text. The left plot is for regular mass ordering and
 right hand side is for reversed mass ordering.  
Figure includes no detector effects 
such as efficiencies, resolution, or backgrounds.  
  \label{fwblehalf} }
\end{figure}

\begin{table}
\label{tab_compare_5}
\begin{tabular}{|l|c|c|c|c|c||c|c|c|c|}
\hline
\multicolumn{2}{|c}{}
& \multicolumn{4}{|c|}{Neutrino Rates} & \multicolumn{4}{||c|}{Anti
  Neutrino Rates} \\
\hline
Beam (mass ordering) & $\sin^2 2 \theta_{13}$  &
\multicolumn{8}{c|}{$\delta_{CP}$ deg. } \\ \cline{3-10}
 &      & ~~0$^\circ$ & -90$^\circ$ & 180$^\circ$ & +90$^\circ$ & ~~0$^\circ$ & -90$^\circ$ & 180$^\circ$ & +90$^\circ$ \\
\hline
NuMI LE 12 km offaxs (+) & 0.02 & 76  & 108  & 69  & 36  & 20 & 7.7 & 17 & 30\\
NuMI LE 12 km offaxs (-) & 0.02 & 46  & 77  & 52  & 21   & 28 & 14 & 28 & 42\\
\hline
NuMI LE 12 km offaxs (+) & 0.1 & 336  & 408  & 320  & 248  & 86 & 57 & 78 & 106\\
NuMI LE 12 km offaxs (-) & 0.1 & 210  & 280  & 224 & 153  & 125 & 95 & 126 & 157\\
\hline
\hline
NuMI LE 40 km offaxs (+) & 0.02 & 5.7  & 8.8  & 5.1  & 2.2  & 2.5 & 1.6 & 0.7 & 3.3\\
NuMI LE 40 km offaxs (-) & 0.02 & 4.2  & 8.0 & 5.7  &  2.0   & 2.3 & 2.2 & 0.8 & 3.6\\
\hline
NuMI LE 40 km offaxs (+) & 0.1  & 17 & 24  & 15  & 9.4   & 6.7 & 2.8 & 4.6 & 8.5\\
NuMI LE 40 km offaxs (-) &  0.1  & 12  & 21  & 16 & 7.7   & 6.6 & 3.4 & 6.4 & 9.6\\
\hline
\hline
WBLE 1300 km (+) & 0.02  & 141  & 192  & 128  & 77 & 19 & 11 & 18 & 36 \\
WBLE 1300 km (-) & 0.02  & 58  & 111  & 88 & 35 & 45 & 25 & 45 & 64 \\
\hline
WBLE 1300 km (+) & 0.1  & 607   & 720  & 579   & 467  & 106 & 67 & 83 & 122\\
WBLE 1300 km (-) & 0.1  & 269  & 388  & 335  & 216 & 196 & 154 & 196 & 240\\
\hline
\hline
WBLE 2500 km (+) &0.02 & 61 & 103  & 88  & 46 & 11 & 4.6 & 4.7 & 11\\
WBLE 2500 km (-) &0.02 & 16 & 36  & 33  & 13 & 28 & 15 & 18 & 31 \\
\hline
WBLE 2500 km (+) &0.1 & 270  & 361  & 328  & 238 & 27 & 13 & 13 & 28  \\
WBLE 2500 km (-) &0.1 & 47  & 92 & 85  &  39  & 103 & 74 & 80 & 109 \\
\hline
\end{tabular}
\caption{
This table contains signal
event rates   after
$\nu_\mu \to \nu_e$ (also for anti-neutrinos) conversion
for the various scenarios described.
The event rates here have no detector model or backgrounds.
The units are charged current events per 100 kTon of detector
mass for 1 MW of beam for $10^7 sec$ of operation.
For NuMI running we assume 120 GeV protons in the LE tune and for WBLE
we have assumed 60 GeV protons.
The charged current cross sections applied as well as the oscillation
parameters used  are described in the text.
}
\label{rate1}
\end{table}

We will now explore the above observations in further detail including 
the feasibility of beams and detectors, current best knowledge on the
performance of detectors, and requirements for 
other physics related applications of 
these very large detector facilities.

\section{Accelerator Requirements}
\label{accsec} 

All phases of the envisioned US neutrino accelerator 
program, Phase-I(NO$\nu$A), Phase-II(option A), or Phase-II(option B),
{\it require} upgrades to the existing proton accelerator 
infrastructure in the US.  Phase-I upgrades, already planned at FNAL, 
will  
increase  the Main Injector extracted beam power to 
0.7 MW  at  120 GeV (this is  called ``proton plan-2'' and has been 
incorporated in the NO$\nu$A project).
The plan to further upgrade the Main Injector to 1.2 MW is 
called ``the SNuMI Project'' \cite{snumicdr}.    
Phase-II will benefit from these upgrades.

We have used beam power in the range of $\sim$0.5 to 2 MW for high
energy protons ($>$30 GeV) in our calculations 
because this level of beam power is now
considered the next frontier for current accelerator technology
\cite{acc1,mcginnis,foster,marchionni,trgt,acc2} and also necessary to obtain sufficient event rate
to perform the next stage of neutrino oscillation physics.  The
technical limitations arise from the need to control radiation losses,
limit the radiation exposure of ground water and other materials, and
the feasibility of constructing a target and horn system that can
survive the mechanical and radiation damage due to high intensity
proton pulses \cite{trgt}.

We quote event rates either in units of $MW\times 10^{7} sec$ or
number of protons on target (POT). A convenient formula for conversion
is below.  $$POT (10^{20}) = {1000\times Beam Power (MW) \times T
(10^7 s) \over 1.602 \times E_p (GeV)}$$ where $T$ is the amount of
exposure time in units of $10^7 s$ and $E_p$ is the proton energy. We
now briefly summarize the understanding of high energy proton beam
power at the two US accelerator laboratories where high intensity
proton synchrotrons are operational, Fermilab and Brookhaven.

{\bf FNAL Main injector (MI):} Discussion is currently 
underway to increase the total power
from the 120~GeV  Main Injector (MI) complex after the Tevatron program
ends~\cite{snumicdr, acc1}. In this scheme protons from the 8~GeV booster,
operating at 15~Hz, will be stored in the recycler
(which becomes available after the
shutdown of the Tevatron program)  while the MI
completes its acceleration cycle, which is shortened from 
the current 2.2 sec to 1.33 sec. In a further upgrade 
 the techniques of
momentum stacking
using the antiproton accumulator, and slip-stacking using the recycler 
will raise the total intensity in the MI to
$\sim 1.2$~MW at 120~GeV \cite{mcginnis}. In the rest of this report 
this will be called {\it the SNuMI plan}.
In the ideal case,  the length of the acceleration cycle is proportional
to the proton energy, making the average beam power proportional to
the final proton energy.
However, fixed time intervals in the beginning and the
end of the acceleration cycle are required for stable operation. These become
important at low
energies and reduce the performance below the ideal. Current
projections suggest that
$\sim 0.5$ MW operation between $40-60$~GeV and $\gsim 1$~MW operation at
120~GeV is possible.

More ambitious plans at FNAL call for
replacing the 8 GeV booster with a new 
 super-conducting LINAC
that can provide $1.5\times 10^{14}$ $H^-$ ions at 10~Hz corresponding
to 2~MW of total beam power~\cite{foster}. Some of the 8~GeV ions
could be injected into the MI to provide high proton beam power at any
energy between 30 and 120 GeV; {\it e.g.}, 40~GeV at $\sim 2$~Hz or
120~GeV at $\sim0.67$~Hz.  Such a plan allows for flexibility in the
choice of proton energy for neutrino production.  This plan will be
called the {\it high intensity neutrino source} upgrade (HINS).

\begin{figure}
\centering\leavevmode
\includegraphics[angle=0,width=0.9\textwidth]{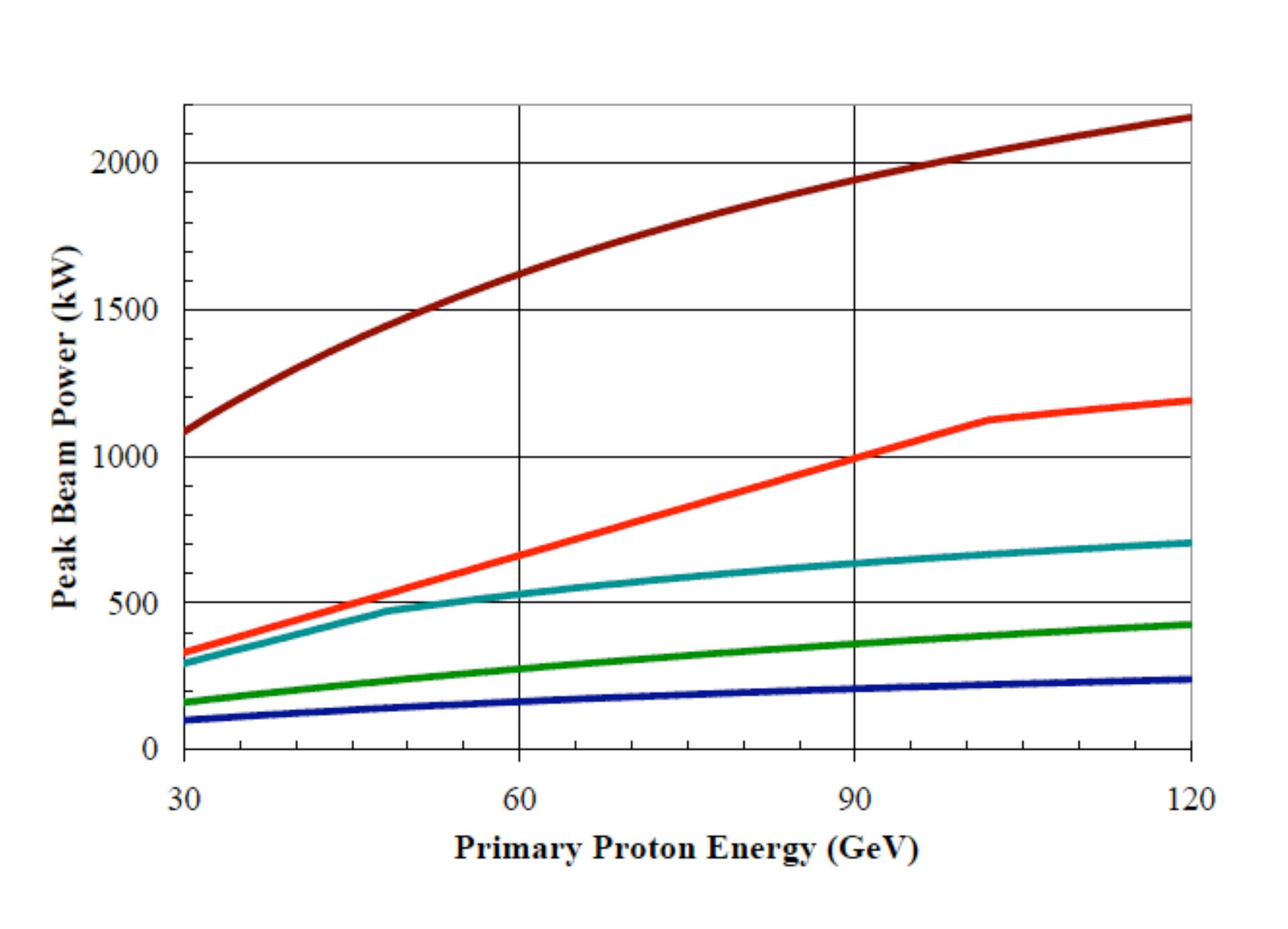}
 \caption{Proton beam power from the Fermilab main injector
 as a function proton energy for various 
scenarios.  Lowest (blue) curve is for the current complex running 
concurrently with the Tevatron. Second (green) curve is for 
the proton plan upgrades, third (light blue) curve is for SNuMI 
recycler stage  which will take place after the 
termination of the Tevatron program, fourth (red) is for the accumulator
 stage upgrades,
the uppermost (brown) is for the HINS upgrade which calls for a new 8 GeV 
LINAC injector.  
   \label{fnalpow} }
\end{figure}

The projected proton intensity from the main injector for the
successive upgrades at FNAL is shown in Figure \ref{fnalpow}\cite{marchionni}.
 A reviewed  cost estimate 
that has been included in the NO$\nu$A project 
for the 700 MW (proton plan-2) 
upgrade is \$60M. 
The cost of the complete SNuMI plan (to 1.2 MW)   
is at the moment very preliminary at $\sim \$54M$   
(without overhead or contingency factors). 
The HINS upgrade is estimated to be approximately  $>$\$300M.

{\bf BNL AGS:} The BNL Alternating
Gradient Synchrotron
(AGS)  operating at 28~GeV
currently can provide about 1/6~MW of beam power. This corresponds to
an intensity of about $7\times 10^{13}$ protons in a 2.5 microsecond
pulse every 2~seconds. The AGS complex can be upgraded
 to provide a total  proton beam
power of 1~MW~\cite{acc2}.
The main components of the accelerator upgrade at BNL are
a new 1.2~GeV super-conducting LINAC to provide protons to the
existing AGS, and new magnet power supplies to increase the ramp rate of the
 AGS magnetic field from about 0.5 Hz to  2.5 Hz.
For 1 MW operation  28 GeV  protons from the accelerator will be delivered
in pulses of $9\times 10^{13}$ protons at 2.5 Hz.
It has been determined that 2 MW operation of the AGS is also possible by
further upgrading the synchrotron to 5 Hz repetition rate and with
further modifications to the LINAC and the RF systems.  The AGS 1 MW upgrade 
is estimated to cost \$343M (TEC) including contingency and overhead costs. 
This cost has been reviewed internally at BNL.

\section{Target and horn development}

All phases of the envisioned US neutrino accelerator 
program, Phase-I (NO$\nu$A), Phase-II(option A), or Phase-II(option B),
{\it require} substantial development for a  new target
capable of operating at high proton intensities 
and perhaps new  focusing horn optics. 

Current understanding of targets, and R\&D in progress is summarized
in \cite{trgt, hylentrgt}. The neutrino event rate is approximately
proportional to the total proton beam power (energy times current)
incident on the target.  The parameters for target design to be
considered for a given power level are proton energy, pulse duration,
and repetition rate.  In addition to these the shape and size of the
beam spot on the target, and the angle of incidence could also be
varied. Studies over the last few years have come to the acceptance
that with optimal choice of the above variables the upper limit for a
solid target operation is $\sim$ 2 MW. For a given accelerator
facility these parameters tend to be correlated and constrained, and
therefore a practical limit for a solid target with current technology
is probably between 1 and 2 MW.  Nevertheless, considerable work is
needed to achieve a practical design for such a high power solid
target and its integration into a focusing horn system.  Above 2 MW,
liquid targets are likely the better choice, but these devices will
require considerable R\&D and testing before they can be considered
practical.

Target R\&D which includes understanding of materials as well as
engineering issues of integration is a critical item for the physics
program considered in this report.

\section{Neutrino beam-lines}

There is currently good experience in building and operating high
intensity neutrino beam-lines in the US. 
The study group has concluded that  it is possible to use  an existing
 or  build a 
new super neutrino beam-line based on
current technology or extensions of current technology 
and operate it for the physics program described in this report.

In the following we summarize the status of US high energy accelerator
neutrino beam-lines.
There are two additional accelerator neutrino beam-lines in the world
with comparable technical requirements: the CERN to Gran Sasso neutrino beam 
which is now operating, 
and the JPARC to Super-Kamiokande neutrino beam which will 
start operation in a few years.  
We will not report on these in this report, but  this study 
included presentations from these facilities. It is clear that 
there is plenty of communication and 
 shared technical information between these centers and the US.

\subsection{NuMI}  
\label{numisec} 

The design and operation of the NuMI beam-line was reported in 
\cite{hylentrgt}. In the NuMI beam 120 GeV protons from the Main Injector, 
in a single turn extraction of $\sim 10 \mu s$ duration every $\sim 2
sec$, are targeted onto a $ 94 cm$ long graphite
target. A conventional 2 horn system is used to charge select and focus
the meson beam into a 675 m long, 2 meter diameter evacuated decay
tunnel. The NuMI beam-line is built starting at a depth of $\sim 50 m$
and is aimed at a downwards angle of 3.3 deg towards the MINOS
detector in Minnesota at a distance of 735 km from the production
target.  The flux of the resulting neutrino beam is well known and
will be described in a separate section below.

NuMI beam transport, target, horns, and shielding were designed for
operation with $4\times 10^{13} protons/pulse$ with a beam power of
400 kW. The goal is to average $3.7\times 10^{20} protons/year$. The
first year run of NuMI achieved typical beam intensities of $2.5\times
10^{13} protons/pulse$ or 200 kW. The total integrated exposure was
$1.4\times 10^{20}$ protons on target for the period from March 2005
to March 2006.  A number of technical problems were encountered and
solved during this time: at the start of the run the cooling water
line to the target failed, one of the horns had a ground fault, and
most notably a detailed study of the tritium production from the
beam-line had to be carried out. Various monitoring systems as well
systems to collect tritiated water were installed to eliminate the
amount of tritium going into cooling water and the environment.  The
experience gained from NuMI operations is indeed invaluable for future
operation of neutrino beams.

NuMI beam-line was built at a total cost of \$109M (TEC).  The
construction time was approximately $\sim$5 years. The beam-line became
operational in March of 2005.  Upgrade and operation of the NuMI
beam-line for higher intensity for Phase-I is included  in the new
SNuMI conceptual design report  at
FNAL\cite{snumicdr, marchionni}.  It is anticipated that for operation at 1 MW,
the primary proton beam-line, the target and horns, and cooling
systems in the target hall will require upgrades. New He bags and
upgrades to the high radiation work areas will also be installed. The
total preliminary cost of this upgrade ($\sim 10M$)  is 
included in the cost of the proton plan-2 upgrade described in 
Section \ref{accsec}.

\subsection{Beam towards DUSEL} 
\label{duselbeam} 

Members of this study group \cite{marchionni} 
have examined the possible siting and 
construction of a new beam-line towards one of the site candidates for 
DUSEL, either Henderson mine in Colorado or Homestake mine in South Dakota.  
The study group has concluded that there are no technical limitations 
to building such a beam-line on the Fermilab site using the same extraction 
line from the main injector as the NuMI beam-line.  The study group has found 
significant advantage for lower energy neutrino flux 
in making the diameter of the decay tunnel for the new 
beam-line up to 4 meters.

The new beam-line at FNAL would use the same extraction from the 
Main Injector into the NuMI line; a new tunnel would pick up the
proton beam from the present tunnel and transport it in the western
direction with the same radius of curvature as the Main Injector so
that up to 120 GeV protons can be used with conventional magnets.
There is adequate space on the Fermilab site to allow a new target
hall with 45 m length and a decay tunnel of length 400 meter and
diameter of 4 meters. This will allow the location of a near detector
with $\sim 300$ meters of length from the end of the decay pipe.  The
new decay pipe would point downwards at an angle of 5.84$^o$ to
Homestake (1289km from FNAL) or 6.66$^o$  to Henderson (1495 km from
FNAL). The diameter of the decay tunnel is a crucial parameter for
both  the neutrino beam intensity and the cost and feasibility of
the beam-line; it will require detailed optimization.  With our present
understanding, construction of a 4 meter diameter decay tunnel with
adequate shielding for eventually 2 MW of operation is possible. 
If the additional concrete shielding is found to be inadequate then
the decay pipe would have to be reduced to 3 meter diameter because of the 
maximum possible span of excavation in the rock under FNAL\cite{bogert}. 
The thickness of the shielding has been scaled from the NuMI experience, but the 
implications of the wider diameter for radiation issues (in particular, 
tritium production) will need careful study. 
After  optimization, the cost of such a project can be reliably
estimated from the known cost of the NuMI project.

A new beam-line from BNL-AGS to either Homestake (2540 km) or Henderson
(2770 km) has also been examined in a BNL report\cite{acc2}. They have
made the choice of building the beam-line on a specially constructed
hill where the shielded target station is located on top of the hill
and the meson decay tunnel is on the downward slope of the hill
pointing towards DUSEL at an angle of 11.7$^o$ (Homestake) or 13.0$^o$ 
(Henderson). Due to the limitations on the height of the hill,
the decay tunnel length is restricted to be $\sim$200 meters with a diameter
of 4 meters. The cost of such a beam-line including construction of the hill
and  proton transport to the top of the 
hill was estimated to be  \$64M (TEC) including contingency and overhead; 
this cost has been  reviewed internally at BNL. Further work on this option
has not been part of this study.

\section{Event rate calculations}
\label{erc} 

The neutrino flux and the numbers of expected events with and without
oscillations were calculated for both the NuMI off-axis beam and a new
broadband beam towards DUSEL. This calculation assumes no detector
resolution model or background rejection capability. Both calculations
were performed using the same GEANT based GNuMI code.  This code has
been extensively tested as part of the MINOS collaboration. It has
been verified against recent data in the MINOS near detector.  The
code and associated cross section model is known to produce agreement
with the MINOS near detector event rate per proton to about 10\% at
the peak of the spectrum and of the order of 20-30\% in the tails of
the spectrum with no adjustments.  
We have also calculated anti-neutrino event rates. The accuracy here is 
worse simply because of the lack of data from the NuMI beam-line. 
The anti-neutrino spectra have disagreements between various production 
codes of $\sim 30\%$. 
We believe this is sufficient
accuracy for the purposes of this study.

It is very likely that neither the specific  off-axis configuration nor the 
broad-band configuration is highly optimized for the physics under 
consideration. Such optimization could result in modest gains, especially 
at low energies. At this stage there is good confidence that 
the possible improvements will not change the overall picture and sensitivity
outlined in this report.  

\subsection{NuMI off-axis locations } 

We have calculated the neutrino flux and event rates at various
off-axis distances from the NuMI beam-line.  NuMI was assumed to be
configured in the medium energy (ME)  or low energy (LE) 
beam configuration for the results 
quoted here.
The low energy configuration provides 
 better event rate at the 40 km off-axis  location in the low energy peak. 
There is, however, 
event rate loss at the 12 km   location.  

The details
of the calculation, as well as the spectra
are in \cite{offaxs}. Tables \ref{tab:rates1} for neutrino running and 
\ref{tab:arates1} for antineutrino running 
summarizes these event rates. The normalization is per $MW\times 10^7 sec$
protons of 120 GeV and for 1 kTon of efficient detector mass. There are no
corrections for the type of target nucleus in the detector. There are
no efficiencies for reconstruction or fiducial cuts in this
calculation.


We have used tabulated cross sections to calculate the 
event rates in the various columns. The column labeled 
``$\nu_\mu$ CC'' is the total charged current muon neutrino 
event rate. ``$\nu_\mu$ CC osc'' is the charged current muon neutrino 
event rate after oscillations. ``$\nu_e$ CC beam'' is the charged current rate 
of electron neutrino contamination in the beam.  ``$\nu_e$ QE beam'' is 
the charged current quasi-elastic event  rate 
of electron neutrino contamination in the beam.  ``NC-1$\pi^0$'' is the 
rate of neutral current single pion production integrated over the 
noted energy range; no detector related rejection is assumed in this 
table. ``$\nu_\mu \to \nu_e$ CC'' is the charged current event rate of 
electron neutrinos after oscillations using the oscillations parameters 
described in Section \ref{sfp2}. ``$\nu_\mu \to \nu_e$ QE'' is the 
quasi-elastic  rate of 
electron neutrinos after oscillations using the oscillations parameters 
described in Section \ref{sfp2}.
For example, the total
$\nu_\mu$ CC event rate in 5 years with $1.7\times 10^7$ sec/yr 
 in a 100 kton detector without oscillations at
40 km (LE) off axis can be calculated to be $5.38\times 100\times 5\times 1.7= 4573$.
 This event count includes events from both the pion and the kaon peaks at
about 0.5 and 4 GeV, respectively.

\begin{table}[!h]
\caption{Signal and background interaction rates for various NuMI beam
  configurations, baselines and off-axis distances. Rates are given
  per MW.10$^7$s.kT. The rates are integrated over the range
  0-20 GeV. For $\nu_{\mu} \rightarrow \nu_e$ oscillations a value of
$\sin^2 2\theta_{13} =0.04$ and $\Delta m^2_{31} = 2.5 \times
  10^{-3} \ {\rm eV}^2$ is used. No detector model is used.}
\begin{center}
\label{tab:rates1}
   \begin{tabular}{|r|r@{.}l|r@{.}l|r@{.}l|r@{.}l|r@{.}l|r@{.}l|r@{.}l|}
     \hline
     Distance off-axis & \multicolumn{2}{c|}{$\nu_{\mu}$ CC} &
\multicolumn{2}{c|}{$\nu_{\mu}$ CC osc} &
  \multicolumn{2}{c|}{$\nu_e$ CC beam} & \multicolumn{2}{c|}{$\nu_e$ QE
beam} & \multicolumn{2}{c|}{NC-$1\pi^0$} &
\multicolumn{2}{c|}{$\nu_{\mu} \rightarrow \nu_e$  CC} &
     \multicolumn{2}{c|}{$\nu_{\mu} \rightarrow \nu_e$ QE} \\
     \hline \hline
\multicolumn{15}{|c|}{NuMI LE tune at 700 km} \\ \hline \hline
     0 km       & 400&2      & 267&6          & 4&55           & 0&444
      & 21&2
      & 3&66          & 0&676 \\
     40 km     & 4&81      & 2&66          & 0&190         & 0&047
      & 0&525
      & 0&071        & 0&038 \\ \hline \hline
\multicolumn{15}{|c|}{NuMI LE tune at 810 km} \\ \hline \hline
     0 km       & 299&0      & 187&4          & 3&40           & 0&332
      & 15&8
      & 3&10          & 0&551 \\
     6 km      & 198&6       & 107&0            & 2&59          & 0&275
      & 11&9
      & 2&53         & 0&506 \\
     12 km & 84&4   & 31&9      & 1&57       & 0&193        & 6&79
      & 1&41
      & 0&367 \\
     30 km     & 11&6       & 8&38           & 0&353         & 0&070
      & 1&32
      & 0&107       & 0&046 \\
     40 km     & 5&38       & 2&91           & 0&195         & 0&045
      & 0&596
     & 0&084        & 0&045 \\
\hline \hline
\multicolumn{15}{|c|}{NuMI ME tune at 810 km} \\ \hline \hline
    0 km       & 949&1      & 781&1          & 7&14           & 0&485
      & 30&6
      & 4&71          & 0&527 \\
     6 km      & 304&9       & 191&4            & 3&83          & 0&313
      & 14&9
      & 3&19         & 0&491 \\
     12 km & 80&5   & 32&0      & 1&81       & 0&174         & 5&74
      & 1&33
      & 0&330 \\
     30 km     & 8&59       & 5&52          & 0&321         & 0&051
      & 0&81
      & 0&094        & 0&038 \\
     40 km     & 4&14      & 2&40          & 0&168         & 0&032
      & 0&427
     & 0&054        & 0&022 \\
     \hline
\end{tabular}
\end{center}

\end{table}

\begin{table}[!h]
\caption{Signal and background interaction rates for various NuMI
  anti-neutrino beam
  configurations, baselines and off-axis distances. Rates are given
  per MW.10$^7$s.kT. The rates are integrated over the range
  0-20 GeV. For $\nu_{\mu} \rightarrow \nu_e$ oscillations a value of
  $\sin^2 2\theta_{13} =0.04$ and $\Delta m^2_{31} = 2.5 \times
  10^{-3} \ {\rm eV}^2$ is used. No detector model is used.}
\begin{center}
\label{tab:arates1}
   \begin{tabular}{|r|r@{.}l|r@{.}l|r@{.}l|r@{.}l|r@{.}l|r@{.}l|r@{.}l|}
     \hline
     Distance off-axis & \multicolumn{2}{c|}{$\bar{\nu}_{\mu}$ CC} &
\multicolumn{2}{c|}{$\bar{\nu}_{\mu}$ CC osc} &
  \multicolumn{2}{c|}{$\bar{\nu}_e$ CC beam} & \multicolumn{2}{c|}{$\bar{\nu}_e$ QE
beam} & \multicolumn{2}{c|}{NC-$1\pi^0$} &
\multicolumn{2}{c|}{$\bar{\nu}_{\mu} \rightarrow \bar{\nu}_e$  CC} &
     \multicolumn{2}{c|}{$\bar{\nu}_{\mu} \rightarrow \bar{\nu}_e$ QE} \\
     \hline \hline
\multicolumn{15}{|c|}{NuMI LE tune at 700 km} \\ \hline \hline
     0 km       & 157&6      & 102&3          & 1&69           & 0&306
      & 19&3    & 1&25       & 0&306 \\
     40 km      & 1&64       & 0&905          & 0&063        & 0&021
      & 0&544   & 0&024      & 0&016 \\ \hline \hline
\multicolumn{15}{|c|}{NuMI LE tune at 810 km} \\ \hline \hline
     0 km       & 117&7      & 71&0          & 1&26           & 0&229
      & 14&4    & 1&026          & 0&285 \\
     6 km       & 77&6       & 39&8            & 0&925          & 0&179
      & 10&8    & 0&800         & 0&241 \\
     12 km & 31&7  & 10&9      & 0&545       & 0&116        & 6&29
      & 0&388      & 0&145 \\
     30 km     & 3&87       & 2&69          & 0&122         & 0&035
      & 1&31
      & 0&043       & 0&025 \\
     40 km     & 1&81       & 0&97           & 0&066         & 0&021
      & 0&609
     & 0&029        & 0&018 \\
\hline \hline
\multicolumn{15}{|c|}{NuMI ME tune at 810 km} \\ \hline \hline
    0 km       & 350&6      & 285&1          & 2&53           & 0&349
      & 23&6
      & 1&59          & 0&316 \\
     6 km      & 112&8       & 69&0            & 1&28         & 0&208
      & 11&9
      & 1&011         & 0&259 \\
     12 km & 27&7   & 9&83      & 0&601       & 0&105         & 4&76
      & 0&348
      & 0&125 \\
     30 km     & 2&66       & 1&67          & 0&109         & 0&027
      & 0&70
      & 0&027        & 0&014 \\
     40 km     & 1&27      & 0&73          & 0&057         & 0&016
      & 0&376
     & 0&015        & 0&008 \\
     \hline
\end{tabular}
\end{center}
\end{table}

\subsection{Wide band beam towards DUSEL} 
\label{wbledu}

The spectra and the event rate for a  beam towards DUSEL
 were calculated 
by using the same GNuMI framework but the geometry of the target, horns, and the decay tunnel 
was changed. The full calculation and the resulting spectra are described in 
\cite{wble}.  The integrated event rates are shown in 
Table \ref{tab:rates2} and Table \ref{tab:arates2}.   There are 
a few of comments of importance: 

\begin{itemize}

\item The calculation in the table is for 1300 km (the FNAL to Homestake distance),
 but it could be easily converted to 1500 km (the distance to Henderson). 
The unoscillated rate scales as $1/r^2$, but the 
oscillated event rate scales according to the oscillation function. 
When we demonstrate the 
full sensitivity 
calculation later in this report we include the variation with distance. For 
1300 versus 1500 km this variation is small.

\item 
Ealier work on sensitivity  used  a 
28 GeV proton beam \cite{hql04}. 
The total $\nu_\mu$ CC event rate in 100 kTon efficient fiducial mass 
 after 5 years at $1.7\times 10^7$ sec/yr  without oscillations using 
$E_p = 28$ GeV protons with 1 MW running is 44625  events integrated over 1-20 GeV.   
It should be kept in mind, however, that according to \cite{acc1}, 
the available beam power is less for lower energies (see Fig. \ref{fnalpow}).  
In the technical note \cite{wble} it has been shown that the 40-60 GeV 
spectrum could be  very similar to the 28 GeV with considerable increase in 
event rate per unit beam power.  It has also been shown that it is possible to 
run at the full energy of 120 GeV and still obtain essentially the same 
spectrum as the 28 GeV one with a small 0.5$^o$ off-axis angle. With such  a choice 
the neutrino (antineutrino) event rate is 76415 (28475)  for 100 kTon and 5 yrs 
for 1 MW and $1.7\times 10^7$ sec/yr.

\item Tables \ref{tab:rates2} and \ref{tab:arates2} represent our present 
understanding of creating such a beam. When optimization is performed coupled to 
the complete understanding detector performance versus energy, the spectrum could be 
adjusted to give the best signal/background performance. 
  This could be accomplished by optimizing the 
 horn optics and/or inserting secondary targets (plugs) that remove 
high energy pions from the beams (see \cite{mbishaisep16}).

\item  We have integrated the rates of various types of events over the same 
energy interval 0-20 GeV for Tables \ref{tab:rates1} to \ref{tab:arates2}. 
It should be understood that there is considerable variation in the signal to 
background ratio as a function of energy. To get a full appreciation of this
we recommend the reader to explore the spectra at the 
study web-site \cite{website}. The variation also depends on oscillation parameters.
In particular, it should be noted that the CP violating phase
as well as the mass hierarchy  is responsible for 
moving the peak of the oscillation probability by as much as $\sim$0.5 (0.7) GeV for 
the 810 (1300) km baseline. This variation coupled to the width of the useful 
spectrum and the detector energy resolution 
 has an impact on the parameter sensitivity of the program.    

\end{itemize}

\begin{table}[!h]
\caption{Signal and background interaction rates at 1300 Km
  (Fermilab-HOMESTAKE) using different WBLE beam energies and off-axis
  angles. The rates integrated over the neutrino energy range of 0 -
  20 GeV. Rates are given per MW.10$^7$s.kT. For $\nu_{\mu}
  \rightarrow \nu_e$ oscillations a value of $\sin^2 2\theta_{13}
  =0.04$ and $\Delta m^2_{31} = 2.5 \times 10^{-3} \ {\rm eV}^2$ is
  used.  No detector model is used.}
\begin{center}
\label{tab:rates2}
   \begin{tabular}{|r|r@{.}l|r@{.}l|r@{.}l|r@{.}l|r@{.}l|r@{.}l|r@{.}l|}
     \hline
     Degrees off-axis & \multicolumn{2}{c|}{$\nu_{\mu}$ CC} &
\multicolumn{2}{c|}{$\nu_{\mu}$ CC osc} &
  \multicolumn{2}{c|}{$\nu_e$ CC beam} & \multicolumn{2}{c|}{$\nu_e$ QE
beam} & \multicolumn{2}{c|}{NC-$1\pi^0$} &
\multicolumn{2}{c|}{$\nu_{\mu} \rightarrow \nu_e$  CC} &
     \multicolumn{2}{c|}{$\nu_{\mu} \rightarrow \nu_e$ QE} \\
\hline \hline
\multicolumn{15}{|c|}{WBLE 120 GeV at 1300 km with decay pipe 2m radius 380 m length} \\ \hline \hline
     0$^{\circ}$     & 198&2      & 104&9          & 1&89           & 0&179
      & 9&11
      & 2&85          & 0&408 \\
     0.5$^{\circ}$     & 89&9      & 37&9          & 1&22           & 0&140
      & 5&62
      & 1&62          & 0&300 \\
     1.0$^{\circ}$     & 34&2      & 19&5          & 0&621           & 0&095
      & 2&95
      & 0&470          & 0&129 \\
     2.5$^{\circ}$     & 4&66      & 2&36          & 0&116           & 0&032
      & 0&550
      & 0&094          & 0&049 \\
\hline \hline
\multicolumn{15}{|c|}{WBLE 60 GeV at 1300 km with decay pipe 2m radius 380 m length} \\ \hline \hline
     0$^{\circ}$      & 151&0       & 69&2           & 1&34          & 0&169
      & 7&83
      & 2&53         & 0&403 \\
     0.5$^{\circ}$      & 77&2       & 28&7           & 0&906          & 0&134
      & 5&33
      & 1&52         & 0&305 \\
     1.0$^{\circ}$      & 33&3       & 18&4           & 0&520          & 0&098
      & 3&08
      & 0&480         & 0&141 \\
     2.5$^{\circ}$      & 5&05       & 2&56           & 0&120          & 0&035
      & 0&611
      & 0&105         & 0&058 \\
\hline \hline
\multicolumn{15}{|c|}{WBLE 40 GeV at 1300 km with decay pipe 2m radius 380 m length} \\ \hline \hline
     0$^{\circ}$     & 110&4       & 44&4           & 1&02          & 0&159
      & 6&50
      & 2&05         & 0&357 \\ \hline \hline
\multicolumn{15}{|c|}{WBLE 28 GeV at 1300 km with decay pipe 2m
     radius 180 m length} \\ \hline \hline
     0$^{\circ}$     & 52&5       & 19&4           & 0&374          & 0&074
      & 3&87
      & 1&05         & 0&223 \\ \hline \hline
\end{tabular}
\end{center}
\end{table}

\begin{table}[!h]
\caption{Signal and background anti-neutrino interaction rates at 1300 Km
  (Fermilab-HOMESTAKE) using different WBLE beam energies and off-axis
  angles. The rates integrated over the neutrino energy range of 0 -
  20 GeV. Rates are given per MW.10$^7$s.kT. For $\nu_{\mu}
  \rightarrow \nu_e$ oscillations a value of $\sin^2 2\theta_{13}
  =0.04$ and $\Delta m^2_{31} = 2.5 \times 10^{-3} \ {\rm eV}^2$ is
  used.  No detector model is used.}

\begin{center}
\label{tab:arates2}
   \begin{tabular}{|r|r@{.}l|r@{.}l|r@{.}l|r@{.}l|r@{.}l|r@{.}l|r@{.}l|}
     \hline
     Degrees off-axis & \multicolumn{2}{c|}{$\bar{\nu}_{\mu}$ CC} &
\multicolumn{2}{c|}{$\bar{\nu}_{\mu}$ CC osc} &
  \multicolumn{2}{c|}{$\bar{\nu}_e$ CC beam} & \multicolumn{2}{c|}{$\bar{\nu}_e$ QE
beam} & \multicolumn{2}{c|}{NC-$1\pi^0$} &
\multicolumn{2}{c|}{$\bar{\nu}_{\mu} \rightarrow \bar{\nu}_e$  CC} &
     \multicolumn{2}{c|}{$\bar{\nu}_{\mu} \rightarrow \bar{\nu}_e$ QE}   \\
\hline \hline
\multicolumn{15}{|c|}{WBLE 120 GeV at 1300 km with decay pipe 2m radius 380 m length} \\ \hline \hline
     0$^{\circ}$     & 75&0      & 37&7          & 0&570           & 0&106
     & 7&79          & 0&669      & 0&160 \\
     0.5$^{\circ}$   & 33&5      & 13&0          & 0&356           & 0&077
     & 4&90          & 0&332     & 0&103 \\
     1.0$^{\circ}$   & 12&0      & 6&47          & 0&185           & 0&056
      & 2&64         & 0&122     & 0&056 \\
     2.5$^{\circ}$   & 1&41      & 0&694         & 0&037           & 0&013
      & 0&499        & 0&033     & 0&022 \\
\hline \hline
\multicolumn{15}{|c|}{WBLE 60 GeV at 1300 km with decay pipe 2m radius 380 m length} \\ \hline \hline
     0$^{\circ}$      & 50&5       & 21&3           & 0&373          & 0&088
      & 6&05          & 0&507      & 0&137 \\
     0.5$^{\circ}$    & 25&4       & 8&52           & 0&248          & 0&066
      & 4&23          & 0&272      & 0&094 \\
     1.0$^{\circ}$    & 10&3       & 5&38           & 0&144          & 0&045
      & 2&52          & 0&116      & 0&058 \\
     2.5$^{\circ}$    & 1&36       & 0&667          & 0&031          & 0&013
      & 0&518         & 0&035      & 0&024 \\
\hline \hline
\multicolumn{15}{|c|}{WBLE 40 GeV at 1300 km with decay pipe 2m radius 380 m length} \\ \hline \hline
     0$^{\circ}$     & 33&8       & 12&5           & 0&270          & 0&069
      & 4&70         & 0&366      & 0&110 \\ \hline \hline
\multicolumn{15}{|c|}{WBLE 28 GeV at 1300 km with decay pipe 2m
     radius 180 m length} \\ \hline \hline
     0$^{\circ}$     & 14&6       & 4&94           & 0&076          & 0&026
      & 2&64         & 0&172      & 0&065 \\ \hline \hline
\end{tabular}
\end{center}
\end{table}

\section{Detector Requirements}

The detector requirements for a detector in a beam towards DUSEL and a
detector in the NuMI off-axis beam are quite different. Although the
physics goal of measuring $\theta_{13}$, mass hierarchy, and, above
all, CP violation is the same, the obstacles to obtain sufficient
sensitivity to this physics are very different for the two techniques.
We will describe the understanding reached in the process of this
study.

Both techniques are attempting to obtain sensitivity to CP violation
in the neutrino sector by collecting sufficient numbers of $\nu_\mu
\to \nu_e$ appearance events. 
By obtaining appearance events at difference oscillation
phases and energy, matter effects and CP effects can be disentangled
to measure oscillation parameters without correlations or ambiguities.
Regardless of the technique the most important experimental parameters are
the numbers of events at or near the oscillation peaks
versus the numbers of irreducible and reducible backgrounds.  The numbers
of events in either technique are roughly proportional to the exposure
defined as the beam power in MW (at some chosen proton energy) times
the total detector efficient fiducial size in kTon times the running time in units of
$10^7$ sec. 
In the following, to set the rough scale for detectors, we will assume that
a few hundred $\nu_\mu \to \nu_e$ events after accounting for 
detector efficiency are needed at $\sin^2 2 \theta_{13} =0.1$ per year.  
As pointed out in
Section \ref{accsec}, accelerator power of $\sim$ 1 MW can be obtained
and handled with current technology; this sets the scale for the
detector size, efficiency,  and running times.

\subsection{Off-axis} In the off-axis technique,
we have considered  two large detectors
 at two different locations. On the NuMI beam-line, the
places considered for the placement of these detectors are: 1)
baseline length of 810 km and off-axis distance of 12 km, 2) baseline
length of 810 km and off-axis distance of 40 km. At a length of 810 km
(which is close to the maximum possible on the NuMI baseline), the
first and second oscillation maxima for the physics under
consideration are at neutrino energy of 1.64 GeV and 0.54 GeV,
respectively, for $\delta m^2_{32}=0.0025 eV^2$.
 The off-axis distances were chosen to obtain a narrow
band neutrino beam at or near these oscillation maxima. These spectra
and the event rates can be seen in \cite{offaxs}.

Shorter baseline lengths for NuMI off-axis detectors have been considered
in the literature \cite{olga2}.  We have commented 
on this approach as part of the answers to 
questions in Appendix A. We will not consider this approach here 
because of the practical difficulties noted.

The main detector requirements for off-axis detectors are: 

\begin{itemize} 
\item {Size:} 
To approach the exposure criteria of 
few hundred events per year for $\sin^2 2 \theta_{13}=0.1$
 the total efficient 
fiducial mass of the detectors at the first and second oscillation maxima 
needs to be  $\sim 100$ kT.  This could be deployed  with $50$ kT at the first location
(12 km off-axis) and $50$kT at the second location (40km off-axis) or all of the 
mass in one location.  

\item {Cosmic ray rejection:} NuMI based off-axis detectors will likely be 
on the surface or have a small amount of overburden. 
Surface or near-surface capability is essential for the NuMI based off-axis 
program because of the geographic nature of the area.
As pointed out in
Section \ref{depth}, a surface detector needs to a) have sufficient data
acquisition bandwidth to collect all events near the beam spill time,
b) eliminate cosmic ray tracks so that the beam events can remain
pure, c) tag events due to cosmic rays so that no cosmic ray induced
events mimic an in-time beam event. These requirements force the surface
detector to be a highly segmented detector with active cosmic ray veto
shielding.

\item {Background rejection:}
There are two contributions to the background from the neutrino beam:
neutral current events and contamination of electron neutrino
events.  The narrow band nature of the neutrino beam is important for
rejection of both of these backgrounds. The neutral current events
which tend to have a falling energy distribution can come from both
the main peak of the neutrino spectrum and the tails. In the case of
the second location, 40 km off-axis,  the large kaon peak will contribute background. The
$\nu_e$ contamination has a broad distribution for both off-axis
locations \cite{offaxs}. To use the narrow band nature of the beam
effectively to suppress backgrounds, the detector must have the
capability to measure neutrino energy (total charged current event
energy) with good resolution, which is approximately the same 
as the width of the narrow band beam.   It should also be able to reject
$\pi^0$ or photon induced showers.

\end{itemize} 

\subsection{\bf Detectors at DUSEL} The two sites for DUSEL
that made a presentation to this study  are 1290
(Homestake) and 1495 (Henderson) km from FNAL.
The study has considered distances as far as $\sim$2500 km and concluded that 
the physics capability, with some exceptions, is roughly the same 
for same sized detector. 
 The first and second
oscillation maxima for 1290 km are at 2.6 GeV, and 0.87 GeV;
for 1495 km, they are at 3.0 GeV and 1.0 GeV, for $\Delta m^2_{32}=0.0025 eV^2$. 
A new  neutrino beam at 0$^o$ or at small off-axis angles 
has been simulated 
\cite{wble} to show that a spectrum could be made to cover 
these energies; the critical parameter in the flux at low energies will be 
the decay tunnel diameter which must be kept to be $\sim 3-4 m$, which is 
a factor of 1.5-2 larger than the NuMI decay tunnel. 
 The beam-line could be operated at any energy between 
30 to 120 GeV proton energy. For higher proton energies work is in progress
to remove high energy neutrinos ($>4 GeV$) that  produce background.  
The beam-line could also be operated at a slight off-axis angle if the background 
can be lowered by modest amount while operating 
at the highest power level possible at 120 GeV. 
 For the purposes of setting broad detector 
requirements we will assume that the spectrum 
is  similar to  Figures \ref{fwble1300} or \ref{fwblehalf}.

Detectors at DUSEL (at either Homestake or Henderson) could be placed 
either on the surface or at a deep site. If placed on the surface the detector 
considerations would be approximately the 
same as those for off-axis detectors because the 
primary design issue would be rejection of cosmic ray background.  
The availability of deep sites at the appropriate baseline distance 
for a very large detector are the main reason
for locating the detector at DUSEL.  Both Henderson and Homestake 
are planning on large detector caverns at a depth of $\sim 5000$ ft. 
We will enumerate the detector requirements assuming this depth.

\begin{itemize} 
\item{Size:} To approach the exposure criteria of 
a few hundred $\nu_\mu \to \nu_e$ appearance events per year 
at $\sin^2 2 \theta_{13}=0.1$, 
the efficient fiducial mass of the detector needs to be $\sim 100 kT$.
In the case of DUSEL all of this mass can be in the same place exposed to 
a beam that contains both oscillation maxima.  

\item{Cavern:} Because of the size required for the detectors, a stable 
large cavity (or cavities) that can house $\sim 100 kT$ of efficient 
 fiducial 
mass will be needed. 
For a water Cherenkov detector, which is well suited for deep operation, 
the efficiency is expected to be $\sim 20\%$ indicating a real detector size
of several hundred kTon. 
 From preliminary studies it appears that both 
Henderson and Homestake satisfy this criteria.

\item{Cosmic ray  rejection:} Since the cosmic ray rate at the deep sites 
proposed for DUSEL detectors is very low, it will not be a major factor in 
detector design. A cosmic ray veto for such a detector might be needed for 
physics other than accelerator neutrino
 physics; for example, detection of solar neutrinos. But it is not required 
for the physics discussed here.   

\item{Surface location for a detector:} 
For a liquid argon TPC, the efficiency and background rejection 
could be high and therefore the detector 
could be $\sim 100 kT$. However, for an underground  liquid argon TPC the 
requirements on the cavern will be dominated by safety concerns regarding 
storage of such a large amount of cryogenic liquid in a deep laboratory.  
If the liquid argon detector is placed on the surface, the requirements
are approximately the same as for the NuMI based off-axis detectors. 
The dominant requirement will be rejection of cosmic ray background.

\item{Background rejection:} There are two  main contributions to the 
in-time background from the beam: neutral current events, and electron 
neutrino contamination in the beam. 
It is expected that the majority of 
the NC background at low energies will  be from single $\pi^0$ events that will 
have to be rejected. 
In the case of using a wide band beam,
there are two tools for signal extraction. 
Pattern recognition with good 
capability will be needed to reduce neutral currents, especially single $\pi^0$ events. 
The oscillation pattern in the energy spectrum will also be used to
extract the signal. The first oscillation node, in particular, will form a peak 
above 2 GeV with a well known shape. To allow such a signal extraction, the detector 
must have good energy resolution for neutrino energy. 
From the work reported here
$\sim 10\%$ energy resolution above 0.5 GeV 
 including Fermi motion effects will be needed. 
For a water Cherenkov detector there is new work on pattern recognition 
to reduce the NC backgrounds and obtain the needed energy resolution. 
For a liquid argon detector, it has been shown 
that the NC background can be suppressed to very low levels 
for low multiplicity events (such as quasi-elastics) while 
maintaining good resolution.

\end{itemize}

\section{Status of detector simulations}

\subsection{Water Cherenkov Detector} 
\label{wcsim}

As part of this work, we have  studied the background rejection and neutrino 
energy resolution (from charged current events) 
  of a large  water Cherenkov detector instrumented in 
the same manner as Super-Kamiokande. Although considerable further work is needed
the capabilities appear to be sufficient for the neutrino oscillation 
program under consideration. The total mass and exposure 
needed to achieve good sensitivity to CP violation in 
neutrino oscillations was also determined.

The technique of water Cherenkov detectors 
with non-focusing optics 
is well understood. 
In particular, the light yield and  the fraction of scattered light
can be modeled accurately. Software techniques exist 
 that  use the pattern of light 
and the time sequence of photons to reconstruct vertices and trajectories of 
charged particles. The vertex resolution depends on the timing accuracy of 
the PMTs. The energy resolution and the energy threshold depends on the 
total amount of detected light. Both of these have been extensively discussed 
in technical articles and Ph.D. theses \cite{sknim, kasuga}. 
Considering the substantial existing knowledge and information about this 
technology, we decided to focus only on the additional new requirements 
imposed by the accelerator neutrino physics under consideration.

For the program considered here an essential problem is to separate 
electron shower events from neutral current  events, especially events 
containing a single $\pi^0$ in the final state. The goal is to search for 
$\nu_e$ charged current induced showering events in the 0.5 to 4 GeV range.  
For example, single $\pi^0$ particles with energies of 1, 2, 3, and 4~GeV 
decay to two photons 
with a minimum (which is also the most probable) 
opening angle of 16, 8,  5, and 4 degrees, respectively. 
The probability of a decay with an opening angle of more than $20^\circ$
for 1, 2, 3, and 4 GeV $\pi^0$'s is 40\%, 8.2\%, 3.6\%, and 2.0\%, 
respectively.   
In a water Cherenkov detector the position where the $\pi^0$ photons 
convert cannot be measured with sufficient precision from the pattern of 
Cherenkov light, which tends to 
be two overlapping showering rings.  At low $\pi^0$ energies the opening 
angle is sufficiently 
large compared to the Cherenkov angle ($\sim 42^\circ$) that single $\pi^0$'s 
can be separated 
quite effectively.  At energies greater than 2 GeV, however, the small 
angular separation between 
the two photons makes such separation difficult. 
 It is well known that resonant single pion
production in neutrino reactions has a rapidly falling cross section
as a function of momentum transfer, $q^2$, up to the
kinematically allowed value~\cite{adler}. 
This characteristic alone suppresses the
background by more than 2 orders of magnitude for $\pi^0$ (or shower)
energies above 2 GeV. 
Therefore a modest $\pi^0$ background suppression (by a factor of
$\sim 15$ below 2 GeV and $\sim 2$ above 2 GeV) should   make
the $\pi^0$ background manageable  over the entire spectrum.

As part of this study 
such background suppression has  been demonstrated using 
complete simulation and reconstruction using the Super-Kamiokande 
detector as the benchmark~\cite{chiaki1, chiaki2}. 
Similar suppression has also been obtained independently by 
another group\cite{fanny1, ednote}.  
In both studies the rejection of backgrounds 
was enhanced beyond the currently well known capabilities 
of a Super-Kamiokande like 
detector by using a combined likelihood method. In this method a number of 
event observables (a complete list can be obtained from the talks in 
\cite{chiaki1} and \cite{fanny1}) with low background discriminating power were combined
in a single likelihood cut. 
The work in \cite{chiaki1} chose to cut on this likelihood as a function of 
reconstructed  energy so that the efficiency of this additional cut
 for charged current 
electron neutrino events was  $\sim$40\%.   The additional rejection for 
neutral current events ranges from a factor of 30 at 300 MeV to a factor 
of 4 at 3 GeV (page 34 in \cite{chiaki1}).
Table \ref{cyrej} shows the rejection power achieved by this method
as a function of energy. 
The table is divided in two parts: before the event energy can be 
reconstructed  the rejection can only be given in terms of 
true  quantities such as ``true energy''. 
After the event is reconstructed the rejection 
is given in terms of ``reconstructed energy''.    
It should be remarked that the reconstructed energy for an NC event 
is considerably lower than the true neutrino energy. 
The total integrated efficiency for signal in this calculation using the 
28 GeV spectrum is $37\%$ from the 
traditional cuts multiplied by $40\%$ efficiency of the likelihood cut. 
As explained in \cite{chiaki1} the likelihood 
cut could be adjusted to have higher efficiency at a cost of higher background.
The total integrated rejection of neutral current background 
is $\sim 13$  for the traditional cuts
multiplied by  $24.0$ using the subsequent likelihood cuts\cite{ncvis}.

The work reported in \cite{fanny1}   compares her results to \cite{chiaki1} 
(page 31-31 \cite{fanny1}).  
In the work reported in \cite{fanny1} it was chosen 
 to retain high efficiency (above 70\%)
to electron neutrino charged current events; she obtained rejection factors of 
10 at 300 MeV declining to 2 at 3 GeV. The two calculations are in 
good agreement  if compared at the same efficiency 
considering that the simulated event sample  and methods of
discrimination were quite different.  

\begin{table}
\begin{tabular}{|l|c|c|c|c|c|c|}  
\hline 
Cut &  \multicolumn{6}{c|}{Energy Bin (GeV)}  \\ \cline{2-7}
    & 0-0.5 & 0.5-1.0 & 1.0-1.5 & 1.5-2.0 & 2.0-3.0 & $>$3 \\ 
\hline  
SK cuts/$E_{true}$ &    &   &  &  & &  \\ 
 $\nu_e$ signal & 74\% &     74\%  &      62\% &     44\% &     36\% &     27\% \\
 CC$\nu_\mu$ bkg. & 0.17\% & 0.44\% & 0.75\% & 0.76\% &  0.90\% &  0.45\% \\
 NC bckg. & 1.7\%   &  4.43\%   &  5.3\% &     7.0\% &     7.7\% &     8.6\% \\
 Beam $\nu_e$ bkg. & 86\% &    75\% &     57\% &     46\% &     36\% &     23\% \\
\hline  
Likelihood cuts/$E_{rec}$ &  & & & & &  \\
$\nu_e$ signal & 40\% &     40\% &     40\% &      39\% &     40\% &     40\% \\
CC$\nu_\mu$ bkg. &  6.8\% & 13.6\% &    6.3\% & 8.0\% & 6.5\% & 2.2\% \\
NC bkg. &  0.72\% & 4.5\% & 6.3\% & 3.9\% & 8.3\% & 7.0\% \\
Beam $\nu_e$ bkg. & 37\% &     41\% &     40\% &     37\% &     39\% &     34\% \\ 
\hline 
\end{tabular} 
\caption{ Simulation and analysis results on the 
fraction of events kept after the traditional cuts (top part of the table) and
the additional efficiency  
after the newly developed likelihood cuts (bottom part of the table). 
The events are 
divided in 4 parts: signal from $\nu_e$ charged current events (of which a small 
part are quasi-elastics), charged current $\nu_\mu$ events, neutral current (NC) events, 
and background due to $\nu_e$ contamination in the beam. 
There is no entry for background from charged current $\nu_\tau$ because 
the beam spectrum is dominantly below $\tau$ production threshold ($\sim 3.5 GeV$) and this 
background is estimated to be low.  
The efficiency for the 
signal and $\nu_e$ background should be the same except for the statistical fluctuations in 
the Monte Carlo due to small statistics of the $\nu_e$ background.
 We have retained the numbers in the table to demonstrate consistency. 
\label{cyrej}}
\end{table}  

The other important component of this study is the neutrino energy resolution 
for charged current electron neutrino events.
The selection procedure described in the previous paragraphs attempts to 
select clean events with a single lepton in the final state. To measure the 
neutrino energy we assume that this selected event is a quasi-elastic 
scattering event.  We then calculate the neutrino energy using the following 
formula: 
$$E_{rec} = {2 M_p E_{lepton} - m^2_{lepton}\over 
		2(M_p -E_{lepton} +P_{lepton} \cos\theta_{lepton})}$$
to reconstruct the neutrino energy ($E_{rec}$)  from the measured 
electron energy ($E_{lepton}$) and electron angle 
($\theta_{lepton}$)  with respect to the neutrino 
direction. 
 The energy resolution using this method  has four components. 
The energy resolution of the electron  has been demonstrated to be  4\% at 
500 MeV improving to 2 \% above 2 GeV (page 84 in \cite{kasuga}).
The  angle of the electron with respect to the beam 
 must be measured to calculate the energy of the 
incoming neutrino. The angular resolution ranges from 3 deg at low energies to 
1.5 deg at high energies (page 81 in \cite{kasuga}). 
The third component to the neutrino energy resolution is the Fermi motion
of the struck nucleon inside the oxygen nucleus. This is often modeled 
using either data from electron scattering or using a simple Fermi gas model.
It adds a contribution of $\sim$100-200 MeV to the resolution.  Finally,
the selected events have a contamination of non-quasielastic events in 
which the  extra particles (such as charged mesons or photons)  
in the final state are either invisible because they are below 
Cherenkov threshold or are  missed because of poor reconstruction.  
The final energy resolution including all these effects has been calculated 
(page 17 in \cite{chiaki3}) to be about $\sim 10\%$ at 1 GeV  with significant non-gaussian
and asymmetric 
tails.  These tails are due to the nuclear effects 
and non-quasielastic contributions. 
The resolution improves at higher energies. 
The effect of the resolution is  that 
the oscillation pattern remains visible although somewhat degrades. 
The resolution will have to be modeled 
well to extract the oscillation signal 
and the oscillation 
parameters from the far detector data; this is true regardless of 
the detector type.

\begin{figure}
\centering\leavevmode
\includegraphics[angle=0,width=1\textwidth]{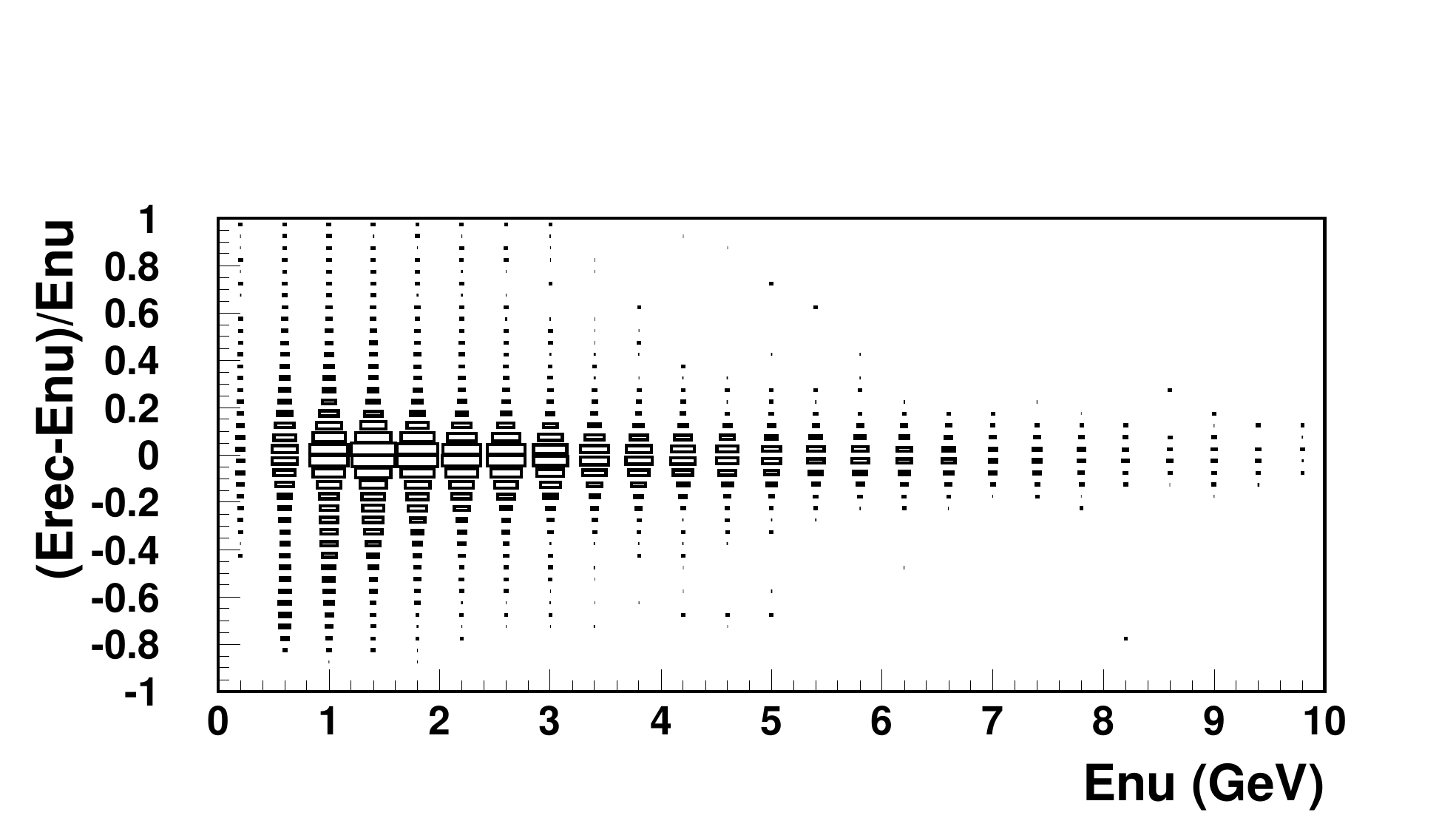}
\includegraphics[angle=0,width=0.93\textwidth]{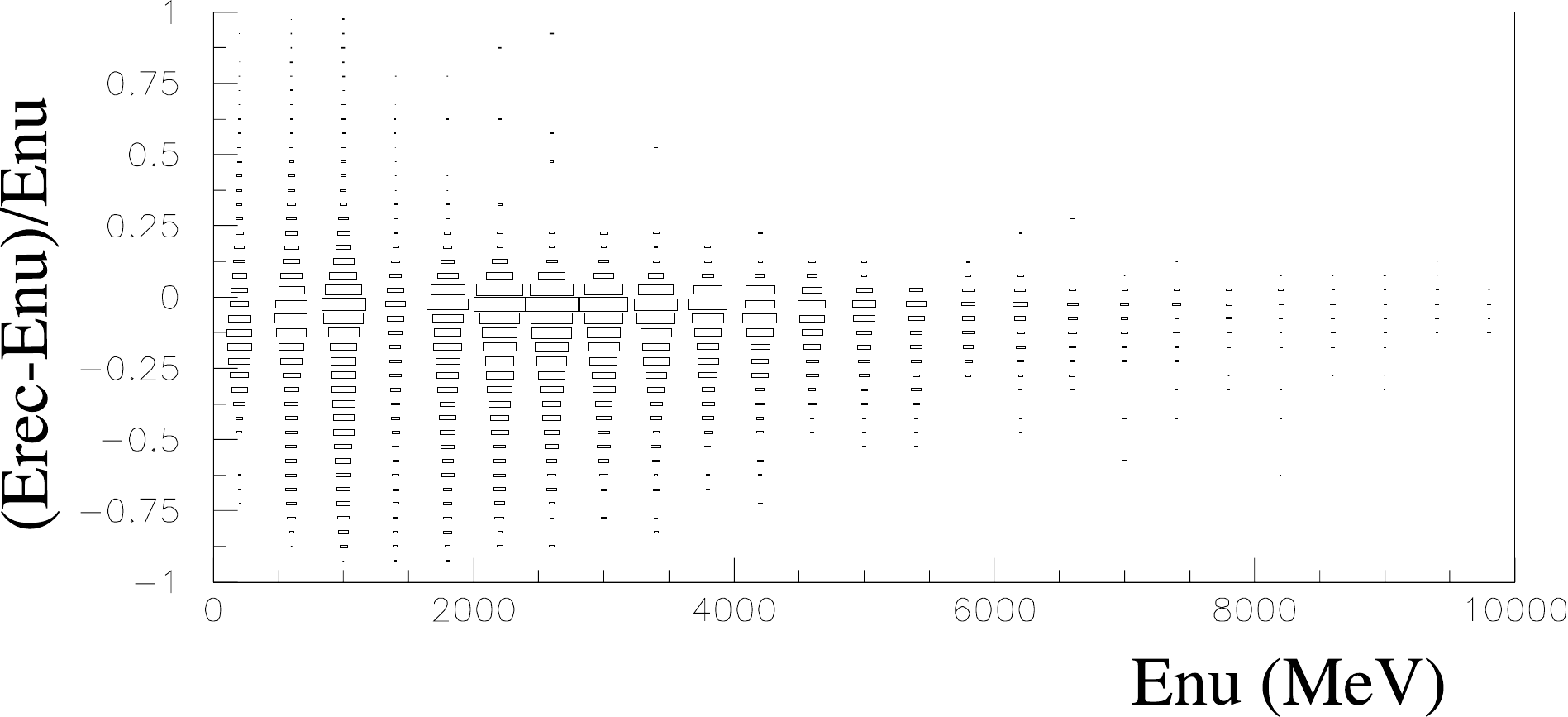}
 \caption{
Top is the parameterized resolution function used in the  
calculation of sensitivity. 
Bottom is the energy resolution on selected electron neutrino events after all 
cuts as described in \cite{chiaki2, chiaki3}.
The tail where reconstructed energy is lower than true energy is due to 
non-quasielastic events that are selected. These 
events have missing particles and therefore have missing energy. 
The bottom plot
 was made from  simulated  events including the effects of oscillations 
over 1480 km, and therefore there is 
a depletion of events in the plot around the oscillation minimum.  
   \label{eres1} }
\end{figure}

 The sensitivity calculations described later in this report were performed
 using 
the GLoBES framework \cite{globes}. For this calculation the 
detector response was parameterized and adjusted to  correspond to 
the full simulation described above.  There are 
some differences that should be 
kept in mind to allow comparison between calculations.  
The first difference is that in the sensitivity 
calculation no events below 
0.5 GeV are used.
 The work in \cite{chiaki1} and \cite{fanny1} includes 
events to lower energies.   
The second  difference is in the energy resolution. The energy resolution obtained 
after complete simulation and reconstruction is shown in  Figure \ref{eres1} bottom 
plot. 
The sensitivity calculation has a parameterized resolution function that includes 
effects of Fermi motion, resolution on  the lepton energy and 
angle, and  non-quasielastic contamination. 
  The two resolution functions are shown  in Figure \ref{eres1}. 
The parameterization has somewhat worse resolution in the core than the 
full simulation, but less tail than the full simulation.  
 The 
input to the calculation is firmly based on full simulation.  
But the parameterized background and resolution allows for fast calculation of 
signal and background rates for different oscillation scenarios; we are also 
 able to change the beam spectrum while keeping the detector performance fixed using 
this tool.

We note that comparison of signal rates between various authors needs to be done with some care
because they may have used different conventions for the sign of the CP violating phase 
$\delta_{CP}$.  The calculation here 
uses the standard convention from PDG (with $e^{-i\delta}$
 in the $U_{e3}$ matrix element), 
but some of the simulations reported in the course of this work have  the 
 convention with the opposite sign on $\delta$. 


We have explored  the feasibility of 
using a water Cherenkov detector for this science with promising results,
 but considerable further 
work is needed to optimize the detector and gain complete understanding 
of its limitations.   
One question that needs exploration is 
what are the intrinsic limitations of the water Cherenkov technique 
in terms of pattern recognition. The literature on the 
 subject is broad and general conclusions can be drawn \cite{yps1,glasl,
anto1,bug1,shioz1}.  The question can be quantified in terms of 
the vertex and angular resolution for single tracks and the ability to 
separate two tracks that are close in angle and have a common vertex.
In the references cited above it has been shown that the photon
detection resolution (time and position) in a  Super-Kamikande 
style water Cherenkov detector (proximity focused) does not 
approach  the multiple scattering contribution to that resolution. 
Therefore, the capabilities of the detector could be improved 
by modest improvements to the timing and granularity of the PMTs or 
by addition of ring imaging techniques.  In other words, the current 
capabilities are dominated by  the characteristics and geometry of
the photo-multiplier array, and there is room for improvement.

\underline{\bf Photomultiplier coverage and optimization}:  
The optimization of PMT coverage and granularity has not been
addressed in our study. This is an appropriate goal for the proponents
of the water Cherenkov technique when they write the full proposal.
Some hints of the effects of high granularity can, however, be found
in the work of Yanagisawa and Dufour. For example, in \cite{chiaki1}
it is shown that the efficiency for detecting a $\pi^0$ particle
increases by 20\% as the event moves away from the wall of the
detector; this indicates that a larger detector or a detector with
more granularity with the same PMT coverage will have better
background suppression. In \cite{fanny2}, the likelihood based
background suppression is shown to have weak dependence on the PMT
coverage (either 20\% or 40\%). These preliminary results indicate  that
as long as the collected numbers of photoelectrons is reasonably
large, the granularity of the PMT readout will have more impact on
pattern recognition.

\underline{\bf Optimizing beam spectra}: In this study we have not made extensive 
attempts at optimizing  the
beam spectra versus the detector performance for backgrounds. In
\cite{wble}, a cross section model  was used to calculate
signal and background shapes for neutrino spectra produced with
different energy proton beams.  It is clear that the background will
increase  for proton energies above 60 GeV because of the
high energy tails. However, for proton energies below 60 GeV, the
signal versus background performance is approximately constant: there
is reduction of background for lower energies, but there is
significant statistical gain (for constant power) at higher energies.
The intensity at low neutrino energies as well
as the elimination of high energy tails will continue to need
examination.

There are three ways to optimize the beam spectrum to reduce 
backgrounds from high energy neutrinos: 1) The optics of the target horn 
assembly need to be 
optimized to increase the flux below 3 GeV and reduce it at higher energy,
2) The beam towards DUSEL could certainly be made slightly off-axis. 
Reference \cite{wble}  shows that 
a $0.5^o$ off-axis angle lowers  
the event rate $>$ 5 GeV by about a factor of 3. The $0.5^0$ off-axis beam 
angle can be accommodated 
in a large 4 meter diameter tunnel, but the implications for shielding,
beam dump, as well as future flexibility of the program must be carefully considered. 
3) The third option for reducing neutrino flux at higher energies 
is to introduce a second target (or beam plug) between the two focusing horns.
This has been studied for NuMI and found to be effective at reducing tail events
by as much as 70\% \cite{numi719}. Such an option needs careful engineering design 
because it will affect the radiation environment in the target area.

\underline{\bf Near Detector}: Lastly, there has been very little study of 
near detector issues for a
 beam towards DUSEL. There are 3 issues that need to be
discussed. 1) \underline{The availability of space}: the beam design discussed in
Section \ref{duselbeam} leaves 300 m of earth shielding from the end of the
decay tunnel to a potential near detector site located within FNAL
boundaries. The depth at this site would be 192~m for a beam to
Henderson and 176~m for Homestake. The NuMI near detector is at a
depth of 105~meters. The feasibility and cost of a near detector
cavern of about 30~meter width/length will need to be examined.
2) \underline{The main requirement for the near detector}: 
The most important function of the near detector is the measurement of 
the neutrino spectrum and backgrounds before oscillations. 
As explained later for the
sensitivity calculation we have assumed that the background
 will be known to about 10\%. This includes the effects
of beam and detector modeling as well as nuclear effects which might
be different between the near and the far detectors if they are
composed of different materials.  A harsher requirement on the near
versus far energy scale of $<1\%$ might come from the need to measure
$\Delta m^2_{32}$ with high precision. 3) \underline{The detector granularity, mass,
and data acquisition}: the event rate at the near site (few events per
beam pulse for a $\sim$100 ton detector) will be much higher compared to the
far detector. Reconstruction of these events will likely require a
fine grain detector (perhaps a modest sized liquid argon TPC) 
with electronics that can separate events within
the 10 $\mu s$ pulse.  How the design impacts the requirements
outlined above will need to be examined.  There is now considerable
experience from the NuMI-MINOS  on how to use the near
detector to perform the appropriate extrapolations from near to far
site.  From that experience \cite{minos} the above requirements do not
appear to be particularly difficult, but the issue should not be treated 
cavalierly.

\subsection{Liquid Argon Time Projection Chamber} 
\label{larsim} 

Liquid Argon TPC detectors, with fine grained tracking and total
absorption calorimetry capability, suggest great promise for
sensitivity to long baseline oscillation physics.  Hand scanning
studies indicate efficiency for charged current quasi-elastic electron
neutrino interactions ($\nu_e$ QE) greater than 80\% and background
rejection of neutral current $\pi^0$ events by a factor of 70~\cite{nusagdoc}.  
Studies from European groups
are consistent with these results~\cite{GLACIER}.  

As part of this study, tools have been developed to simulate and
reconstruct events in the E$_\nu$ = 0.5-5 GeV energy region.  Studies
using these new tools confirm the efficiencies and background
rejection from the hand scanning work.  Sensitivity calculations
folding in these efficiencies, background rejection factors, and
resolutions indicate LArTPCs are $\sim$3 times more sensitive than an
equal mass of Water Cherenkov detector (See Section \ref{sensi}).

It is primarily the imaging capability that enables LArTPCs to distinguish
different event classes from each other. Specifically, while
conventional detectors can typically identify only the outgoing lepton
in QE interactions, LArTPCs can tag both the outgoing lepton and the
recoil proton.  Furthermore, a LArTPC can easily and unambiguously
identify the interaction point of energetic gamma-rays, for example
from $\pi^0$ decay, if the separation from the primary vertex is
larger than 2 cm~\cite{alessandro}.

As part of this study, a GEANT3 simulation of a Liquid Argon TPC was
studied and developed to best quantify 
the detector performance
\cite{stevenstalk}.  The Monte Carlo used the 
NUANCE event generator as input and simulated events in a $7\times10\times10 m^3$
box, roughly equivalent to 1 kTon.
 Events are digitized using standard GEANT libraries, and
Monte Carlo truth studies performed on this output.  Given the imaging
capability of a LArTPC, this is an acceptable approximation of an
actual event.  The criteria to tag a $\nu_e$ QE interaction are
first to see an electron shower as distinct from a muon track.  This
is assumed to be 100\% efficient.  The second criteria is to see a
recoil proton coming from the same vertex as the electron.  The well
established low energy threshold for this is a proton with kinetic
energy $>$ 40 MeV~\cite{40mev}.  Imposing this requirement, the
efficiency for $\nu_e$ QE events is $>$90\%. As these are first pass
studies, we default  to the more conservative 80\%
efficiency determination from the hand scanning.  Neutral
current $\pi^0$ backgrounds, with subsequent $\pi^0 \rightarrow \gamma
\gamma$, arise from both $\nu_{\mu} n \rightarrow \nu_{\mu} n \pi^0$
and $\nu_{\mu} p \rightarrow \nu_{\mu} p \pi^0$ interactions.  The
first,$\nu_{\mu} n \rightarrow \nu_{\mu} n \pi^0$, is rejected because
of the lack of any recoil proton.  The second,$\nu_{\mu} p \rightarrow
\nu_{\mu} p \pi^0$, is tagged by observation of a 2cm or larger gap
between the vertex of the recoil proton and at least one of the gammas
from the decaying $\pi^0$ which converts into an $e^+ e^-$ shower.
Combining these requirements, only 0.5\% of the NC $\pi^0$ backgrounds
are not rejected.  Further rejection factors are expected by looking
at the energy deposited in the first few cm of tracks
initiating electron showers versus gamma showers.  The overlapping
$e^+e^-$ from the gamma shower deposits twice the energy at the
beginning of the track as the single electron.

These studies have been performed using the WBLE flux 
generated with 40 GeV protons  used in this study.  
Further study is needed to understand reconstruction for the other  beam
options and the NuMI option. Nevertheless, these results  are
relevant across a broad range of energies.  In particular, for NC
$\pi^0$ rejection, the separation between the primary vertex and the
closest gamma conversion point is roughly independent of the incoming
neutrino energy~\cite{alessandro}.
High
multiplicity events in the deep inelastic scattering (DIS) 
 region may be very challenging to
reconstruct.  Efficiency and background rejection for 
  DIS events for the different flux configurations is also needed.

Advances in automated reconstruction were also pursued as part of this
study. 
 The Hough transform based fit algorithm was
designed to reconstruct linear tracks from a quasi-elastic 
event through a parameterization by
angle.
 It efficiently identifies both primary and secondary vertices
and reconstructs tracks with resolution of  $\sim 2^\circ$ 
(RMS)~\cite{alessandro,colinstalk}.  This fitter suffices for 
events with linear tracks and
low multiplicity such as   quasi-elastics and 
resonance events.  A  study of the capability to automatically identify 
and reconstruct
electromagnetic showers is in the early stages.

In the future, this simulation and reconstruction package can be used
to study energy resolution for different classes of events.

 For the results in this report, we use the 
 energy resolution from previous work.  For
QE events, a 5\% energy resolution was assumed.  This is valid
down to $\sim$1 GeV, below which few events contribute to the
oscillation signals.  For non-QE events, a 20\% neutrino energy
resolution was assumed.  This is likely too conservative in the
resonance region where low multiplicity events can still be well
measured  by LArTPCs, but likely too optimistic  for DIS events
above  2-3 GeV.  Understanding these resolutions as
a function of energy is part of the ongoing program of simulation and
reconstruction studies.

\section{Status of detector design and technology} 

\subsection{Water Cherenkov conceptual Design}

The water Cherenkov detectors discussed in this study were largely 
conventional based on the well known technology developed and perfected 
over the last  three decades. The main difference is the 
factor of $\sim 10$ increase in fiducial mass compared to the largest existing 
detector (Super-Kamiokande).  This large increase can be accomplished either by 
increasing the size of the detector or by building several detectors (or both). 
The second important parameter for this detector is the number and size 
of the  photo-multiplier tubes (PMTs). 

Two conceptual designs were reported for this study. 
They were specifically for the two possible DUSEL
locations of Homestake or Henderson, but the authors 
have acknowledged that their ideas could be adapted to either site
with appropriate considerations for site dependent cost factors.  
 
The design reported for the Henderson site (UNO \cite{uno}) 
has a single cavity of dimension 60 m wide, 60 m high, and 180 m long.  
The 180 meter length is divided in 3 sections. Each section is a separate 
optical volume with photo-cathode coverage of 10\% for the end sections and 
40\% for the central section using the 
20 inch diameter PMT developed by Hamamatsu. 
 Each section has fiducial volume (depending on
specific physics cuts) of about 150 kTon.  The depth of the 
detector in Henderson 
will be approximately 5000 ft.

The design reported for Homestake houses the detector at 4850 ft depth
in 3 separate large caverns or modules \cite{300kt}. 
The size of the caverns will be  
cylindrical  with diameter/height of $\sim 53$ meters. 
The location is at the 4850 ft level of Homestake which is proposed as
the Early Implementation Plan for the Homestake lab. 
The collaboration proposes that  the same level be used to 
 accommodate several more cavities to take the total detector mass to 
megaton over a long period, but the baseline detector is 3 modules.  
 The exact dimensions of the cavities 
will be determined by the need to maintain  
fiducial mass of 100 
kTon for 
accelerator neutrino events.  Each detector will be instrumented by 
10 to 13 inch diameter PMTs with photo-cathode coverage of 25\%. 
At this stage of simulation and understanding  of PMT performance,
the Homestake proponents consider  the 
choice of smaller but larger numbers of tubes for granularity 
 adequate for reconstructing the accelerator neutrino events. 
The  concept  for the Homestake detector 
including the physics and a rough estimate for the cost 
was presented to a program 
committee for the Homestake interim laboratory. 
The review  can be obtained at 
\cite{hspac}. 

The main concerns for both designs is the cost and time 
required to build stable
and safe  cavern(s)\cite{laughton}  
and the manufacturing of the necessary 
number of photo-multiplier tubes.

For the single cavity (UNO) concept
 an estimate based on the cost of Super-Kamiokande has been made for 
the cavity excavation of \$168M; the engineering 
and stability of the cavity needs detailed 
examination. The total cost including 56000 
large 20 inch PMTs and 15000 smaller 8 inch tubes for 
outer veto volume was estimated to be \$437M. 
The total construction time will be approximately 10 yrs 
dominated by the PMT  manufacturing time\cite{uno}.

For the multi-cavity Homestake design, the proponents 
have performed an initial engineering design for the 
cavity construction and a  stability 
study\cite{stable}. The cost for constructing 
3 cavities is estimated to be approximately  \$70M
 which includes  contingency factors. 
The time scale for constructing the first 
 cavern is 4 yrs and each additional cavern is readied 
6 months after the completion of the previous one. 
 The total cost including approximately 
50000 PMTs for each  detector module is \$309M.
 The impact on this cost if the size of the 
module is increased for additional fiducial 
volume is explained in Appendix A. 
Based on the Super-Kamiokande experience, the installation time 
for the PMTs will be about 1 yr for each module.

The  largest unknown at present for both designs is the schedule
 for manufacturing the large numbers of PMTs.  For the 20 inch PMT
option, there appears to be only one vendor at present with a labor 
intensive manual process. For the 
smaller
PMTs there could be multiple vendors with more automated 
manufacturing processes. We will comment on this issue again at the end 
of the report. 

There are other technical concerns for such a large water Cherenkov detector:
the handling, temperature  and purification 
of such a large amount of water, the engineering for mounting the 
PMTs and the cabling of the large number of channels,
 maintenance of the PMTs and associated electronics,  and 
the radiation environment in the deep site which can affect  the data rate 
and the energy threshold of the detector. There  is no detailed engineering design for these 
items, however these issues have been examined by previous 
generations of these types of detectors. Based on that previous experience both detector designs 
have included approximate costs in their estimates.

\subsection{Liquid Argon TPC Conceptual Design}

While LArTPCs show great promise with excellent efficiencies and
background rejection for a variety of physics goals, they have not yet
been demonstrated on scales larger than few hundred tons in size.  An
active R\&D program culminating in  the T600
program~\cite{t600} has illustrated the capabilities of the detector,
however, further R\&D is necessary to consider massive detectors, on the scale
of tens of ktons.

There are several different design ideas for massive detectors
including a modularized detector~\cite{cline}, a single detector but
with modularized drift regions~\cite{nusagdoc}, and a single open
volume, very long drift detector combining charge and light
collection~\cite{GLACIER}.  The technical issues described here are
relevant primarily for single massive detectors with modularized drift
regions, the design studied by the contributors to this study.  For
these, there are no major obstacles  to scaling to detectors on the
scale of 50-100 kTon, however, there is an R\&D path that must be
realized in order to consider massive detector construction,
operation, and data analysis.  Details
of this path and major R\&D goals can be found in
~\cite{lar218}.  The major challenges for scaling to a large
detector include:
\begin{itemize}
\item Argon purity
\item Signal to noise in a massive TPC
\item Understanding Cost and Schedule
\end{itemize}
Progress and path for each of these is described below.

For ionization electrons to drift 3 m in a LArTPC,  10ms electron
lifetime must be achieved and maintained.  Studies from the T600 run
suggest this is possible, but, for a massive detector,
modifications must be made to the purification system, and 
the ability to reach purity
levels necessary in an industrial environment must be demonstrated. 
Over the past year,
Fermilab has embarked on purity testing towards this goal.  They have
developed a new non-proprietary Trigon filter 
(unlike those used in the T600) that can be regenerated in-line. 
 With
this filter system, Fermilab has achieved 12 ms lifetimes in a small
test vessel.  Over the next year, purity studies will continue with a
materials test stand~\cite{teststand} at Fermilab where argon will be
re-purified after being exposed to contaminants expected in a massive
LArTPC.  An additional challenge to purity is a consequence of the 
inability to
achieve vacuum before the initial argon fill in a massive detector.
An idea to purge the vessel with clean argon gas prior to liquid fill
is being tested at Fermilab now with studies continuing in the
upcoming year~\cite{argonpurging}.  

A very massive detector will have signal wires as long as tens of
meters.  Long wires present challenges related to wire breakage, wire
assembly and stringing, and electronics noise.  Existing R\&D work at
Fermilab focuses on assembly techniques and noise pickup using a long
wire test stand~\cite{longwires}.  Work on electronics design
to maximize signal to noise specifically by employing cold electronics,
is underway at Fermilab and Michigan State University.  A new idea for
internal wire configuration, a cellular design, avoids many of the
stringing and assembly problems of long wires by stringing wires onto
pre-assembled ladders before installation\cite{cellulardesign}.

There are two cost drivers for a liquid argon TPC which have some
certainty at this point.  The first of these was given by the LArTPC
group in its September 2005 report to NuSAG\cite{nusagdoc}.  
 There, the cost of liquid argon alone (without
a purification system) is reported as about \$1~M  per kton.
Subsequent to that report, a simple scaling relationship has been
developed based on information from two vendors for tanks appropriate
for containing liquid argon (but without modifications required to put
a TPC inside it).  This relationship, which is expected to be valid
between 5 kTon and 50 kTon, is \$2.72M + 0.306 \$M/kTon.  Thus
taking a 50 kTon detector as an example, these two cost drivers (the
liquid argon plus a containment tank) would cost about $ \$50M +
\$18M = \$68M$. 

There are many other costs, both technically driven and project
driven, but the design of the TPC itself needs to be specified in more
detail before such a  complete costing exercise can 
converge.  For example, the recently developed cellular design for
the TPC  significantly changes the requirements on the
containment tank compared to the design in the September 2005 LArTPC
report to NuSAG.  Since this design allows for fabrication of
the TPC wire planes  at the same time as the containment tank
is being constructed, 
the schedule for construction of the detector is shorter.  If electronics
are used in the liquid argon, the cellular design will change 
and the requirements on pattern recognition will become easier.
Finally, the idea of using several smaller tanks to achieve a large
mass will impact the cost of the purification system as well as the
cost of the containment tank(s).  These are some of the design choices
for the TPC that need to be made before a cost estimate of the
technical components, other than the liquid argon and the cost
for a single containment tank, can be made.

In addition to the major challenges for scaling to large detectors as
described above, issues relating to detector siting have  been
studied.  Water Cherenkov detectors  must
be located deep underground due to cosmic ray backgrounds.  By contrast,
liquid argon detectors could  be located on or near the surface.
  As part of this study, cosmic rates in
a massive LArTPC detector were calculated and their impact on the
physics program was considered~\cite{cosmicswriteup} and is discussed
in more detail in Section~\ref{depth}.  If massive LArTPCs are
sited with some overburden, such as at the 300~ft drive-in site at
Homestake, cavern construction for these detectors must still be
understood.  As part of this study, cavern designs modeled after
liquefied natural gas vessels built within ships hulls were
considered~\cite{brombergnote}.  This design is promising and studies
on this are ongoing.

The R\&D path towards a massive detector includes small scale tests
and studies as described above.  Construction of a significantly
larger prototype, $\sim$1kTon, is  necessary before
embarking on the massive detector project.  The details of this R\&D
path at Fermilab will be addressed within the next year.

\section{Overburden and shielding } 
\label{depth} 

In this section we briefly discuss the overburden issue in the context
of accelerator  neutrinos. For non-accelerator physics 
the issue is discussed in Section \ref{nonacc}. 

In summary, the background rates in a large detector due to cosmic
rays have been calculated for both surface and underground
locations. A preliminary evaluation of the consequences for both data
acquisition and background to accelerator neutrino events suggests:
 1) It is not possible to operate a water Cherenkov detector of
size $> 50 kT$ on the surface. 2) A fine grained tracking detector
such as a liquid argon TPC could be operated on the surface to take
data within the short ($\sim 10\mu S$ at FNAL Main Injector, $\sim 2.5
\mu S$ for BNL AGS) accelerator spill\cite{cosmicswriteup}, however background
rejection of $\sim 10^8$ ($\sim 10^3-10^4$) will be needed against
cosmic muons (photons) by either active veto or pattern recognition to
reduce the background rate to acceptable levels; this rejection is
in addition to the rejection obtained by the 
timing requirement. We provide a few more details of the
calculations below.

A cylindrical tank of size 50 m height/diameter (approximately 100kT
of water) will have a rate of cosmic muons (with momentum $>0.5
GeV/c$) 250 kHz from the top and 250 kHz from the sides. For a 10 $\mu
s$ beam spill this corresponds to 5 muon tracks in the detector. For a
single volume water Cherenkov detector in which the photo-multipliers
are mounted on the walls looking inwards, each muon on the average
will produce a hit in more than 50\% of the PMTs. Therefore, each
cosmic ray should be assumed to deaden the entire detector for a
period of time which is dependent on the dwell time of the muon track and 
the light inside the detector, the pulse shapes from PMTs, and 
 the data acquisition electronics. All the effects are of order $1
\mu s$ and therefore make the detector unworkable at the surface. For
example, for a detector similar in technology to Super-Kamiokande, the
dead-time from the above event rates will exceed 50\% \cite{sknim}.  To reduce
this dead-time using fast pulse digitizers is costly, and requires
significant software and hardware R\&D to resolve overlapping pulses
to reconstruct events with contained vertices. The consequences on
background rejection and resolution are at present unknown. The depth
required to reduce the number of in-time cosmics to various levels is
given in Table \ref{intime}.  A depth of at least $\sim 1000$ meters water
equivalent is needed to reduce the muon rate to a level comparable to
the rate of events from the neutrino beam so that minimal dependence
on pattern recognition (and a modest active veto capability) is needed
to separate beam related events.

\begin{table}  
\begin{center} 
\begin{tabular}{rr}  
Intime cosmics/yr & ~~~Depth (mwe) \\
\hline  
$5\times 10^7$ &  0 \\ 
4230 & 1050 \\
462 &	2000 \\
77 &	3000 \\
15 &	4400  \\
\hline 
\end{tabular}
\end{center} 
\caption{Number of cosmic ray muons in a 50 m height/diameter detector 
in a 10$\mu s$ pulse for $ 10^7$ pulses, corresponding
 to approximately 
1 year of running, versus depth in meters water equivalent.  
}
\label{intime}  
\end{table}  

A 50 kT liquid argon TPC can be contained in a cylindrical tank of
size 35.5 m height/diameter; such a detector will have a cosmic ray
muon rate of 125 kHz from the top and 125 kHz from the sides.
An examination of cosmic rays \cite{cosmicswriteup} in a liquid argon TPC has
considered their effects on data acquisition and event reconstruction,
and as a source of background.  The rate of cosmic rays was shown to be 
tolerable  with the proposed drift-time and data
acquisition system for cycles up to 5 Hz. 
In this scheme the detector takes data in a short time interval
(currently proposed to be 3 drift times) near the beam time.
 This is sufficient to cover
most possible accelerator cycle times discussed above.
 The high
granularity of the detector should allow removal of cosmic muons from
the data introducing a small ($<0.1\%$) inefficiency to the active
detector volume, so that most of the accelerator induced events are
unobscured. If a cosmic ray muon (photon) event mimics a contained
in-time neutrino event it must be rejected based on pattern
recognition. The rejection required is $\sim 10^8$ for muon cosmics
and $\sim 10^3-10^4$ for photon cosmics; given the fine grained nature
of the detector this rejection is considered achievable, but still
needs to be demonstrated by detailed simulations.

\section{Analysis of sensitivity to oscillation parameters}

\label{sensi}

In this section we will combine the information from the previous 
sections on the intensity of the accelerator beam and detector performance
to calculate the sensitivity to oscillation parameters.  
The main features of the accelerator and detector performance can be 
summarized as follows: 

For the sensitivity calculations we will assume that we can obtain 
a total of $60\times 10^{20}$ protons at 120 GeV. This total is to be 
divided between neutrino and anti-neutrino running.  
To convert this luminosity to running time we will assume 
that the accelerator can produce proton intensity 
according to Figure \ref{fnalpow}  in the accumulator 
upgrade scenario. In the accumulator upgrade scenario a power level of 
1.2 MW  is expected at 120 GeV. 
This running scenario will be used for both the off-axis  and 
the DUSEL based options for the Phase II program.  
We will make comments on running at
lower proton energies as well as more exposure. The impact on 
the running time will be according to the power curve in 
Figure \ref{fnalpow}.  The raw event rates can be obtained from Tables 
\ref{tab:rates1} to \ref{tab:arates2}.

For the first  DUSEL based calculation
 we have assumed a water Cherenkov detector 
with a total fiducial mass of 300 kTon with the performance described in 
Section \ref{wcsim}. The calculation was performed with the GLoBES package\cite{globes}
with the beam spectra and detector performance specified according to the work 
in this report.  For the DUSEL baseline we have performed calculations ranging 
from 500 km to 2500 km with various beam configurations. 
We cannot display all calculations in this 
report due to length considerations, but they can be obtained from the 
study website\cite{website}. Differences in parameter sensitivity  due to 
baseline will be discussed. Most of the calculations shown here will be   
for the 1300 km  distance.

For the second DUSEL based calculation we have assumed a 100 kTon liquid argon
time projection chamber with the performance indicated in section \ref{larsim}.
Briefly, we assume 80\% efficiency for electron neutrino events with 
very little  background from other sources.

For the off-axis calculations several different combinations were calculated.  
First, for comparison purposes 
the calculation is performed for NO$\nu$A with the detector performance 
obtained from the NO$\nu$A collaboration.  Second, a 100 kTon total mass for a
liquid argon detector TPC was assumed for phase II. The performance was 
evaluated for setting the entire detector mass at the same location as 
NO$\nu$A and also for setting 50 kT  at the NO$v$A site (12 km off-axis) 
and 50 kT  at 
the site (40 km off-axis) where the second oscillation maximum
 can be observed.  Doubling of the total mass at the two sites was 
also examined.

Lastly, we note that unless otherwise noted the oscillation parameters 
used for the calculations are as follows: 
$$\Delta m^2_{21}=8.6\times 10^{-5} eV^2$$ 
$$\sin^2 2 \theta_{12}= 0.86$$, 
$$\Delta m^2_{32}=2.7\times 10^{-3} eV^2$$
$$\sin^2 2 \theta_{23}=1$$

The parameters, 
$\theta_{13}$, $\delta_{CP}$ and the mass hierarchy (normal or reversed) 
are left free in the calculation.  

Before describing the sensitivity to $\nu_e$ appearance, we first 
make a few comments on the $\nu_\mu$ disappearance measurement. 
Either of the two experimental concepts (a new beam to a DUSEL location or 
new off axis detectors on the surface in the NuMI beam-line) have sufficient 
statistical reach to make a very precise measurement of the atmospheric 
oscillation parameters ($\Delta m^2_{32}$ and $\sin^2 2 \theta_{23}$.  
 For a 100 kTon detector, $\sim$ 10000 $\nu_\mu$ CC events per year 
are expected 
in either scenario with approximately 1/2 disappearing due to oscillations
(see Section \ref{erc} for exact numbers for specific beam configurations).
 The statistical precision
after several years of running, therefore, will be $<1\%$ for both
$\Delta m^2_{32}$ and $\sin^2 2\theta_{23}$.  A discussion of this 
measurement for a DUSEL based detector can be seen in 
\cite{hql04}. A similar discussion for the off-axis scenario is in 
\cite{nova}. To obtain the best measurement, both the 
neutrino event energy resolution (including nuclear target effects
 due to Fermi motion and re-scattering) and the absolute energy scale 
need to be well modeled. With current knowledge of these limitations 
the measurement will most likely be systematically limited
to about 1\% for both $\Delta m^2_{32}$ and $\sin^2 2 \theta_{23}$. 

There is an important difference between the off-axis measurement 
and the broad band measurement. The oscillation shape including a nodal pattern,
if the baseline distance is sufficient, can be measured with a DUSEL based 
detector. Such a measurement will exhibit less correlation between 
the two parameters $\Delta m^2_{32}$, which determines the position of 
the node in energy, and $\sin^2 2 \theta_{23}$, which determines the 
depth (or amplitude) of the node. A precise measurement of the shape 
could also limit non-standard physics models of decay, decoherence, 
extra-dimensions, etc.

\subsection{Sensitivity of a FNAL to DUSEL based program} 
\label{sensi_1}

For the calculations reported here we have used the 120 GeV 
beam with 380 meter decay tunnel with a 0.5 deg off-axis angle. 
As explained above, the spectrum from such a configuration is
well matched to the physics at this current time. The 
energy of the proton beam and the horn optics need to be optimized further.

\subsubsection{Water Cherenkov Detector} 
\label{sensi_1_1}

\begin{figure}[h]
\includegraphics[angle=0,width=0.49\textwidth]{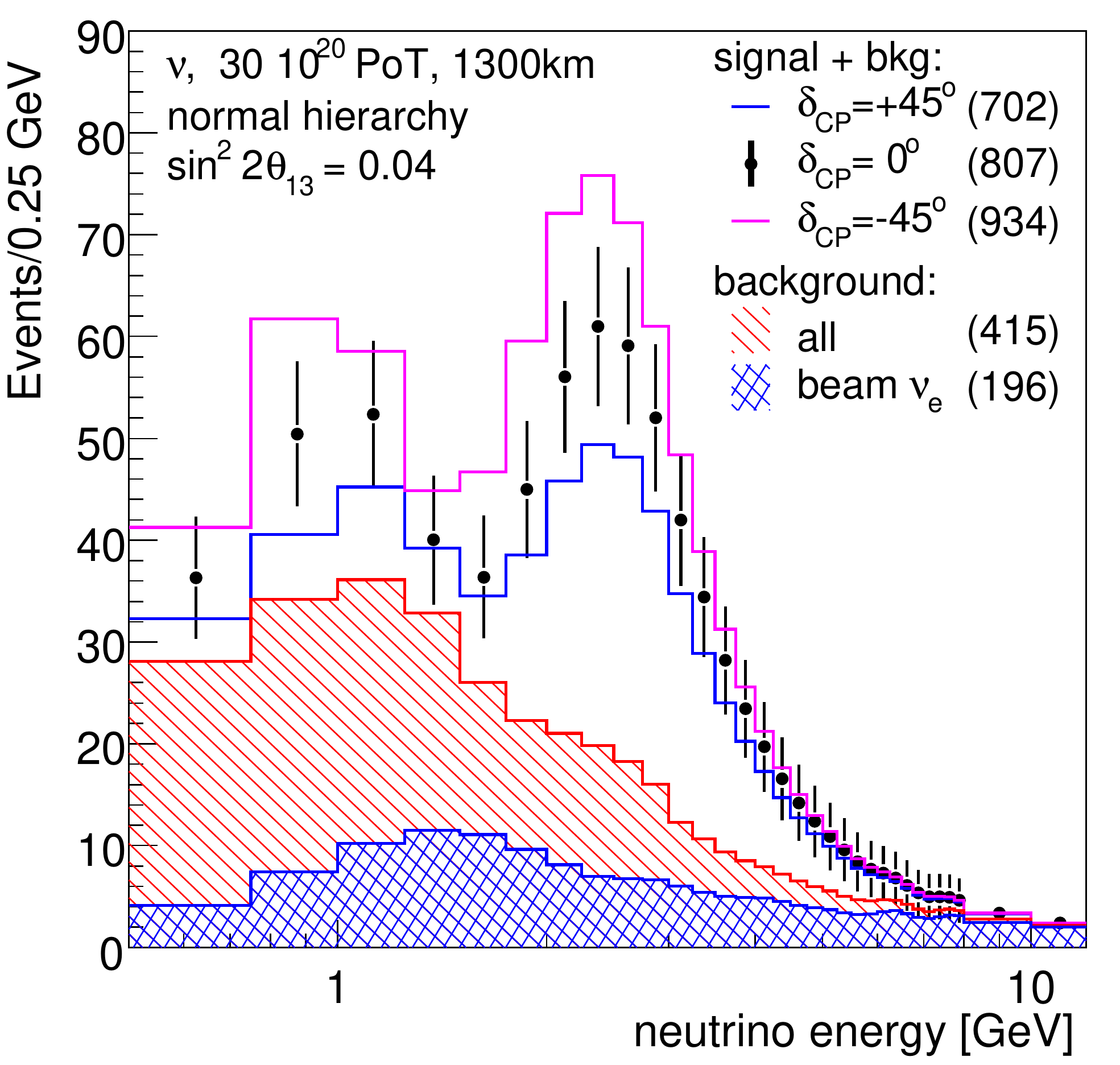}
\includegraphics[angle=0,width=0.49\textwidth]{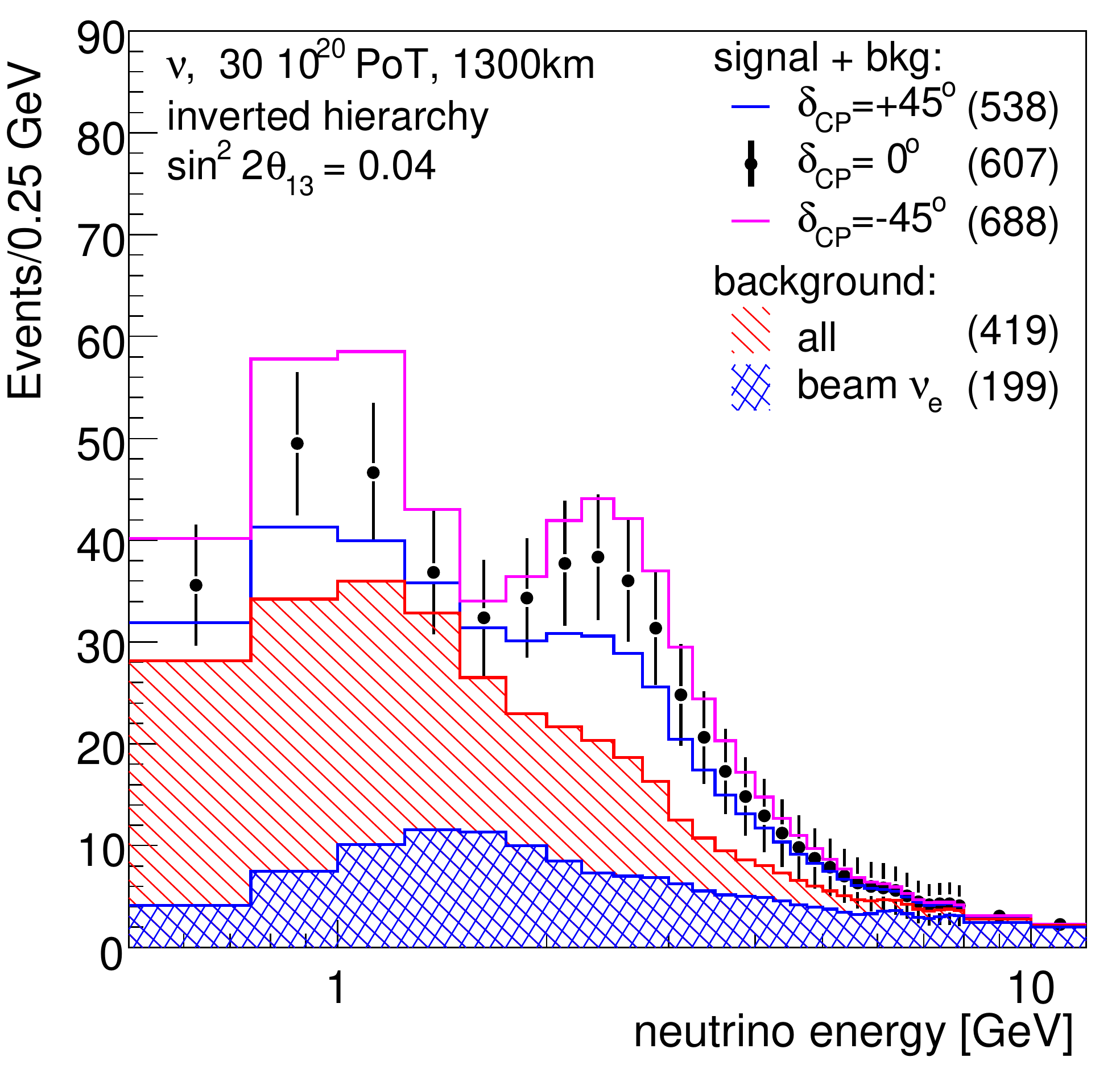} \\
\includegraphics[angle=0,width=0.49\textwidth]{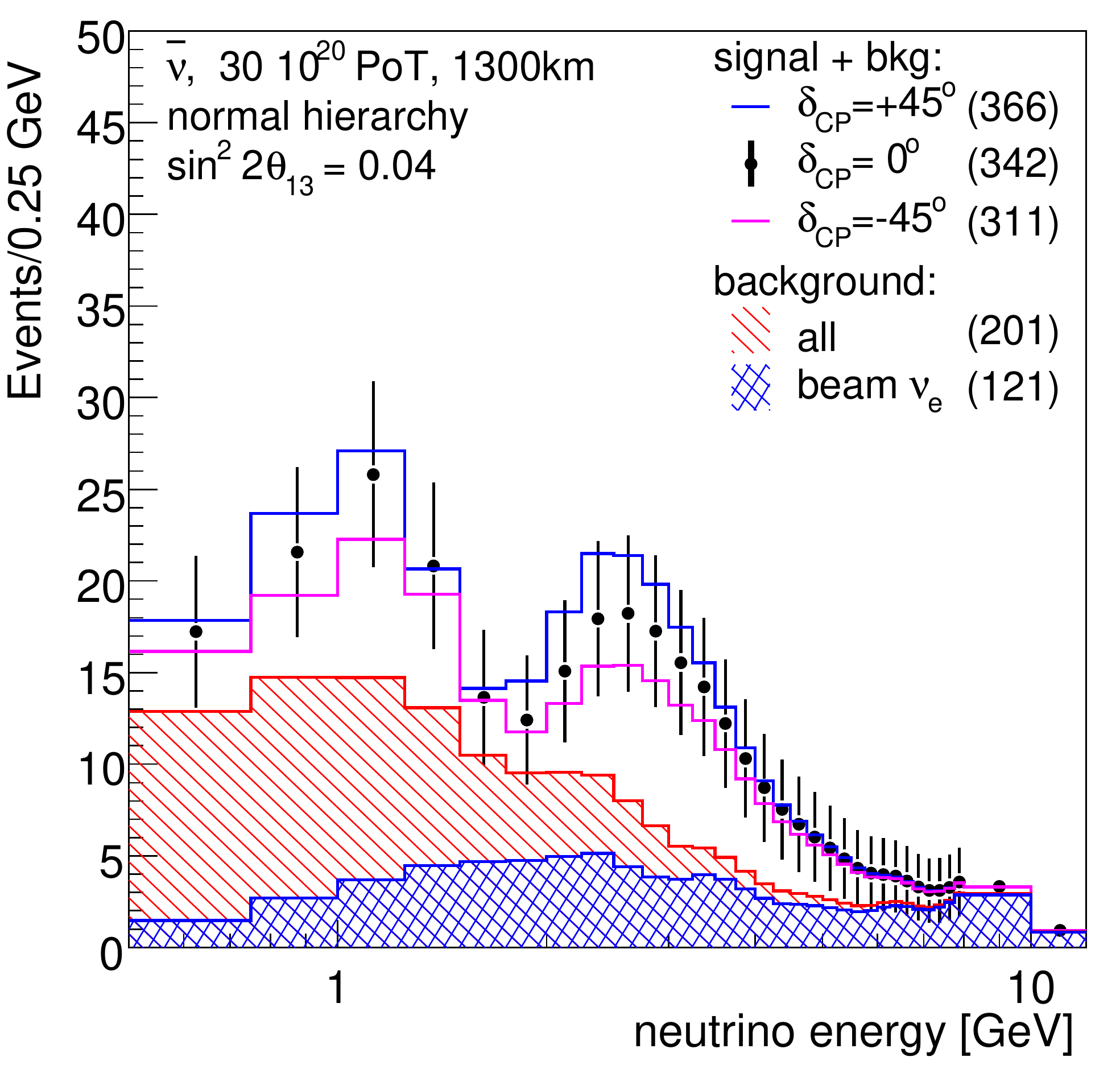}
\includegraphics[angle=0,width=0.49\textwidth]{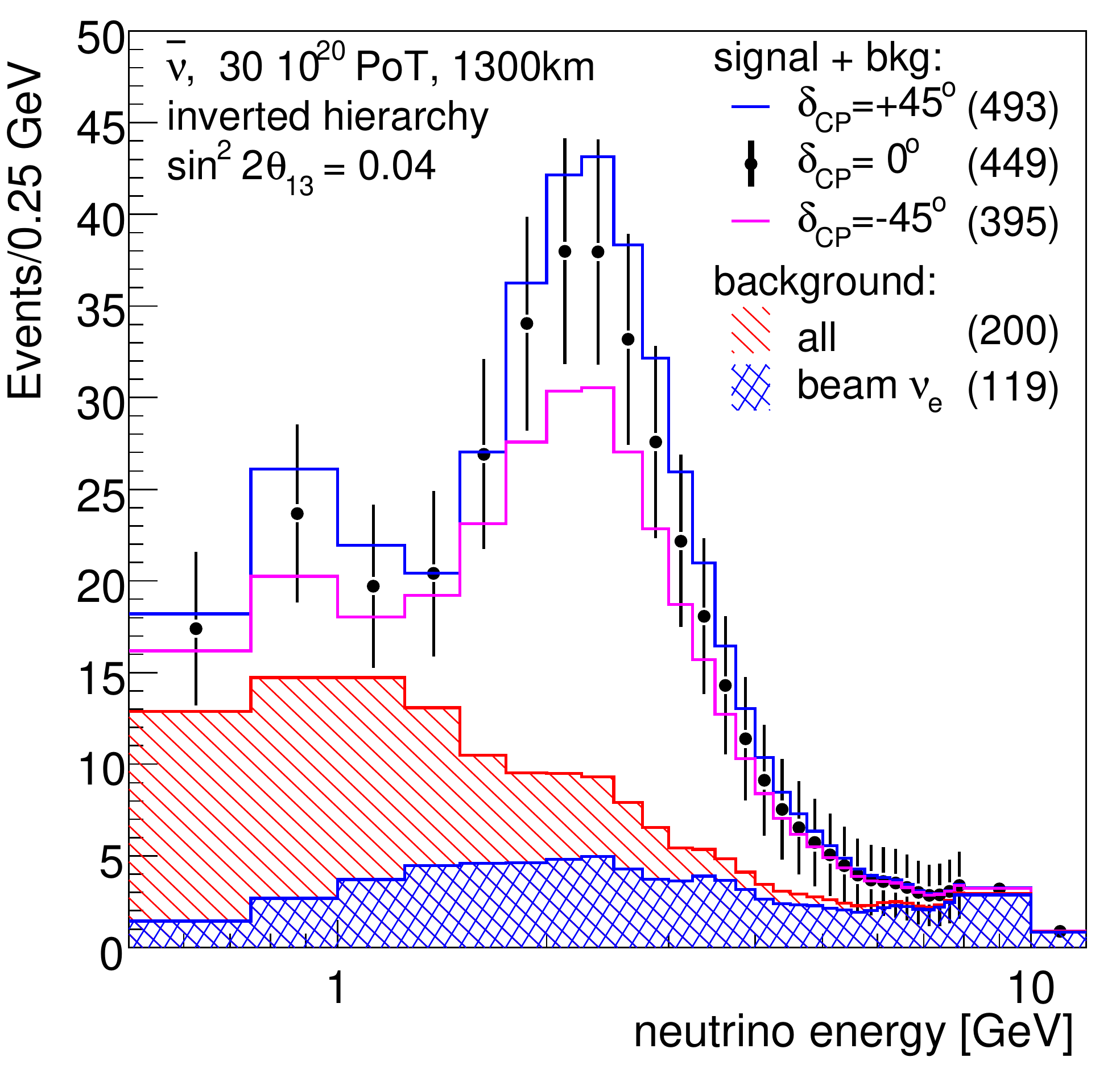}
 \caption{\it Simulation of detected electron neutrino (top plots) and
 anti-neutrino (bottom plots) spectrum (left for normal hierarchy, right
 for reversed hierarchy) for 3 values of the CP parameter $\delta_{CP}$,
 $-45^o$, $0^o$, and $-45^o$, including background
 contamination.  
This simulation is for 300 kT of water Cherenkov detector with 
the performance described in Section \ref{wcsim}.  
This is for an 
exposure of $30\times 10^{20}$ POT for each neutrino and anti-neutrino running. 
 The
 hatched histogram shows the total background. The $\nu_e$ beam
 background is also shown. The other  parameters and
 running conditions are shown in the figure.   \label{nueap} }
\end{figure}

The reconstructed electron neutrino spectrum with 300 kT of fiducial mass 
and a total exposure of $60\times 10^{20}$ protons
(divided equally between neutrinos and antineutrino running)  is shown in 
Figure \ref{nueap}. This spectrum includes effects of 
nuclear motion, detector resolution, detector signal efficiency, and background 
rejection using the performance as described  in 
Section \ref{wcsim}.  The plot is made for  $\sin^2 2 \theta_{13}=0.04$
and a baseline of 1300 km. The plots for 1480 km can be obtained from  \cite{website}.  
The left plots are for the normal mass hierarchy ($m_1 < m_2 < m_3$). 
The right hand plots are for the reversed  hierarchy ($m_3< m_1 < m2$). 
The top plots are for neutrino running and bottom plots are for anti-neutrino running. 

By fitting the spectra in Figure \ref{nueap} we can extract 
the parameters $\theta_{13}$, $\delta_{CP}$, and the mass hierarchy. 
We calculate a $\chi^2$ function and extract the confidence levels 
for a simultaneous fit to these three parameters. 
For the input values of the other oscillation parameters we assume 
1 sigma errors as follows: 

$$ \theta_{12} =0.59\pm 10\%, \Delta m^2_{21}=(0.86\pm 10\%)\times 10^{-5}$$
$$\theta_{23} = \pi/4 \pm 5\%, \Delta m^2_{31}=(2.7\pm 5\%) \times 10^{-3}$$ 

We also include 5\% error on the matter density. The calculation includes 
correlations between all parameters and accounts for possible degeneracies.   
The spectra were fit with statistical errors and with 10\% systematic error on 
the background and 1\% systematic error on the normalization with no correlations 
between neutrino and anti-neutrino channels.  
Details of the analysis method  are in \cite{wble-glb} where the same analysis 
was performed with a different spectrum and detector performance.

\begin{figure}[h] 
\includegraphics[angle=0,width=0.45\textwidth]{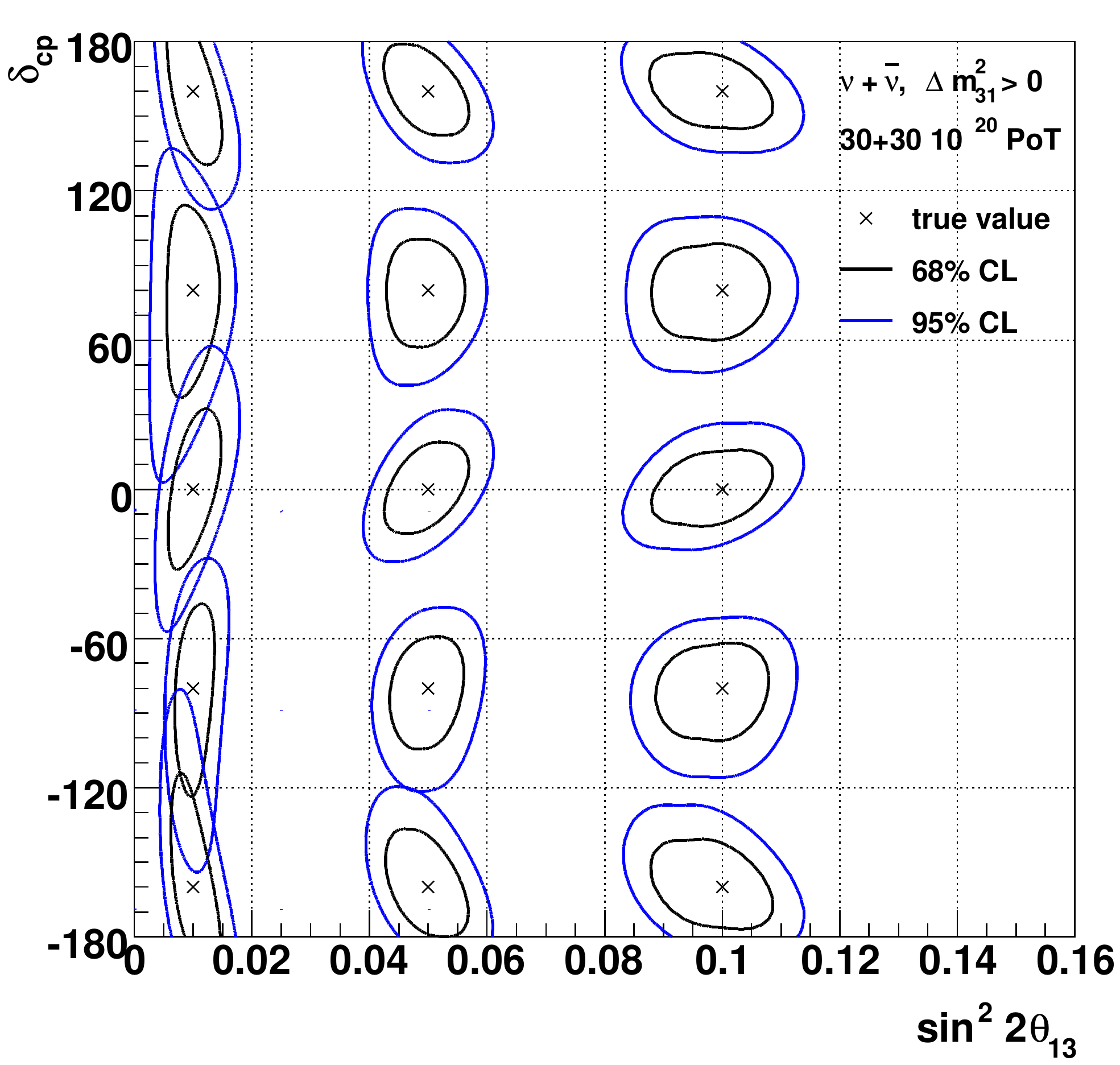} 
\includegraphics[angle=0,width=0.45\textwidth]{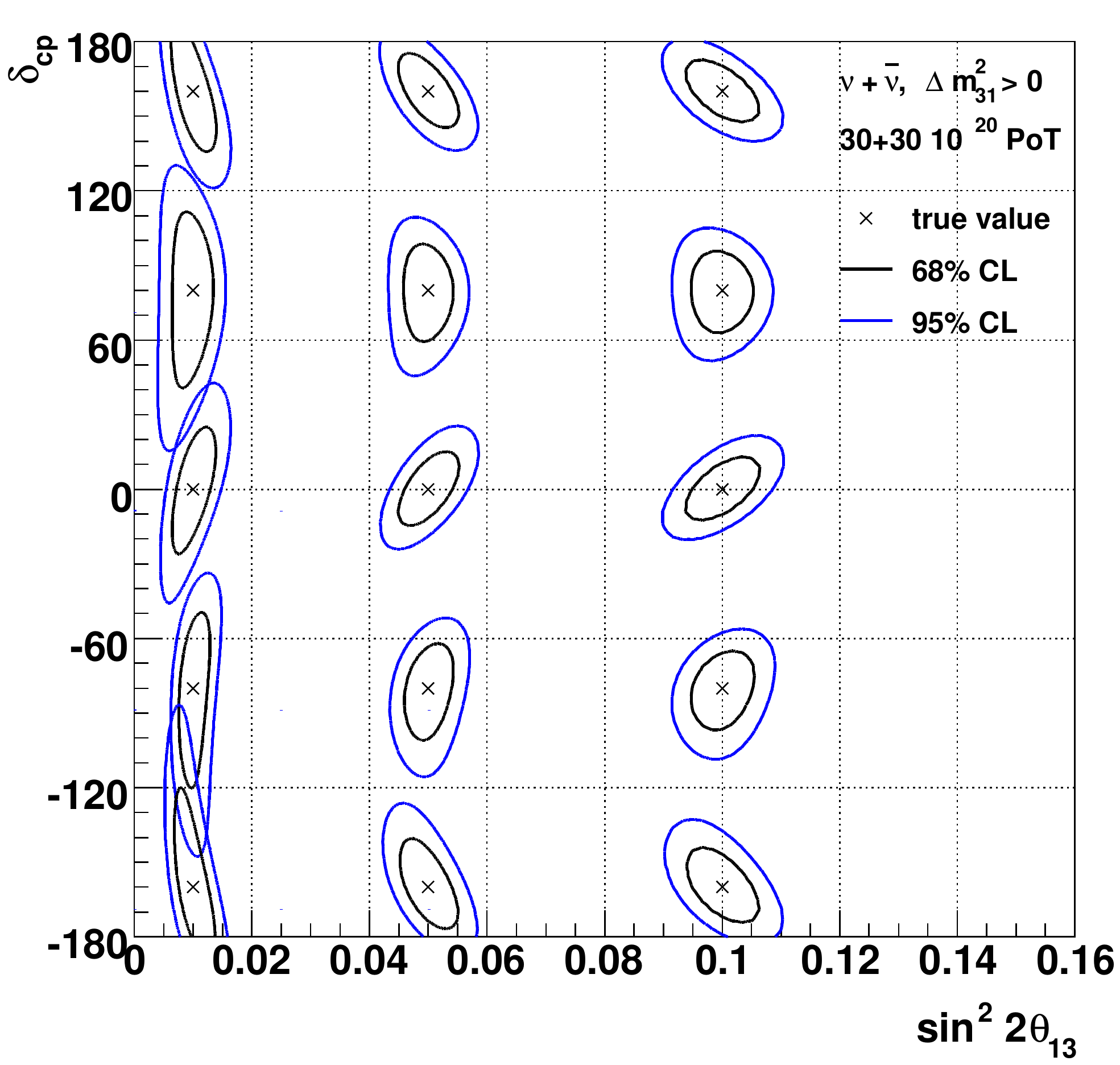} 
 \caption{\it 
 90\% and 95\%  confidence level error contours in  $\sin^2  2 \theta_{13}$
versus $\delta_{CP}$ for statistical and systematic errors (left hand plot)
 for 15 test points.
This is for a 300 kT water Cherenkov detector with a total 
exposure of $60\times 10^{20}$ POT.
The right hand side is for statistical errors alone. 
This plot was made for normal mass hierarchy. 
We assume 10\% systematic errors on the background for this plot.
  \label{bubble} }
\end{figure}

In Figure \ref{bubble} we show the confidence level contours for measuring 
the pair of parameters ($\theta_{13}$ and $\delta_{CP}$). This calculation 
was performed for normal mass hierarchy, a baseline of 1300 km, and 
a total exposure of $60\times 10^{20}$ protons equally divided between 
neutrino and anti-neutrino running.
  The result for 1480 km 
is approximately the same.   
In the case of normal hierarchy, the neutrino data alone can be used 
to measure the parameters over a large range of parameter space. But if the 
mass hierarchy is reversed, anti-neutrino data has to be used. 
 The resolution obtained after combining both neutrino and anti-neutrino 
data  is approximately independent of mass hierarchy.   
It is clear that the parameter measurement will 
suffer from background below $\sin^2 2 \theta_{13}=0.01$, but above this value 
 the resolution on the CP phase of about $\pm 20^o$ (1 sigma) is approximately 
independent of $\theta_{13}$.

\begin{figure} 
\includegraphics[angle=0,width=0.45\textwidth]{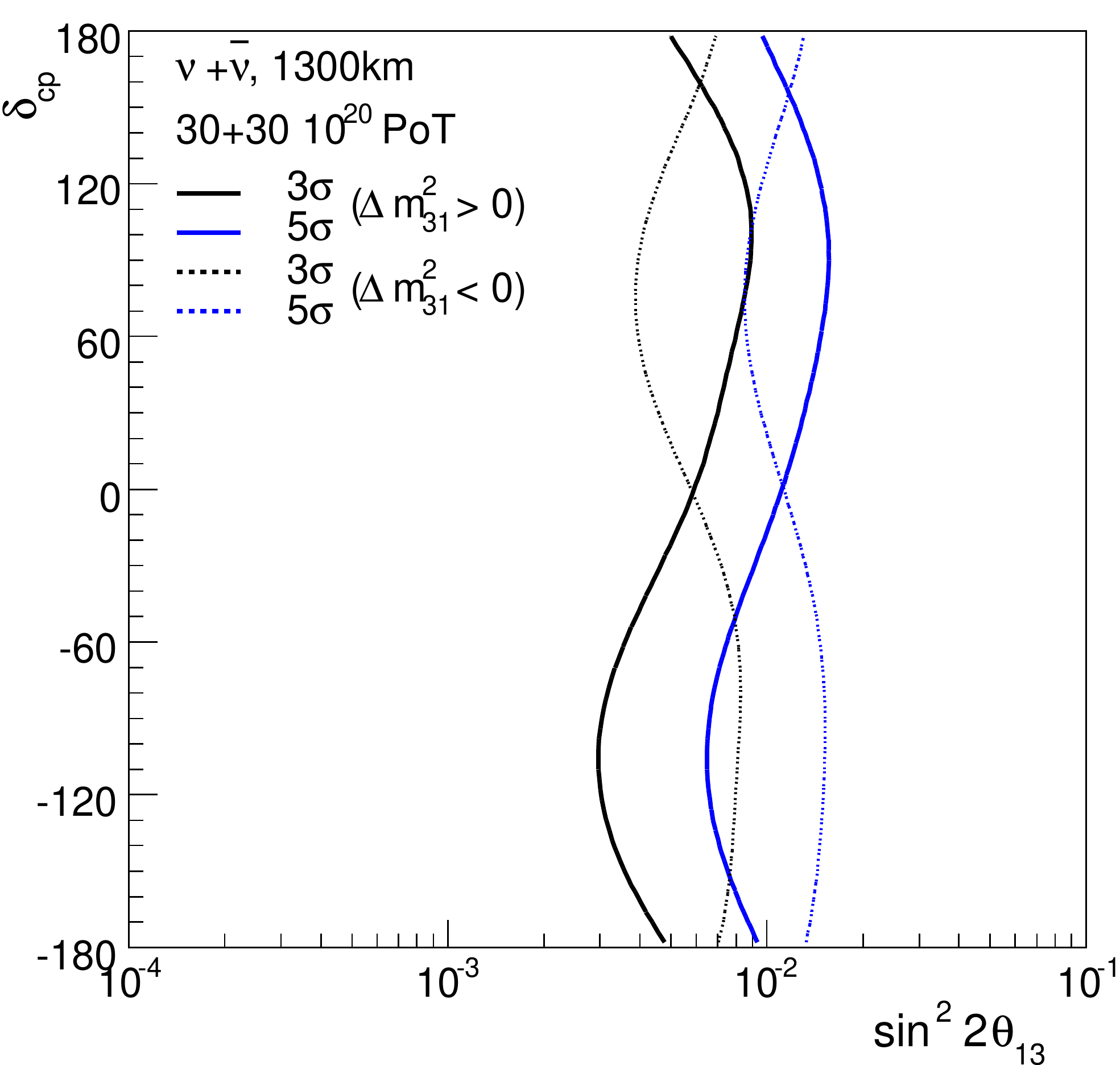} 
\includegraphics[angle=0,width=0.45\textwidth]{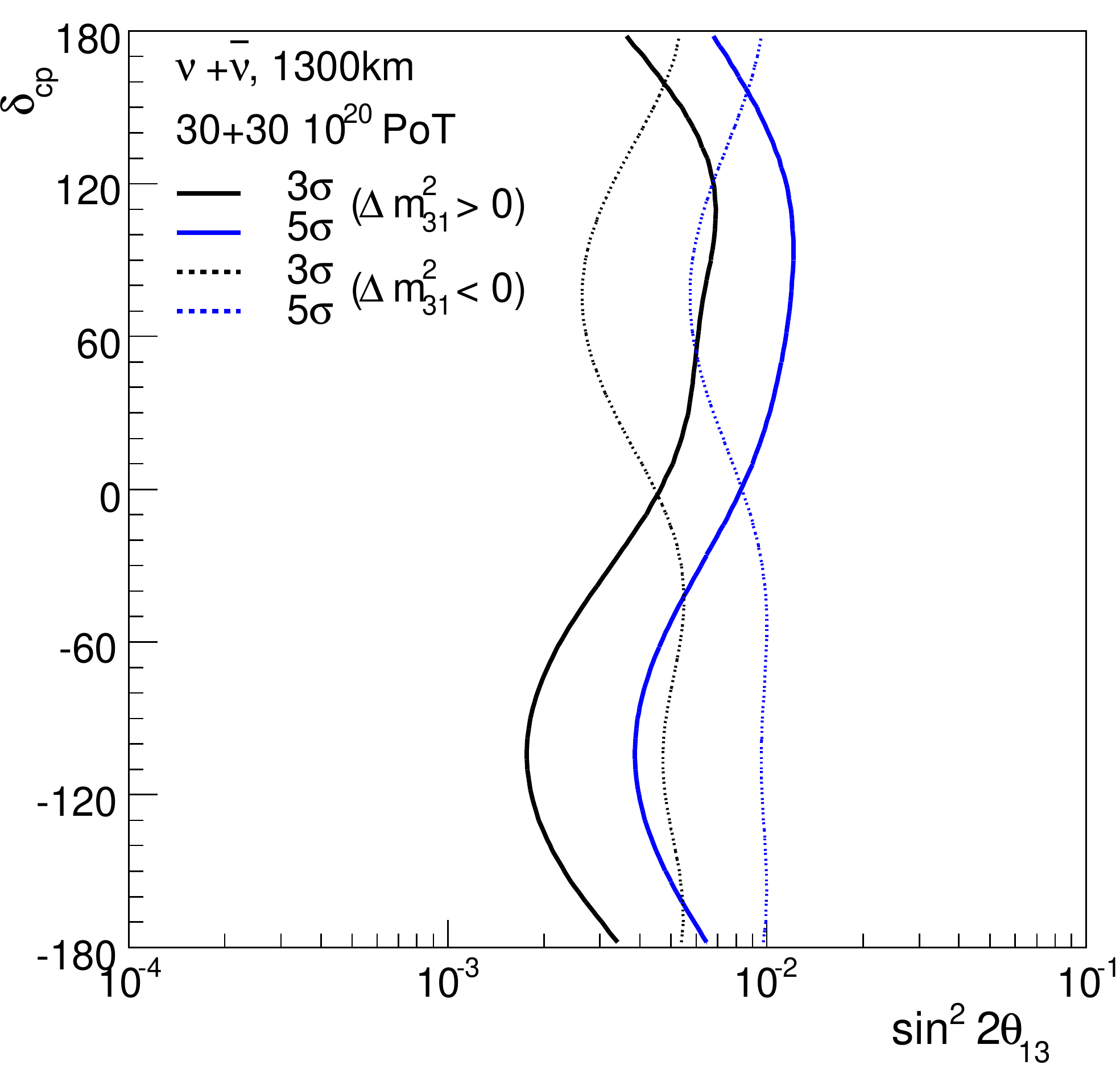} 
 \caption{\it 
 3 sigma and 5 sigma  confidence level exclusion limits for determining a non-zero value for 
$\theta_{13}$  in  $\sin^2  2 \theta_{13}$
versus $\delta_{CP}$ for statistical and systematic errors (left hand plot).
This is for a 300 kT water Cherenkov detector with a total 
exposure of $60\times 10^{20}$ POT.
The right hand side is for statistical errors alone. 
The solid (dashed) lines are for normal (reversed)  mass ordering.
We assume 10\% systematic errors on the background for this plot.
  \label{limit1} }
\end{figure}

\begin{figure}[h] 
\includegraphics[angle=0,width=0.45\textwidth]{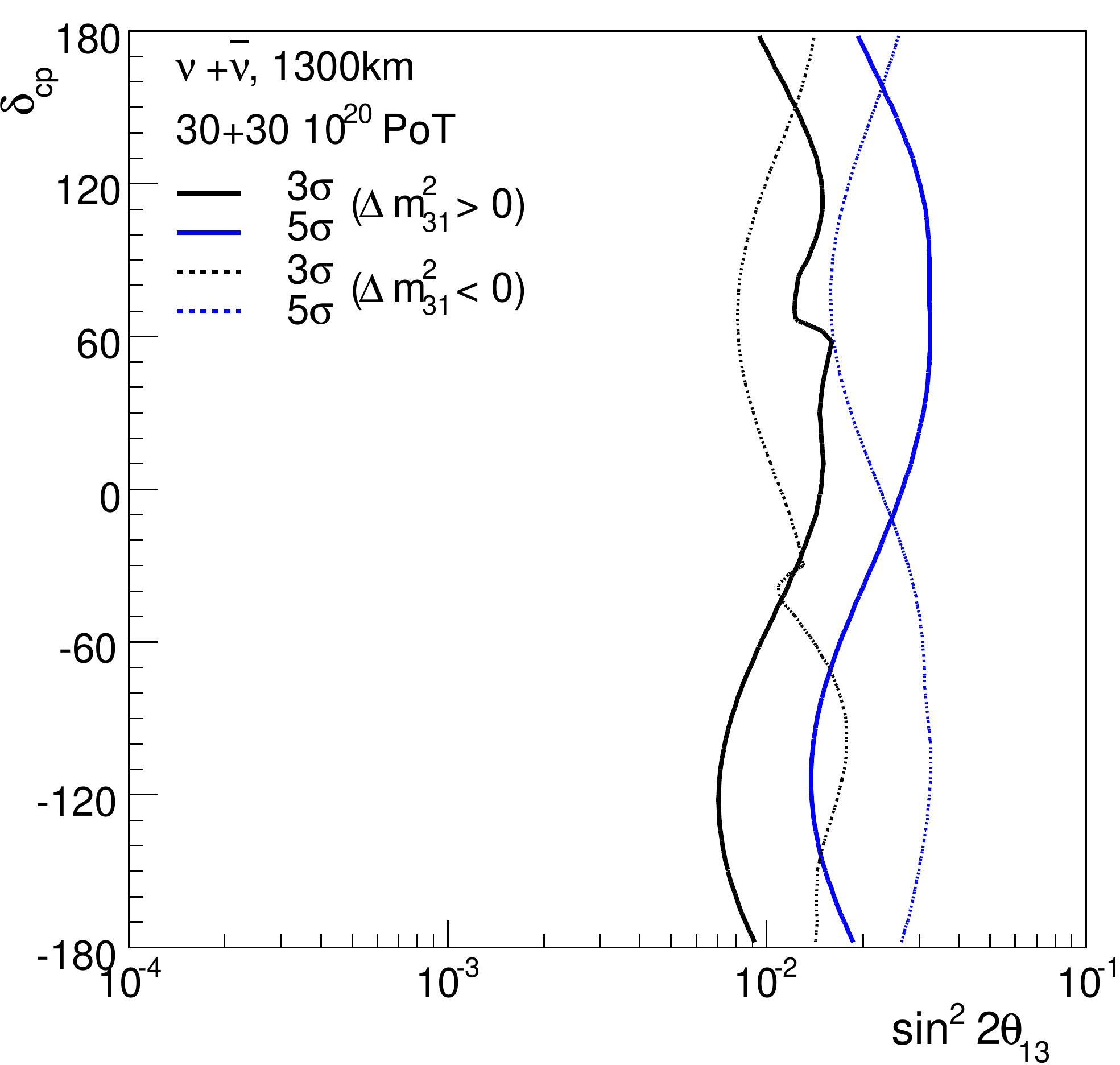} 
\includegraphics[angle=0,width=0.45\textwidth]{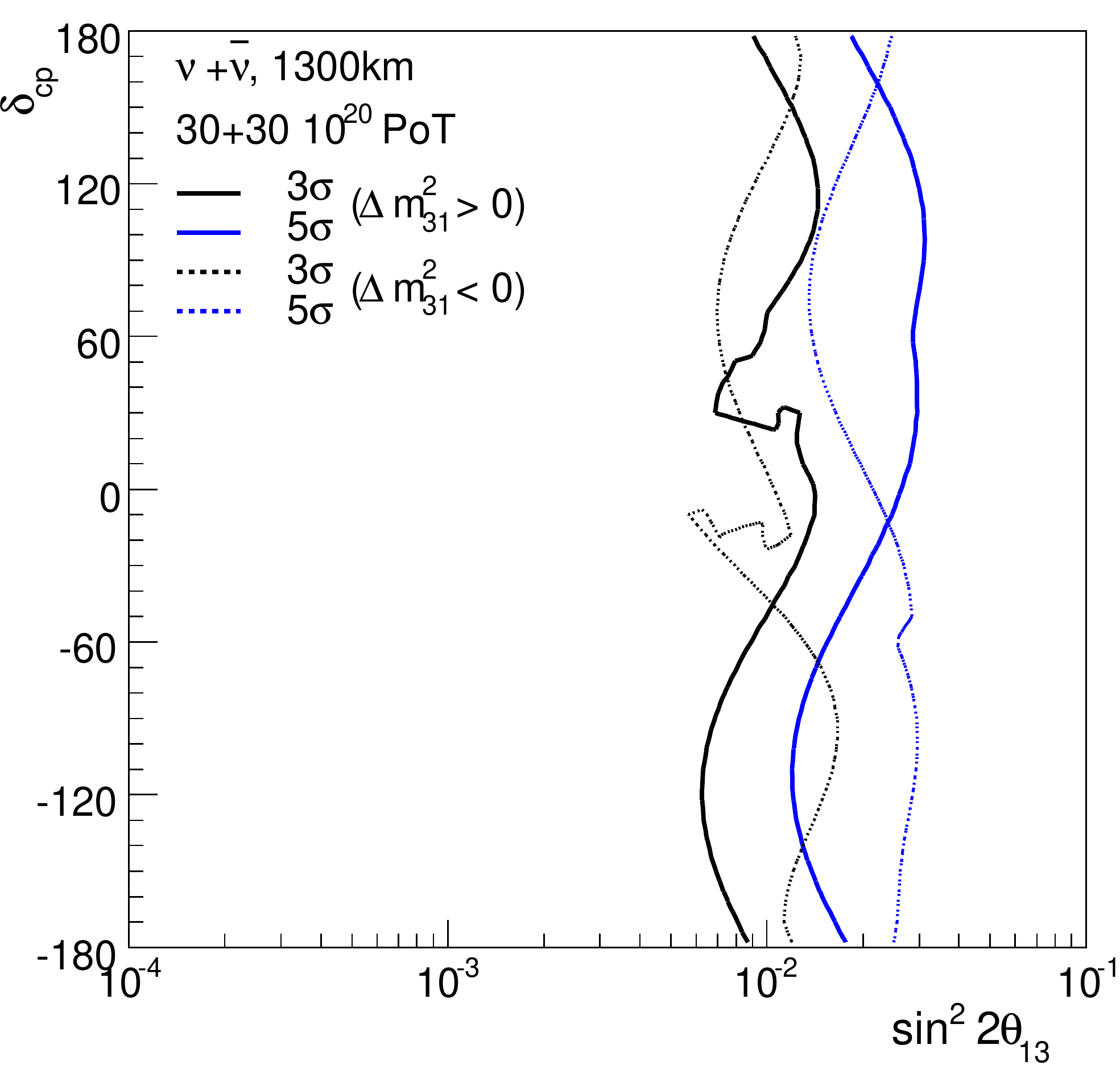} 
 \caption{\it 
 3 sigma and 5 sigma  confidence level exclusion limits for determining the 
mass hierarchy in  $\sin^2  2 \theta_{13}$
versus $\delta_{CP}$ for statistical and systematic errors (left hand plot).
This is for a 300 kT water Cherenkov detector with a total 
exposure of $60\times 10^{20}$ POT.
The right hand side is for statistical errors alone. 
The solid (dashed) lines are for normal (reversed)  mass ordering.
We assume 10\% systematic errors on the background for this plot.
  \label{limit2} }
\end{figure}

\begin{figure}[h] 
\includegraphics[angle=0,width=0.45\textwidth]{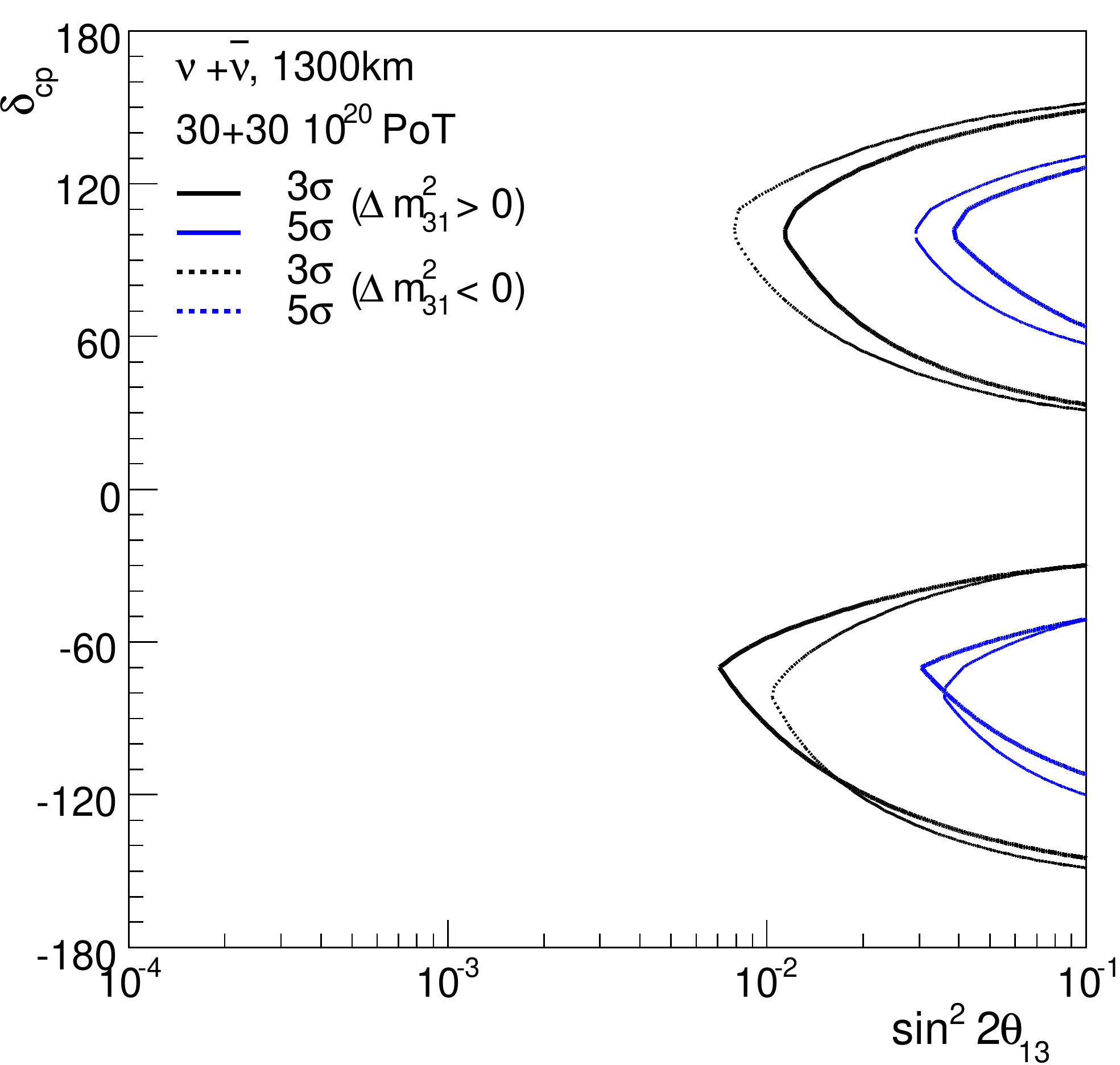} 
\includegraphics[angle=0,width=0.45\textwidth]{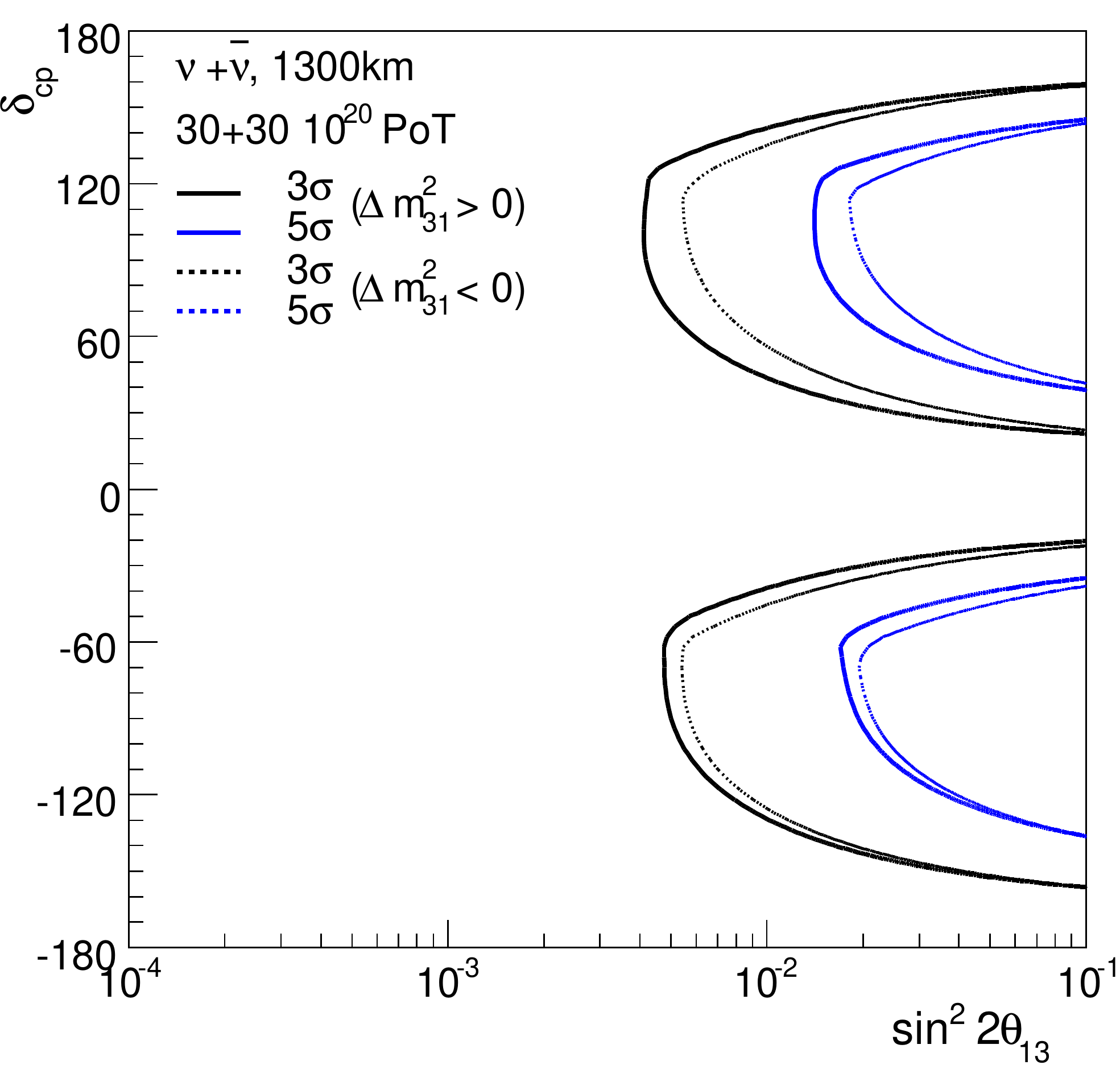} 
 \caption{\it 
 3 sigma and 5 sigma  confidence level exclusion limits for determining 
CP violation  in  $\sin^2  2 \theta_{13}$
versus $\delta_{CP}$ for statistical and systematic errors (left hand plot).
This is for a 300 kT water Cherenkov detector with a total 
exposure of $60\times 10^{20}$ POT.
The right hand side is for statistical errors alone. 
The solid (dashed) lines are for normal (reversed)  mass ordering.
We assume 10\% systematic errors on the background for this plot.
  \label{limit3} }
\end{figure}

If there is no excess of electron events observed then we can set a
limit on the value of $\sin^2 2 \theta_{13}$ as a function of
$\delta_{CP}$.  Such  sensitivity limits are shown in
Figure \ref{limit1}. 
The range of parameters over which the mass hierarchy can be resolved 
is shown in Figure \ref{limit2}. We have chosen to display the limits separately 
for the two mass hierarchies.
Some of the structure in the 3 sigma lines is due to the limited number of 
bins used in the calculation. 
The region to the right hand side of each curve excludes the opposite mass 
hierarchy at the respective confidence level. 
Similarly the range of parameters over which CP violation can be established 
(i.e. determine that $\delta_{CP}$ is not 0 or $\pi$) is displayed in Figure 
\ref{limit3}.

\clearpage

\clearpage  

\begin{figure}[h] 
\includegraphics[angle=0,width=0.45\textwidth]{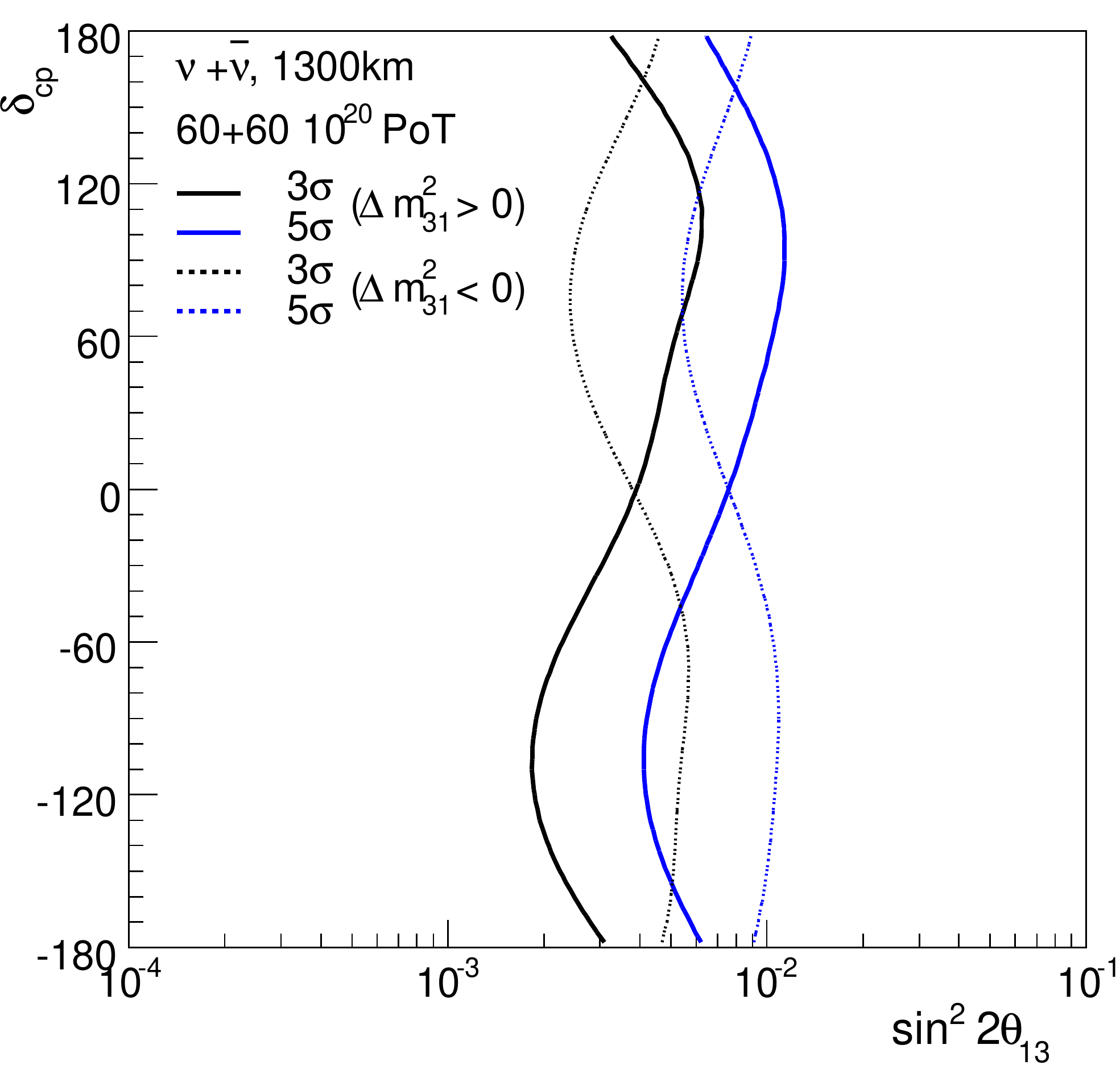} 
\includegraphics[angle=0,width=0.45\textwidth]{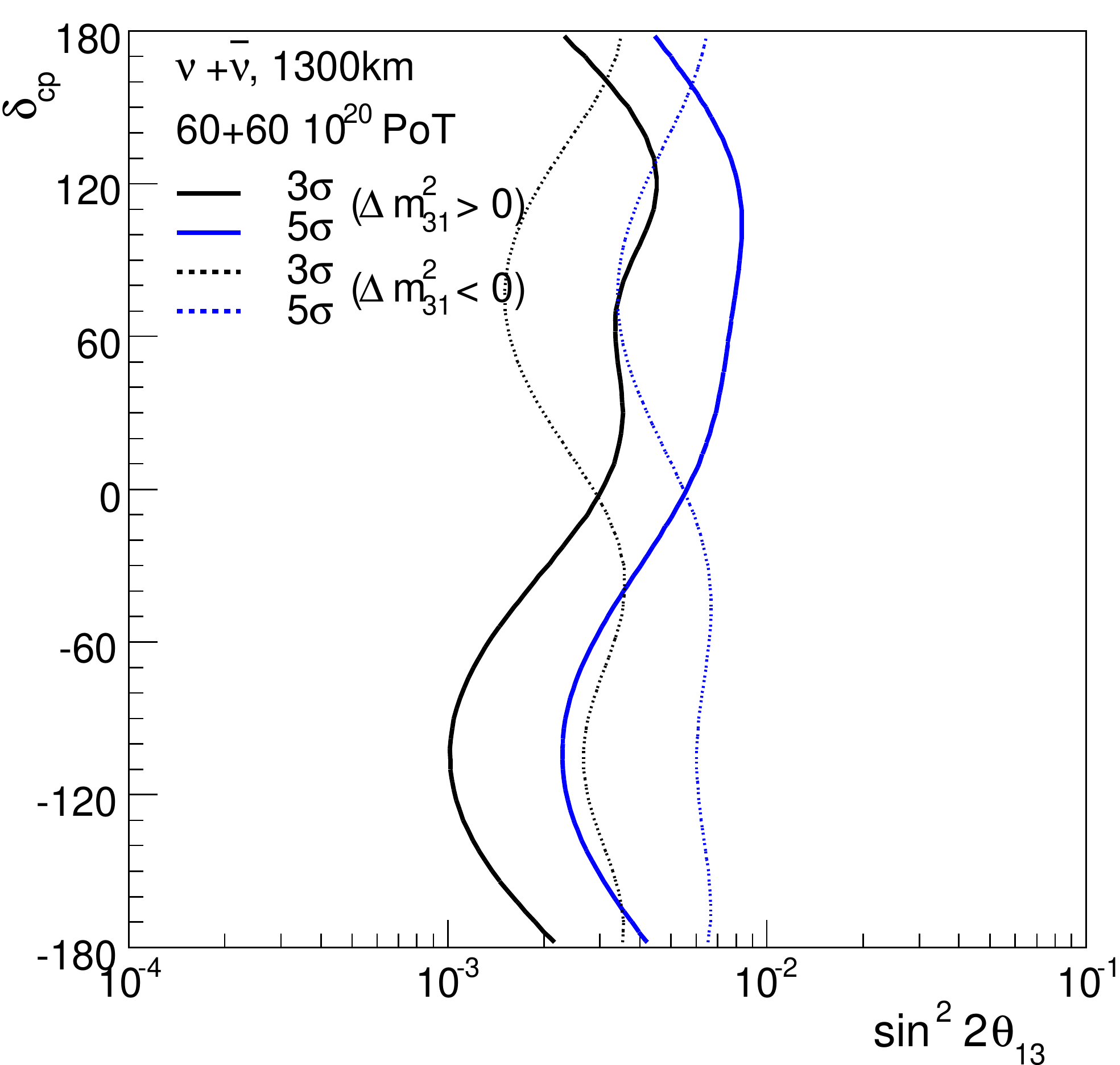} 
 \caption{\it 
 3 sigma and 5 sigma  confidence level exclusion limits for determining a non-zero value for 
$\theta_{13}$  in  $\sin^2  2 \theta_{13}$
versus $\delta_{CP}$ for statistical and systematic errors (left hand plot).
This is for a 300 kT water Cherenkov detector with a total 
exposure of $120\times 10^{20}$ POT.
The right hand side is for statistical errors alone. 
The solid (dashed) lines are for normal (reversed)  mass ordering.
We assume 10\% systematic errors on the background for this plot.
  \label{limit160} }
\end{figure}

\begin{figure}[h] 
\includegraphics[angle=0,width=0.45\textwidth]{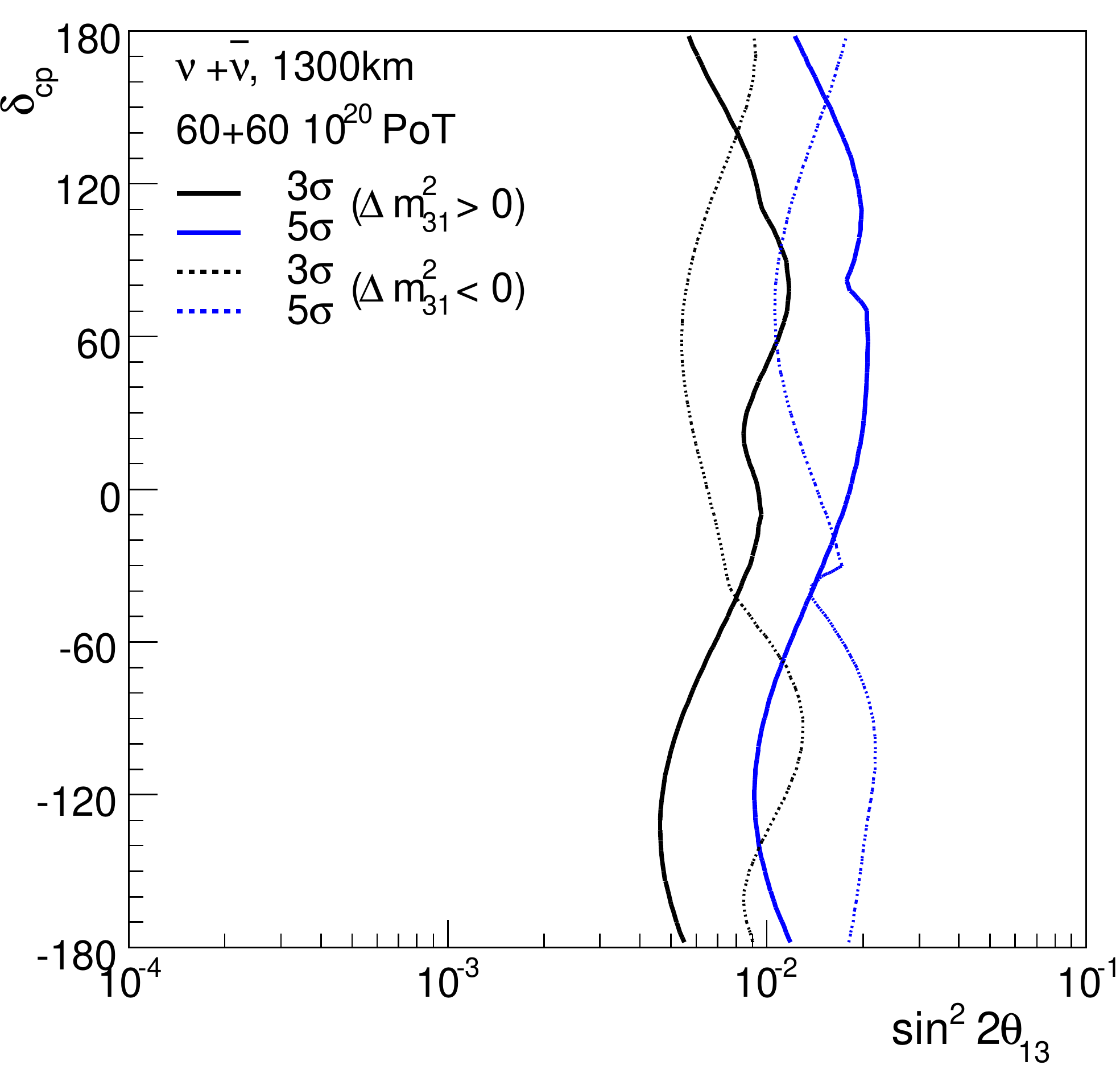} 
\includegraphics[angle=0,width=0.45\textwidth]{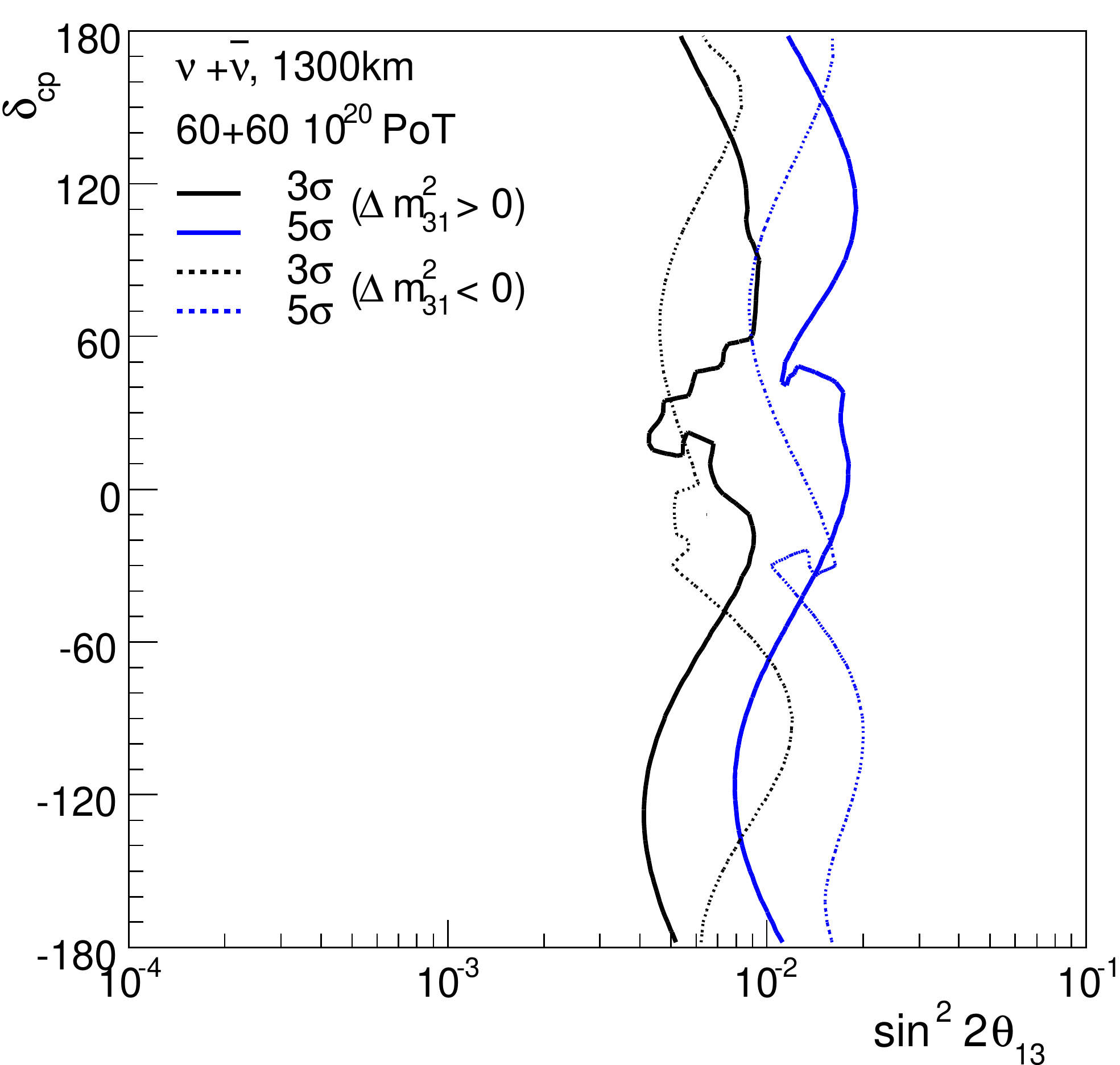} 
 \caption{\it 
 3 sigma and 5 sigma  confidence level exclusion limits for determining the 
mass hierarchy in  $\sin^2  2 \theta_{13}$
versus $\delta_{CP}$ for statistical and systematic errors (left hand plot).
This is for a 300 kT water Cherenkov detector with a total 
exposure of $120\times 10^{20}$ POT.
The right hand side is for statistical errors alone. 
The solid (dashed) lines are for normal (reversed)  mass ordering.
We assume 10\% systematic errors on the background for this plot.
  \label{limit260} }
\end{figure}

\begin{figure}[h] 
\includegraphics[angle=0,width=0.45\textwidth]{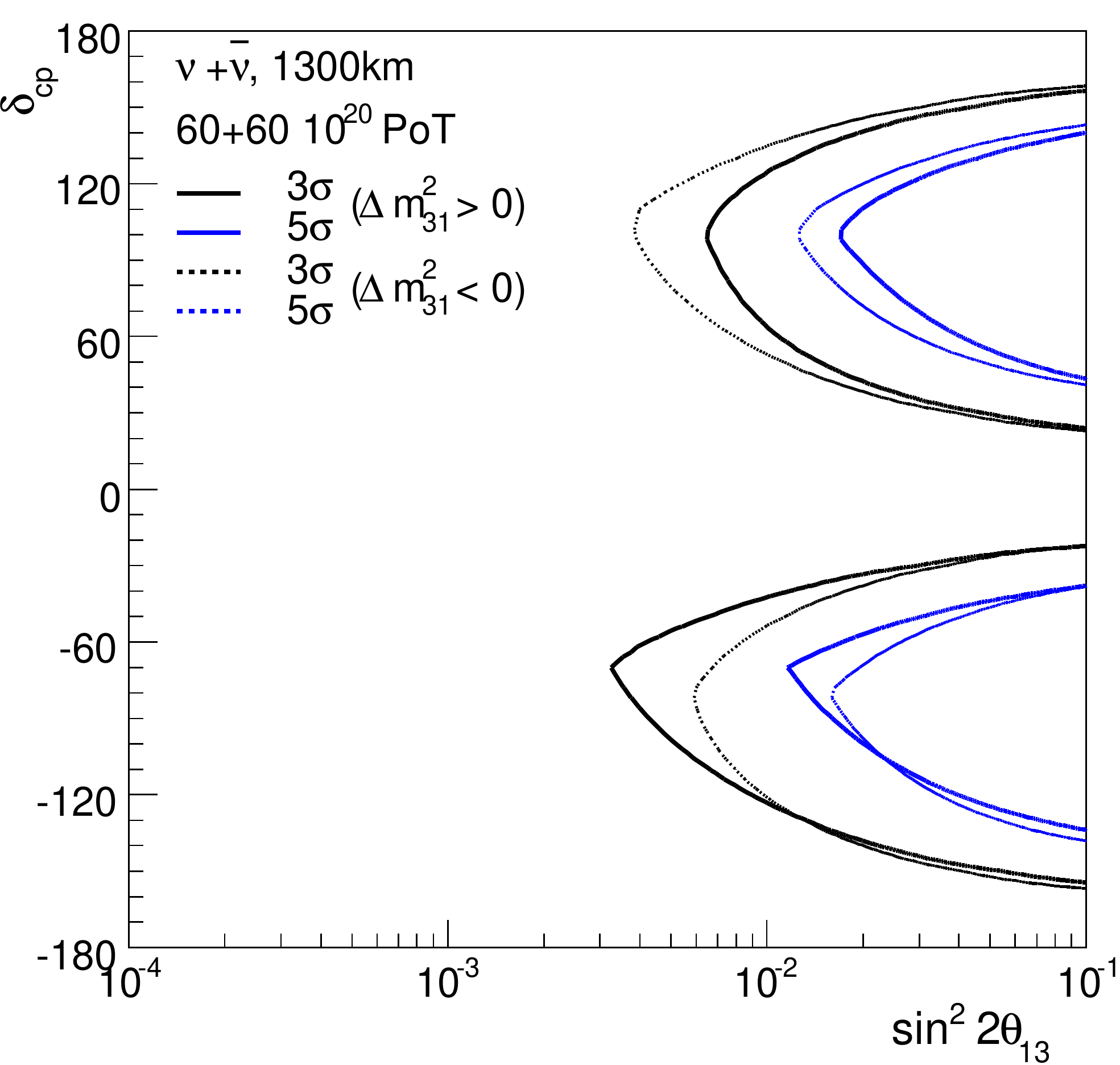} 
\includegraphics[angle=0,width=0.45\textwidth]{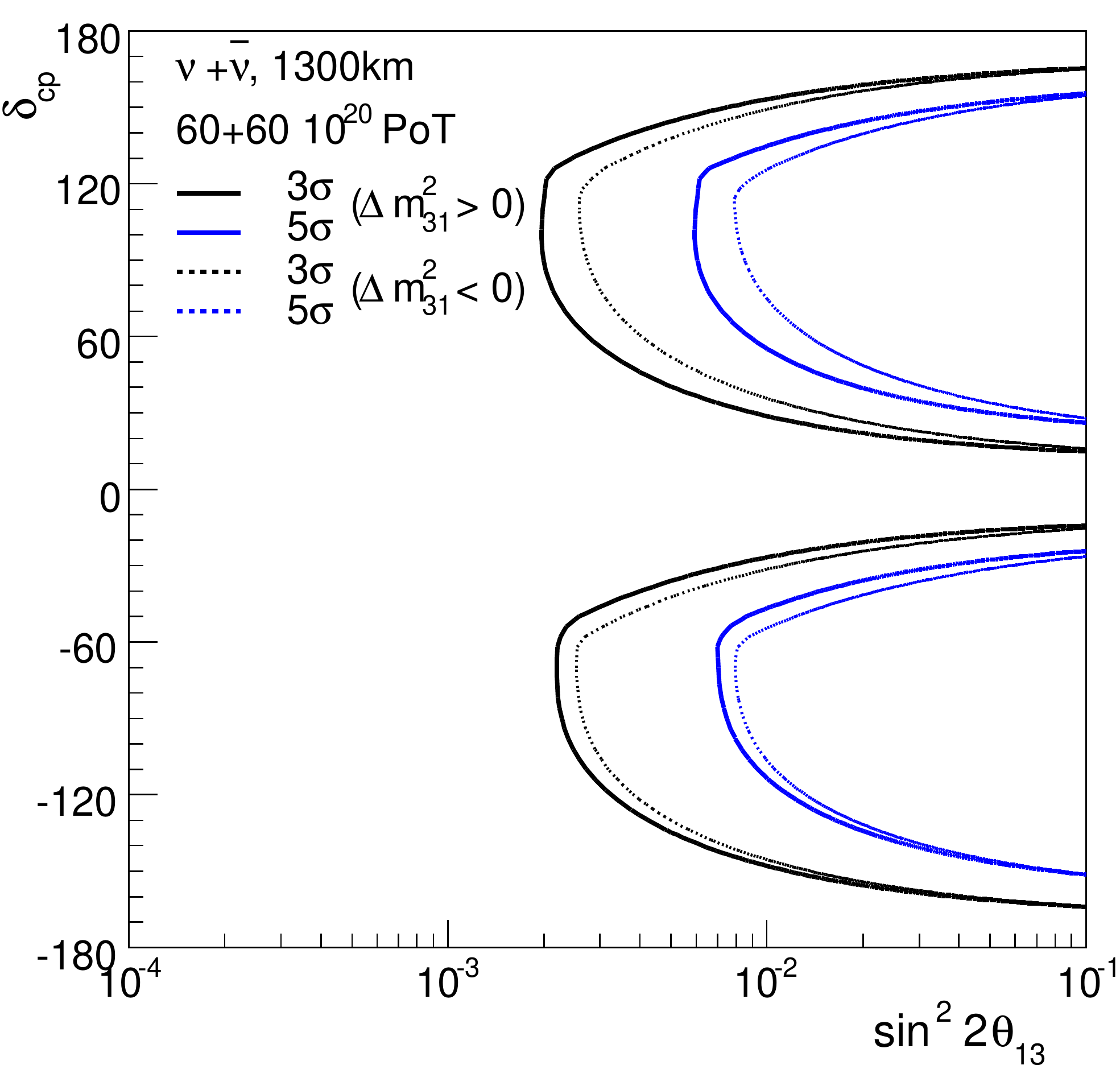} 
 \caption{\it 
 3 sigma and 5 sigma  confidence level exclusion limits for determining 
CP violation  in  $\sin^2  2 \theta_{13}$
versus $\delta_{CP}$ for statistical and systematic errors (left hand plot).
This is for a 300 kT water Cherenkov detector with a total 
exposure of $120\times 10^{20}$ POT.
The right hand side is for statistical errors alone. 
The solid (dashed) lines are for normal (reversed)  mass ordering.
We assume 10\% systematic errors on the background for this plot.
  \label{limit360} }
\end{figure}

{\bf Sensitivity variation with exposure}: 
The exposure assumed in the above plots was 
$30\times 10^{20}$ protons on target for each neutrino 
and antineutrino running.  This corresponds to 3 years of running for 
each polarity for 1.2  MW of beam power  and $1.7\times 10^7$ sec per year of 
running at 120 GeV.
If we were to run the antineutrino beam for twice the exposure of neutrino, then the 
the number of events is approximately balanced between $\nu$ and $\bar\nu$.
 Such unequal running 
is advantageous in the case of the reversed hierarchy. 
 An analysis of the total exposure was performed in  \cite{wble-glb}.
It was found that  longer exposures will have relatively modest effect on
the sensitivity to $\sin^2 2 \theta_{13}$ and the mass hierarchy resolution, 
but could be important for improving the precision on the CP violation measurement.  
Exclusion contours for twice the exposure 
(total exposure of $120 \times 10^{20}$ protons) are shown in 
Figures \ref{limit160} (for determining non-zero $\theta_{13}$), 
\ref{limit260} (for determining mass hierarchy), and \ref{limit360}
(for determining CP violation).

\begin{figure}[h]
\includegraphics[width=0.7\textwidth]{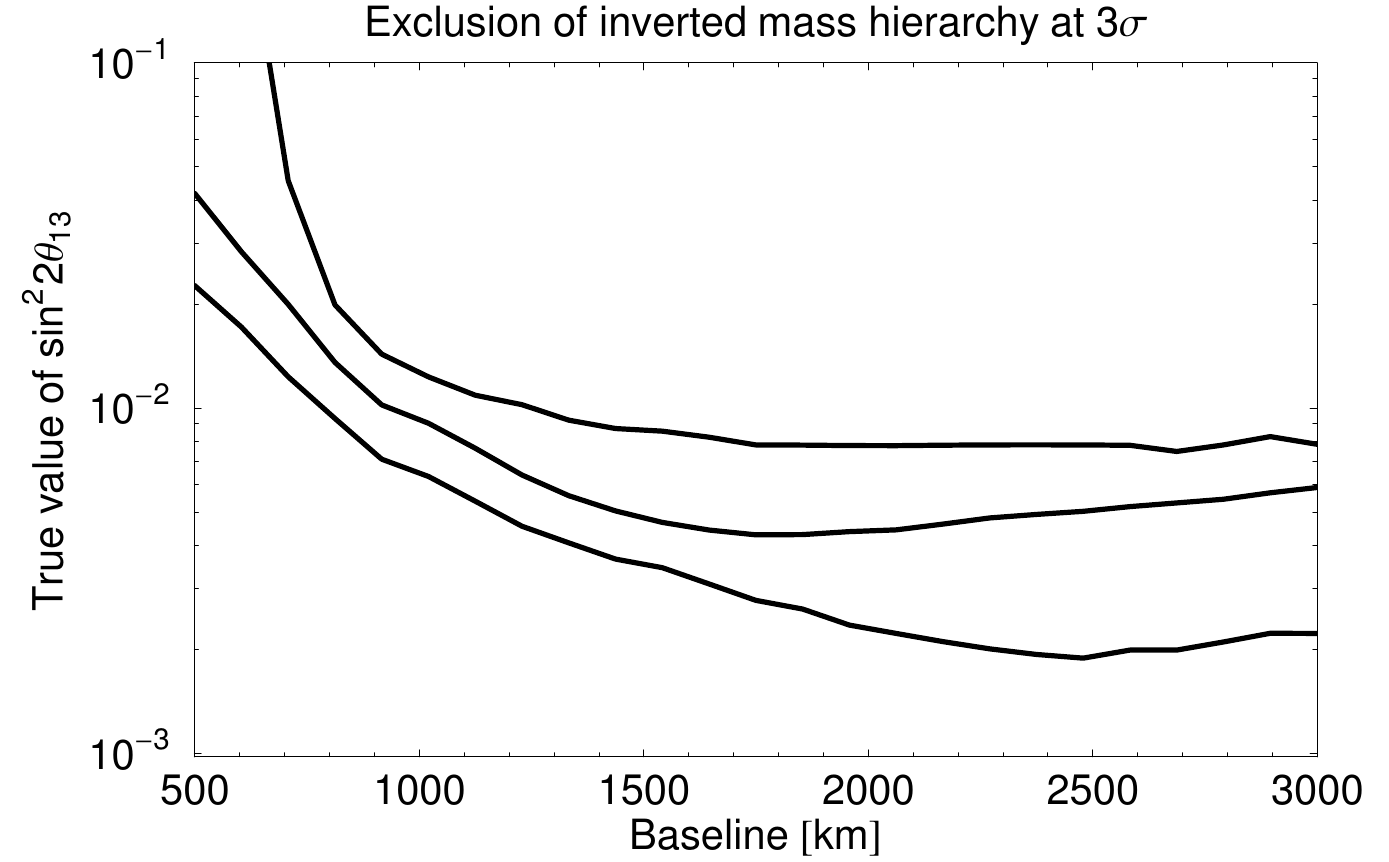}
 \caption{\label{MHbase}
   Discovery reach for a normal mass hierarchy at $3\sigma$ for CP
   fractions 0 (lower-most line, best case), 0.5 (middle line) and 1
   (uppermost line, worst case) as a function of the baseline. The
   detector mass, beam power and exposure are kept the same  for all
baselines. For further explanation of the plot please see \cite{wble-glb}. }
\end{figure}

{\bf Sensitivity variation with distance:}
Analysis in \cite{wble-glb} showed that there is significant variation in 
sensitivity up to 1500 km for determination of the mass hierarchy. This is 
reproduced in Figure \ref{MHbase}.   The variation in sensitivity to 
$\theta_{13}$ was found to be mild partly because of the larger matter 
enhancements (in neutrino (antineutrino) mode for normal (reversed) mass hierarchy)
at longer distances.  There is a slow decrease in the sensitivity to CP violation 
at longer distances, but this is attributed to the shape of the spectrum used for 
this calculation.  At distances below 1000 km, there is a  degradation in the 
CP sensitivity  because of the need to resolve the 
ambiguity due to the mass hierarchy with the same data.
Complete calculations with same assumptions for detector size and performance 
and spectra can be obtained from the \cite{website} for distances up to 2600 km; 
this covers the various options for the DUSEL locations.


{\bf Sensitivity variation due to systematics and parameter variation:} 

The sensitivity calculation reported in this section follows 
the prescription from \cite{globes, wble-glb}. They include the parameter 
variation as described above (Section \ref{sensi_1_1}).  
In addition, we assume a 10\% systematic error on the total background.
Considering recent and past experience with background determination 
in long baseline experiments, the 10\% systematic error is very likely a 
pessimistic assumption \cite{miniboone, e776}, especially with a planned near detector. 
Sensitivity estimates with other assumptions for the systematic error 
can be obtained from the website \cite{website}. 
The conclusion from these studies is that the background systematic
 error will most likely dominate over the parameter variation.  
Therefore, in Figures \ref{bubble} to \ref{limit360}, we have chosen to 
show the 
sensitivity with and without systematic errors. 
Our  conclusion is that the wide band technique 
which leads to  a spectrum measurement is robust against parameter changes and 
background systematic errors over a reasonable range \cite{wble-glb}.

\clearpage 

\subsubsection{Liquid Argon Detector} 
\label{sensi_1_2}

\begin{figure}[h]
\includegraphics[angle=0,width=0.49\textwidth]{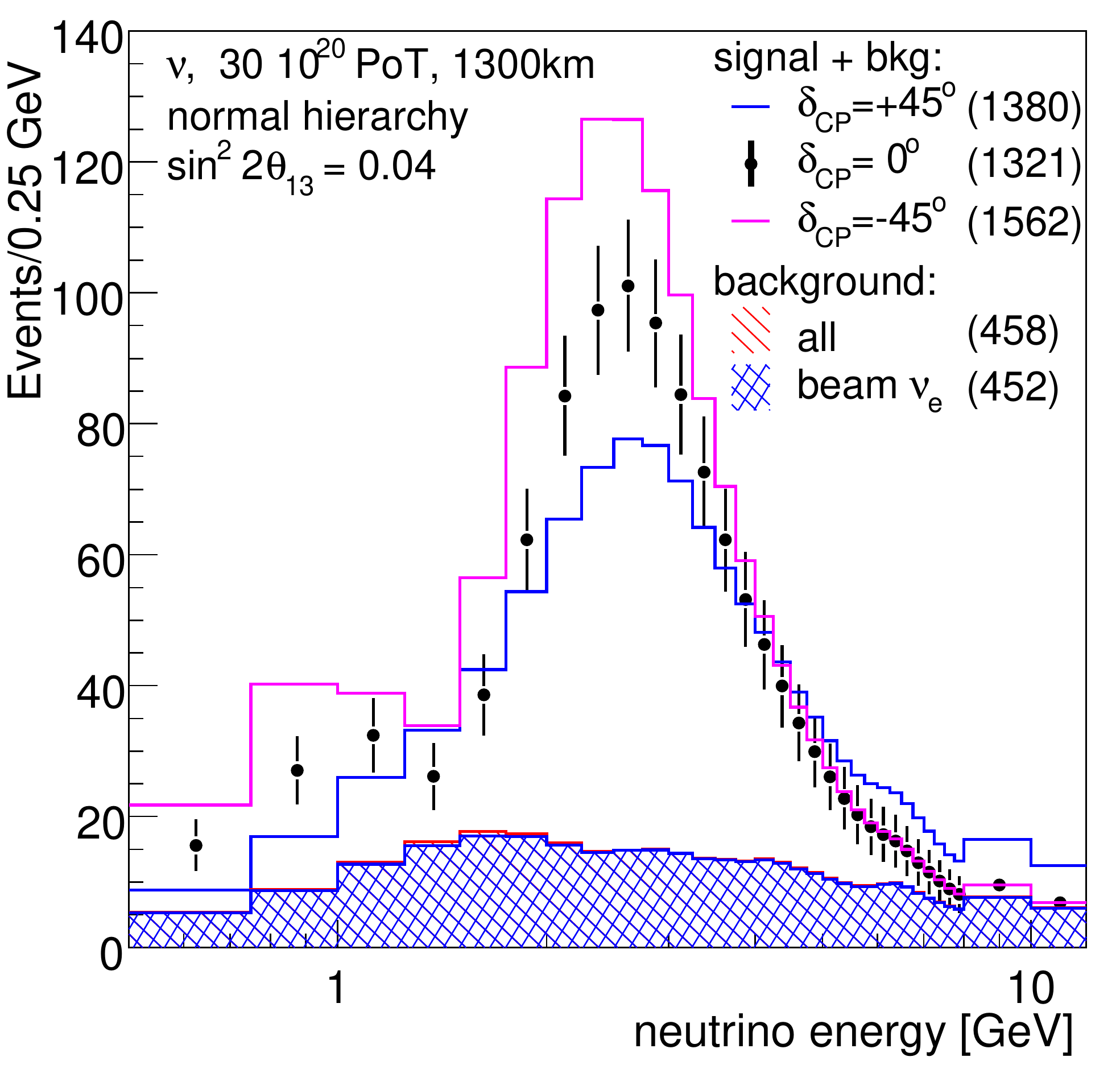}
\includegraphics[angle=0,width=0.49\textwidth]{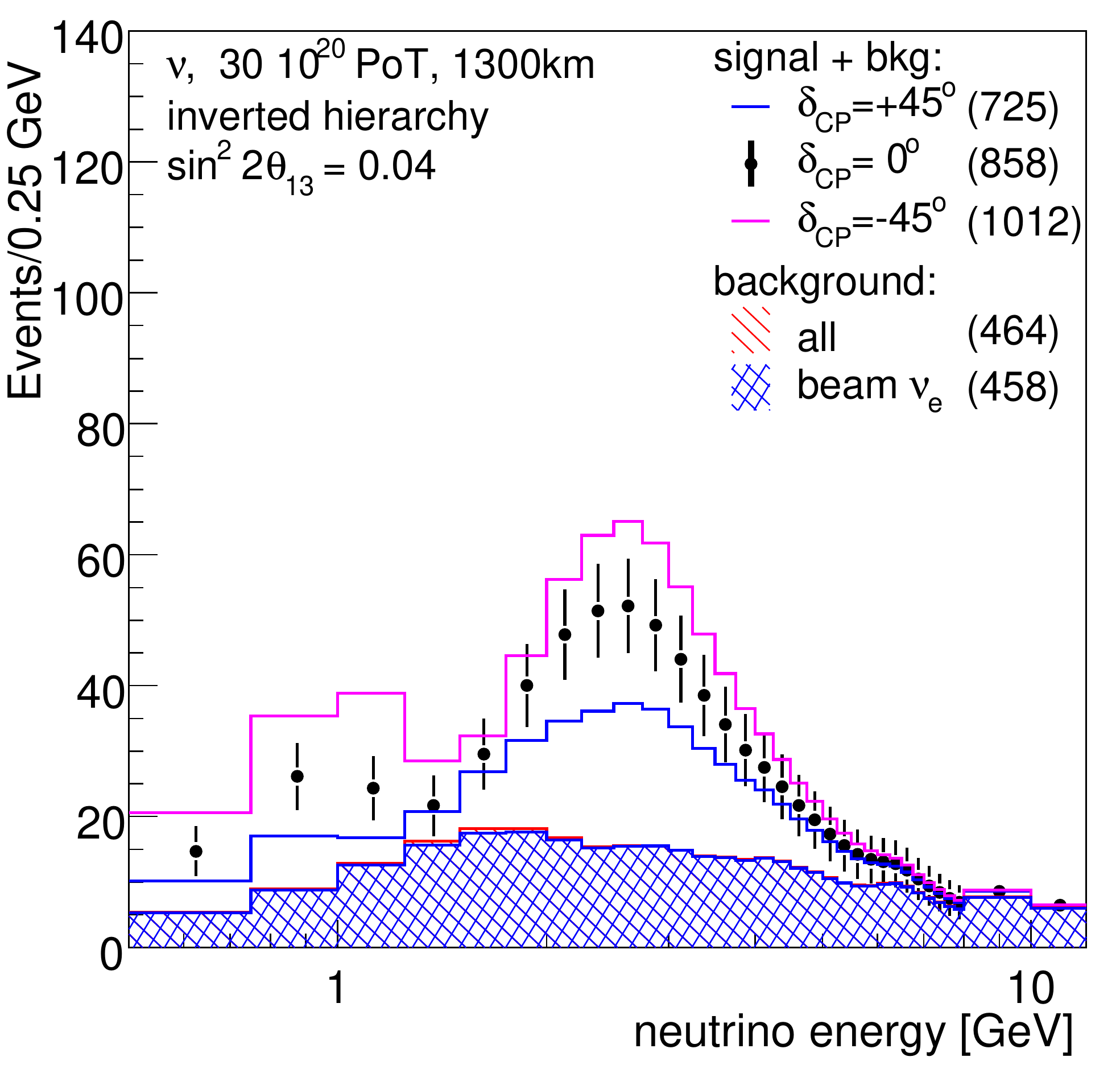} \\
\includegraphics[angle=0,width=0.49\textwidth]{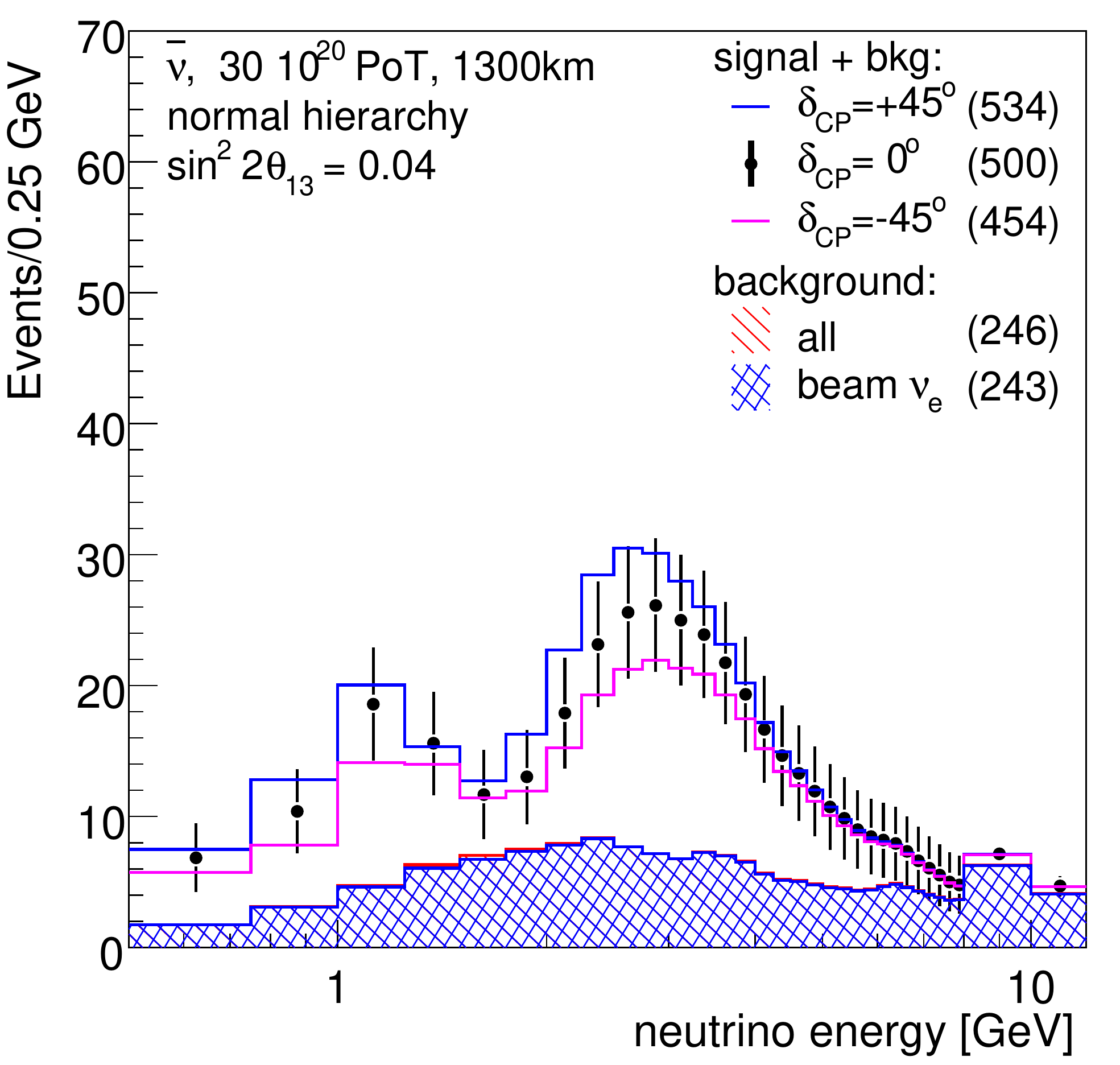}
\includegraphics[angle=0,width=0.49\textwidth]{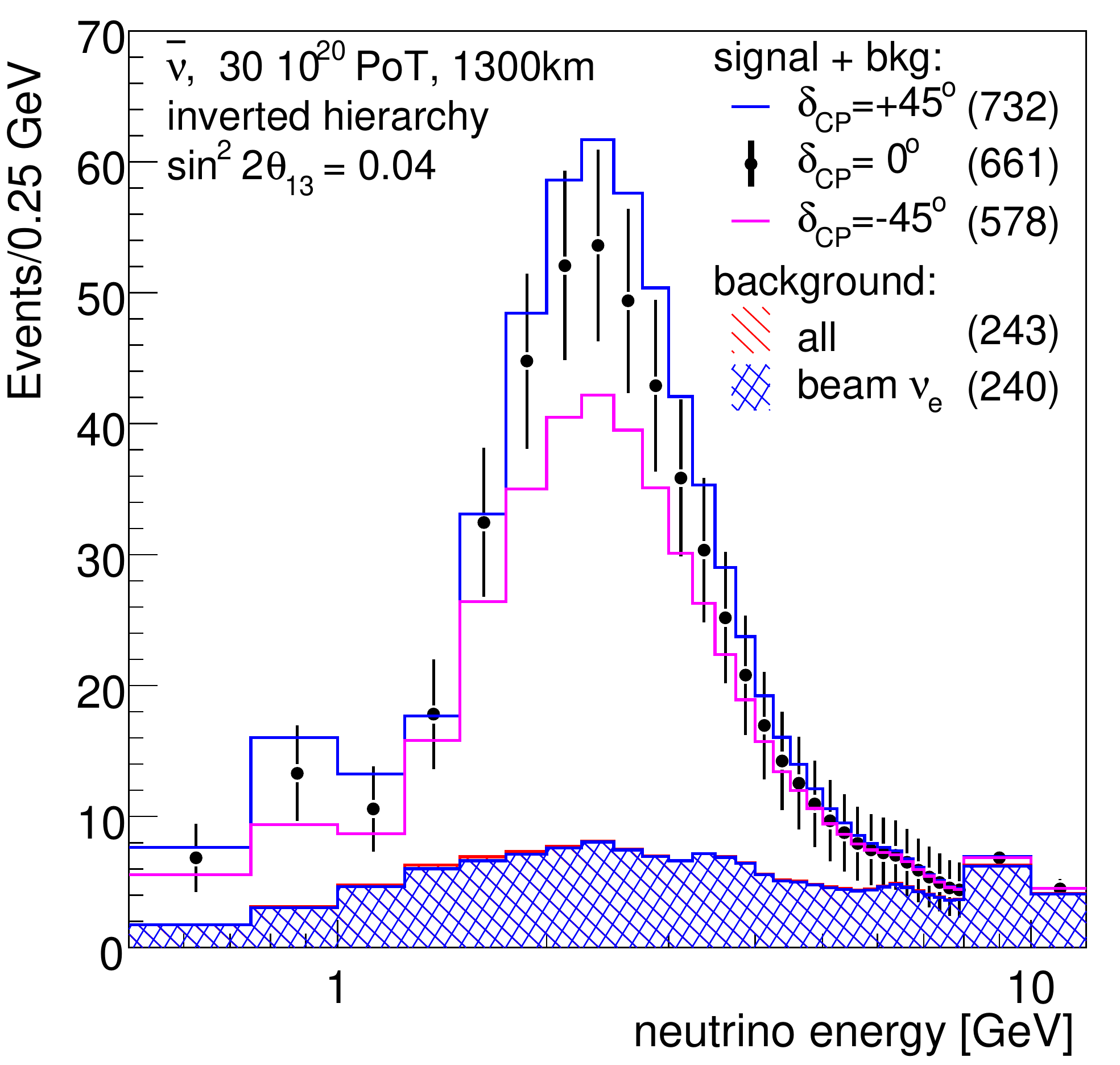}
 \caption{\it Simulation of detected electron neutrino (top plots) and
 anti-neutrino (bottom plots) spectrum (left for normal hierarchy, right
 for reversed hierarchy) for 3 values of the CP parameter $\delta_{CP}$,
 $-45^o$, $0^o$, and $-45^o$, including background
 contamination.  
This simulation is for 100 kT of LAr  detector (with 
the performance described in the text) placed at DUSEL, 1300 km away from FNAL.     The
 hatched histogram shows the total background, which is dominated by the $\nu_e$ beam
 background. The other  parameters and
 running conditions are shown in the figure.   \label{nueaplar} }
\end{figure}

If a 100 kTon fiducial mass liquid argon time projection chamber can be 
built, then it can be placed 
 at one of the DUSEL  sites and  used as a long baseline 
neutrino oscillation detector. We have assumed that such a detector 
can have 80\% efficiency for all charged current electron neutrino 
events and has background rejection capability  
(Section \ref{larsim}) that virtually rejects all NC and CC backgrounds.
We further assume that the detector will have resolution characterized by 
$20\%/\sqrt{E/GeV}$ for non-quasielastic events and 
$5\%/\sqrt{E/GeV}$ for quasielastic events.  
The spectra that result from these assumptions are displayed in 
Figure \ref{nueaplar} for the same parameters and exposure as 
Figure \ref{nueap}.  

\begin{figure}[h] 
\includegraphics[angle=0,width=0.45\textwidth]{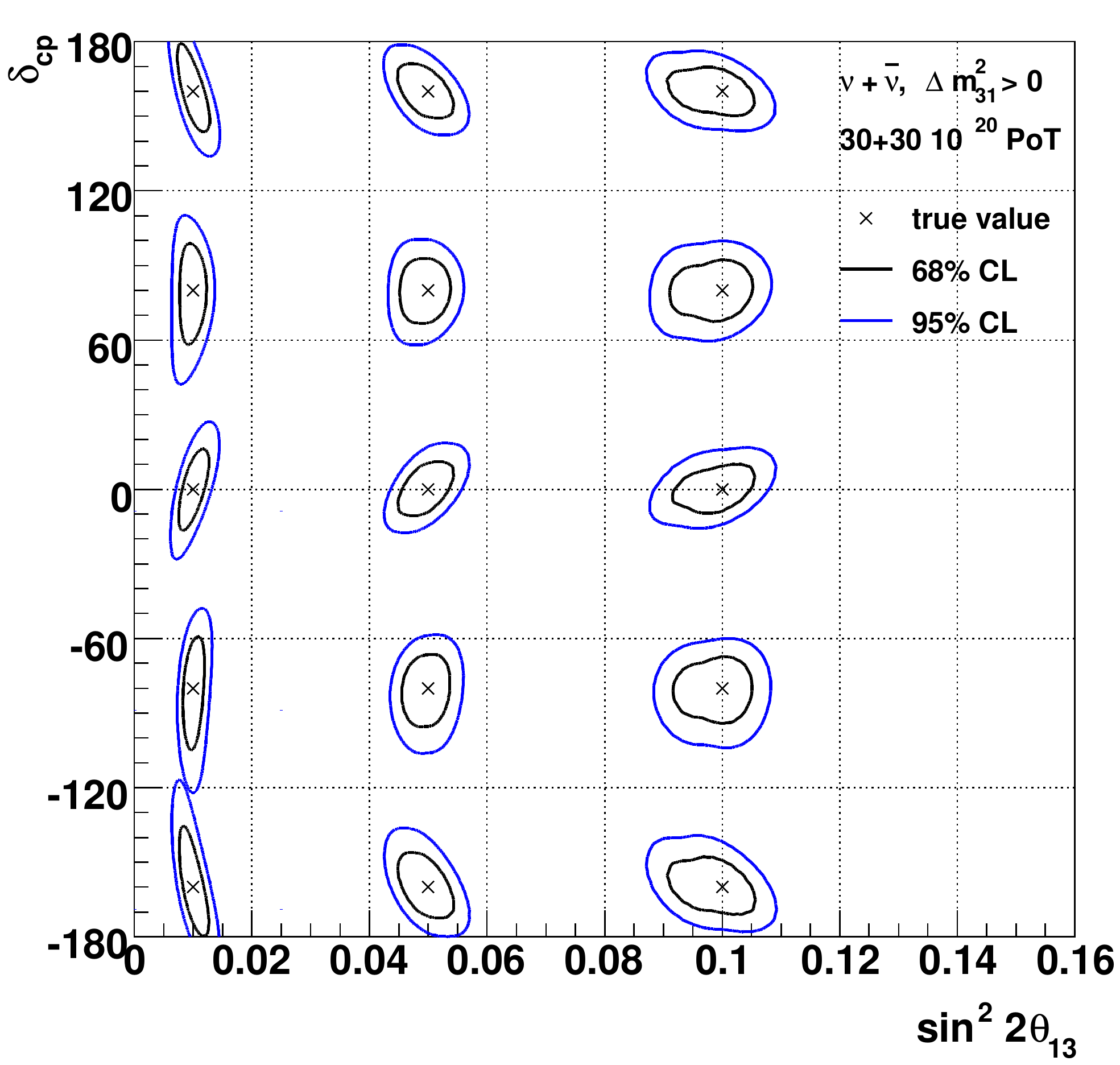} 
\includegraphics[angle=0,width=0.45\textwidth]{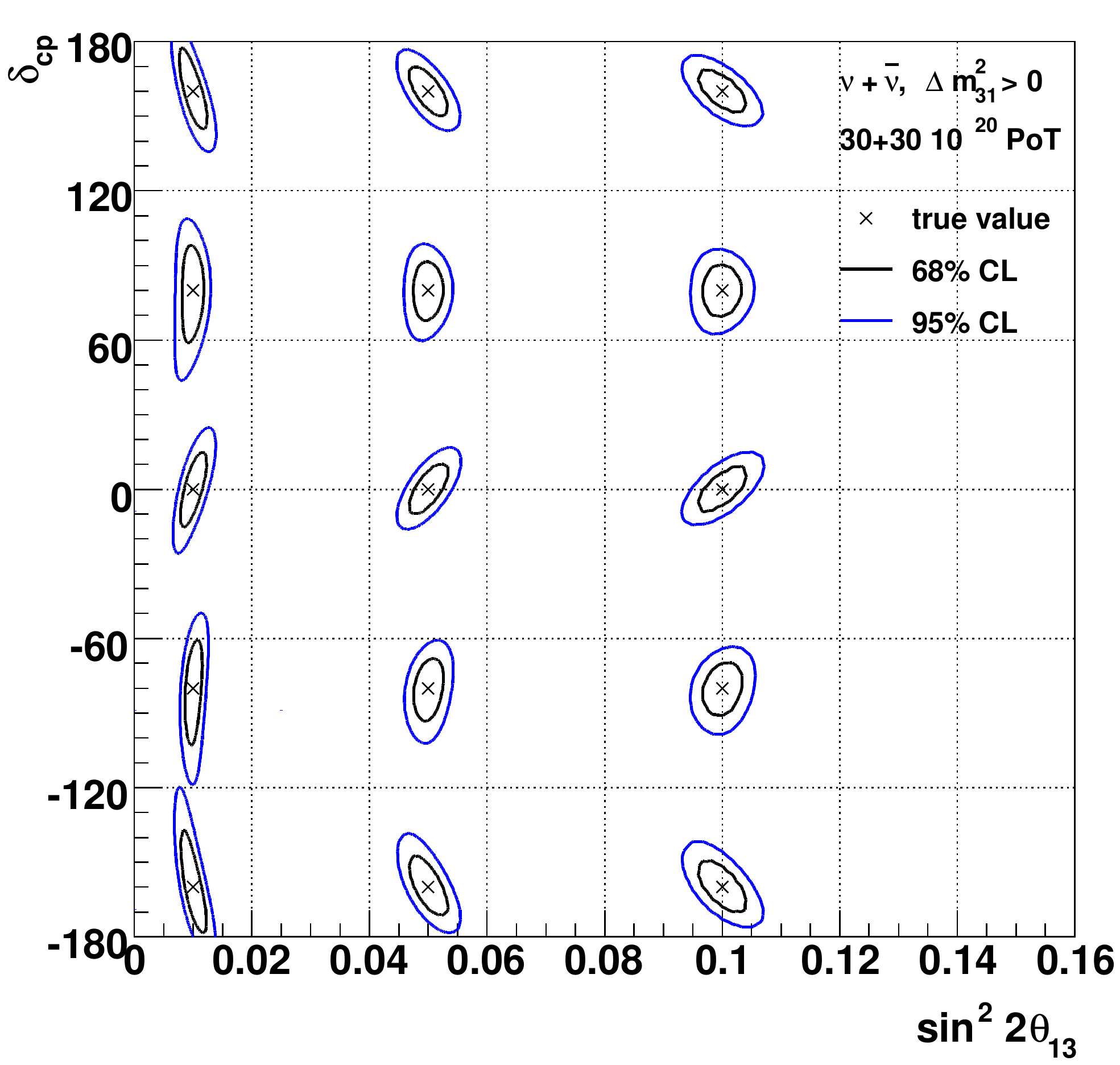} 
 \caption{\it 
 90\% and 95\%  confidence level error contours in  $\sin^2  2 \theta_{13}$
versus $\delta_{CP}$ for statistical and systematic errors (left hand plot)
 for 15 test points. 
This plot is for a 100 kTon liquid Argon TPC placed at DUSEL 1300 km away from FNAL.  
This is for combining both  neutrino and anti-neutrino data.
The right hand side is for statistical errors alone. 
This plot was made for normal mass hierarchy. 
We assume 10\% systematic errors on the background for this plot.
  \label{bubblelar} }
\end{figure}

\begin{figure}[h] 
\includegraphics[angle=0,width=0.45\textwidth]{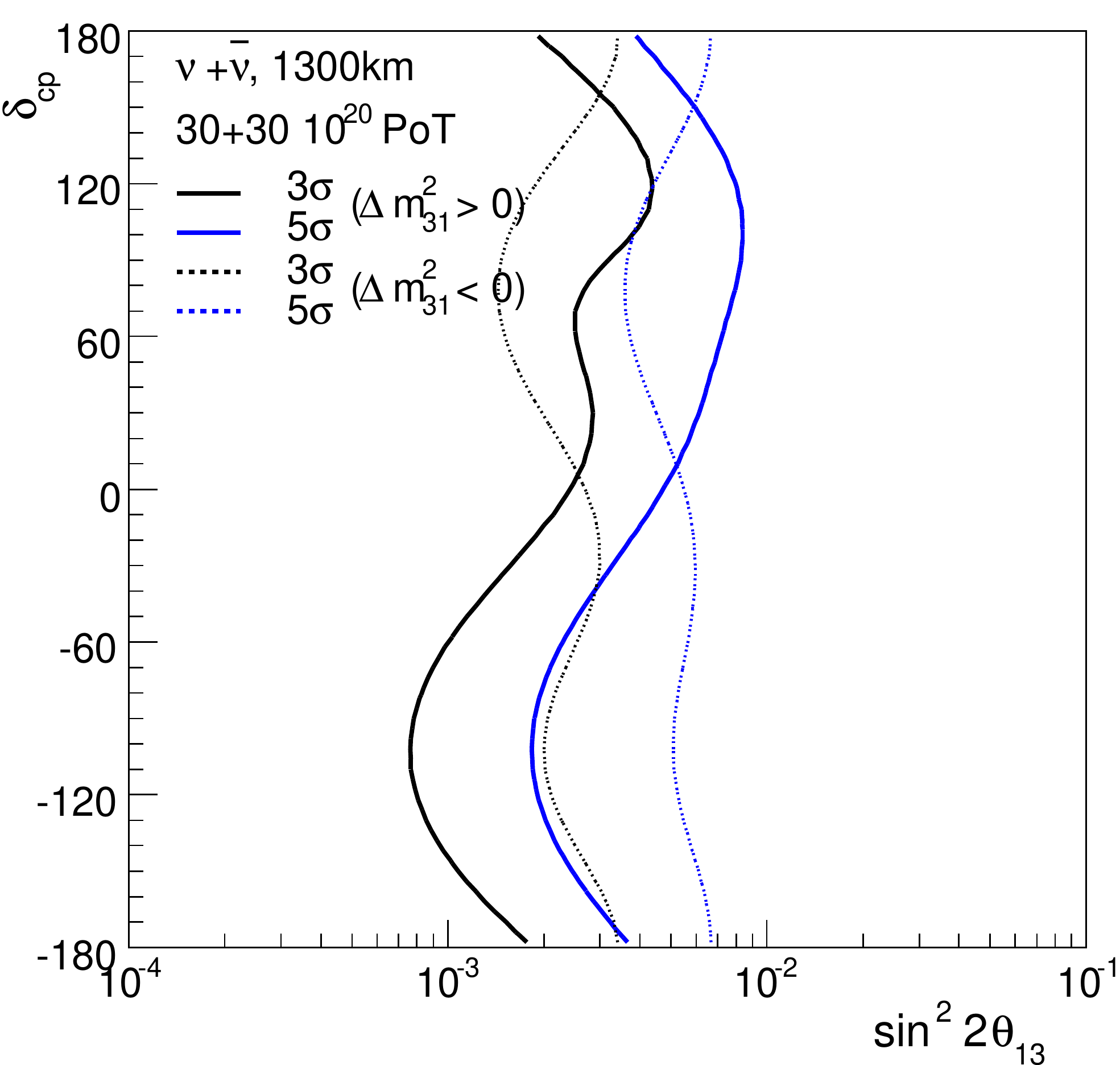} 
\includegraphics[angle=0,width=0.45\textwidth]{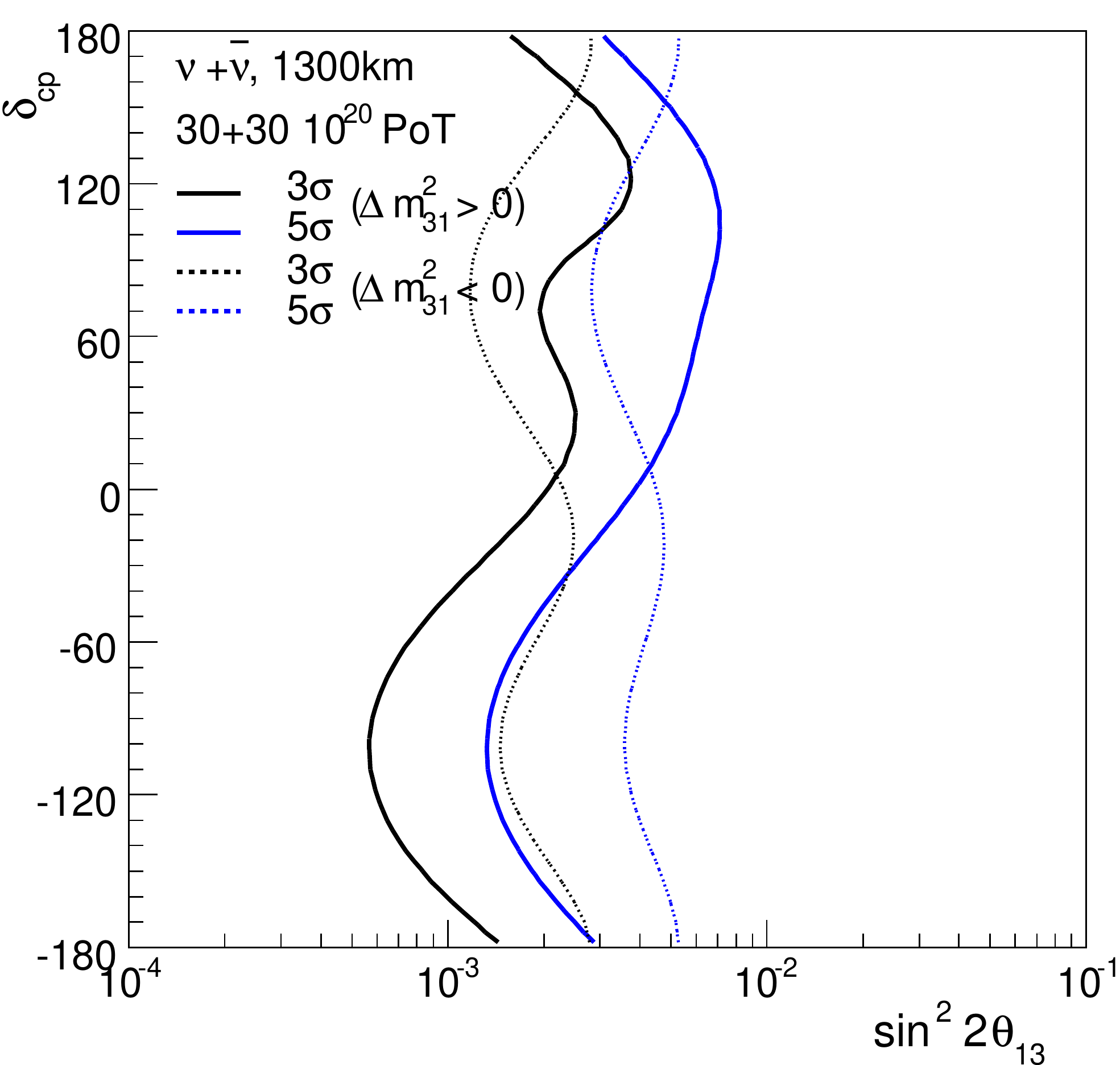} 
 \caption{\it 
 3 sigma and 5 sigma  confidence level exclusion limits for determining a non-zero value for 
$\theta_13$  in  $\sin^2  2 \theta_{13}$
versus $\delta_{CP}$ for statistical and systematic errors (left hand plot).
This plot is for a 100 kTon liquid Argon TPC placed at DUSEL 1300 km away from FNAL.  
This is for combining both  neutrino and anti-neutrino data.
The right hand side is for statistical errors alone. 
The solid (dashed) lines are for normal (reversed)  mass ordering.
We assume 10\% systematic errors on the background for this plot.
  \label{limit1lar} }
\end{figure}

\begin{figure}[h] 
\includegraphics[angle=0,width=0.45\textwidth]{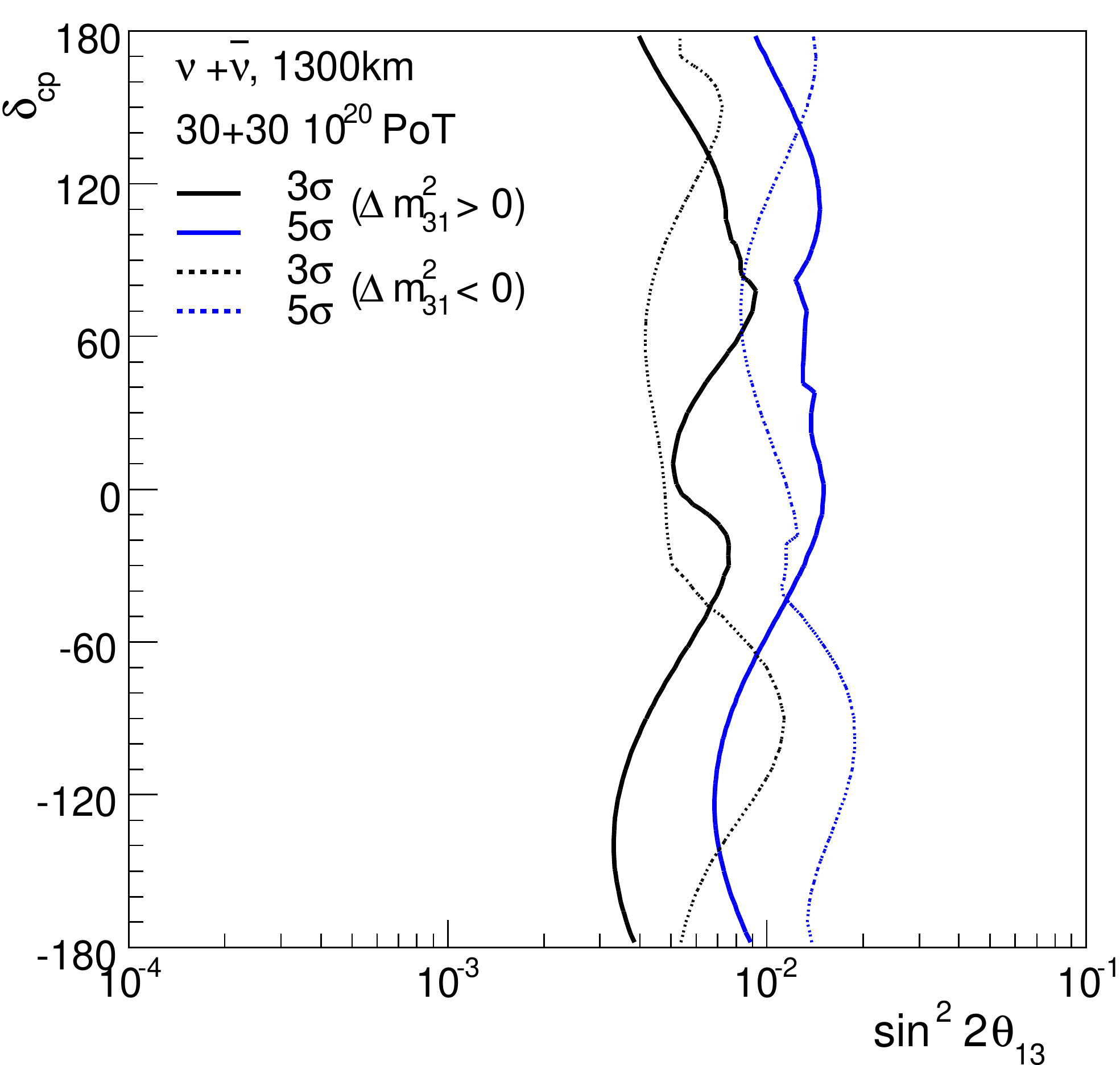} 
\includegraphics[angle=0,width=0.45\textwidth]{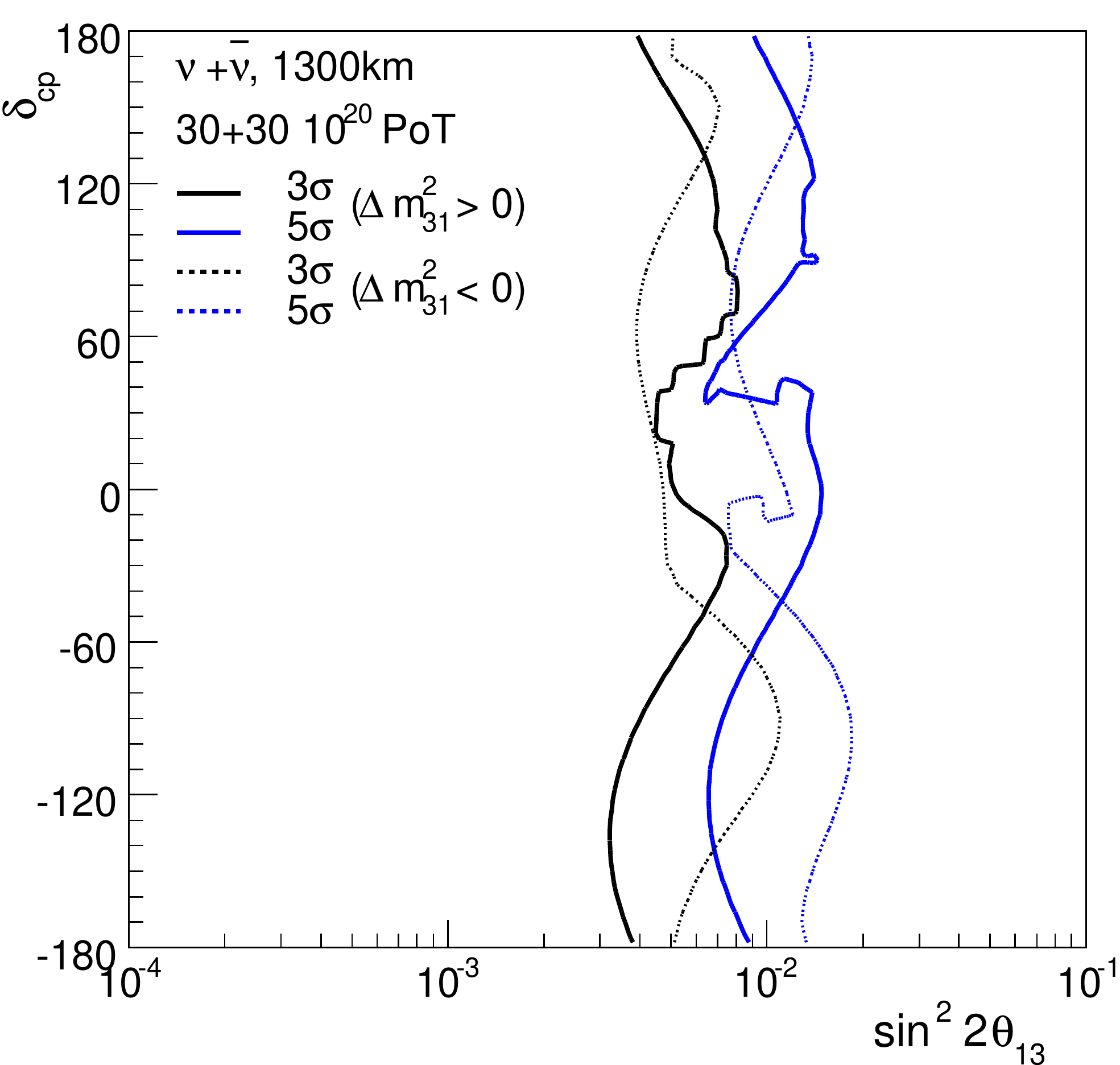} 
 \caption{\it 
 3 sigma and 5 sigma  confidence level exclusion limits for determining the 
mass hierarchy in  $\sin^2  2 \theta_{13}$
versus $\delta_{CP}$ for statistical and systematic errors (left hand plot).
This plot is for a 100 kTon liquid Argon TPC placed at DUSEL 1300 km away from FNAL.  
This is for combining both  neutrino and anti-neutrino data.
The right hand side is for statistical errors alone. 
The solid (dashed) lines are for normal (reversed)  mass ordering.
We assume 10\% systematic errors on the background for this plot.
  \label{limit2lar} }
\end{figure}

\begin{figure}[h] 
\includegraphics[angle=0,width=0.45\textwidth]{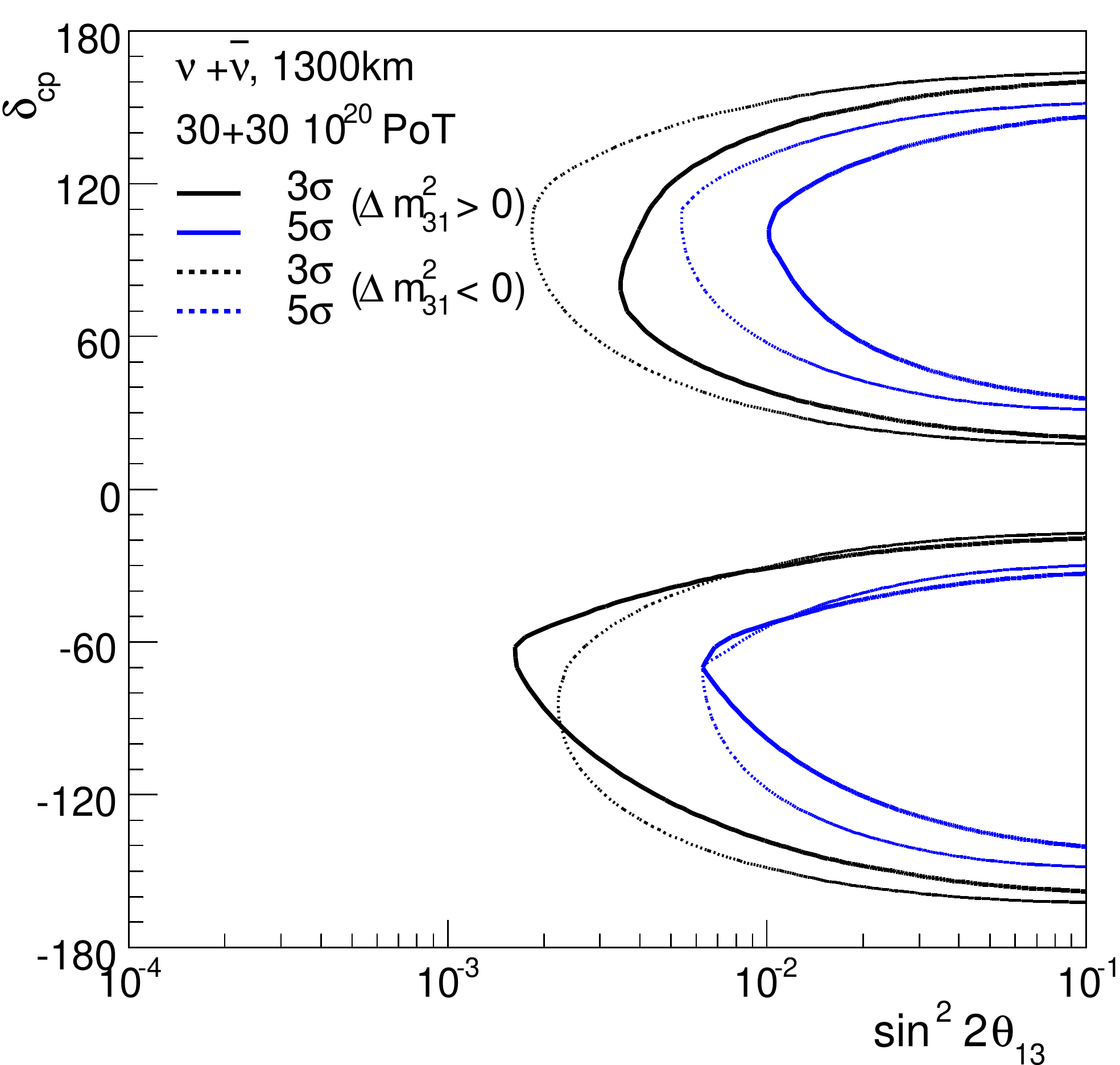} 
\includegraphics[angle=0,width=0.45\textwidth]{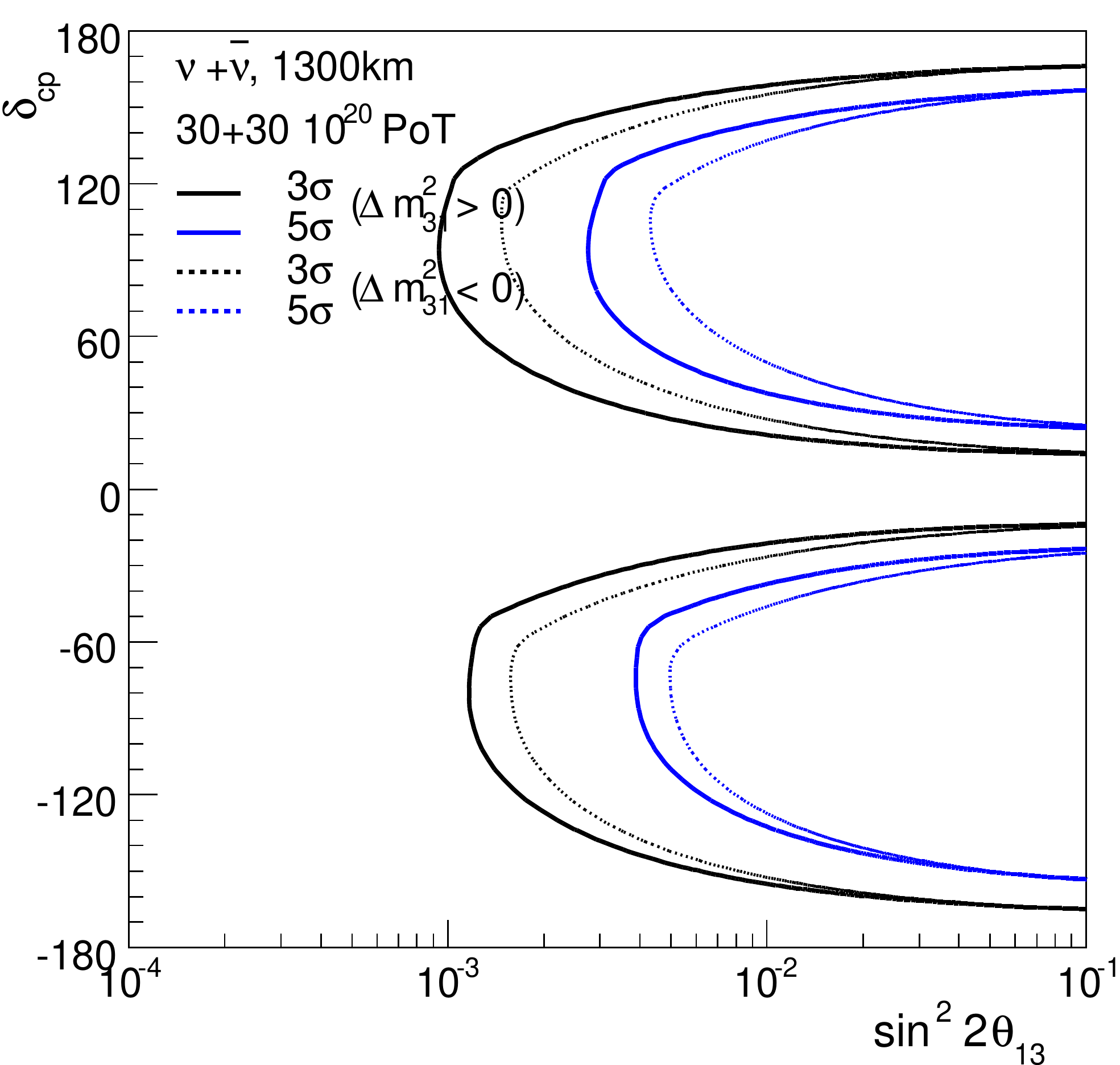} 
 \caption{\it 
 3 sigma and 5 sigma  confidence level exclusion limits for determining 
CP violation  in  $\sin^2  2 \theta_{13}$
versus $\delta_{CP}$ for statistical and systematic errors (left hand plot).
This plot is for a 100 kTon liquid Argon TPC placed at DUSEL 1300 km away from FNAL.  
This is for combining both  neutrino and anti-neutrino data.
The right hand side is for statistical errors alone. 
The solid (dashed) lines are for normal (reversed)  mass ordering.
We assume 10\% systematic errors on the background for this plot.
  \label{limit3lar} }
\end{figure}

The parameter resolutions and sensitivity limits for the 100 kT liquid Argon TPC at 
DUSEL are shown in Figures \ref{bubblelar} to \ref{limit3lar}.  
If there is no excess of electron events observed then we can set a
limit on the value of $\sin^2 2 \theta_{13}$ as a function of
$\delta_{CP}$.  Such  sensitivity limits are shown in
Figure \ref{limit1lar}.
The range of parameters over which the mass hierarchy can be resolved 
is shown in Figure \ref{limit2lar}. We have chosen to display the limits separately 
for the two mass hierarchies.
The region to the right hand side of each curve excludes the opposite mass 
hierarchy at the respective confidence level. 
Similarly the range of parameters over which CP violation can be established 
(i.e. determine that $\delta_{CP}$ is not 0 or $\pi$) is displayed in Figure 
\ref{limit3lar}.  
The comments we made regarding dependence on exposure, baseline, and oscillation 
parameters for the water Cherenkov 
detector are equally applicable to the sensitivities one would obtain with the 
liquid argon detector placed at DUSEL.

\clearpage 

\subsection{Sensitivity of a NuMI based off axis program} 
\label{sensi_2}

In this section we present the sensitivity results for several of the
scenarios using NuMI off-axis neutrinos which we studied.  The
following assumptions were made for this part of the study.
We have attempted to make the calculations as directly comparable to the previous 
section as possible. But there are small differences. These are also listed below.  

\begin {itemize}
\item The source of the proton beam used to create neutrinos will be the Fermilab Main Injector.
\item Current planning at Fermilab to maximize the proton intensity from the Main Injector via a series
of staged upgrades, will in fact occur over the next decade.
\item The possibility of a new source, such as the HINS (High Intensity Neutrino Source\cite{acc1}),
 or alternative ideas to improve the Main Injector, is considered 
 as the final stage of the upgrade path, ultimately providing an annual
 proton intensity of 2 $\times$ 10$^{21}$ protons per year.
\item Though the Fermilab Main Injector can be operated with extracted
 proton energies of less than
120 GeV, we have assumed that the optimum operation is at 120 GeV,
based on total delivered beam power.
\item The neutrino beam which we are considering here is the 
existing NuMI beam, which is a conventional 
horn focusing beam, capable of producing both neutrino and
anti-neutrino beams (by reversing the current in the horns). We do not
consider any reconfiguration or modification to the existing 2-meter
diameter, 675-meter long decay pipe.
\item We  assume that upgrades to targets, horns,  shielding 
and cooling systems will be required to accommodate proton intensities
significantly higher than that for which the facility was designed.
\item We do assume that the target and horn configuration can be adjusted to optimize neutrino rates.
\item We assume that detectors situated in NuMI off-axis locations will most likely be sited on or near the surface. We 
do not discuss the detector designs required to reject the backgrounds from cosmic ray interactions associated with a surface location.
\item  We have used an efficiency  of 80\% for all charged current electron neutrino events
and a neutral current rejection factor of 0.001, consistent with the parameters
 of a  Liquid Argon detector. 
\item We assume that the background is known with a systematic error of 5\%. 
\item The assumption on oscillation parameters is stated at the beginning of this section. For the calculation 
here these parameters as well as the matter density are assumed to be fixed.  
\item In our first pass analysis our sensitivities have been generated assuming no prior knowledge 
 of either  \sthetae, the mass hierarchy  or \deltacp. Iteration of the sensitivity calculation for the
 mass hierarchy and \deltacp have also been done such that  for values of   \sthetae  $\underline{>}$ 0.02,
 the angle is known (as will be the case from Phase I experiments). For values $\underline{<}$ 0.02 we assume the angle is unknown.

\end{itemize}

\begin{figure}[h] 
\includegraphics[angle=0,width=0.5\textwidth]{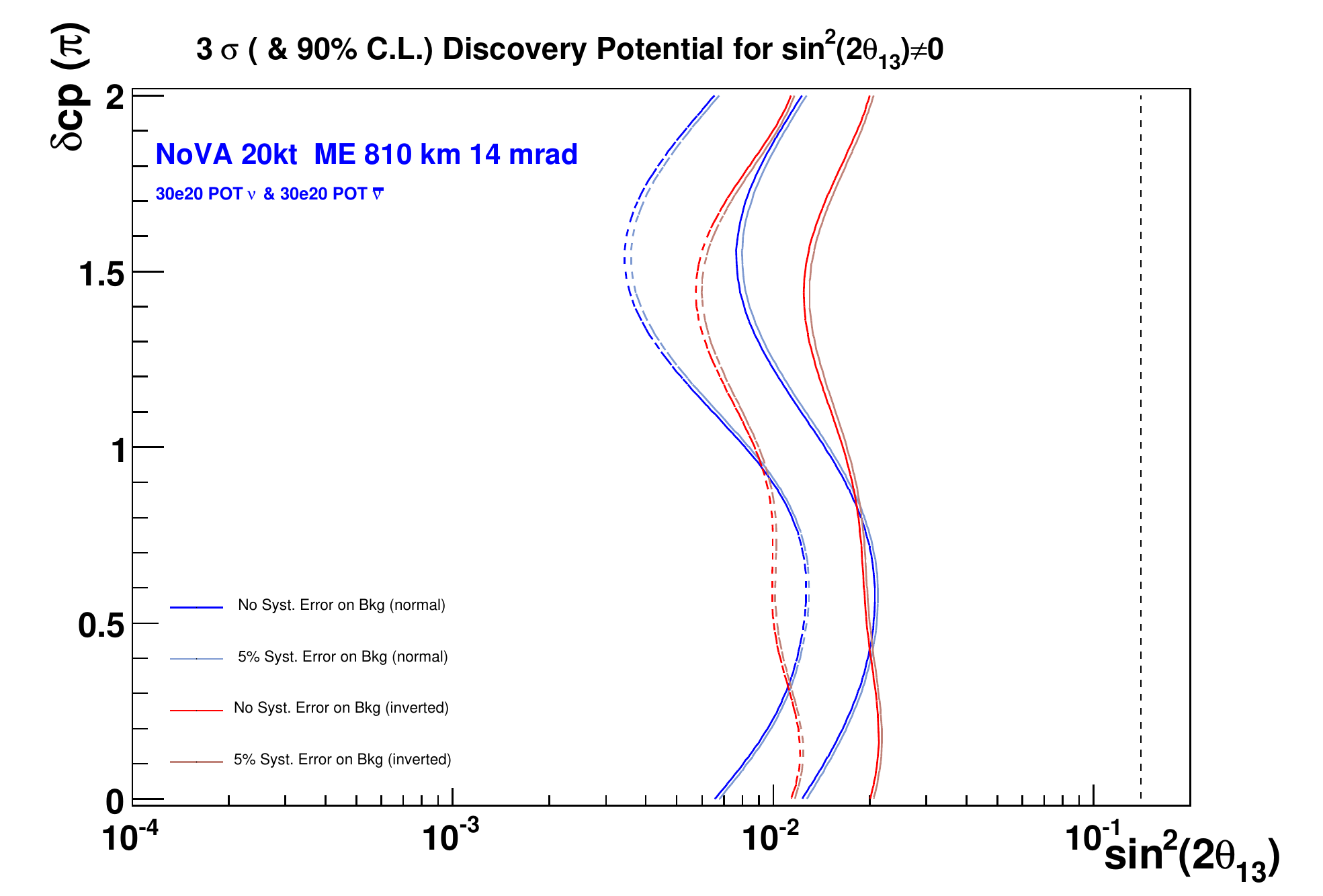} \\
\includegraphics[angle=0,width=0.5\textwidth]{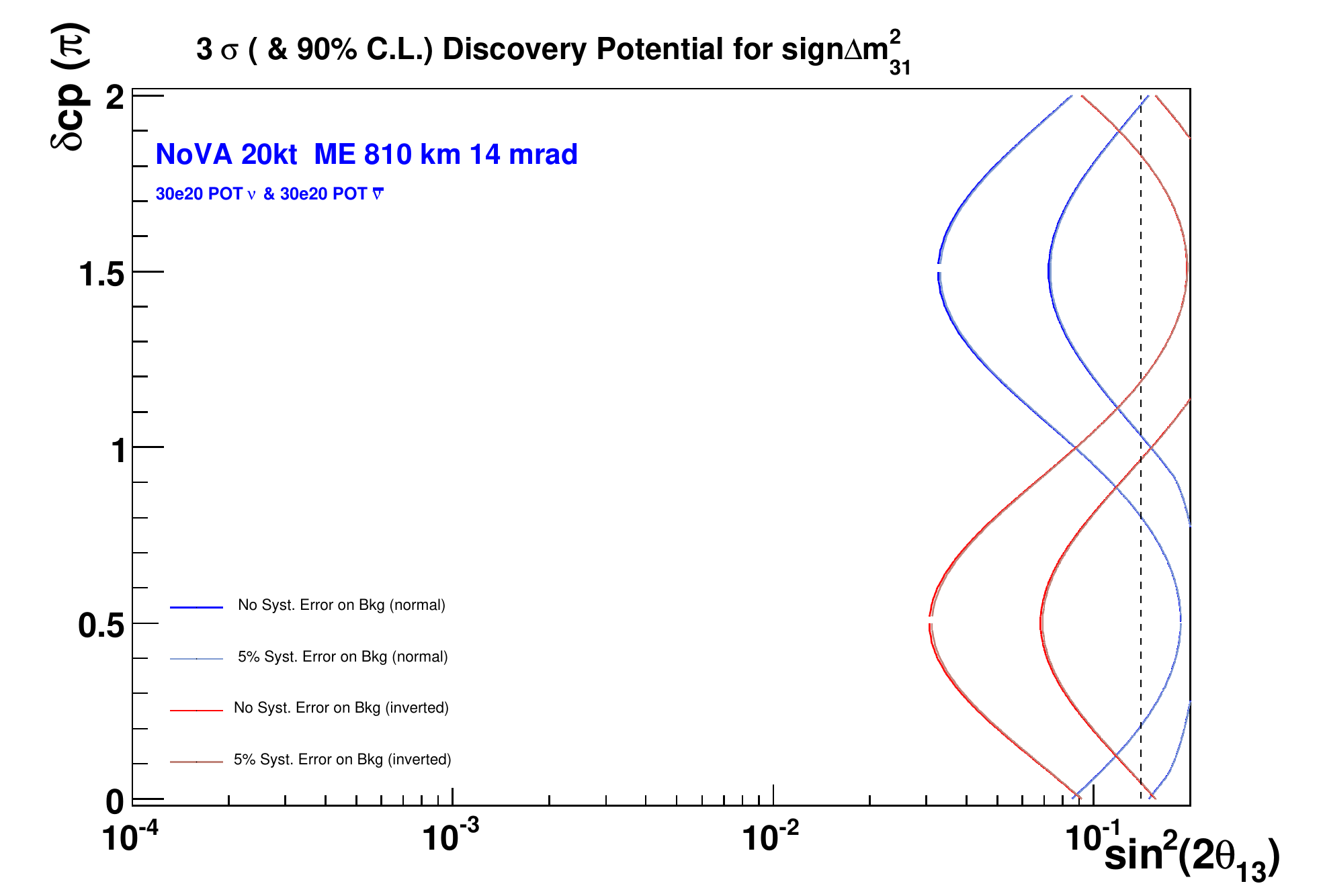}\\
\includegraphics[angle=0,width=0.5\textwidth]{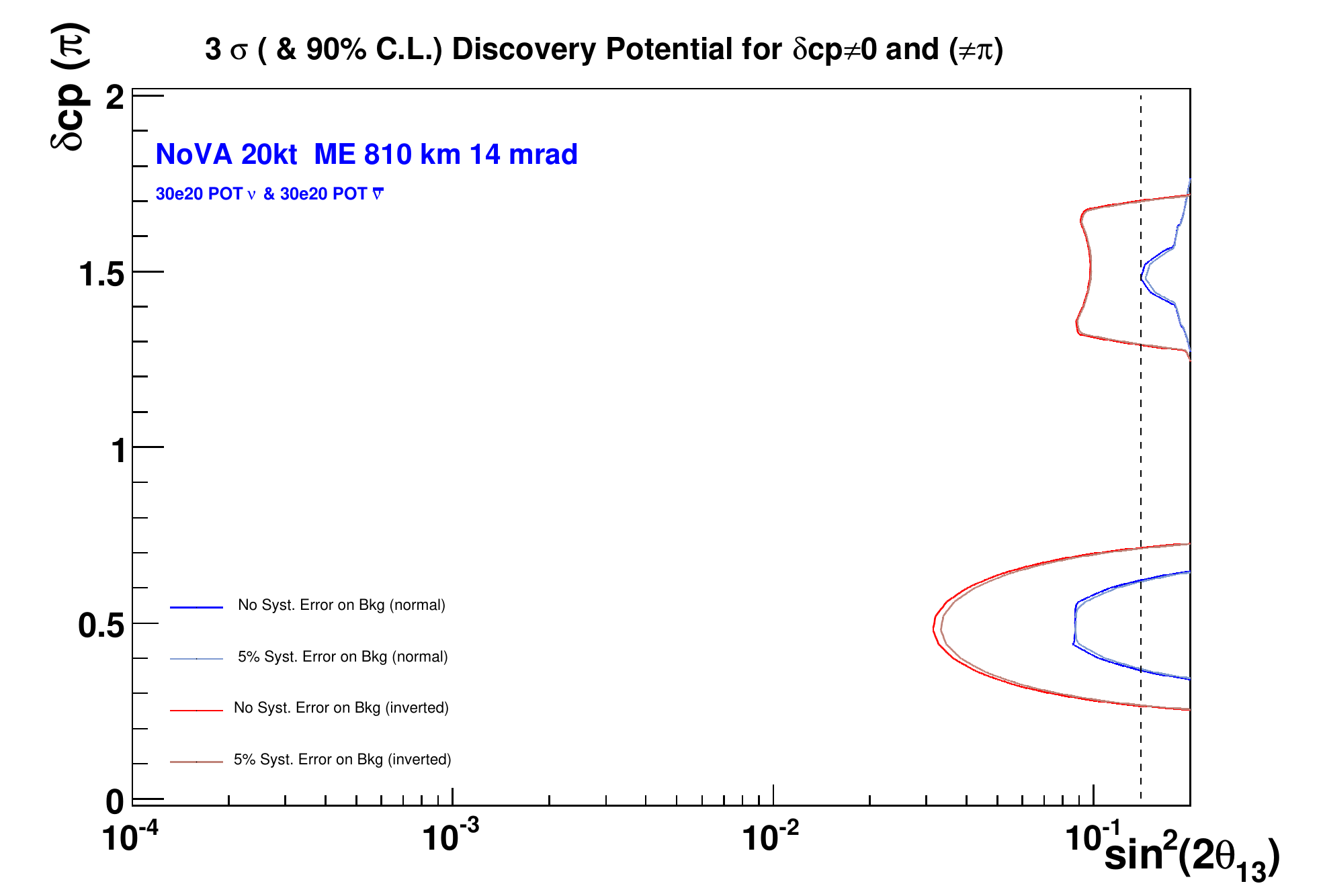} 
 \caption{\it 90\%, 3 sigma confidence level exclusion limits for
 determining a non-zero value for $\theta_{13}$ (top), for excluding the
 opposite mass hierarchy (middle), and for excluding CP violation
 (bottom) in $\sin^2 2 \theta_{13}$ versus $\delta_{CP}$.  These plots
(blue for normal and red for reversed hierarchy)
 are for a 20 kTon NO$\nu$A detector placed at the off-axis location
 on the NuMI beam-line with a total exposure of $60\times 10^{20}$
 protons, and for combining both neutrino and anti-neutrino data.
 5\% background systematic errors are assumed. \label{limitnova} }
\end{figure}

As a cross-check and starting point for our study, we have calculated
the sensitivities for a 20 kton \nova detector (see Figure
\ref{limitnova}).  Our results are consistent with those produced for
the \nova project Technical Design Report.  Note, for this and all
subsequent figures, the dashed line is placed at the current Chooz
limit.

For each scenario we assumed that the Phase II program consisted of
running for an equal time in neutrino and anti-neutrino mode. We show
the plots for an integrated proton intensity of 30$\times$10$^{20}$ in
each mode. We also assume that the \nova detector continues to take
data during Phase II. The new detectors are all Liquid Argon
technology.

\begin{figure}
\includegraphics[width=1.05\textwidth]{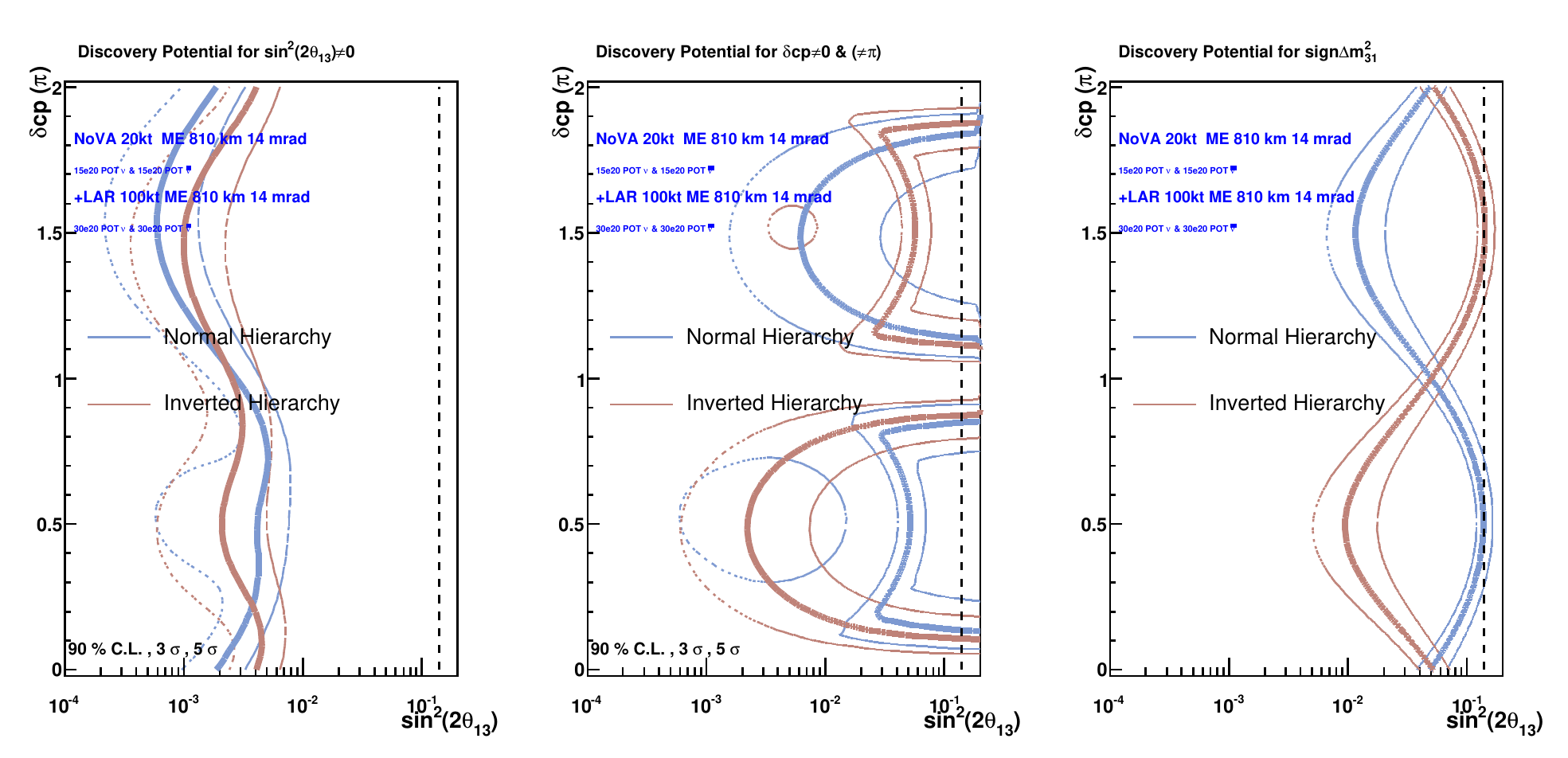}
\caption{\label{Case6}Scenario 1 : 90\%, 3$\sigma$, and 5$\sigma$ confidence level exclusion limits for  determining a non-zero value of \thetae (left), 
 for excluding CP violation (center), and
for excluding the opposite mass hierarchy (right), in \sthetae versus \deltacp.  
 }
\end{figure}

The \emph{first} scenario considered was placing a 100 kton detector
at the 1st maximum, i.e. simply increasing the mass and efficiency of
the \nova configuration. The sensitivities are shown in
Figure~\ref{Case6}.

\begin{figure}
\includegraphics[width=1.05\textwidth]{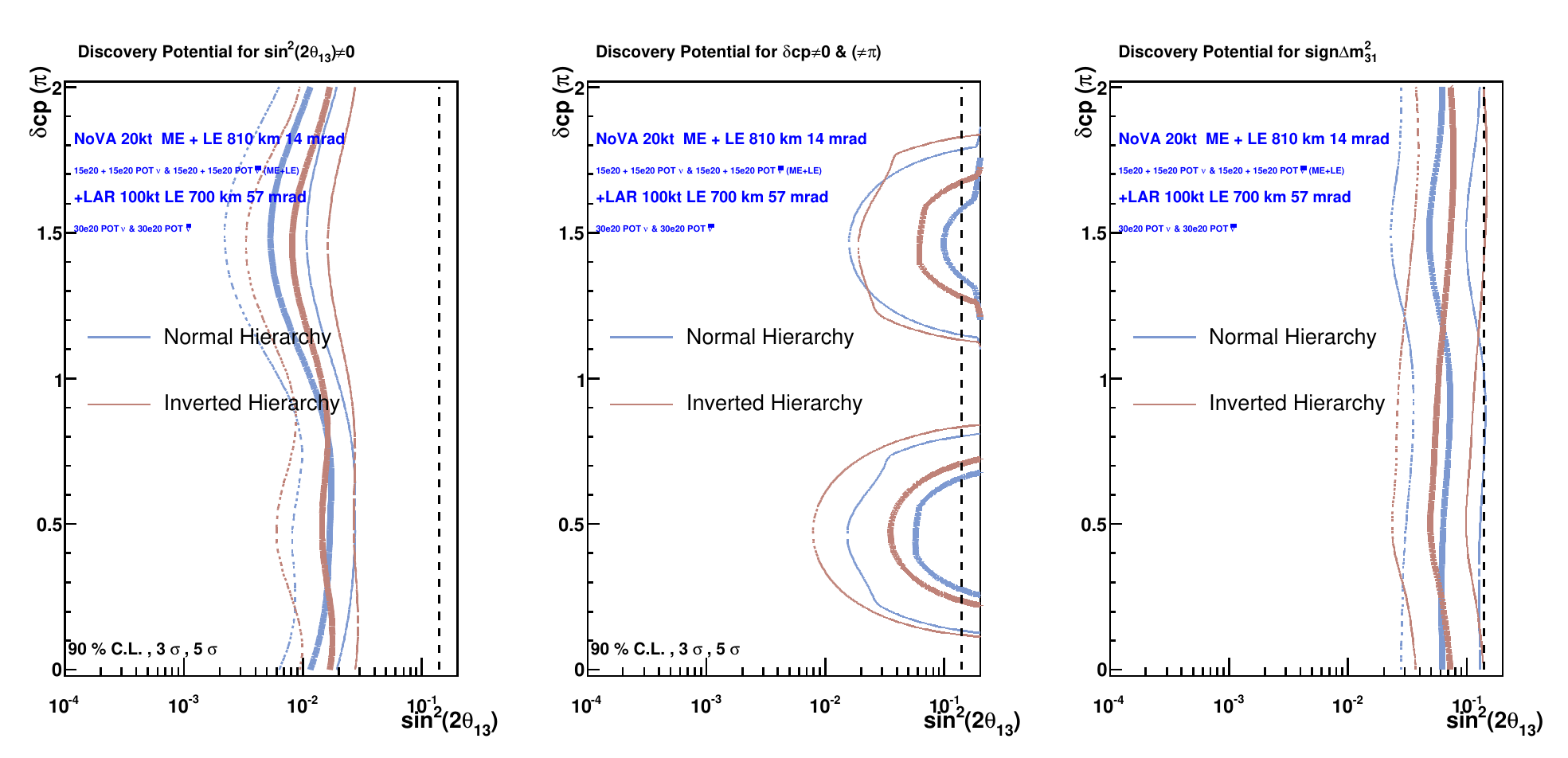}
\caption{\label{Case7}Scenario 2 : 90\%, 3$\sigma$, and 5$\sigma$ confidence level exclusion limits for  determining a non-zero value of \thetae (left), 
for excluding CP violation (center), and
for excluding the opposite mass hierarchy (right), in \sthetae versus \deltacp.  
 }
\end{figure}

The \emph{second} scenario we studied, was to place a 100 kton
detector at a baseline of 700 km and an off-axis angle of 57 mrad (40
km). This location corresponds to the second oscillation maximum,
where the matter effects are small (due to the lower energy of the
neutrinos), but the CP effects are large.  
The \nova detector is the only detector at the first
maximum site (L = 810, 14 mrad off axis).  These results are shown in
Figure~\ref{Case7}. We find that with this configuration, running
neutrinos and anti-neutrinos, the sensitivity to the mass hierarchy
flattens over the range of possible \deltacp values, but the discovery
potential is limited to values of \thetae relatively close to the
current limit.

\begin{figure}
\includegraphics[width=1.05\textwidth]{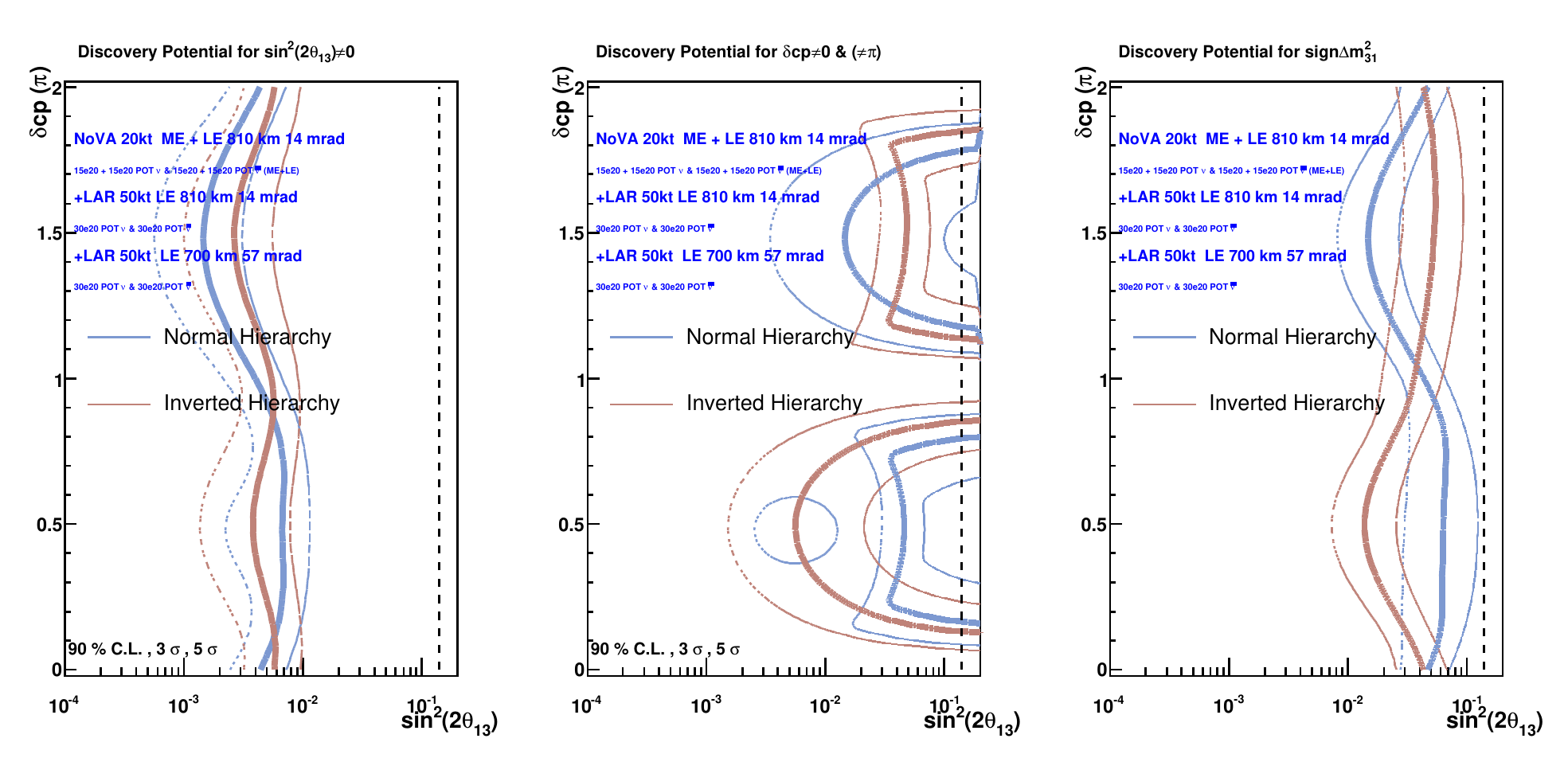}
\caption{\label{Case8}Scenario 3 : 90\%, 3$\sigma$, and 5$\sigma$ confidence level exclusion limits for  determining a non-zero value of \thetae (left), 
for excluding CP violation (center), and
for excluding the opposite mass hierarchy (right), in \sthetae versus \deltacp.  
 }
\end{figure}

A \emph{third} scenario was to split the mass between the two
locations, 50 ktons at each the first and second maximum. These
results are in Figure~\ref{Case8}.

Finally, a \emph{fourth} scenario was the same as the third except
that the detectors at each site were 100 ktons.

We summarize our studies in  Table~\ref{compare}. 
Because some of the scenarios studied have the benefit of
"flattening" the sensitivity over \deltacp, we have included the
sensitivity limits for both 50\% and 100\% coverage of the \deltacp
space. We have also included in this table the limits which can be
reached as the Phase I program evolves, and as the Liquid Argon
technology also evolves.

From these studies we conclude the following:
\begin{itemize}
\item \sthetae down to 0.02  can be measured by the Phase I (NO$\nu$A) experiment.  
Phase I experiments however, have limited or no sensitivity to 
determining the mass hierarchy, and essentially no sensitivity to \deltacp. 
\item If  \sthetae is large, i.e. $>$0.04 a Phase II experiment using the NuMI beam can be designed specifically  to determine the mass hierarchy.  Such an experiment is like our 4th scenario (see Table~\ref{compare}), with  two massive  detectors  placed  at the first and second maximum sites. In this experiment  the mass hierarchy can be resolved for all values of \deltacp. The experiment also has sensitivity to \deltacp.

\item  The most interesting and complex situation to plan for,  is if Phase I experiments indicate that 0.02 $<$ \sthetae $<$ 0.04.  In this case we find that the configurations studied for the NuMI Off-Axis option can have relatively good sensitivity to determine the mass hierarchy, as well as  some sensitivity to $CP$. 

\item If Phase I experiments  conclude  that  \sthetae $<$ 0.02  the Phase II program can continue the search.  Continued running, more protons and larger more efficient detectors, placed  at the 1st maximum (the \nova site), allows one to reach sensitivities to well below 10$^{-2}$, (of the order $\sim$0.003) as can be seen in our first scenario.

\end{itemize}

\begin{figure}
\includegraphics[width=1.05\textwidth]{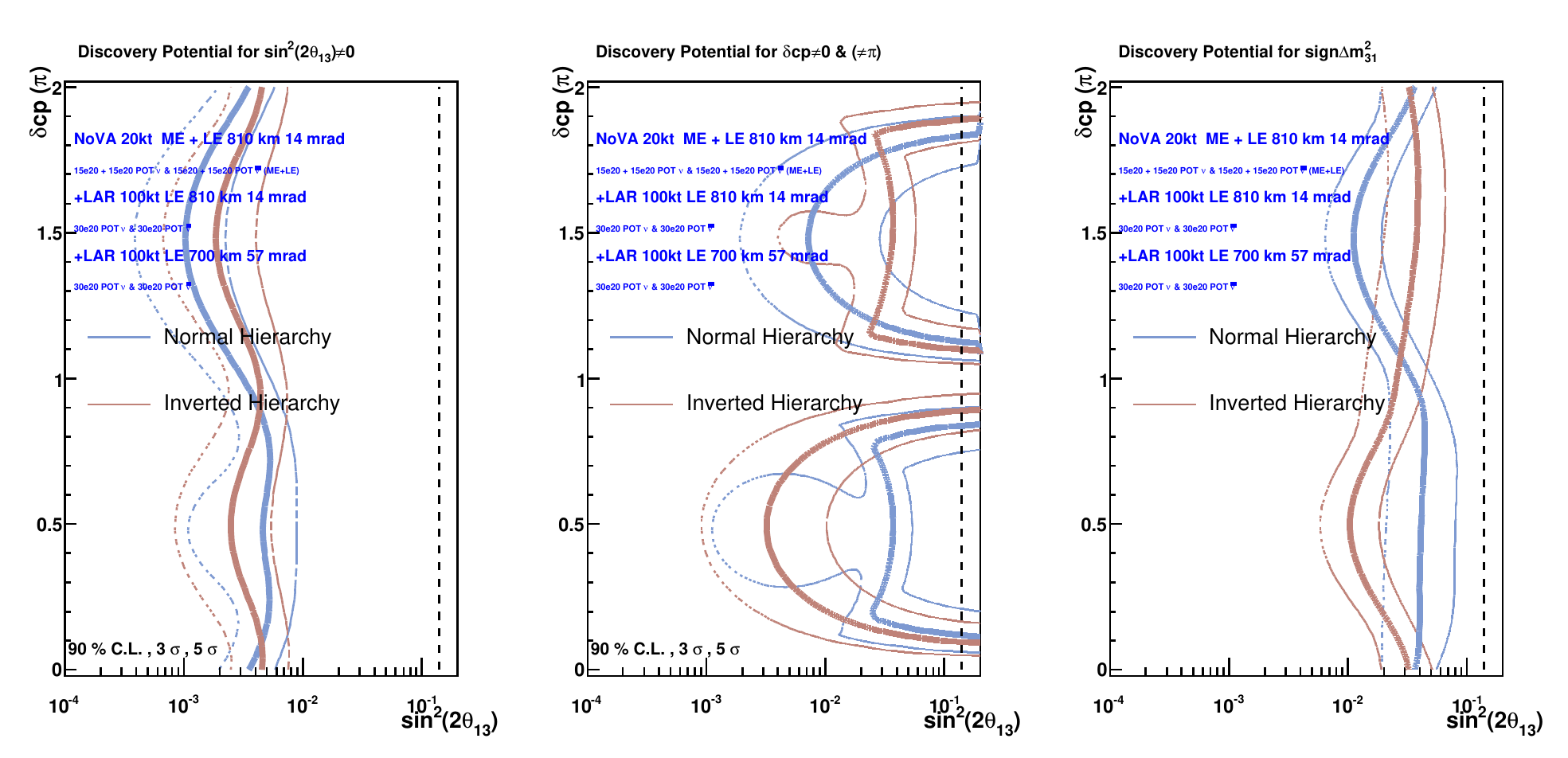}
\caption{\label{Case81}Scenario 4 : 90\%, 3$\sigma$, and 5$\sigma$ confidence level exclusion limits for  determining a non-zero value of \thetae (left), 
for excluding CP violation (center), and
for excluding the opposite mass hierarchy (right), in \sthetae versus \deltacp.  
 }
\end{figure}

\begin{table}[h]
\centering
\begin{tabular}{|c|c|c|c|c|c|}
\hline
Detector Location         &  Detector                   &   Exposure                              &                        &      &           \\
 (L(km), $\theta$(mr))   &  (technology/mass) &   POT ($\nu$) &   \sthetae       & sgn(\delma)          & CPV \\      
      (beam tune)                                   &                                    &   /POT(\nubar)   &                   &                            &   \\
\hline
810, 14(ME)                         & \nova/20kt	             &    15/-               &  0.018/0.030   & 0.17/NA       & NA  \\
\hline         
810, 14 (ME)                       & \nova/20kt                   &   15/15                      &   0.018/0.024    & 0.16/NA       & NA  \\
\hline                                       
810, 14(ME)                      & \nova/20kt                   &   30/30               &   0.012/0.020    & 0.10/NA       & NA \\
\hline
\hline
810, 14(ME)                     & \nova/20kt                   &   30/30                   &   0.007/0.013    & 0.08/0.20       & NA \\
810,14(ME)                            & + LAr/5kt                      &   30/30               &                             &                        &         \\
\hline
810, 14(ME)                           & \nova/20kt                     &   30/30                 &   0.004/0.009    & 0.05/0.15    & 0.07/NA \\   
810,14(ME)                            &  + LAr/20kt                   &   30/30           &                             &                       &               \\                           
\hline
\hline
Scenario 1                  & & & & & \\
810, 14(ME)                           & \nova/20kt                &  30/30                   & 0.0018/0.005  & 0.03/0.12    & 0.03/NA \\
810, 14(ME)                           & + LAr/100kt                  &  30/30                &                           &                      &                 \\
\hline
Scenario 2               & & & &  & \\
810, 14(ME/LE)                         & \nova/20kt                &  30/30               & 0.011/0.018  & 0.05/0.07    & 0.07/NA \\
700, 57 (LE)                        & + LAr/100kt                  &  30/30           &                           &                      &                 \\
\hline
Scenario 3         & & & &   & \\
 810, 14(ME/LE)                         & \nova/20kt                   &  30/30        &                                  &                             &                         \\
 810, 14 (LE)                      & + LAr/ 50kt                     &   30/30        &      0.0035/0.006     &     0.033/0.06    &  0.035/NA       \\
 700, 57 (LE)                        & + LAr/ 50kt                     &  30/30           &                                   &                            &                       \\
\hline
Scenario 4      & & & &  & \\
810, 14(ME/LE)                         & \nova/20kt                   &   30/30        &                                   &                             &            \\
810, 14(LE)                 & + LAr/100kt                   &   30/30        &      0.0027/0.0046     &     0.030/0.042    &  0.022/NA  \\
700, 57 (LE)                         & + LAr/100kt                   &   30/30       &        &               &                           \\
\hline
                                     
\end{tabular}
\caption{\label{compare}Sensitivity comparisons for all NuMI Off-axis scenarios that were evaluated.
These numbers were calculated with the normal hierarchy assumption. 
 The first three cases represent three possible stages of the Phase I (\nova) program. The values given represent the value of \sthetae  where a 3 $\sigma$ determination of the parameter can be made for  50\%(/ 100\%)  of the possible values of \deltacp. Note that for determining the sensitivity  to mass hierarchy and \deltacp, for values of \sthetae $>$0.02 we assume that the angle is known (i.e. from Phase I experiments.) }
\end{table}

\subsection{Comparison of sensitivity estimates} 
\label{comp_1}

\begin{table}

\begin{center}
\begin{tabular}{|c|l|c|c|c|c|c|c|}
\hline
Option & Beam & Baseline & Detector & Exposure (MW.yr$^{*}$) &
$\theta_{13} \neq 0$  & CPV &$sgn(\Delta m_{31}^2)$ \\ \hline \hline
(1) & NuMI ME, 0.9$^{\circ}$     & 810 km  & NO$\nu$A 20 kT & 6.8 & 0.015 & $>$ 0.2 & 0.15\\ \hline
(2) & NuMI ME, 0.9$^{\circ}$     & 810 km  & LAr 100 kT & 6.8  & 0.002 & 0.03  & 0.05 \\ 
(3) & NuMI LE, 0.9$^{\circ}$,  3.3$^{\circ}$,    & 810,700 km  & LAr 2 $\times$ 50 kT & 6.8  & 0.005 & 0.04  & 0.04 \\ 
(4) & WBLE 120GeV, 0.5$^{\circ}$ & 1300km  & LAr 100 kT & 6.8  & 0.0025 & 0.005 & 0.006 \\ \hline

(5) & WBLE 120GeV, 0.5$^{\circ}$ & 1300km  & WCe 300 kT  & 6.8  & 0.006 & 0.03  & 0.011 \\
(6) & WBLE 120GeV, 0.5$^{\circ}$ & 1300km  & WCe 300 kT  & 13.6 & 0.004 & 0.012 & 0.008 \\ \hline
\end{tabular}
\end{center}
\caption{Comparison of the sensitivity reach of different long
  baseline experiments. The sensitivity is given as the value of
  $\sin^2 2\theta_{13}$ at which 50\% of $\delta_{cp}$ values will have
  $\geq 3\sigma$ reach for the choice of mass hierarchy with worst sensitivity. 
We assume equal amounts of 
  $\nu$ and $\bar{\nu}$ running in the total exposure. The assumption on running time is 
  $1.7 \times 10^7$ seconds of running per year.
Also see Table \ref{compare}.
\label{tab_compare_sens}
}
\end{table}

\begin{figure}[tp!]
\begin{center}
\includegraphics[height=18cm]{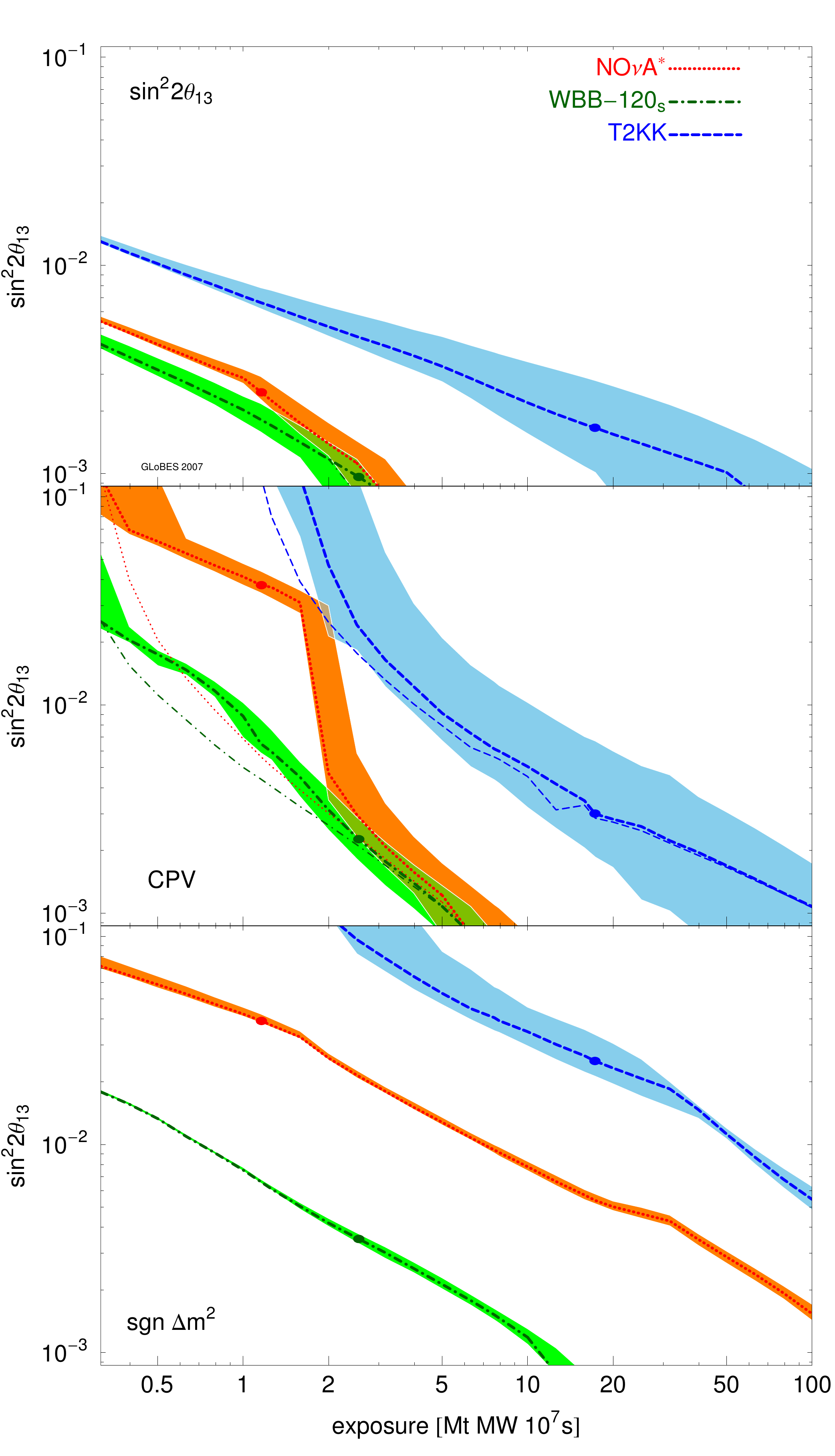}
\end{center}
\caption{The $\sin^2 2 \theta_{13}$ reach at $3\sigma$ 
for the discovery of nonzero $\sin^2 2 \theta_{13}$ (top),
 CP violation(middle), and the normal hierarchy (bottom)
 as a function of exposure. The curves are for 
a fraction of $\delta_{CP}$ of 0.5, 
which means that the performance will be better for 50\% of all 
values of $\delta_{CP}$, and worse for the other 50\%. 
The light curves in the CPV panel are made under the assumption that the mass hierarchy 
is known to be normal. 
 The shaded regions result by varying the systematic uncertainties from 
2\% (lower edge) to 10\% (upper edge). 
This figure is reproduced from \cite{wble-glb2}. 
\label{fig:lumiscale}}
\end{figure}

A summary of the sensitivity reach for non-zero $\theta_{13}$, CP
  violation and the sign of $\Delta m_{31}^2$ for 6 different
  combinations of beams, baselines, detector technologies, and
  exposure is presented in Table ~\ref{tab_compare_sens}.
Several more configurations for the off-axis scenario are presented in 
Table \ref{compare}.
 The
  sensitivity reach is given as the lowest $\sin^2 2\theta_{13}$ value
  at which \underline{at least 50\%} of $\delta_{cp}$ values will have
  $\geq 3\sigma$ reach. For this table we use the mass hierarchy with the worst
  sensitivity to determine the minimal value of $\sin^2 2\theta_{13}$
  for which $\geq$ 50\% of $\delta_{cp}$ values will have $\geq
  3\sigma$ sensitivity to a particular measurement. We estimated these
  values of $\sin^2 2\theta_{13}$ from the studies and plots discussed
  in Sections ~\ref{sensi_1} and ~\ref{sensi_2}. We note that
  different options are sensitive to different values of
  $\delta_{cp}$, such that being sensitive to 50\% $\delta_{cp}$
  values does not necessarily imply that a given experimental option
  is sensitive to the same region of oscillation parameter phase space
  as another.

We compare the wide-band FNAL to DUSEL program, option (4), with the
  narrow-band off-axis NuMI-based program, option (2), for the same
  exposure of 6.8 MW.yr (1 experimental year is defined as
  $1.7 \times 10^7$ seconds). This is equivalent to an integrated
  exposure of $60 \times 10^{20}$ protons-on-target for proton beam
  energies of 120 GeV. We assume equal amounts of exposure for
  neutrinos and anti-neutrino (reverse horn current) running.  A
  liquid Argon TPC with a total mass of 100 kT is assumed as the
  detector technology of choice for the purpose of the comparison. We
  note that slightly different assumptions on the systematic
  uncertainties on the oscillation parameters and backgrounds went
  into the sensitivity estimates for NuMI off-axis (5\% uncertainty on
  the background) and the wide-band FNAL to DUSEL options (10\%
  uncertainty on the background). The effect of the different
  assumptions is $\leq$ 15\% variation on the value of $\sin^2
  2\theta_{13}$ at which the sensitivity reaches 50\% of
  $\delta_{CP}$. We find that for the same exposure of 6.8 MW.yr, and
  the same  liquid Argon TPC detector technology (size and same performance), 
the wide-band FNAL to
  DUSEL approach has significantly better sensitivity to CP violation,
  the sign of $\Delta m_{31}^2$, and comparable sensitivity to
  non-zero values of $\theta_{13}$. To illustrate the improvement in
  sensitivity over the existing program, the sensitivities of the
  current NO$\nu$A experiment (as shown in Figure ~\ref{limitnova}) at
  the same exposure, are summarized as option (1) in Table
  ~\ref{tab_compare_sens} These are the sensitivity limits
  expected from the NO$\nu$A experiment only, before combination with
  T2K\cite{jhfsk}. Analysis of combining with T2K has been performed elsewhere\cite{olga1}.

The  value of $\sin^2 2\theta_{13}$ at which at least 50\% of
  $\delta_{cp}$ values will have $\geq 3\sigma$ reach as a function of
  exposure for the NuMI ME beam at 810km (labeled NO$\nu$A$^*$), and
  the wide-band 120 GeV beam at 1300km (labeled WBB-120$_{\rm s}$) is
  summarized in Figure ~\ref{fig:lumiscale} from reference
  ~\cite{wble-glb2}.  A LAr TPC is the detector technology assumed for
  NuMI off axis (labeled NO$\nu$A$^*$) and WBB-120$_{\rm s}$.
 We find that after reaching an
  exposure of 2 MT.MW.$10^7$ seconds (for 100kT LAr and a 120
  GeV beam, this is an exposure of $10^{22}$ protons-on-target), the
  mass hierarchy-$\delta_{cp}$ degeneracy is sufficiently resolved for the
  NO$\nu$A$^*$ approach (option (2)) - and the sensitivity to CP
  violation approaches that of the wide-band beam at the
  1300km baseline. For the mass hierarchy, the wide-band FNAL to DUSEL
  approach always has significantly better sensitivity independent of
  the exposure. Option (3) in Table ~\ref{tab_compare_sens} is a
  NuMI-based program with a 50 kT detector at the 1st oscillation
  maximum running concurrently with another 50kT module placed at the
  2nd oscillation maxima. We find that option (3) has worse
  sensitivity to non-zero values of $\theta_{13}$ when compared to
  option (2) and slightly better sensitivity to the sign of $\Delta
  m_{31}^2$.

Option (5) summarizes the FNAL to DUSEL sensitivity when the 100 kT
LAr TPC of option (4) is replaced by a 300 kT water
Cherenkov detector. We find that the sensitivity worsens due to the
lower signal statistics and higher NC backgrounds in a water Cherenkov
detector. We can recover some of the lost sensitivity by doubling the
exposure of the water Cherenkov detector as shown in option (6). For
the same exposure, the FNAL to DUSEL program with a 300 kT water
Cherenkov detector, option (5), has the same sensitivity to CP
violation as the NuMI based program with a 100 kT LAr TPC in options
(2) and (3)  and significantly better sensitivity to 
the sign of $\Delta m_{31}^2$. We find the FNAL to DUSEL program with a 300 kT
water Cherenkov detector 
has similar sensitivity to non-zero $\theta_{13}$ as the NuMI based
program with two 50 kT LAr TPC's at the 1st and 2nd oscillation
maxima, option (3). 

We summarize the comparison studies as follows:

\begin{itemize}

\item \sthetae down to 0.02  can be measured by the Phase I (NO$\nu$A) experiments.  
Phase I (NO$\nu$A) experiment,  however, have limited or no sensitivity to 
determining the mass hierarchy, and essentially no sensitivity to \deltacp. 

\item All Phase II experimental options will 
improve the sensitivity to CP violation by
at least an order of magnitude over the existing Phase I program.

\item Given the same exposure and detector technology (LAr TPC), the FNAL to DUSEL 
program with a wide band beam has
significantly better overall  sensitivity to neutrino
oscillations when compared to a shorter baseline NuMI based program
with an off-axis beam. 

\item The FNAL to DUSEL program with a  300 kT water Cherenkov
detector has similar sensitivity to CP violation when compared to a
NuMI off-axis program with a 100 kT LAr TPC, and significantly better
sensitivity to the sign of $\Delta m_{31}^2$.

\item A NuMI off-axis program with two 50 kT LAr TPCs at the 1st
and 2nd oscillation maxima at baselines of 810 and 700 km respectively
has marginally better sensitivity to the sign of $\Delta
  m_{31}^2$ but significantly worse sensitivity to non-zero
  $\theta_{13}$ when compared with putting the full 100 kT mass at the
  1st oscillation maxima.

\end{itemize}

\clearpage

\section{Sensitivity to non-accelerator physics}
\label{nonacc} 

A well instrumented very large detector, in addition to its
accelerator based neutrino program, could be sensitive to nucleon
decay which is one of the top priorities in fundamental science.  All
of the detector technologies we consider will lead to enhanced
detection and study of neutrinos from natural sources such as the Sun,
Earth's atmosphere and lithosphere, and past and current supernova
explosions. To achieve these goals, the key issues are cosmogenic
backgrounds and low energy thresholds ($\sim 5$ MeV); the first
primarily depends on the depth of the detector and the second depends
on the depth, the radioactivity from the materials used in the
detector and in the surrounding rock, and the detector noise
(photosensor noise in the case of a water Cherenkov detector, and
electronic noise in the case of a liquid argon TPC).

In this section we briefly summarize the potential of a large detector for 
nucleon decay and astrophysical sources of neutrinos. We also comment on the 
technical requirements on the detector.    For each topic we attempt to 
identify where the requirements of the accelerator program match and
 where they diverge.

\subsection{ Improved Search for Nucleon Decay}
\underline{Theoretical Motivation:} While current experiments show that the proton lifetime exceeds about
$10^{33}$ years, its ultimate stability has been questioned since the
early 1970's in the context of theoretical attempts to arrive at a
unified picture of the fundamental particles - the quarks and leptons
- and of their three forces: the strong, electromagnetic and
weak. These attempts of unification, commonly referred to as ``Grand
Unification'', have turned out to be supported empirically by the
dramatic meeting of the strengths of the three forces that is found
to occur at high energies in the context of so-called ``Supersymmetry'',
as well as by the magnitude of neutrino masses that is suggested by
the discovery of atmospheric and solar neutrino oscillations. One of
the most crucial and generic predictions of grand unification,
however, is that the proton must ultimately decay into leptonic matter
such as a positron and a meson, revealing quark-lepton unity. 
A class of well-motivated theories of
grand unification, based on the symmetry of SO(10) and Supersymmetry,
which have the virtue that they successfully describe the masses and
mixings of all quarks and leptons including neutrinos, and which also
explain the origin of the excess of matter over anti-matter through a
process called ``leptogenesis'', provide a conservative (theoretical)
upper limit on the proton lifetime which is within a factor of ten of
the current lower limit. This makes the discovery potential for proton
decay in a next-generation experiment high.

From a broader viewpoint, proton decay, if found, would provide us
with a unique window to view physics at truly short distances - less
than $10^{-30}$ cm., corresponding to energies greater than $10^{16}$
GeV - a feature that cannot be achieved by any other means. It would
provide the missing link of Grand Unification. Last, but not least, it
would help ascertain our ideas about the origin of an excess of matter
over anti-matter that is crucial to the origin of life itself. In this
sense, and given that the predictions of a well-motivated class of
Grand Unified theories for proton lifetime are not far above the
current limit, the need for an improved search for proton decay
through a next-generation detector seems compelling. 
The theoretical
guidance provided by some promising models
 points  towards the need for
improved searches for proton decaying into $\bar\nu K^+$ and
$e^+\pi^0$ modes with lifetimes less than about $2\times 10^{34}$ and
$10^{35}$ years, respectively. Should proton decay be discovered in
these modes, valuable insight would be gained by searches for other
related modes including $\mu^+ \pi^0$ and $\mu^+ K^0$.

\underline{Current status of experimentation:} 
The ``classical'' proton decay mode, $p\to e^+\pi^0$, can be efficiently
detected with low background. At present, the best limit on this mode
( $> 5.4\times 10^{33}$ yr, 90\% CL) comes from a 92 kTon-yr exposure
of Super-Kamiokande. The detection efficiency of 44\% dominated by
final-state $\pi^0$ absorption or charge-exchange in the nucleus, and
the expected background is 2.2 events/Mton-yr.  The mode $p\to \bar\nu
K^+$, is experimentally more difficult in water Cherenkov detectors
due to the unobservable neutrino and the fact that the kaon is below
Cherenkov threshold. The present limit from Super-Kamiokande is the
result of combining several channels, the most sensitive of which is
$K^+\to \mu^+ \nu$ accompanied by a de-excitation signature from the
remnant $~^{15}N$ nucleus. Monte Carlo studies suggest that this mode
should remain background free for the foreseeable future. The present
limit on this mode is $> 2.2\times 10^{33}$ yr (90\% CL).

\underline{Requirements for the next stage of experimentation:} 
Since the lifetime of the nucleon is unknown, 
and could range from just above present
limits to many orders of magnitude greater,
increases in sensitivity by factors
of a few are insufficient to motivate new experiments. Thus, continued
progress in the search for nucleon decay inevitably requires much
larger detectors than Super-Kamiokande. The efficiency for 
detection of the $e^+ \pi^0$ mode
is dominated by pion absorption effects in the nucleus, and cannot be
improved significantly. An order of magnitude improvement in this mode
can only be achieved by running Super-Kamiokande for an additional
30-40 more years, or by constructing an order of magnitude larger
experiment.  The decay modes of the nucleon are also unknown, 
and produce quite different experimental signatures, so future
detectors must be sensitive to most or all of the kinematically
allowed channels. Moreover, the enormous mass and exposure required to
improve significantly on existing limits (and the unknowable prospects
for positive detection) underline the importance of any future
experiment's ability to address other important physics questions
while waiting for the proton to decay. 

\underline{New facilities under consideration:}
A variety of technologies for discovery of nucleon decay have been
discussed. Of these, underground water Cherenkov appears to be the
only one capable of reaching lifetimes of $10^{35}$ years or
greater. Cooperative, parallel studies of a future underground water
Cherenkov proton decay experiment are underway in the U.S. and
Japan. The proposed designs range from 300 kTon (14 times
Super-Kamiokande) to 1 Mton. Liquid Argon or scintillation techniques
have also been discussed in the proton decay community and may have
significant efficiency advantages for certain modes that are dominant
in a certain broad class of SUSY theories. Liquid Argon time
projection chambers potentially offer very detailed measurements of
particle physics events with superb resolution and particle
identification. Liquid Argon feasibility will be demonstrated in the
near future with the operation of a 600-ton ICARUS detector. If
expectations are correct, it should have a sensitivity that is
equivalent to a 6000-ton water Cherenkov detector in the $p\to \bar\nu
K^+$ mode. The liquid scintillator approach is presently being
explored with the 1~kTon KamLAND experiment. It should also have enhanced
sensitivity to this mode by directly observing the $K^+$ by $dE/dx$
and observing the subsequent $K^+\to \mu^+\nu$ decay.


\underline{Performance and feasibility:}
Detailed Monte Carlo studies, including full reconstruction of
simulated data, indicate that the water detectors could reach the goal
of an order of magnitude improvement on anticipated nucleon decay
limits from Super-Kamiokande. With sufficient exposure, clear
discovery of nucleon decay into $e^+ \pi^0$ would be possible even at
lifetimes of (few) $\times 10^{35}$ years where present analyses would
be background-limited, by tightening the selection criteria. For
instance, with a detection efficiency of 18\%, the expected background
is only 0.15 events/Mton-yr, ensuring a signal:noise of 4:1 even for a
proton lifetime of $10^{35}$ years. A water Cherenkov detector would
also provide a decisive test of super-symmetric SO(10) grand unified
theory by reaching a sensitivity of a (few)$\times 10^{34}$ years for
the $\bar\nu K^+$ mode.

As we have discussed, a much smaller liquid argon could do
particularly well on the mode $\bar\nu K^+$ as the efficiency could be
as much as 10 times larger than that in the water Cherenkov detectors
due to the extraordinary bubble chamber-like pattern recognition
capabilities. Due to this, a single observed event could be powerful
evidence for a discovery. The $e^+\pi^0$ mode however would be limited
by the smaller size of these detectors.

The search for n-nbar oscillation is another test of baryon
non-conservation. While this is not one of the favorite predictions of
conventional SUSY grand unification, this process, taking place in the
nuclear potential, can reach an equivalent sensitivity to baryon
non-conservation of $10^{35}$ years.

\subsection{Observation of Natural Sources of Neutrinos} 

All of the detector technologies we consider will lead to enhanced
detection and study of neutrinos from natural sources such as the Sun,
Earth's atmosphere and lithosphere, and past and current supernova
explosions.  There may also be previously unsuspected, natural
neutrino sources that appear when the detector mass reaches the
hundreds of kilotons scale.  The liquid scintillator technique is of
particular note here because it could allow the detection of low
energy antineutrinos from Earth's lithosphere.  This physics, however,
requires low energy thresholds which are difficult to obtain without
eliminating cosmogenic background by locating the detector at great
depth and with careful selection of materials with low intrinsic
radioactivity for the detector construction.  The low activity
concerns become important if we attempt to push the threshold to below
$\sim 5 MeV$.  The low activity requirement is not essential for accelerator 
physics.

Solar neutrinos have already been observed in the Super-Kamiokande and
SNO detectors. If the large detector concepts discussed here result in
construction of the underground experiment, it may become possible to
increase the observable event rate enough to clearly observe spectral
distortion in the $<5$ to $14$ MeV region.  One could also measure the
as yet undetected hep solar neutrinos (with an endpoint of 18.8 MeV)
well beyond the $~^8$B endpoint (~14 MeV).  These measurements would
require a very comprehensive understanding of the detector systematics
and energy resolution, but a better determination of the solar
spectrum as well as detection of the day-night effect with high
statistics would represent a significant advance in the evolution of
solar nuclear physics measurements.

The observation of supernova neutrino events in a large neutrino
detector of the type being discussed in this report is straightforward
and has historical precedent.  The SN 1987A supernova, in fact, was
seen by two large water Cherenkov detectors (11 events in Kamiokande-II
(total mass~3kT) and 8 events in IMB (total mass~7kT)) that were
active in proton decay searches at that time.  The predicted
occurrence rate for neutrino- observable supernovas (from our own
galaxy and of order 10 kpc distant) is about 1 per 20 years, so events
will be very rare.  However, the information from a single event,
incorporating measured energies and time sequence for tens of
thousands of neutrino interactions, obtained by a very large neutrino
detector, could provide significantly more information than has ever
been obtained before about the time evolution of a supernova.  In
addition to obtaining information about supernova processes, the small
numbers of SN1987a neutrino events have been extensively used to limit
fundamental neutrino properties.  Supernova processes continue to have
very high interest because of the recent detection of the acceleration
of the rate of expansion of the universe using type Ia supernova.
Recent work has shown that diffuse neutrino events from past core
collapse supernova (which produce neutrino bursts) could be used  to
gain independent knowledge on the cosmological evolution parameters\cite{Hall:2006br}.
Therefore detection of supernova neutrinos, either as a burst from a
single supernova or as a diffuse source from past supernovas, should
be a key mission of the multipurpose detector facility.

Recently, there has been substantial progress in the detection of
relic supernova neutrinos using the inverse beta decay reaction
$\bar\nu_e + p \to e^+ + n$. This ability could be obtained with some
futher investment as described below. Neutrons are presently invisible
in water Cherenkov detectors.  After thermalizing, they are captured
by free protons in the water, emitting a 2.2 MeV gamma which is well
below a typical threshold and which is also overwhelmed by the large
radon backgrounds at such energies.  However, if we were to dissolve
gadolinium in the form of gadolinium (tri)chloride, $GdCl_3$, in the
water (the price of gadolinium has dropped three orders of magnitude
in recent years, making such a detector affordable) then the
experiment would become  sensitive to the neutron capture
gamma cascade (total energy = 8.0 MeV) produced by Gd following
positron emission from the inverse beta reaction \cite{vagins}. With a
concentration of 0.1\% Gd (0.2\% $GdCl_3$) by mass, over 90\% of the
neutrons will be visibly captured on Gd rather than on protons.
 
By requiring coincident signals, i.e., a positron's Cherenkov light
followed shortly thereafter ($<100 \mu s$) and very close to the same
spot by the gamma cascade of a captured neutron, backgrounds to the
diffuse supernova neutrino signal could be greatly reduced. Diffuse
supernova neutrino background (DSNB) models vary, but with the Gd 
in the water the 50kton Super-Kamiokande should see about five DSNB events each
year above 10 MeV with essentially no background. One can
easily imagine a next-generation water Cherenkov detector seeing $>
100$ supernova relic neutrinos every year.  
Adding gadolinium would
greatly improve the response to a supernova within our own galaxy as
well, allowing the deconvolution of the various neutrino signals
(charged current, neutral current, elastic scattering) and, among
other things, doubling the pointing accuracy back to the progenitor
star.  Such a detector would also be sensitive to late black hole
formation to much longer times than  at present, since the
distinctive coincident inverse beta signals can be distinguished from
the usual singles backgrounds.  An abrupt cutoff of these coincident
signals would be the unmistakable signature of a singularity being
born.

The continued study of atmospheric neutrinos  in the large
underground detector will provide useful additions to the program
carried out so successfully by the Super-Kamiokande Experiment.  A
detector with mass ($\sim$1 Mton) would be a powerful tool for studying
neutrino physics from atmospheric neutrinos. Thanks to the larger
dimensions of the detector, higher energy neutrino-induced muons can
be fully contained and their energy can be measured. Using the
atmospheric neutrino flux, the distinctive oscillatory pattern as a
function of L/E could be directly observed.  The factor of 20 increase
in detector fiducial mass will allow statistical improvements in all
the topics studied and, perhaps, the emergence of new scientific
topics. 
Other natural sources of neutrinos, such as 
lithospheric neutrinos, have not yet been studied
extensively and could, in principle, be observed by the new detector
concepts.
  An initial result in this area has recently been announced
by KamLAND.  Typically, the neutrino energies for these processes
are below 10 MeV and are sensitively dependent upon the low-energy
threshold capability of the new detectors.  The liquid scintillator
detector concepts are likely to have the best opportunities for
advancing these topics, but liquid argon detectors could also
contribute.

Finally, we note that there may be galactic sources of neutrinos that
are of lower energy and greater abundance than the ultra high-energy
neutrino sources to be explored by detectors such as the Ice Cube
Cherenkov detector now being constructed deep under the Antarctic ice
sheet by an NSF sponsored collaboration.  Galactic neutrinos have a
natural source in inelastic nuclear collisions through the leptonic
decays of charged secondary pions.  This source is expected to be of
comparable intensity and energy distribution to the high-energy
photons that are born from neutral pion decays in the same collisions.
Such neutrino sources, currently not detectable with Super-Kamiokande,
could be seen by a megaton-class neutrino detector that runs for
several decades.

\subsection{Depth requirements for non-accelerator physics} 

It is difficult to consider all possible non-accelerator physics
channels and precisely predict the most optimum depth for either water
Cherenkov detector or a liquid argon detector.  The answer could easily
depend on various technical assumptions, but it is certainly clear
that depth comparable to  or larger than  present detectors
(Super-Kamiokande is at 1000 m of rock or 2700 meter-water-equivalent
depth) is needed  for the best physics reach. A quantitative
summary of depth considerations can be seen in
\cite{hsobeldeep}.

Nucleon decay modes can be divided in two classes: ones where all of
the nucleon energy is visible and ones where some of the nucleon
energy escapes detection. In the first case, the total momentum and
energy balance is a powerful tool for background reduction, and it has
been often argued that these modes should require only modest
shielding from cosmic rays. Indeed, most of the decay modes that were
searched for in the first generation detectors required only modest
depth. IMB operated successfully at a depth of 2000 feet.  However, in
a very large water Cherenkov detector, cosmics not only produce
background, but also reduce the live-time of the experiment by keeping
the detector occupied by frequent large energy deposits. If we require
live-time to be more than 90\%, a shallow depth of few tens of meters
appears sufficient. This conclusion does not include consideration of
the data rate, which is continuous for non-accelerator physics, and
could be unmanageably high near the surface.  The requirement of a
reasonable data rate ($< 10 Hz$ of muons) increases the depth
required to approximately the  Super-Kamiokande depth.

For the second class of nucleon decays in a water Cherenkov detector,
 a low energy tag from
dexcitation photons may need to be used (For example $p\to \bar\nu
K^+$ with a $\sim 6.3$ MeV gamma from $~^{15}N$ de-excitation followed
by $K^+ \to \mu^+ \nu$ with lifetime of 12 ns).  These require low
energy thresholds for photons. This is difficult with a background of
fast-neutron (spallation products from muons in the detector or in the
surrounding rock) induced low energy background events at shallow
depths. Nevertheless,  since the tagging photon is in-time to the main event
(with time window of $< 50 ns$), one could conclude that these events
also may not require much more than Super-Kamiokande depths. A subclass
of events are, however, subject to fast neutron backgrounds.  As an
example of this, the mode $n\to \bar\nu \bar\nu \nu$ can be searched
for by observing the de-excitation of the residual nucleus. The
proposed ultimate DUSEL depth (about 6500 mwe) would reduce the muon
background by about a factor of 100 with respect to Super-Kamiokande and
certainly help in the observation of these modes with a low energy
component.

For a liquid argon calorimeter, much higher resolution may permit
relaxation of these issues.  In particular, the $\bar\nu K^+$ mode
could be much easier to detect because the kaon could be identified 
by its energy deposit (dE/dx). Nevertheless, some minimum 
depth will very likely be necessary
to reduce backgrounds from fast neutrons and to  reduce the data
rate to manageable levels.

For solar neutrinos in a water Cherenkov detector, the important issue
is dead-time introduced by spallation induced fast neutron backgrounds.
At Super-Kamiokande this dead-time is $\sim 20\%$. To maintain the same
level of dead-time for a much larger detector, depth similar to or
greater than Super-Kamiokande (2700 mwe) will be needed.  For a liquid
argon detector, this requirement could be relaxed 
because the dead volume around a cosmic muon could be better defined.

For a supernova in our galaxy (10kpc), the signal level is so large
($\sim 10000/sec$ over a 10 sec burst), that the spallation background
at depths as shallow as 500 mwe are manageable.  For detection of
supernova in neighboring Andromeda ($\sim$750 kpc), however, greater
depth ($>1300$ mwe) is needed. Optimizing depth for diffuse relic
supernova neutrino search needs to take into account the deadtime loss
as well as background from spallation products such as $~^9$Li which
beta decays and then ejects a neutron. The analysis in
\cite{hsobeldeep} suggests that this search may require depths similar
to Super-Kamiokande even if one could get the enhancement in signal to
background from gadolinium  loading.

In summary, the driving issues for depth consideration for future large
water Cherenkov or liquid argon detectors will be backgrounds to low
energy events from spallation products and data rates.  
If one wants to maintain 
sensitivity to specific important physics channels such as $p\to
\bar\nu K^+$ in a water detector, and solar and supernova neutrinos in
either technology, depth in the same range as the current
Super-Kamiokande detector is needed. Greater depth will enhance the
physics reach of the detector.

\section{Results and Conclusions}

The following summary results and 
conclusions were discussed  
at the Sep. 17 2006  meeting of the study group.  
The broad conclusions have been refined by 
significant additional numerical work 
since then. We have first listed the broad conclusions from the study.
A summary of comparisons for the various experimental approaches follows.

\begin{itemize}

\item Very massive detectors with efficient fiducial mass of
$>100$ kTon (in the case of water Cherenkov several hundred kTon and 
in the case of a liquid argon detector $\sim 100$ kTon; for accelerator
based neutrino physics these two would be roughly equivalent in sensitivity) 
could be key shared research facilities for the future particle,
nuclear and astrophysics research programs.  Such a detectors can
be used with a long baseline neutrino beam from an accelerator
laboratory to determine (or bound) leptonic CP violation and measure
all parameters of 3 generation neutrino oscillations.  At the same time,
 if located in a
low background underground environment, they would have additional
physics capabilities for proton decay and continuous observation of
natural sources of neutrinos such as supernova or other astrophysical
sources of neutrinos.


\item  The Phase-II program will need  considerable upgrades to the 
current accelerator intensity from FNAL.  
Main Injector accelerator intensity
upgrade to $\sim$ 700 kW (from the current $\sim$200 kW) 
is already planned for Phase-I of the program
(NO$\nu$A). A further upgrade to 1.2 MW is under design and discussion  as 
described briefly in this report.  
The phase-II program could be carried out with the 
these planned upgrades. Any further improvements, perhaps with a new intense
source of protons, will obviously increase the statistical sensitivity and 
measurement precision.  
Such an upgrade could significantly reduce the running times
(especially in antineutrino mode) and increase 
statistical precision.  

\item A water Cherenkov detector of multi-100kTon size
 is needed 
to obtain sufficient statistical power to reach good sensitivity to 
CP violation. This requirement is independent of whether one uses
the off-axis technique or the broadband technique in which the detector is 
housed in one of the DUSEL sites.  

\item  High signal efficiency at high energies and excellent background
reduction in a liquid argon TPC allows the size of such a detector to
be smaller by a factor of 3 to 4 compared to a water Cherenkov detector for
equal sensitivity. Such a detector is still quite large.

\item The water Cherenkov technology is well-known.
The issues of signal extraction and background reduction were 
discussed and documented at length in this study. The needed  
background reduction and energy resolution is   achievable and 
well understood for the broadband beam approach, 
but not yet fully optimized.    
Key issues for scaling up the current generation 
of water Cherenkov detectors 
(Super-Kamiokande, SNO, etc.)  and locating such 
detectors in underground locations 
in DUSEL have been investigated. 
The cost and schedule for such a detector could be created with high 
degree of confidence. A first approximation for this was reported to 
this study.

\item For a very large liquid argon time projection detector
key technical issues have been identified for the building of the
detector.  A possible development path includes understanding argon
purity in large industrial tanks, 
mechanical and electronics issues associated
with long wires, and construction of at least one prototype in the
mass range of 1 kTon.

\item In the course of this study  we have examined the surface operation of 
the proposed massive detectors for accelerator neutrino physics.  
  Water Cherenkov detector are suitable for 
deep underground locations only. Surface or near surface operation of 
liquid argon TPCs is possible, but 
requires that adequate rejection of cosmic rays be  demonstrated.
Surface or near surface operation capability is essential for the 
off-axis program based on the existing NuMI beam-line because of the 
geographic area through which the beam travels.     

\item For an off-axis program based on the NuMI beam-line, baselines of 
about 800 km and off-axis distances of 10 to 40 km were considered for 
CP violation physics. Since the detector location is on the surface 
the best choice appears to be a fine grained detector such as a 
large liquid argon TPC. 
The scenarios considered for this program were: a) 100 kTon LArTPC at the
 2nd oscillation maximum (40-60 mrad) in conjunction with the Phase I NO$\nu$A
detector.  b) 100 kTon LArTPC at the Phase I NO$\nu$A site. 
For scenario a) we find that the simple addition of a 2nd detector does not 
have significant sensitivity for CP. Scenario b) does have sensitivity 
as shown in Figure \ref{Case6}.
 A third scenario, using two detectors of 
50 kTon each at the first and second maximum has also been analysed (see Figure \ref{Case8}). 
Additional scenarios are presented in Table \ref{compare}.


\item For a wideband program to DUSEL (either at Henderson (1495 km) or
Homestake(1290 km)), two choices for detector technology were considered:
a deep sited large water Cherenkov detector with fiducial mass of 
$\sim 300$ kT or a 100 kT liquid argon TPC (which may be located 
either on the surface or underground). These were found to have good
 sensitivity for CP violation after exposure to 
the same amount of beam. 
The better signal to background ratio for the liquid Argon TPC 
allows for better sensitivity which can be compensated by increased exposure 
or a larger water Cherenkov detector.   
 The sensitivity for 1300 km location 
and its variation for  exposure are shown in Figures \ref{limit1}
to \ref{limit3lar}.  
The sensitivity was found to be about the same for 1495 km.


\item Baselines shorter than 500 km on the NuMI beamline from FNAL have severe
 technical limitations for performing the CP violation science
 because of the low energy of the oscillated events, 
difficulty of separating the ambiguities due to mass hierarchy, and  
 the surface location of the massive detectors.

\end{itemize}

\subsection{Brief comparison of experimental approaches}

In the course of this year long study we have been able to draw several very clear conclusions. Regardless of which options evolve into a future program, the following will be required.
\begin{enumerate}
\item A proton source capable of delivering 1 - 2 MW to the neutrino production target.
\item Neutrino beam devices (targets and focusing horns) capable of efficient operation at high intensity.
\item Neutrino beam enclosures which provide the required level of environmental and personnel radiological protection.
\item Massive ($>>$100 kton) detectors which have have high efficiency, resolution and background rejection.
\item For each of the above items, significant investment in R and/or D is required and needs to be an important aspect of the current  program.
\end{enumerate}

We have found that the main areas of this study can be discussed
relatively simply if we divide them into two broad categories : 1) The 
neutrino beam configuration and 2) The detector technology. Further, we
are able to summarize our conclusions in two  tables which show
the pros and cons of the various options.

In Table \ref{ctable1} we compare the pros and cons of using the existing NuMI
beam and locating detectors at various locations, 
versus a new wide
band neutrino beam, from Fermilab but directed to a new laboratory
located at one of the potential DUSEL sites, i.e. at a baseline of
1300 to 2600 km.

In Table \ref{ctable2} we compare the pros and cons of constructing massive
detectors (~100 - 300 kT total fiducial mass) using either water
Cherenkov or liquid argon technology.

\begin{table}[h]
\centering
\begin{tabular}{|l|l|l|}
\hline
 & Pro & Con \\
 \hline
NuMI On-axis & Beam exists; & L $\sim$ 735 km \\
                          & Tunable spectrum; &  Sensitivity to mass hierarchy is limited \\
                          &                  & Difficult to get flux $<3$ GeV \\
\hline                          
NuMI Off-axis   & Beam exists ;                             &  L $\sim$ 800 km  \\
(1st maximum) & Optimized energy;          &  Limited sensitivity to mass hierarchy \\
                           & Optimized location  for  & \\
                           & 1st detector;                    & \\
                           & Site will exist from \nova project; & \\
\hline
NuMI Off-axis & Beam exists; &  L $\sim$ 700-800 km; \\
(2nd maximum) & Optimized energy; & Extremely low event rate;  \\   
                             & Improves mass hierarchy  &  A new site is needed; \\
                             & sensitivity if $\theta_{13}$ is large; &  Energy of events is $\sim 500 MeV$;  \\
                             &   & Spectrum is very narrow  \\
\hline
WBB to DUSEL & More optimum (longer) baseline;   & New beam construction project $>$\$100M;  \\
                             & Can fit oscillation parameters    &   Multi-year beam construction; \\   
                             & using energy spectrum;                 &  \\
                             & Underground DUSEL site for detector; & \\
                             & Detector can be multi-purpose; & \\
\hline
\end{tabular}
\caption{\label{ctable1}Comparison of the existing NuMI beam to a possible new wide band low energy (WBLE) beam to DUSEL }
\end{table}

\begin{table}[h]
\centering
\begin{tabular}{|l|l|l|}
\hline
 & Pro & Con \\
 \hline
Water  & Well understood and proven technology;      &  Must operate underground;  \\ 
Cherenkov		&  Technique demonstrated by SuperK (50kT);  &  Scale up factor is $<10$;    \\
                &                                          & Cavern stability must be assured  \\ 
                &			&  and could add cost uncertainty; 			\\
                &    New  background rejection techniques  & NC background depends on spectrum \\
	         &  available; 		& 		and comparable to instrinsic background; \\ 
                   &  Signal energy resolution $\sim10\%$;   &  Low $\nu_e$ signal efficiency (15-20\%);   \\
                                 & Underground location      &                                    \\
                                 & makes it a multi-purpose detector;      &       \\
				 & Cosmic ray rate at 5000ft is $\sim$0.1 Hz.   &                \\
 & Excellent sensitivity to $p\to \pi^0 e^+$ & Low efficiency to $p\to K^+ \bar\nu$ \\ 
\hline      
Liquid   & Technology demonstrated by & Scale up factor of $\sim$300 is needed;   \\   
Argon         &  ICARUS (0.3kT); &   \\
TPC     &    &  Needs considerable R\&D for costing; \\ 
       & Promises high efficiency and &  Not yet demonstrated  by  \\ 
                                 & background rejection;     & simulation of a large detector; \\
                                 &  Has potential to operate     &  Needs detailed safety design for   \\
                                 & on (or near) surface;          &  deep location in a cavern; \\
                                 & Could be placed on surface  & Needs detailed  demonstration            \\
			&	either at NuMI Offaxis or DUSEL;          &   of cosmic ray rejection;    \\
				&				& Surface cosmic rate $\sim$500kHz; \\ 
                                 & Better sensitivity to            & Surface operation limits \\
                                 & $p\to K^+ \bar\nu$       & physics program; \\
\hline

\end{tabular}
\caption{\label{ctable2} Comparison of Water Cherenkov to Liquid Argon detector technologies }
\end{table}

\subsection{Project timescales}  

In the following  we briefly comment on the possible timelines for 
the different components of the program we have described in the report.  
At this stage it is difficult to understand the funding, manpower, 
and  other constraints to the program, therefore the 
study group has decided to comment only on technically driven schedules. 

\begin{itemize} 

\item {\bf The FNAL proton upgrade timeline:} 
The SNuMI project which aims to upgrade the Fermilab accelerator complex to 
deliver higher intensity  from the Main Injector,   submitted a conceptual 
design report (CDR) in the fall of 2006. 
The timescale for the project will become clearer after the review process is completed.
A preliminary timeline has been provided to this study. 
The complete upgrade  will be carried out in 
two steps. In the first step, the recycler 
based upgrade (proton plan phase-II)
 will bring the total beam power to 700 kW by 
early 2011. In the second step, the accumulator upgrade 
(the complete SNuMI project) 
will bring  the total intensity to 
1.2 MW. An aggressive plan calls for performing the 
complete upgrade 
up to 1.2 MW by 2012. 
But this will depend on the outcome of reviews 
and discussions that will take place in 
the next year.

\item {\bf Construction of a new beam towards DUSEL:} 
Construction of a possible  new beam towards is not part of the SNuMI project. 
Only  preliminary discussions, cost, and schedule estimates exists. 
The scope of the project is similar to the NuMI project which was described 
in Section \ref{numisec}. Based on the NuMI experience, a rough outline for the 
project could be: 1 to 2 years for preparation and geological site investigations,  
2.5 to 3 years for civil construction, and  1 year for installation of 
technical systems: a total of 4.5 to 6 years for construction of the beam-line. 
There are a number of issues that are different between NuMI and a new beam-line to DUSEL.
These are related to the greater downwards angle of the DUSEL 
beam-line and the proximity of the DUSEL 
beam-line to the FNAL site boundaries. 
These issues and their mitigation will be addressed in a 
separate note \cite{laughton}.

\item {\bf Construction of a deep large water Cherenkov detector:} 
There are two well recognized considerations that define the time scale over
 which a large water Cherenkov detector could be built: the underground cavern 
construction and manufacturing of large numbers of photo-multiplier tubes. 
For both the single cavern  and the multiple cavern concepts of 
the detector a significant period of exploratory excavations and bore holes 
will be needed. After this period ($\sim 1-2$ yrs) approximately 
5 to 6 yrs of excavation is needed to reach the needed 
total volume. The PMT manufacturing period depends on the choice of 
the PMT, which is different for the two different concepts for the detector. 
 For 20 inch PMTs,  the UNO plan calls for manufacturing 56000 tubes in 
about 8-10 yrs. For the Homestake multiple module proposal, the plan calls
 for manufacturing 10-12 inch PMTs with a rate of about 150000 tubes 
in 6-7 yrs.   The collaborations are communicating with two large manufacturers of 
hemispherical PMTs.  Preliminary conclusions are that each of the two 
manufactures have sufficient capacity currently  to produce 10-12 inch diameter 
PMTs at about 1/2 the rate 
that is needed for these projects. For either choice,  smaller or larger diameter, 
the production capacity 
needs to be  enhanced to meet the need, 
but the investment needed in not considered 
extraordinary.   There could be bottlenecks in production of 
materials (for example,  glass) that need to be fully understood.  

\item {\bf Construction of a very large LARTPC:} 
The cost and schedule estimate for a very large liquid argon TPC of
size (50 to 100 kTon) must be preceded by a series of development
steps.  Although the viability of the technique has been established
by the ICARUS group, a factor of 10 cost reduction is required to make
a very large detector economically possible.  The
development program is outlined in \cite{lar218} and contains three
projects. One project involves techniques for the
purification of liquid argon to achieve long electron drift times, low
noise electronics design, and materials qualification. A second project
is the construction of a $\sim$3 ton module to test design concepts
for the very large detector; and the third project is the design and
construction of a $\sim$ 1 kT detector to be constructed using the techniques 
proposed for the very large detector.  

The first project is in progress at FNAL and
Yale. Long (many millisecond) electron drift lifetimes have been achieved
and the project is expected to be complete by mid 2007. Dependent on 
funding the second project could
produce results by the end of 2007. The siting and mass of the
detector to be proposed for the third project are under discussion. Once a
choice is made, the group would like to start the design immediately. 
Completion of the design
for project 3 is expected to take 1 year  and  requires successful
completion of the other projects. A preliminary cost
for project 3 at this time is $\sim $ \$10 M. The above program is
essential for a LARTPC detector on the surface or
underground. The cost implications for siting a very large detector
at underground locations are being discussed, but they need
further work.

\end{itemize}

\section{Acknowledgments}

We are grateful for the support that the directorates of both Fermi
National Accelerator Laboratory and Brookhaven National Laboratory
have given to this work. The interaction with the NuSAG committee was
also extremely important for setting the time table to finish the
large number of calculations performed for this report.

\newpage 

\appendix

\section{Answers to questions raised by NUSAG}

{\bf 
1. Noting the existence of discrepant sensitivity calculations even
for the same detector, it would be most useful to have any such
calculations performed with consistent assumptions and methodologies.

a) Fixed, common, stated values of the mixing parameters not
explicitly under study. 

 b) Common, stated and plotted, cross sections
vs. En.  Common, stated nuclear models. 

 c) Stated assumptions about
energy resolution, background rejection.  

d) If appropriate, common
total p.o.t.  If sensible, use a common proton energy and anti-nu
running fraction.  If not, state the optima chosen. 

 e) What methods
are used to extract the oscillation parameters from the final event
sample (counting? fitting the spectrum?)  

f) Standardized, stated
method for defining sensitivity.

2.  Give sufficient detail in tables and/or plots to allow a reader to
understand how the numbers for rates or sensitivities are obtained.
We would expect that many of the results would be easily accessible to
a physicist with a calculator.  Here are some useful inputs that come
to mind (meant as a guide only): 

a) Specify the signal channel(s).
(We will assume here that it is quasi-elastic.) 
 b) What simple cuts
(energy, etc.), if any, do you apply?  

c) The number of INTRINSIC $\nu_e$
events reconstructed as signal, and their reconstructed energy
spectrum (in reconstructed Enu(QE) or Evis, or Ee, or whatever you'll
use.)  

d) What is the purity of the QE selection, that is, for true
$\nu_e$ events, what fraction of those selected as QE are actually QE (as
a function of E)?  

e) The total number of NC $\pi^0$ events, and spectra
vs. true Enu and $\pi^0$ momentum.  

f) The number of NC $\pi^0$ events
reconstructed as signal, and their reconstructed energy spectrum.
What is the true Enu spectrum for the NC pi0 events reconstructed as
signal?  

g) The NC $\pi^0$ rejection assumed, as a function of... ($\pi^0$
momentum?) 

h) The assumed systematic errors on each of the
backgrounds, with any relevant dependence on energy.  How are these
estimates arrived at?  

i) The assumed signal efficiency as a function
of energy.  How are these estimates arrived at?  

j) Provide tables and
spectra (vs. true and reconstructed $E_\nu$) giving the initial population
of events, before cuts, by process (QE, CCpi+, DIS,...), how these
numbers diminish as the cuts are applied, and in the final sample at
the various oscillation parameter test points.  An entry at the 3-s
sensitivity limit would be informative.  Scatter-plots of reconstructed
vs. true En for individual signal and background channels may be
informative.
}

{\bf  3.  Specify the level of simulation that goes into your
currently-generated sensitivity estimates.  For example: 

a) How is
energy resolution treated?  Give a plot of the assumed energy
resolution (electron energy and neutrino energy) vs. energy.  

b) How
is the selection of QE events treated?  

c) How is the rejection of
pi0's modeled?
}

We are grateful to the NUSAG committee to provide questions that
could be used to guide the study. 
The  report was written with the desire to answer these questions. 
Some of the details that these questions ask for are in the supporting 
documents which can be obtained from the study website: 
\underline{http://nwg.phy.bnl.gov/fnal-bnl/}. 
To keep the length of the report minimum we have decided not to repeat 
the material that can be found in the body of the report.

{\bf  
4.  What near detector location/size/technology/performance/cost is
assumed/needed to achieve the assumed systematic errors?
}

In section \ref{wcsim}  we have summarized the thoughts on the near detector 
issues for the Phase-II(B) DUSEL based broadband approach.  The requirement on 
systematic error on the background are relatively modest (10\%). The harshest 
requirement might be on the energy scale systematic of 1\% which is needed to 
achieve the best precision on the atmospheric parameters of $\Delta m^2_{32}$
and $\sin^2 2 \theta_{23}$.  The main technical issues for the near detector
are the location  for its deployment, the deviation from $1/r^2$ behavior
of the flux due to the close location of the detector, and the high event rate 
at the near site. The study did not look at these problems in detail.  
Fortunately, there is now rich experience on these issues from  the running NuMI-MINOS experiment. Most of this experience can be 
applied directly to the future project. 

For Phase-II(A) approach using the NuMI offaxis beam, the near detector 
requirements have not been studied. The location of such a detector 
could be in the existing tunnel that connects the NuMI near detector site to 
the beam tunnel.  The study has not looked at the event rates or potential 
difficulties due to the deviation from both $1/r^2$ behavior
and from having a source with a wider angular acceptance at the 
near detector than the far detector.

{\bf  
5.  If possible, for comparison purposes, use the same methodologies
to make parallel sensitivity estimates for NoVA (single detector) and
T2K.  What sensitivity for NoVA do you calculate for the same number
of p.o.t. assumed in question 1?
}
Please see Section \ref{sensi_2}.

{\bf 
6.  All sensitivity calculations for off-axis configurations must
include events from neutrinos in the high-energy peak from kaon decay.
}

The detector performance criteria are in Section \ref{sensi_2}. 

{\bf 
7.  What detector technologies are still worth pursuing for a 2nd
off-axis detector -- Liquid scintillator?  Water Cerenkov?  Liquid
Argon?  Other?
}

Over the past several years, three potential detector technologies 
have been considered for a next generation experiment: 
 liquid scintillator (similar to NO$\nu$A), water Cherenkov and 
a liquid argon TPC. Here, we summarize the conclusions which have
 been made to date in regard to the detector technology that would be best suited to the off-axis beam.

Studies of a massive liquid scintillator 
detector using the simulations developed for NO$\nu$A have shown that  
 the backgrounds (mostly neutral current) would be approximately 1:1
 with the signal at the second maximum and this option was not considered further.

A water Cherenkov detector of the size proposed for DUSEL could give sufficient 
rate in the NuMI beam, though there might again be a question of background rejection. 
 However it has been concluded that this size of detector must be sited deep
 underground to avoid being swamped by cosmic ray muons and there is no existing 
deep site available along the NuMI beam, and so we do not consider this a viable option.

A liquid argon TPC has the advantages of high efficiency  and high background 
rejection for neutral current events, using the high  spatial resolution.  Thus 
for the same sensitivity in the same beam it can be factors of around $~$3 
smaller than a water Cherenkov detector.  For the sensitivity studies  we have assumed 
 liquid argon detector(s) with a total fiducial mass of 100 kTon.  

{\bf 
8.  There were several references to the possibility of a detector at
$\sim$250 km in the NuMI beam.  
Is this being pursued by the Working Group?
What are the general properties of this approach?
}

Shorter baseline lengths for NuMI off-axis detectors have been
considered in the literature \cite{olga2}.  For example, for a
baseline of 250 km, the first and second oscillation maxima are at
0.50 GeV and 0.17 GeV, respectively.  There are two reasons for
considering shorter baselines: small matter effects and larger numbers
of events because of the closer distance. This solution, however has
several difficulties. The main ones are: i) The low energies needed
forces us to consider large off-axis angles ($>40 mrad$) where the flux
of neutrinos is rather poor and the contamination from high energy
neutrinos from kaon decay large.  This largely negates the advantages
of the larger flux because of the closer distance. The event rate can
be easily obtained from \cite{offaxs} by scaling.  ii) Natural choice
for a detector at these energies is a water Cherenkov counter. Since
most of the events at these energies are quasi-elastics for which a
water Cherenkov detector has good efficiency, little is gained by
utilizing a liquid argon TPC. The water Cherenkov detectors needed are
too large for operation on the surface as explained in Section
\ref{depth}.  iii) For the first oscillation maximum,
 an experiment with almost identical
parameters is already being carried out in Japan (T2K). Combining the
results of T2K and Phase-I of the US program is a subject of various
reviews\cite{olga1}.

{\bf 
9.  Provide cost and schedule estimates for the same fiducial mass and
PMT coverage/channel count used for sensitivity estimates.  (We
realize that fiducial/total mass ratios may be hard to estimate, but
the assumptions should be stated.)
}

We hope that the committee understands that the work reported in this
study was carried out in parallel in a very short period of time.  In
addition, members of the study group are considering several options
for detector sites and design.  Therefore it is difficult to obtain
complete consistency in the assumptions that went into simulations
versus detector design and cost estimates, etc. Obviously we will do
the best we can to point out the various points of departure and will
depend on good judgment.

The design and cost for a detector in the Henderson laboratory were
provided in the presentation of Prof. Chang kee Jung at
\cite{uno}. The fiducial volume for UNO was quoted to be 440 kT at a preliminary cost
of \$437M.

A conceptual detector design for 300 kT for Homestake was presented to
the committee in \cite{300kt}. The authors of that report provided
the following answer for their choices:

\noindent 
A single 100 kiloton module will have a cylindrical fiducial volume
with a diameter of 50 meters and a height of 50 meters.  The PMTs on
the vertical face of the cylinder will have their photo-cathodes on the
surface of a 52 meter diameter cylinder.  The top and bottom PMTs will
be separated by 52 meters.  This layout defines a fiducial volume that
begins 1 meter inside the PMT photo-cathode surface.  In addition,
there will be 0.5 meter veto region surrounding the entire detector so
that the chamber walls will be on a 53 meter diameter cylinder.

Our budget estimate for the excavation of the detector chamber was
based on a 50 meter diameter by 50 meter high cylinder.  The change
from 50 meter to 53 meters involves a volume increase of 18\% and a
surface area increase of 12\%.  Although our budget breakdown details
permit us to apply the above scale factors to each of the volume and
surface area budget items, we decided, for this answer to merely use
an average cost increase of 15\%.  When applied to a single module, the
construction cost increases  from \$29.1 million to
\$33.5 million.  Note, that these numbers include a contingency of 30\%.
The total single 100 kiloton detector cost increases from \$116.6
million to \$121 million, an increase of 3.6\%.

Similarly, when this cost increase factor is applied to three
detectors, the three chamber cost increases from \$66.1 million to \$76
million and the total three detector cost increases from \$308.9
million to \$318.8 million.

In the above we not included the effect of moving the PMTs from the
original 50 meter diameter cylinder to a 52 meter diameter cylinder, a
surface area increase of 8\%.  If apply this factor to the previously
assumed PMT and associated electronics cost of \$62.1 million this
creates another \$5 million increase per 100 kiloton detector.  The
final cost including all contingencies is then \$126 million for a
single 100 kiloton detector and \$323.8 million for three such
detectors. The above  increase is less than 10\% for budget that has
a contingency of about 34\%.

Finally, the simulations for the background estimates were reported  in 
 \cite{chiaki1}.
They were performed with the exact geometry of the Super-Kamiokande detector
(with 40\% PMT coverage using 20 inch diameter tubes). The Homestake 
detector cost is for 11 inch tubes and 25\% coverage. 
We are confident that this coverage is sufficient because of several factors.
First, the PMT information both Hamamatsu and Photonis shows
 that smaller diameter 
semi-hemispheric tubes have higher quantum efficiency (QE) than the 20 inch tube (for example, 
the Hamamatsu 10.5 inch tube has  QE of 25\% and the 20 inch tubes has QE of 20\% at 390 nm)
This difference is apparently well-known and documented. Secondly,  
the collection efficiency (efficiency of collecting the photo-electron into the dynode structure)
is also known to be larger for the smaller diameter tubes. The collection efficiency factor 
(an increase of about 13\% for the 10.5 inch tubes versus the 20 inch) is not well documented.  
Therefore, if corrected by these two effects, the 25\% coverage with smaller 10.5 inch tubes 
corresponds to 
35\% coverage ($\sim 25\%\times {0.25\over 0.20}\times 1.13$) with the 
20 inch photomultiplier tubes. 
We also expect that with a larger detector and far larger granularity,
the background rejection will get better requiring less total 
coverage.  Nevertheless,  we understand that 
all of the above  has to be demonstrated with benchtop 
 measurements and  detailed simulations for which we
would like to ask for substantial R\&D funds.  
If we must increase the coverage to 40\% to achieve the physics goals then the cost increase 
will be approximately \$112M which is certainly beyond the contingency we have allowed at this
point.

{\bf  
10.  For the modular water Cherenkov approach, are you defining 3
modules as your baseline detector?
}

The authors of report \cite{300kt} reply:

``Yes, there are three main reasons we believe 3 modules is 
an optimum choice to start with. 
First, because of the long running times possible at FNAL it appears that 
a 300 kTon fiducial mass is sufficient to reach the desired sensitivity for 
neutrino oscillations.  Second, it is clear that for proton decay 
searches a larger detector is needed, but  
for current background projections a few background events are expected in 
favored decay modes after  exposure of 1 MT-yr. 
We believe that proton decay searches will benefit from
further detector and analysis improvements after reaching this level of sensitivity.
Any modules built after the first 3 modules will benefit from this knowledge. 
Third, there is considerable cost saving by starting the simultaneous construction of 
3 cavities in the region of relatively  well-known  Homestake rock near the  Ray Davis Chlorine chamber 
as explained in \cite{300kt}.''

{\bf 11.  For the water Cherenkov counters, we will be eager to hear of
progress in algorithms for rejecting $\pi^0$s (and the testing of them).
What is the increase in $\pi^0$ rejection over that achieved by Super-K
(as a function of $\pi^0$ energy) assumed in your current calculations?
What have you reached with your own simulations/algorithms?  Describe
briefly the algorithmic improvements.  Does this rejection depend more
on total photo-cathode coverage, or on granularity?} 

The detailed account of the $\pi^0$ rejection is described in the 
accompanying report
by C. Yanagisawa et al. \cite{chiaki2} as well as in the presentations by 
Yanagisawa and Dufour \cite{chiaki1, fanny1}.   It is also summarized in 
section \ref{wcsim}. 

The improvement to the signal to background  depends on the 
neutrino spectrum and
neutrino oscillation parameters. In the following we use the same neutrino
energy spectrum and the same neutrino oscillation parameters as used in the
above report with the CP violating phase of 45 degrees.

For the baseline of 1480 km (Fermilab to Henderson), using the new
algorithm  the signal to the background ratio can be improved from 0.30 to 1.9,
while retaining 40\% of the signal events accepted by the current
Super-Kamiokande algorithm.
For the baseline of 2540 km (BNL to Homestake), using the new algorithm 
the signal to the background ratio can be improved from 0.35 to 2.1, while
retaining 40\% of the signal events accepted by the current Super-Kamiokande
algorithm.

Dependence of these results above on the granuarity and photocathode coverage
has not been studied in a systematic fashion, as we extensively used
Super-Kamiokande-I (photocathode coverage of 40\%) Monte Carlo sample. 
It is also found that, given the 40\% photocathode coverage, the signal
to background ratio can be significantly improved for neutrino events with
 reconstructed neutrino energy of $<$1.2 GeV  for a detector with 
better granularity.
 Other observations concerning this issue 
are touched upon in section \ref{wcsim}.

{\bf  12.  Though the worldwide community of proponents of large water
Cerenkov detectors seems to cooperate in simulations, algorithms,
etc., we do not see evidence that there is any global planning
(site-independent design studies or physics programs, etc.) underway
for such a detector.  Please comment. }

One of the most useful results of the NUSAG process has been the 
cooperation in simulations and algorithms for large water Cherenkov 
detectors. This cooperation was most evident in the participation from 
the T2KK group in our discussions. We have also had fruitful interactions 
with the proponents of the Frejus based water Cherenkov detectors.  

The description and calculations for the water Cherenkov approach in this 
report was  a result of cooperation between two US based groups:
the UNO group that wants to develop  a single very large cavern for the 
detector and the Homestake based group which wants to develop the detector in 
multiple modules.  Both groups have worked together to understand and
suppress the backgrounds in the detector  and also have settled on a 
similar physics strategy for addressing CP violation in neutrino 
oscillations. 

There are currently 5 well considered 
 proposals for a very large water Cherenkov 
detector worldwide:    Hyper-kamiokande detector in Japan, 
a possible detector in Korea on the same neutrino beamline as 
JPARC to Super-Kamiokande, a very large detector in the Frejus laboratory in 
France, and the two possible sites for DUSEL (Homestake or Henderson mines) 
 in the U.S with either a large single volume detector(UNO) or 
a detector in multiple modules such as the Homestake proposal. 
A long baseline neutrino oscillation program with emphasis on 
reaching sensitivity to CP violation in neutrino mixing is central to all
these proposals.     
Therefore we believe there is good cooperation and agreement 
 on the issue of the physics  program for such a detector.    

The other two factors for such a detector are a) site development,
and b) photo-sensor and electronics R\&D and  acquisition.
The site development is a very large part of this detector design,
and therefore must be handled locally. There is cooperation and 
communication between these groups to compare  costs and schedule for 
the site development. The costs and schedules appear understandable 
after considering the differences between the engineering and 
accounting practices in these geographic region, but we do not see 
how global site independent planning can be performed here.   
The photo-sensor and electronics R\&D is the dominant item in these 
projects. For the photo-sensor R\&D
  we agree that good cooperation  could be helpful and 
 lower the costs for everyone.  

It should be remarked that each of the above geographical regions 
has a unique virtue for locating this massive detector. For Frejus, it is 
the availability of CERN as a neutrino source and the  deep location 
next to the Frejus highway tunnel.  For both HyperK and the Korea based 
detectors the uniqueness lies in the location on an existing 
neutrino beamline from JPARC.   For the US sites the uniqueness is in
first the distance available from Fermilab or BNL ($>$1000km) which is now 
recognized as essential for performing the next generation experiment with
large  CP and matter effects, and second the depth available at the potential 
DUSEL sites to suppress cosmogenic backgrounds.  
Finally, 
the size of the detector projects  are large but at a scale 
that could be contemplated on a national level.  Therefore, global planning 
for a single such detector and site independent studies (in the manner of 
a very large accelerator project), is perhaps not warranted.

{\bf For Liquid Argon:} 

These questions were answered by the liquid argon subgroup. The
answers  were
coordinated by Prof. Bonnie Fleming.

{\bf   NuSAG recommends that the Liquid Argon group reweight its
emphasis from sensitivity/reconstruction/pattern recognition to
hardware issues and cost estimates.  We realize that a full switch
cannot occur if the LAr group is a big part of the more generic
off-axis calculations in the Working Group, but, for example,
LAr-specific reconstruction and particle ID algorithms seem less
pressing than technical feasibility.} 

{\bf  13.  What has actually been measured on purity of the Ar in a tank
made with industrial technology?  If not yet tried, when will the
first tests be?}

Response:  No tests have yet been performed on purity of Argon in a tank made
with the industrial technology necessary for construction of a massive
detector.  This test will require a large tank, $\sim$1 kTon,
constructed using the same techniques as a large detector.  This
project has been envisaged by the LArTPC group as outlined in their
report to NuSAG in 2005.  A specific plan for this component of the
R\&D path is presently under study and expected to converge within a
year.  In the meantime, small scale tests using the Materials Test
Stand at Fermilab (see writeup for details) will have first results
addressing purity issues within this year. 

{\bf  14.  When do you expect to have tried 3-m drifts and long wires in the
US?  What effect will the capacitance of very long wires have on
electronic noise?}

Response: A program to study 5m drifts using a prototype vessel at Fermilab is
in the design stages.  Depending on funding, results from this project
are expected within the next two years.  As well, long drift tests are
underway in Europe on the same timescale.

A 30m long wire with 4 meters of interconnecting cable to electronics
will have a capacitance of 620pF~\cite{nusagdoc}.  Using commercial
amplifiers, a signal to noise of $\sim$9 can be achieved, adequate for
LArTPCs.  Another configuration that has been considered is to use
cold electronics, eliminating the interconnecting cable.  This option
is under study at Michigan State University in Carl Bromberg's group.

{\bf 15.  What are the R\&D milestones, with an estimated schedule, that
would lead to a first realistic cost estimate for a detector of the
2nd-off-axis or wide-band class?}

Response:  Before developing a realistic cost estimate for a massive
detector, 50-100kTons in size, a reasonably sized, scaled down
version of the massive detector should be constructed and operated.
This detector will test purity in a vessel constructed using the same
industrial techniques envisaged for the large detector, electronics,
ability to handle cosmic ray rate, and cellular design.  As well,
smaller scale tests such as the 5m drift test, long wires test etc, as
described in the summary document, are necessary. However, it is the
1kTon scale test that drives the schedule.  The schedule for this
project is not yet fully fleshed out.

\newpage  

\section{NuSAG Charge}

The charge letter is reproduced on the next two pages 

\newpage

Please obtain the letter  from \underline{http://nwg.phy.bnl.gov/fnal-bnl}. 

\newpage

Please obtain the letter from \underline{http://nwg.phy.bnl.gov/fnal-bnl}. 

\newpage

\section{Charge to this working group}

\begin{verbatim}  

April 5, 2006 
	
Dear Colleague,

This letter is being sent to you as a follow-up to the Long Baseline
Workshop held at Fermilab on March 6-7. This mailing list is composed
of those who attended the study and signed up to receive further
information or have subsequently expressed interest in the
study. Since the kick off meeting we have redrafted the goals of the
study. We have inserted a time scale which we judge to be
achievable. The is goal is described in the attached document. You can
anticipate that within days you will get a further document in which
Milind Diwan and Regina Rameika have attempted to parse the study goals
into a set of work packages. We would like to hear from people who are
prepared to do some work on these issues. Especially we would be very
happy to hear from people new to these studies.

However, as you might expect we do have some likely suspects in mind
and Gina and Milind will be contacting people to help. Finally, we
will also be recruiting an Organising/Advisory Committee to help us
guide this study. We look forward to seeing progress on this study and
would welcome your suggestions for additions, adjustments and
approach.

With Best Regards,

Sally & Mont

Sally Dawson, Chair, Physics Department, Brookhaven National Laboratory.  
Hugh Montgomery, Associate Director, Fermi Natinal Accelerator Laboratory
\end{verbatim}  

\newpage

\includegraphics[angle=0,width=1.2\textwidth]{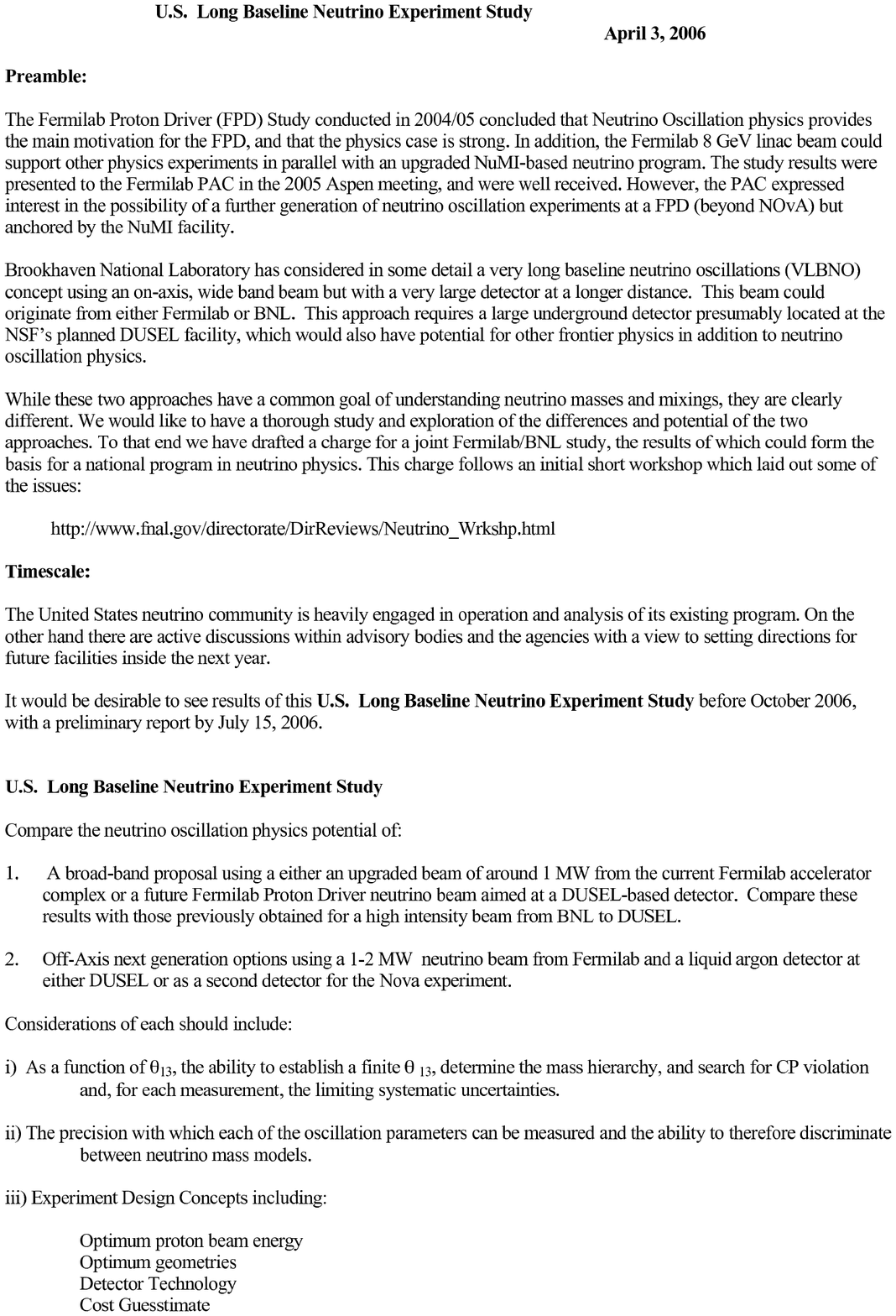}

\newpage 

\section{Study group membership}  
\label{mem} 

Chairs  \\ 
\begin{tabular}{l|l|l} 
\hline 
Sarah Dawson (co-chair) & Brookhaven National Lab. & dawson@bnl.gov \\
Hugh Montgomery (co-chair) & Fermi National Accelerator Lab. & mont@fnal.gov \\
\hline 
\end{tabular}  
\bigskip 
\bigskip 

International Advisory Group \\
\begin{tabular}{l|l|l} 
\hline 
Milind Diwan (co-leader) & Brookhaven National Lab. & diwan@bnl.gov \\
Regina Rameika (co-leader) & Fermi National Accelerator Lab. & rameika@fnal.gov \\
Joshua Klein & University of Texas &  jrk@mail.hep.utexas.edu \\
Franco Cervelli & INFN, Pisa  &  franco.cervelli@pi.infn.it \\ 
Maury Goodman & Argon National Lab. &  maury.goodman@ANL.GOV  \\
Bonnie Fleming & Yale University & bonnie.fleming@yale.edu \\
Karsten Heeger & Lawrence Berkeley Lab. &  	     KMHeeger@LBL.GOV                 \\
Steven Parke &  Fermi National Acc. Lab.          &      parke@FNAL.GOV                   \\
Takaaki Kajita & University of Tokyo & kajita@suketto.icrr.u-tokyo.ac.jp \\
\hline 
\end{tabular}  
\bigskip

A full list of participants is available at 
http://nwg.phy.bnl.gov/\~diwan/nwg/fnal-bnl/folks.txt

\newpage

\section{Relevant resources and URLs for the study group}  

Main websites for this study group are: 

\underline{http://home.fnal.gov/~rameika/LBL\_Study/LBL\_mainframe.htm}

\underline{http://nwg.phy.bnl.gov/fnal-bnl/}

Additional materials can be found at:  

\underline{http://www.fnal.gov/directorate/DirReviews/Neutrino\_Wrkshp.html}

\underline{http://www.hep.net/nusag\_pub/May2006talks.html}

\underline{http://www-numi.fnal.gov/}

\underline{http://nwg.phy.bnl.gov/}

\underline{http://www-lartpc.fnal.gov/LBStudy\_LAr/2006LB.html}

\underline{http://www.dusel.org/}

\underline{http://www.lbl.gov/nsd/homestake/}

\underline{http://nngroup.physics.sunysb.edu/husep/}

\newpage 

\section{Schedule of meetings and report preparation} 
\label{sch}

Most of the work of the working group was carried out by small 
subgroups that worked on the individual documents. 
The work was mostly carried out  by email and telephone. The following meetings 
were very helpful for wider interactions. 

\bigskip 
\begin{tabular}{ll}
November 14, 2005 & FNAL/BNL meeting to explore collaboration, BNL \\
March 3, 2006 & Charge letter for NuSAG to examine APS study recommendation for  \\
	& a next generation neutrino beam and detector configuration \\
March 6-7, 2006	 & First kick-off workshop for organization of the study at FNAL \\
April 5, 2006 & Charge letter to the study from Dawson and Montgomery \\
April 11, 2006  & Preparation of the task list and assignments \\
May 20, 2006 & Presentations to NuSAG committee about the study in Chicago \\
June 27-28, 2006 & Second workshop on detector technologies at FNAL \\ 
July 6, 2006 & Status report to HEPAP from NuSAG, presentation by P. Meyers,   \\
	    & HEPAP meeting in Washington D.C. \\ 
July 15, 2006 &  Deadline for preparation of individual reports from the task list \\
September 16-17, 2006  & Third workshop on preparation of the joint summary report \\
October 16, 2006 & Deadline for presentation of the joint report \\
December, 2006 & Deadline of report from NuSAG to HEPAP \\
\end{tabular} 

\bigskip 

Other meetings of note where interactions took place are 

\begin{tabular}{ll}  
June 13-19, 2006 &  Neutrino 2006, Conference in Santa Fe \\
July 11-21, 2006 &  Neutrino Physics with Liquid Argon TPCs, Yale Univ. \\
August 24-30, 2006 &  NuFact 06 Workshop, UC/Irvine \\
Sep 21-23, 2006 & NNN06, University of Washington  \\
March 29-30, 2007 & Fermilab Physics Advisory Meeting \\
\end{tabular}  

\newpage

\


\begin{thebibliography}{99}
\bibitem{ncvis} 
The results for efficiency
 given in references \cite{chiaki1} or \cite{chiaki2}  are with respect to  
visible events. Most of the NC elastic events are invisible in a 
water Cherenkov detector. We have adjusted the efficiency in table \ref{cyrej} 
for this.
\end{thebibliography}
\end{document}